\documentclass{iagtese_en}	
\hypercolor				
\usepackage{calligra}
\usepackage[T1]{fontenc}
\usepackage{amssymb}
\usepackage{pifont}
\usepackage{pgfornament}

\usepackage{subfig}
\usepackage{pdfpages}

\usepackage{color, colortbl}
\usepackage{rotating}

\usepackage{miama}
\usepackage[T1]{fontenc}

\usepackage{amsmath}	
\usepackage{mathptmx}

\begin{document}			
\institution{University of S\~{a}o Paulo (USP)\\Institute of Astronomy, Geophysics and Atmospheric Sciences\\Department of Astronomy}

\title{A Multi-technique Study of the Dynamical Evolution of the Viscous Disk around the Be Star \\$\omega$ CMa}

\translator{{Tese apresentada ao Departamento de Astronomia do Instituto de Astronomia, Geof\'{i}sica e Ci\^{e}ncias Atmosf\'{e}ricas da Universidade de S\~{a}o Paulo como requisito parcial para a obten\c{c}\~{a}o do t\'{i}tulo de \textbf{Doutor em Ci\^{e}ncias}.\\ \\
\'{A}rea de Concentra\c{c}\~{a}o: Astronomia\\
Orientador:\\Prof. Dr. Alex Cavali\'{e}ri Carciofi}}

\author{\small{Sayyed} \Large{Mohammad Reza Ghoreyshi}}

\date{S\~{a}o Paulo \ano}
						%
\pagestyle{empty}			
						%
\maketitle					%
\Dedicatoria				%
\hfill
\vfill
\fmmfamily

{\begin{center}
    {\Huge To my dear wife, Minoo, who tolerated all difficulties,}
\end{center}}
\par
{\begin{center}
    {\Huge and}
\end{center}}
\par
{\begin{center}
    {\Huge to my dear kids, Arshida \& Arshavir, who endured a difficult time,}
\end{center}}
\par
{\begin{center}
    {\Huge for this thesis becoming true.}
\end{center}} 

\normalfont
\par
\vspace{0.5cm}
\hrulefill \pgfornament[scale=0.25]{77} \pgfornament[scale=0.25,symmetry=v]{77} \hspace{-1cm} \hrulefill
\vspace{2cm}

		%
\Agradecimentos			%

I would like to thank my supervisor, Prof. Dr. {\bf Alex Cavali\'{e}ri Carciofi}, for the patient guidance, encouragement and advice he has provided throughout my time as his student. I have been extremely lucky to have a supervisor who cared so much about my work, and who responded to my questions and queries so promptly. 

I would also like to thank every body at IAG who helped me scientifically or technically. In particular I would like to thank Dr. {\bf Daniel Moser Faes} for his constant helps at different points of this thesis. It would have been very difficult for me to take this work to completion without his incredible support and advice. 

I must express my sincere gratitude to {\bf Minoo}, my wife, for her continued support, encouragement and patience for experiencing all of the ups and downs of my research time.

I gratefully acknowledge the funding received towards my PhD from Coordena\c{c}\~{a}o de Aperfei\c{c}oamento de Pessoal de N\'{i}vel Superior (CAPES) PhD fellowship. 

This research made use of ESO, simbad and vizier databases (CDS, Strasbourg), as well as NASA Astrophysics Data System. This work was possible by the use of the computing facilities of the Laboratory of Astroinformatics (IAG/USP, NAT/Unicsul), whose purchase was done by the Brazilian agency FAPESP (grant 2009/54006-4) and the INCT-A.

At last I wish to thank many other people whose names are not mentioned here but this does not mean that I have forgotten their help. 

\vfill

\begin{flushleft}
\rule{6cm}{0.5pt}\\
{\footnotesize{This thesis was written in \LaTeX{} making use of the {\tt IAGTESE} class developed by students of IAG/USP.}}
\end{flushleft}
						%
\Epigrafe					%
\vfill
\begin{flushright}

``\textit{Capable is he(she) who is wise ------ Happiness from wisdom will arise}''\\

\vspace{0.4cm}

Abolghasem Ferdowsi [940 - 1020]

\end{flushright}

\vspace{0.5cm}

\begin{flushright}

``\textit{Human beings are members of a whole ------ In creation of one essence and soul\\
If one member is afflicted with pain ------ Other members uneasy will remain\\
If you have no sympathy for human pain ------ The name of human you cannot retain}''\\

\vspace{0.4cm}

Saadi Shirazi [1210 - 1292]

\end{flushright}

\vspace{2cm}
\Resumo					%
Estrelas Be s\~{a}o um subtipo espec\'{i}fico de estrelas de sequ\^{e}ncia principal de tipo espectral B. Elas possuem caracter\'{i}sticas \'{u}nicas tais como a presen\c{c}a de linhas de emiss\~{a}o em seu espectro, que se originam de um disco circunstelar. Nos \'{u}ltimos 50 anos, a estrela Be gal\'{a}ctica $\omega$ CMa exibiu erup\c{c}\~{o}es quasi-regulares, a cada 8 anos aproximadamente, onde a estrela torna-se mais brilhante na banda V. Nestas erup\c{c}\~{o}es um novo disco se forma nos primeiros 3-4 anos e depois dissipa-se nos 4-6 anos seguintes. Temos acesso a uma base de dados rica (incluindo fotometria, polarimetria, interferometria e espectroscopia) de $\omega$ CMa desde Mar\c{c}o de 1964, que cobre v\'{a}rios ciclos de erup\c{c}\~{o}es e quiesc\^{e}ncias. Assim, a natureza nos proveu um experimento perfeito para estudar como discos de estrelas Be crescem e dissipam-se.

H\'{a} um corpo de evid\^{e}ncias cada vez maior que sugerem que os discos de estrela Be s\~{a}o bem descritos pelo modelo de decr\'{e}scimo viscoso (VDD), segundo o qual a forma\c{c}\~{a}o e estrutura do disco depende da viscosidade cinem\'{a}tica do g\'{a}s. Entretanto, a maioria dos testes conduzidos com o VDD at\'{e} o momento foram feitos para sistemas que n\~{a}o mostram forte variabilidade temporal. Usamos a rica base de dados de $\omega$ CMa para conduzir o primeiro teste aprofundado do VDD em um sistema fortemente vari\'{a}vel.

Usamos o c\'{o}digo de transporte radiativo HDUST para analisar e interpretar os dados. Desta an\'{a}lise obtemos (1) um modelo fisicamente realista do ambiente circunstelar, (2) a viscosidade do g\'{a}s, e (3) uma estimativa confi\'{a}vel das taxas de perda de massa e momento angular durante os eventos de forma\c{c}\~{a}o do disco.

Nossas simula\c{c}\~{o}es conseguem reproduzir a variabilidade fotom\'{e}trica muito bem, o que sugere que o modelo VDD descreve corretamente a evolu\c{c}\~{a}o estrutural do disco. Mostramos que o par\^{a}metro de viscosidade \'{e} vari\'{a}vel, com valores entre 0.1 e 1. Adicionalmente, as fases de constru\c{c}\~{a}o do disco t\^{e}m valores de viscosidade maior. Contrariamente ao que se acredita, mostramos que durante a dissipa\c{c}\~{a}o a taxa de perda de momento angular n\~{a}o \'{e} necessariamente nula, o que implica que $\omega$ CMa n\~{a}o experimenta uma quiesc\^{e}ncia verdadeira, mas alterna entre uma fase de alta taxa de perda de momento angular (erup\c{c}\~{a}o) e uma fase de baixa taxa (quiesc\^{e}ncia).
Confrontamos as taxas de perda de momento angular com as preditas pelos modelos evolutivos de Genebra, e encontramos que nossas taxas s\~{a}o mais que 10 vezes menores que as taxas de previstas pelos modelos.

O modelo desenvolvido para reproduzir a curva de luz na banda V foi aplicado a v\'{a}rios outros observ\'{a}veis. De forma geral, os resultados desta estudo multi-t\'{e}cnica foram muito positivos, com uma boa concord\^{a}ncia com a fotometria multi-banda, polariza\c{c}\~{a}o, e a maioria das caracter\'{i}sticas espectrais. Este \'{e} um resultado muito relevante, pois prova que um modelo que foi constru\'{i}do apenas apenas a partir de v\'{i}nculos para a interna do disco (a curva de luz na banda V), pode ser extendido para todo o disco e tamb\'{e}m outros processos f\'{i}sicos.		%
\Abstract					%
Be stars are main-sequence stars and a specific subclass of B type stars with the unique characteristic of showing H\,{\sc i} Balmer emission lines in their optical spectra that originates from a circumstellar disk around the star. Over the past 50 years, the Galactic Be star $\omega$ CMa has exhibited quasi-regular outbursts, every 8 years or so, when the star brightens by about half a magnitude in the $V$-band. During these outbursts a new disk is formed during the first 3-4 years, and then dissipates in the following 4-6 years. We have access to a rich dataset (including photometry, polarimetry, interferometry and spectroscopy) of $\omega$ CMa since March 1964 covering several outbursts and quiescence phases. Thus, nature has provided us the perfect experiment to study how Be star disks grow and dissipate.


There is an increasing body of evidence that suggests that Be disks are well described by the Viscous Decretion Disk (VDD) model according to which the formation and structure of the disk depend on the kinematic viscosity of the gas. However, most observational tests of the VDD to-date were done for systems that do not display strong temporal variability. We use the rich dataset available for $\omega$ CMa to perform the first in-depth test of the VDD scenario in a system with strong temporal variability.

We use the radiative transfer code {\tt HDUST} to analyze and interpret the observational dataset. From this analysis we (1) obtain a realistic physical model of the circumstellar environment; (2) measure the viscosity parameter of the gas, both during the formation and dissipation phases of the disk; (3) obtain a reliable estimate of the stellar mass and angular momentum loss rates during outburst.

Our simulations offer a good description of the photometric variability, which suggests that the VDD model adequately describes the structural evolution of the disk. Furthermore, our analysis allowed us to determine the viscosity parameter $\alpha$, as well as the net mass and angular momentum (AM) loss rates. We find that $\alpha$ is variable, ranging from 0.1 to 1.0, not only from cycle to cycle but also within a given cycle. Additionally, build-up phases have larger values of $\alpha$ than the dissipation phases. We also find that, contrary to what is generally assumed, during dissipation the outward AM flux is not necessarily zero, meaning that $\omega$ CMa does not experience a true quiescence but, instead, switches between a high AM loss rate state to a low AM loss rate one during which the disk quickly assumes an overall lower density but never zero. We confront the average AM loss rate with predictions from stellar evolution models for fast-rotating stars, and find that our measurements are smaller by more than one order of magnitude.

The model developed using the $V$-band photometry as a constraint was applied to several other observables. Overall, the results of this multi-technique study were very positive, with a good match for multi-band photometry, polarization, and most spectroscopic characteristics. This is a very relevant result, as it proves that a model that was constructed from constraints only from the very inner part of the disk (the $V$-band light curve), could be extended to the whole disk and to other physical processes.		%
\listoffigures 				
\listoftables 				
\tableofcontents 			
\cleardoublepage			%
\pagestyle{fancy}			
\chapter{Introduction}
\label{chap:intro}

Photons are by far the most common messengers from the celestial objects. Whatever we learned from them came after collecting these messengers in astronomical laboratories, usually called observatories, with research instruments such as photometers, polarimeters, spectrographs, etc., and then analyzing them with computers. 

\begin{figure}[!ht]
\begin{minipage}{0.5\linewidth}
\centering
\subfloat[Hiltner600, B1V type star]{\includegraphics[width=1.0\linewidth]{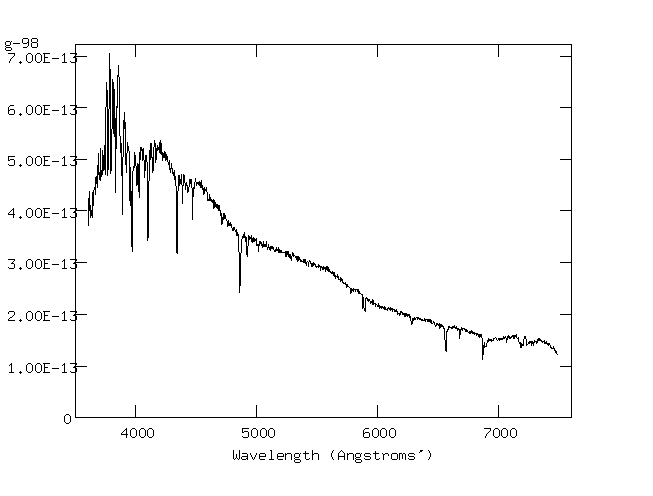}}
\end{minipage}%
\begin{minipage}{0.5\linewidth}
\centering
\subfloat[HD16908, B3V type star]{\includegraphics[width=1.0\linewidth]{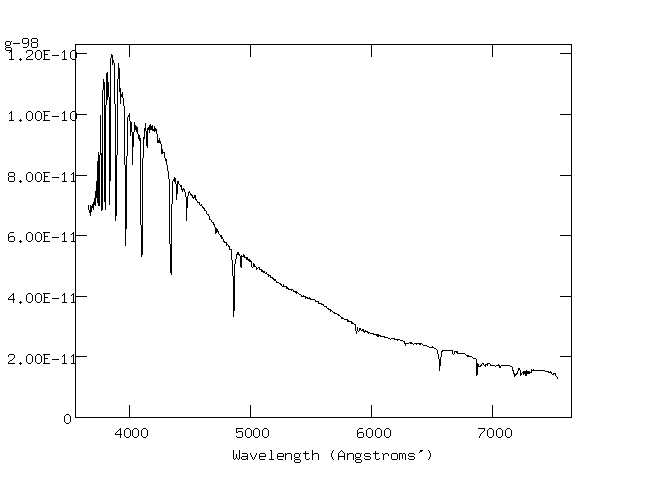}}
\end{minipage}\par\medskip
\begin{minipage}{0.5\linewidth}
\centering
\subfloat[HD14372, B5V type star]{\includegraphics[width=1.0\linewidth]{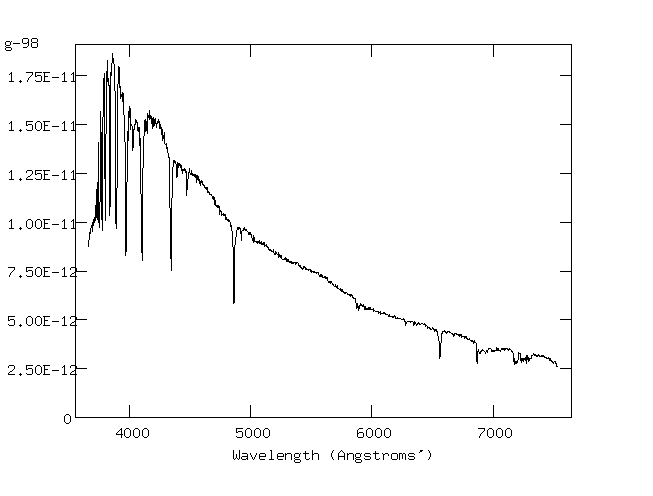}}
\end{minipage}%
\begin{minipage}{0.5\linewidth}
\centering
\subfloat[HD90994, B6V type star]{\includegraphics[width=1.0\linewidth]{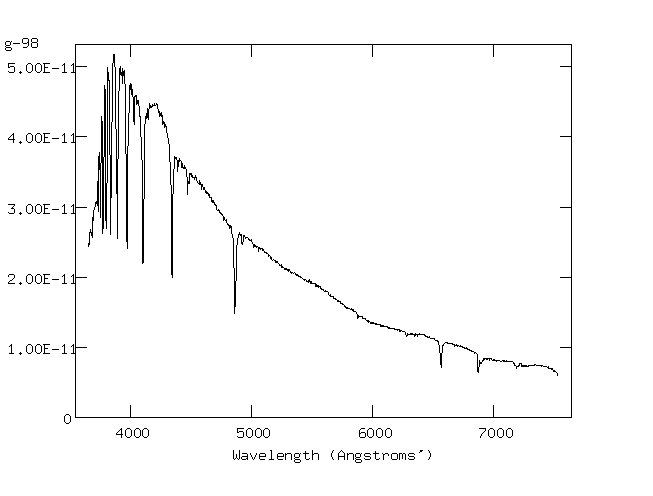}}
\end{minipage}%
\caption[Typical B-type Stellar Spectra]{Typical stellar spectra of few subclasses of B-type stars. B stars are hot stars defined by the presence of hydrogen (H) and neutral helium (He) in the optical spectra. H line strength increases over the B star subclasses while He decreases (Credit: Perry Berlind, Harvard-Smithsonian Center for Astrophysics).}
\label{fig:stellar_spectra}
\end{figure}

When spectroscopy was established as a regular observational technique, it was understood that the stellar spectra are formed by the superposition of a continuum spectrum, frequently resembling that of a black body, and several absorption lines (Fig.~\ref{fig:stellar_spectra}). In this doctoral research we deal with a specific group of massive stars whose spectra show, instead, emission lines in their spectra. This seemingly small difference is actually a key aspect of the subgroup of objects we study, called ``Be stars''.

\begin{figure}
\begin{center}
\includegraphics[width=0.5\linewidth,angle=0]{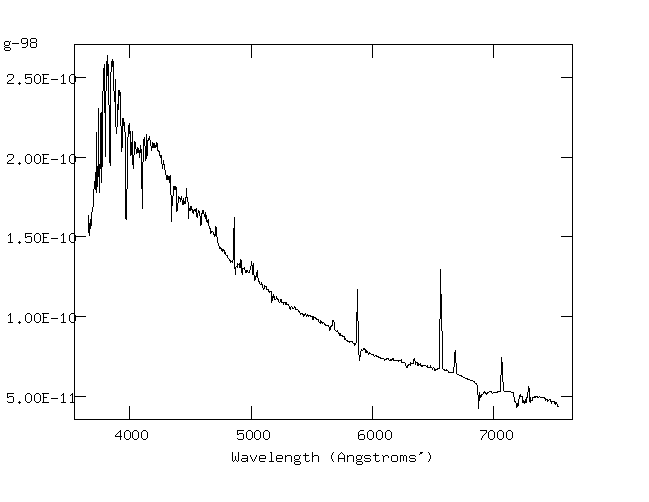}
\caption[Typical Be star spectrum]{Spectrum of $\beta$ Lyrae, a B7Ve star showing emission lines in H$\alpha$ and He lines at 4920\AA\ and 5876\AA\ (Credit: Perry Berlind, Harvard-Smithsonian Center for Astrophysics).} 
\label{fig:beta_lyrae}
\end{center}
\end{figure}

\section{Be stars}
\label{sect:be_star}


Although massive stars make a small fraction of the total stellar populations (approximately 0.1\% of the stars in the Solar neighborhood; \citealt{ledrew2001}), by providing the main source of heavy elements and UV radiation they play a key role in the evolution of the Universe. Through a combination of winds, massive outflows, expanding H\,{\sc ii} regions, and supernova explosions they provide an important source of mixing and turbulence in the interstellar medium (ISM) \citep{maeder1992, smartt2009}. They affect the star and planet formation processes \citep{bally2005} as well as the physical, chemical, and morphological structure of galaxies \citep{kennicutt1998, kennicutt2005}.

With masses in the range of $\sim$3 to $\sim$20M$_{\odot}$, B-type stars can be considered both massive stars ($M\geq8\mathrm{M}_{\odot}$, that may end their lives as a supernova, or intermediate mass stars ($M\leq8\mathrm{M}_{\odot}$) whose end stage of stellar evolution will be a white dwarf. In Fig.~\ref{fig:stellar_spectra}, sample spectra of B stars for different spectral subtypes are shown, revealing their strong UV fluxes and prominent H and He lines. Be stars are a specific subclass of main sequence B-type stars that rotate very fast. They are further characterized by the presence of one or more emission lines in their spectrum \citep{collins1987}. In Fig.~\ref{fig:beta_lyrae} a sample spectrum of the Be star $\beta$ Lyrae is shown, revealing its strong H and He emission lines. The emission includes mainly the first members of the Balmer line series, as well as other species, such as He\,{\sc i} and {\sc ii}, Fe\,{\sc ii} lines, etc. Be stars are not rare objects as about 20\% of B-type stars in Milky Way are Be stars with higher proportion in lower metallicity environments \citep{martayan2011, rivinius2013a}.

Table \ref{table:Be_stellar_parameter} contains the main stellar parameters of different spectral subtypes of B stars from early (B0) to late (B9) spectral types, based on the \cite{schmidt-Kaler1982}.


\begin{table}
\begin{center}
\caption[The main stellar parametrs of the B-type stars]{The main stellar parametrs of the B-type stars.}
\begin{tabular}{|c|c|c|c|c|c|c|c|c|c|c|c|}
\hline
\hline
\cellcolor{black} & $M$(M$_{\odot}$) & $R$(R$_{\odot}$) & $T_\mathrm{eff}$(K) & $L$(L$_{\odot}$) \\
\hline
\rowcolor[HTML]{B2BEB5}
B0 & 17.5 & 7.7 & 30000 & 44000 \\
B0.5 & 14.6 & 7.4 & 28000 & 30000 \\
\rowcolor[HTML]{B2BEB5}
B1 & 12.5 & 6.8 & 26000 & 19000 \\
B1.5 & 10.8 & 5.8 & 25000 & 11000 \\
\rowcolor[HTML]{B2BEB5}
B2 & 9.6 & 5.6 & 23000 & 7800 \\
B2.5 & 8.6 & 5.2 & 22000 & 5400 \\
\rowcolor[HTML]{B2BEB5}
B3 & 7.7 & 4.9 & 20000 & 3700 \\
B4 & 6.4 & 4.4 & 18000 & 1900 \\
\rowcolor[HTML]{B2BEB5}
B5 & 5.5 & 4.1 & 17000 & 1200 \\
B6 & 4.8 & 3.9 & 15000 & 700 \\
\rowcolor[HTML]{B2BEB5}
B7 & 4.2 & 3.7 & 14000 & 400 \\
B8 & 3.8 & 3.5 & 13000 & 300 \\
\rowcolor[HTML]{B2BEB5}
B9 & 3.4 & 3.4 & 12000 & 200 \\
\hline
\end{tabular}
\label{table:Be_stellar_parameter}
\end{center}
\end{table}


\section{The Be phenomenon}
\label{sect:be_phenomenon}

The transient appearance of emission lines in the spectrum of Be stars is known as ``Be phenomenon''. It is believed that the emission lines come from a circumstellar gaseous environment that most likely is in the form of an equatorial, dust-free Keplerian disk. The physical mechanism that brings the stellar materials to an orbit around the star is not well known. Whatever it is, it should be related to the rapid rotation of the star. Recent evidence \citep[e.g.,][]{baade2016} based on space-based photometry, as well as earlier evidence based on ground-based spectroscopy \citep[e.g.,][]{rivinius1998} demonstrate that stellar pulsation also plays a key role on the Be phenomenon. However, we do know what is the fate of the material once it reaches orbit around the star \citep{carciofi2011, rivinius2013a}: the material slowly diffuses outwards by means of viscous torques. Other similar astrophysical systems that are viscosity driven are accretion disks, but there is an important difference: the direction of flow that is from star outwards in the case of Be stars and inwards in the case of accretion disks. Because Be disks are (usually) outflowing, they are referred to by the neologism ``decretion disks''.

The key property that distinguishes Be from B stars is that the former have much higher rotational velocities, sometimes close to the critical rotational speeds \citep{catanzaro2013}. Recent results, summarized in the review paper by \cite{rivinius2013a}, show that Be stars rotate in average at 80\% of their break-up speed. Individual stars may rotate at much higher rates, as, for instance, Achernar, that rotates at about 96\% of the critical limit \citep{carciofi2008a}.

Due to their peculiarities, the study of Be stars provides a unique opportunity to probe and understand several important branches of astrophysics, e.g., asymmetric mass-loss processes, evolution of fast-rotating stars, astroseismology, and, what is most relevant for this research, astrophysical disks.


\section{The disks of Be stars}
\label{sect:be_disk}

As a subclass of B-type stars, Be stars are in the group of hot stars with effective temperature between 12000\,K to 30000\,K (Table \ref{table:Be_stellar_parameter}). Being close to such hot stars, no dust can form in their hot and ionized disks. Therefore, they are completely devoid of dust, with opacities (primarily from hydrogen and free electrons) much simpler than other astrophysical disks where the dominant opacity source is dust. Size distribution, chemical composition, and even spatial distribution of dust grains in such complex astrophysical systems are rather uncertain \citep{glassgold2006}. Moreover, the disks in Be stars are kinematically (relatively) simple. All these together, make Be star disks good laboratories for the study of the physics of astrophysical disks in general.

Several theoretical models were proposed to describe the mechanism of disk formation around Be stars, such as the wind-compressed disk \citep{bjorkman1993}, magnetic wind-compressed disk \citep{cassinelli2002}, and viscous decretion disk (VDD) that was proposed by \cite{lee1991} and further developed by many authors \citep{bjorkman1997, okazaki2001, bjorkman2005, sigut2007, carciofi2011}. Owing to the recent advancements of observational techniques, such as interferometry, we can now safely state that the VDD model is the only one that passes all observational and theoretical tests proposed so far. These recent advancements are described in detail in the review paper by \cite{rivinius2013a}. \cite{carciofi2011} used the VDD model to construct the basis of new paradigm for the interpretation of Be stars observations. Then, it was widely used to successfully model Be stars \citep{jones2008, silaj2010, touhami2011, carciofi2012, klement2015, vieira2015, faes2016, arcos2017, jones2017, klement2017, vieira2017, marr2018, rimulo2018}. More details about the VDD model can be found in Sect.\,\ref{chap:theories}.

How a Be disk is formed? Since Be stars rotate close to their critical velocity, the outer layers of the star are loosely bound to the star and by means of a still unknown mechanism they leave the surface of the star, a process we refer to as an outburst, or mass loss event. Although most of the ejected particles falls back on the stellar surface, some of them have enough angular momentum (AM) to stay in orbit and form a ring-like structure. Thereafter, another mechanism comes into play: viscous shear. Owing to viscous torques some particles are lifted to larger radii to where they carry AM \citep{carciofi2009, carciofi2012} and thereby the disk grows in size. 

\cite{struve1931} suggested that the gaseous dust-free disks around Be stars give them a flattened geometry. Progress in interferometric observations provided the opportunity to verify this hypothesis in 1990s \citep{dougherty1992, stee1995, quirrenbach1997}. 

The disk has a wedge shaped structure with opening angle between 2.5$^\circ$ \citep{wood1997} to 20$^\circ$ \citep{quirrenbach1997}. Recently, more accurate opening angle measurements have been provided by \cite{cyr2015} in the range of 3.7$^\circ$ and 14.0$^\circ$. The reported values of opening angles are not consonant with each other because disk flaring at larger radii affects the measurements. 

Finally, it is worth mentioning that the disks in the late type Be stars are more tenuous and less variable than those in early type ones \citep{vieira2017}.


\section{Variations in the observables}
\label{sect:variations}

Changes in the geometry (e.g., inclination of the disk with respect to the spin axis of the star) and physical parameters of Be stars (including both the central star and its disk; e.g., stellar temperature and disk density) are plausible. Any physical changes within Be stars or their disks reveal themselves in variations on observables such as photometry, spectroscopy, polarimetry, and interferometry, each related to different regions of the system and connected to different physical mechanisms responsible for their origin. For instance, polarization in Be stars arises from scattering of stellar radiation off free electrons in the disk, while emission lines are associated with radiative deexcitation processes (in the visible and IR) or ressonant scattering (in the UV). Therefore, multitechnique follow up observations of Be stars allow for studying the workings of these fascinating systems.

Changes in brightness and spectral line appearance are the typical variations in classical Be stars. Their time scales cover a large range from minutes to years \citep{okazaki1997, floquet2002, kogure2007, haubois2012} that means that various astrophysical phenomena play role in their structure. The origins of these variations are disk formation and dissipation for the longer term variations \citep{okazaki1997, haubois2012, rimulo2018} and pulsations within the B star photosphere for the short ones \citep{baade2000, huat2009}. According to the results of photometric studies, earlier type Be stars are more likely variable \citep{hubert1998, labadie2018, shokry2018}.

\begin{figure}
\begin{center}
\includegraphics[width=\columnwidth,angle=0]{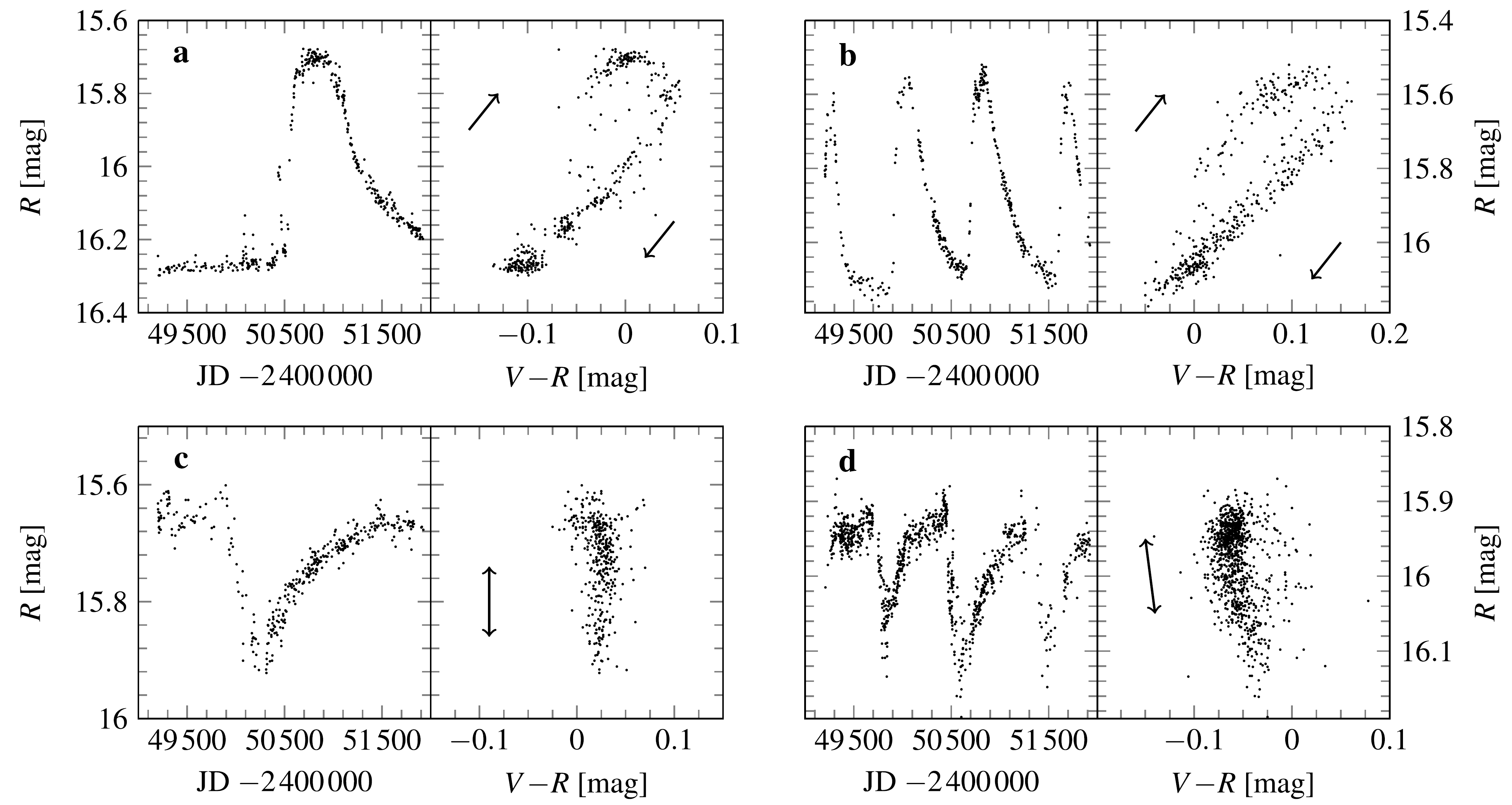}
\caption[Color and magnitude variations of four SMC Be stars]{Color and magnitude variations of four Small Magellanic Cloud (SMC) Be stars. Up left (a): MACHO 23.4148.53 and up right (b): 17.2109.68 showing brightening. Bottom left (c): MACHO 17.2594.208 and bottom right (d): 77.7427.129, showing dimming. The arrows display the temporal sense of the color changes. \citep[Credit:][]{rivinius2013a}.} 
\label{fig:color_mag_variation}
\end{center}
\end{figure}

{\sf Outburst} and {\sf quiescence} are two important phases during the life cycles of a Be star disk, and are behind the long-term, secular variations observed in many light curvers, as well as short-term, low amplitude ones. For a (nearly) pole-on star, an outburst in the visible photometric light curve is typically a sudden rise in the flux of the system that can be the result of disk formation as a consequence of mass being ejected by the star \citep[see,][]{haubois2012}. Conversely, if the star is seen edge-on (i.e., with the disk seen projected against the stellar disk) the outburst will appear as a sudden decline in brightness. Usually, an outburst is followed by a more gradual (relative to the initial change) decay back to baseline that is called quiescence. A quiescence phase is usually associated with the cessation of the mass loss, followed by the dissipation of the disk. Examples of such events are shown in Fig.~\ref{fig:color_mag_variation}, which also illustrates how the colors of the star changes as a result of the varying disk conditions.

The net brightenings and fadings associated with disk events seen pole-on, or, conversely, fadings and brightenings when seen edge-on, are called as bumps and dips \citep{rimulo2018}, respectively. The distinction between a bump or dip is no longer possible for inclination angles close to $\sim$70$^\circ$ because the excess emission will be approximately canceled out by the absorption introduced by the disk \citep{haubois2012, sigut2013}. 

The duration, frequency and amplitude of outbursts are different not only from star to star but even for a given star whose observations may show large variations over time. However, some Be stars possess disk that remain stable for many years or decades while others reveal very dynamic variations in relatively short time (Fig.~\ref{fig:color_mag_variation}). For the Be stars with a stable disk, it is common to have density waves traveling around the disk. The timescales of such movements are hundreds of times longer than that of expected in Keplerian motion at a given radius, typically at a period on the order of 10 years \citep{okazaki1991, papaloizou1992}.

Such global oscillation modes in the disk reveal themselevs in spectroscopic observations as variations in the ratio of the violet-to-red (V/R) peaks  of Be star emission lines \citep{carciofi2009, stefl2009}. The enhancement in the violet (V) peak is seen when the high-density portion of the disk is coming towards the observer, while the enhancement in the red (R) peak is the result of receding of the high-density part from the observer. Also, photometric variations as a line-of sight effect can be produced from the density waves, depending on the inclination angle of the system \citep{escolano2015}.

The variations observed in Be stars can be classified according to their length ranging from hours to decades as the result of different involved astrophysical phenomena. Recurring $\beta$-Cephei type pulsations leads to the ultra-rapid variations \citep{huang1989} on timescales of hours. Non-radial pulsations \citep{baade1982a, rivinius2003}, stellar rotation \citep{balona1990, balona1995}, and localized mass ejections in the inner regions of the disk cause the short-term variations with periods of days \citep{carciofi2007}. Changes in the disk as a result of binarity cause intermediate length variations occurring over periods of weeks to months, as these variations are tied to the orbital period \citep{panoglou2017}. The formation and dissipation of the disk, and perturbations in the circumstellar disk are in connection to the long-term variations with periods of decades. The Be stars $\gamma$ Cas and 59 Cyg \citep{hummel1998} are samples of former case and the disk dissipation observed between 1986 and 1996 in BIII type Be star $\pi$ Aqr \citep{wisniewski2010, draper2011} is a good sample of the latter. The observed variations in $\gamma$ Cas and 59 Cyg can be a sign of binarity connected to a tilted rotating disk. In such a system the companion is misaligned with the equatorial plane of the Be star. Consequently, the tidal interactions inclines the disk respect to the star and aligns it to the companion's orbital plane \citep{cyr2017}. Change in geometry of the system on the plane of the sky and the rotation of such a tilted disk can result in the variation in emission-line width and profiles.

\begin{figure}[!t]
\begin{center}
\includegraphics[height=0.5 \textheight, width=0.5 \columnwidth,angle=0]{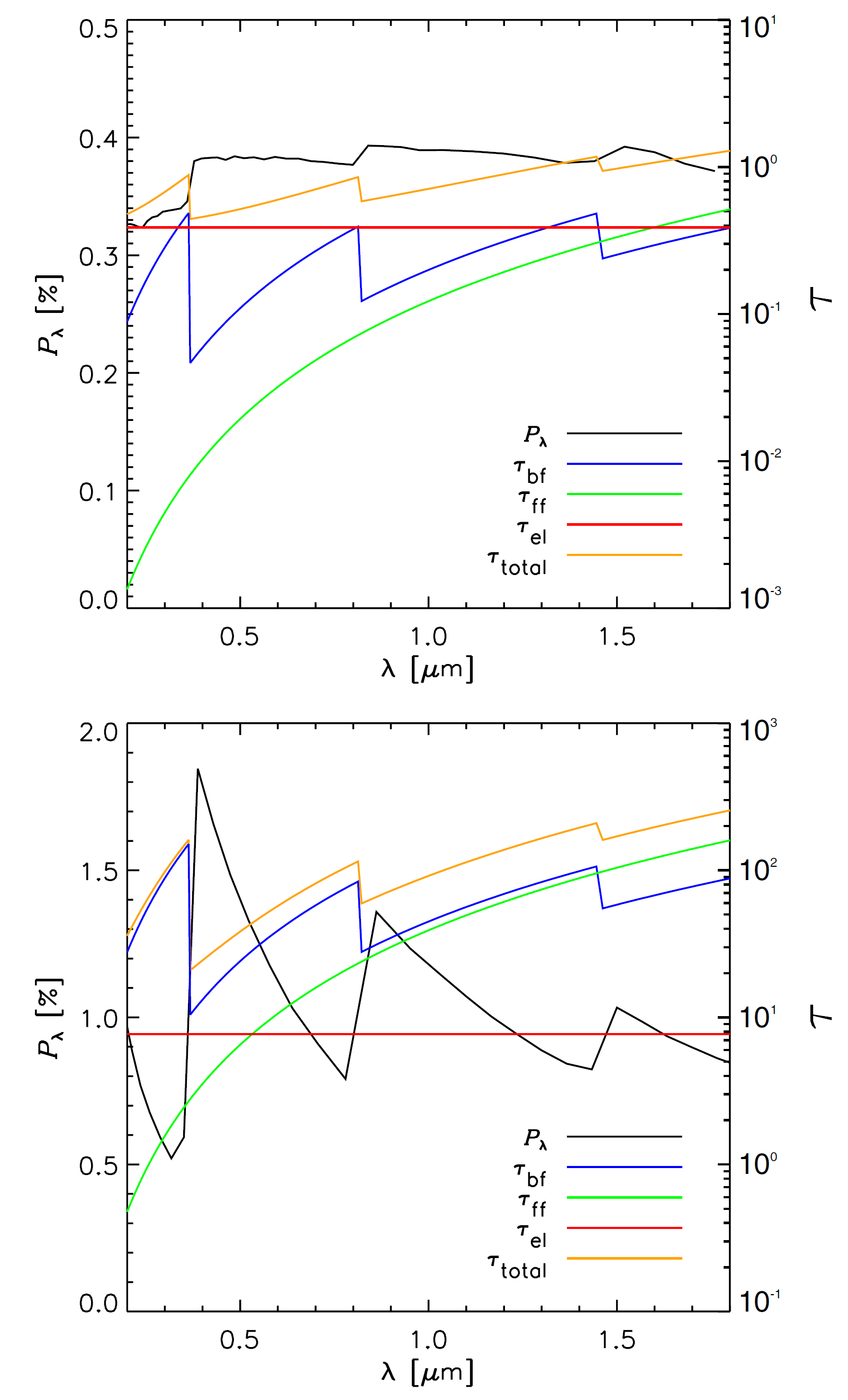}
\caption[Sample polarized spectrum of a B2e star]{Theoretical polarized spectra and total optical depth of two different B2e star disks with two different base densities (Upper panel: 4.2$\times$10$^{-12}$ g\,cm$^{-3}$; lower panel: 8.4$\times$10$^{-11}$ g\,cm$^{-3}$) with inclination angle of $\sim$70$^\circ$. The relative contribution to the total opacity (orange line) of each opacity source is shown with blue (bound-free absorption) green (free-free absorption) and red (Thomson scattering) lines (Credit: \citealt{haubois2014}).} 
\label{fig:polarization_opacity}
\end{center}
\end{figure}

One of the main characteristics of Be stars is the linearly polarized light \citep{hall1950, behr1959} that shows variations in time. Its amplitude can be up to $\sim$2\% of the total light emitted. The polarization level along with data gathered by other techniques provide valuable information on the geometry and physical nature of the disk. The polarization level is typically proportional to the number of scatterers (i.e., total number of free electrons); on the other hand, the slope of the polarization depends on the H continuum opacity, and therefore is a $\rho^2$ diagnostics \citep{haubois2014}. This is illustrated in Fig.~\ref{fig:polarization_opacity}, that shows theoretical polarization spectra for a tenuous disk, whose polarization is almost grey, and a dense disk, that displays the typical ``saw-tooth'' behavior that mimics the spectral dependence of the H opacity.

\section{Classification of the light curves}
\label{sect:lightcurves}

Several studies were performed to classify the light curves of Be stars according to their long-term morphology \citep[e.g.,][]{mennickent2002, sabogal2005}. Such a study for Be star candidates within the SMC performed by \cite{mennickent2002} showed that Be star candidates have light curves in diverse shapes covering those ones with morphologies similar to simpler bumps and dips discussed above and other with much more complex shapes. Their classification include five main types of variability.

\begin{figure}
\begin{center}
\includegraphics[height=0.9 \textheight, width=0.8 \columnwidth,angle=0]{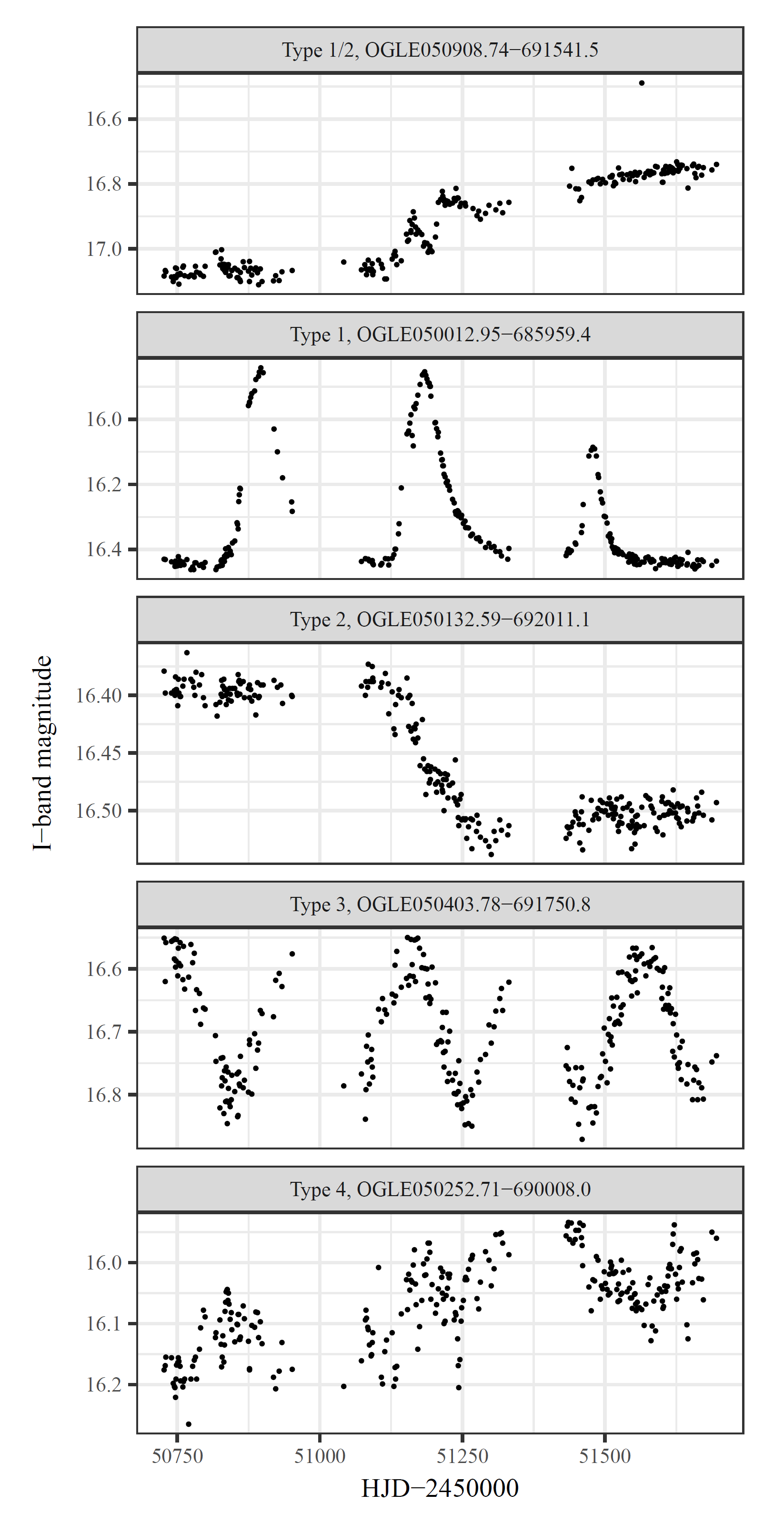}
\caption[Classification of the light curves of Be satrs]{Different types of light curve from five Be star candidates in Large Magellanic Cloud (Credits: \citealt{perezortiz2017}).} 
\label{fig:lightcurve_class}
\end{center}
\end{figure}

\begin{itemize}
    \item{Type-1 which in turn involves two sub-classes. Members of the first sub-class are light curves with rapid and sharp rise of brightness which dims slowly, generally lasting tens of days with amplitudes <0.$^m$2. The second ones have slower brightness increment and slower decrement as well, lasting hundreds of days with amplitudes >0.$^m$2. Their outbursts show hump-like shape and have more symmetric rising and fading timescales. Some stars have light curves with a mix of both sub-classes, and distinction can be hard to be made. The stars of type-1 do not have always a constant level of quiescence brightness. The Galactic Be stars $\omega$ CMa, $\lambda$ Eri, $\mu$ Cen and those stars analyzed in Hipparcos photometry by \cite{hubert1998} and \cite{smith2000} are counterparts of this type.}

    \item{Type-2 are the stars whose light curves consist of long (hundreds of days) phases with different brightness levels of few tenths of magnitude. The outburst in this type is sharp as well. No Galactic star has been observed from this class, yet \citep{sabogal2014}.}

    \item{Type-3 involves stars with periodic or quasi-periodic light curves.}

    \item{Type-4 stars demonstrate fortuitous variability in their light curves with time scales from days to years.}

    \item{Type-1/2 are faint stars found only in the range of 15.5 < V < 17.0 which display brightness jumps over the outbursts simultaneously.}
\end{itemize}

It is important to note that by the time the above study (and other similar ones) were made, dynamical VDD models, such as the ones used in this work, were not available.
The authors, therefore, lacked the interpretation tool that models such as presented by \cite{haubois2012} provide. It is now believed that all variability types described above can be explained by the dynamics of a viscous disk subjected to a disk feeding rate that varies over time. When this variation is simple (such as on and off behavior), the resulting light curvers will be bumps and dips such as the ones of Fig.~\ref{fig:color_mag_variation} and the upper panels of Fig.~\ref{fig:lightcurve_class}. A more erratic disk feeding rate will result in the complex light curves frequently observed.


\section{The viscous decretion disk model in action}
\label{sect:vdd_action}

\begin{figure}[!t]
\begin{center}
\includegraphics[height=0.5 \textheight, width=0.5 \columnwidth,angle=0]{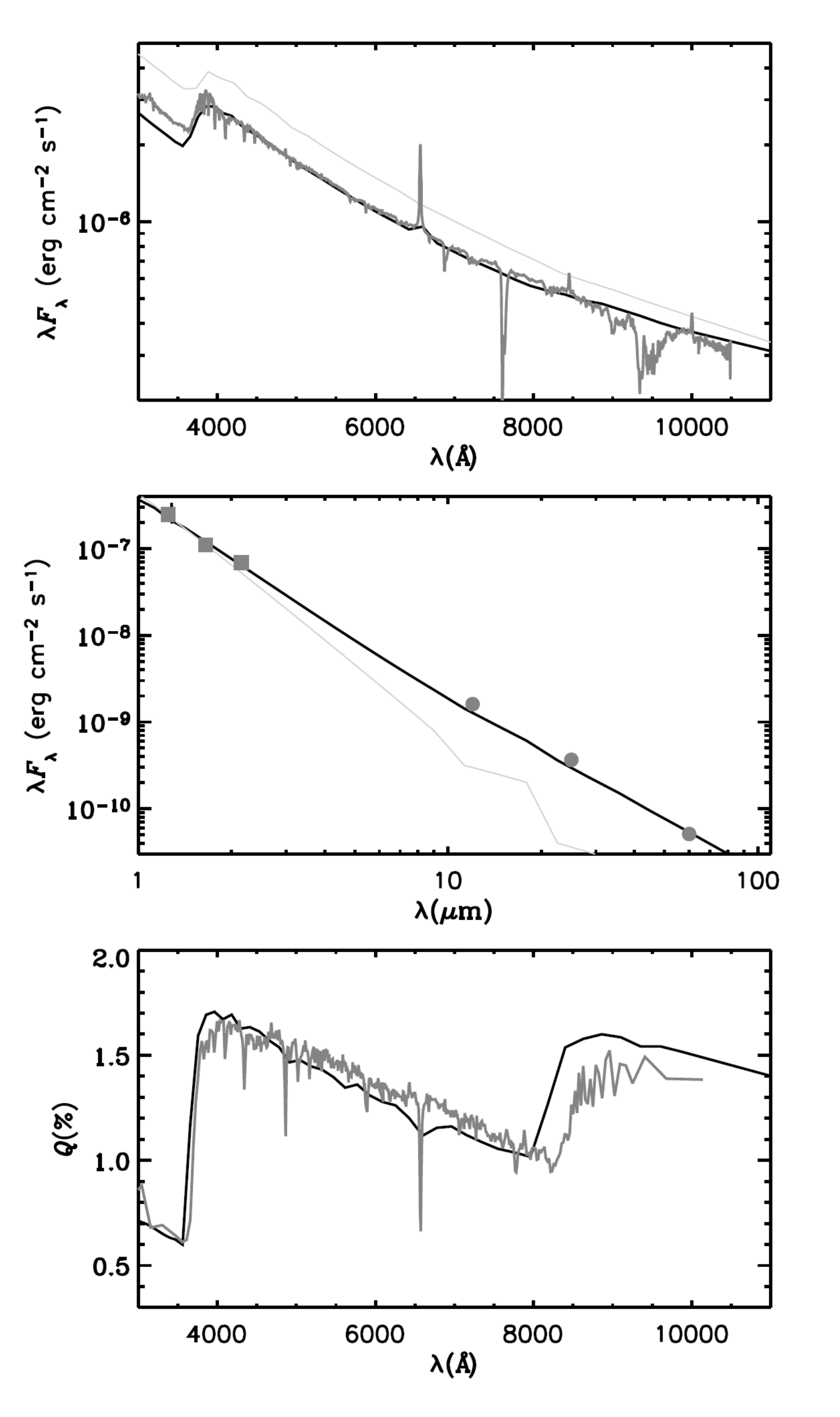}
\caption[Fit of the emergent spectrum and polarization of $\zeta$ Tau]{Fit of the emergent spectrum and polarization of $\zeta$ Tau \citep{carciofi2009}. The dark grey lines and symbols are the observations and the black lines represent the 2D model results. Top: visible SED. Middle: IR SED. Bottom: continuum polarization. In the two upper panels, the light grey line corresponds to the unattenuated stellar SED.} 
\label{fig:zeta_tauri}
\end{center}
\end{figure}

The solution of a VDD model in the near-steady state limit (i.e., a disk fed at a constant rate for an extended period of time) is relatively straightforward to obtain if one assumes the disk to be isothermal \citep[e.g.,][]{bjorkman1997, okazaki2001, bjorkman2005}. In this case, one finds that the disk is in vertical hydrostatic equilibrium and, because the orbital speed is much larger than the sound speed, the disk is geometrically very thin, with opening angles of only a few degrees. By comparing the model with observations, it is possible to test the various model predictions, thereby gaining a better understanding of the nature of Be star disks. As an example, Fig.~\ref{fig:zeta_tauri} shows a comparison between the VDD predictions computed using the radiative transfer code {\tt HDUST} \citep[][see Chap.~\ref{chap:theories}]{carciofi2006a, carciofi2008b} and observations of the Be star $\zeta$ Tau that demonstrates that the model can satisfactorily account for most observational properties.

The VDD model has been successfully applied to study individual stars \citep[for instance,][]{carciofi2006b, jones2008, carciofi2009, carciofi2012, klement2015, klement2017} and samples of Be stars \citep[for instance,][]{silaj2010, touhami2011, vieira2017, rimulo2018}. The main previous results are summarized by \cite{rivinius2013a}.

\begin{figure}
\begin{center}
\includegraphics[width=1.0 \columnwidth,angle=0]{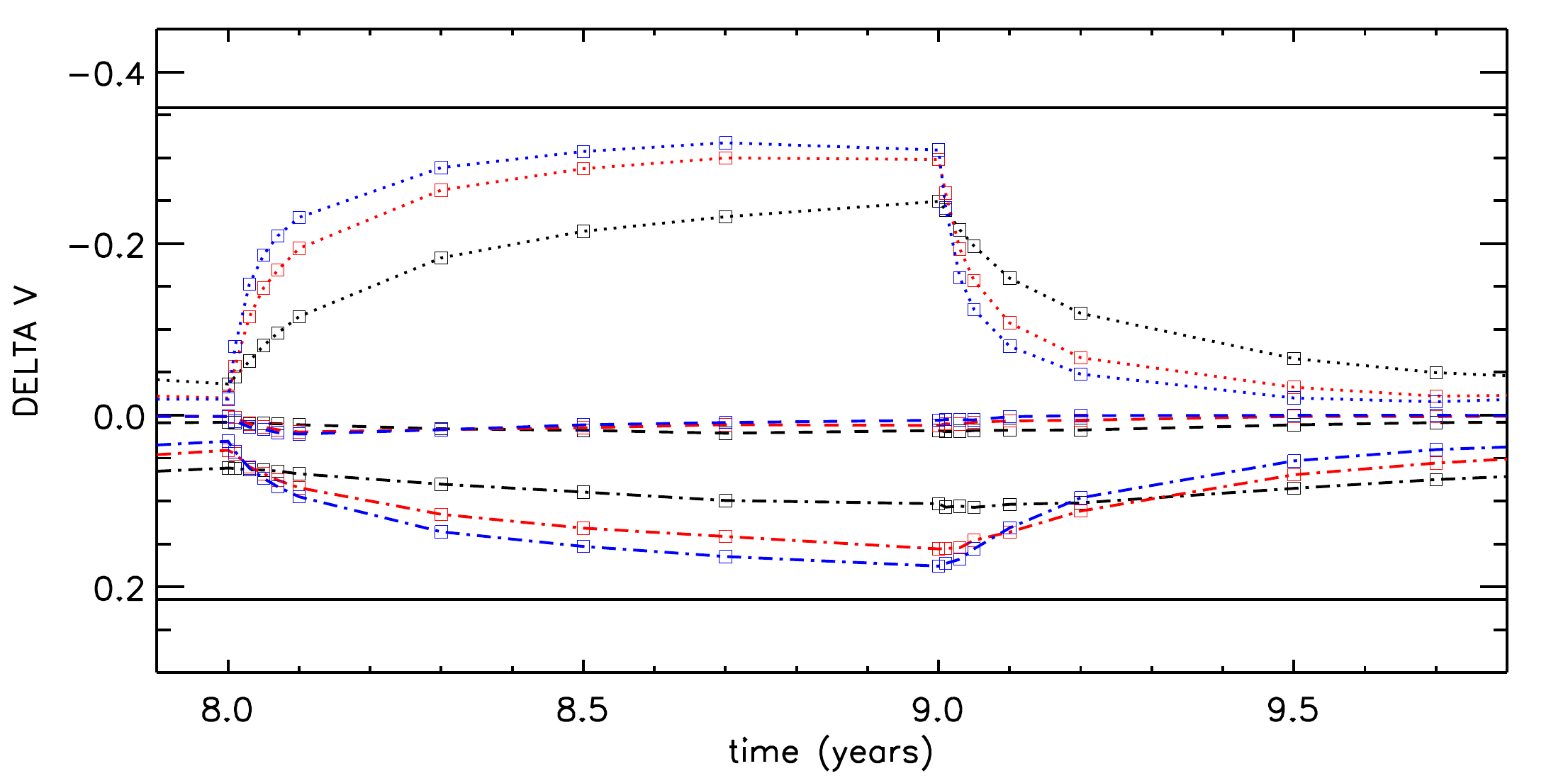}
\caption[$V$-band light curve variation]{$V$-band light curves associated with a periodic one-year long disk build-up ($\dot M>0$) followed by one-year long dissipation ($\dot M=0$). The dotted, dashed and dot-dashed lines represent the light curves for inclination angles of $0^{\circ}$ (face-on), $70^{\circ}$ and $90^{\circ}$ (edge-on), respectively. The black, red and blue colors represent models for $\alpha=0.1$, $0.5$, and $1.0$, respectively. The solid black lines in each graphic indicate the asymptotic value for the band \citep{haubois2012}.} 
\label{fig:v_band_variation}
\end{center}
\end{figure}

One of the main aspects of the Be phenomenon is the intrinsic variability of the Be spectrum, which is attributed to the changing conditions of the circumstellar disk. The VDD model was used not only to reproduce the observables of Be stars in static circumstance but for the dynamic situation. The first effort to understand the dynamical evolution of VDDs around isolated Be stars was done by \cite{jones2008}. This was later followed by a systematic study by \cite{haubois2012} who coupled the the 1-D time-dependent hydrodynamics code {\tt SINGLEBE} \citep[][See Chap.~\ref{chap:theories}]{okazaki2007} and {\tt HDUST} code to study the theoretical effects of time variable mass loss rates ($\dot M$) on the structure of the disk and its consequences on the observed photometry. Figure~\ref{fig:v_band_variation} shows model variations in the $V$-band light curve expected in a disk with periodic one-year long disk build-up ($\dot M>0$) followed by one-year long disk dissipation ($\dot M=0$), for three different values of disk viscosity (see Chap.~\ref{chap:theories}) and three different viewing angles. An important conclusion of their work was that time-dependent VDD calculations can explain well the observed temporal photometric phenomenology of Be stars such as loops in the color-magnitude diagram (Fig.~\ref{fig:color_mag_variation}), believed to track the process of the disk formation, during which the stars become redder and brighter, and dissipation, as the stars move back to their intrinsic colors and brightness. However, the tracks in the color-magnitude diagram are often complex and dependent on the inclination angle, as discussed by \cite{haubois2012}. Moreover, they showed that the observed light curve is strongly affected by the mass injection rate history, and found a relation between the radial slope of the density and the disk dynamical state. The disk formation is associated with steep radial density profiles while disk dissipation results in the flatter density slopes \citep{haubois2012}. The latter also can be the result of accumulation effect by binary interaction \citep{okazaki2002, panoglou2016}.

\section{Main goals}
\label{sect:goals}

\begin{figure}[!t]
\begin{center}
\includegraphics[width=0.8\linewidth, height=0.4\linewidth]{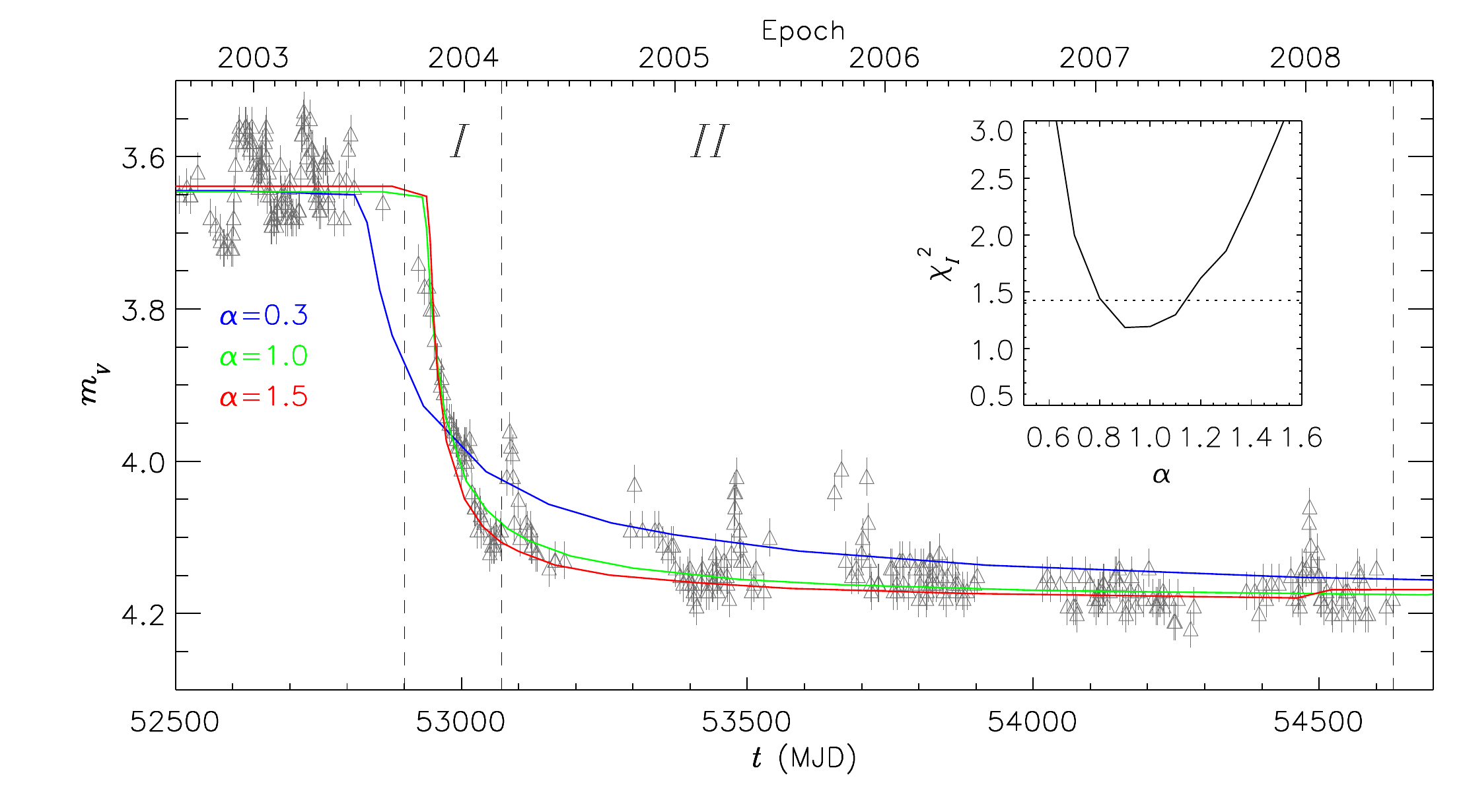}
\caption[First measurment of $\alpha$ parameter by \cite{carciofi2012}]{Fitting the $V$-band dissipation curve of $\omega$ CMa with the VDD model \citep{carciofi2012}. Visual observations of $\omega$ CMa (gray triangles) are shown in comparison to the model fits for different values of $\alpha$. Phase I (MJD = 52900\,--\,53070) is the initial decline, which was used to determine the value of $\alpha$. Phase II (MJD = 53070\,--\,54670) is the slow disk-draining phase. The inset shows the reduced chi-squared of the Phase I fit for different values of $\alpha$. The horizontal dotted line indicates the 90\% confidence level.} 
\label{fig:first_alpha}
\end{center}
\end{figure}

Notwithstanding all successes for the VDD model mentioned above, there are several challenges that remain to be solved. The aim of this PhD thesis is to address some of them. For instance, the VDD model was never tested in both (successive) formation and dissipation states of disk life. The conclusions of \cite{haubois2012}'s work gained further observational support when \cite{carciofi2012} modeled, for the first time, the light curve of a Be star. The chosen star ($\omega$ CMa) passed from an active phase, that lasted from 2000 to 2003, to a quiescent phase at the end of 2003. The model of the dissipation curve (Fig.~\ref{fig:first_alpha}) allowed the authors to measure the viscoity parameter of the disk ($\alpha=1.0\pm 0.2$; for a definition of $\alpha$ please refer to Chap.~\ref{chap:theories}). Their work represented just a first step towards understating the physical conditions of $\omega$ CMa's disk. As such, the study had several limitations, among which we cite the fact that $\alpha$ was determined only during the phase of disk dissipation and by fitting only the $V$-band light curve, which limits the study to the conditions of the inner disk. Lifting these and other limitations was one of the main motivations for the present work. Since the physical conditions of the disk during the disk formation is completely different than dissipation, the possible ability of the VDD model to reproduce this event will represent a strong theoretical support for the VDD theory. 

Additionally, to date the the VDD model was only tested against photometric light curves in the visible \citep{carciofi2012, rimulo2018}. An important part of this research is to test the VDD model with multitechnique observations. The extension to other observables represents a key step forward, as each observables probes both different physical processes (scattering, recombination, etc.) and different physical regions of the disk. 


\section{Strategy and questions}
\label{sect:strategy}

We will study 34 years of photometric observations of the Be star $\omega$ CMa, combined with a rich multitechnique data, using state-of-the-art VDD models combined with detailed radiative transfer calculations. 

The main questions addressed in this PhD thesis are:

\begin{enumerate}
    \item{Is the disk viscosity always constant in time? The modeling of the different outbursts will give us hints to a possible temporal variability of this quantity.}
    \item{Is viscosity at outburst phase the same as dissipation phase (because temperature and density conditions are quite different in these phases)?}
    \item{Is the disk feeding rate ($\dot M$) the same for each cycle?}
\end{enumerate}

Polarization is an important observable for this study because it is produced by scattering of starlight off the disk, within about 10 stellar radii, and is critically dependent on three quantities: the inclination angle of the disk, the geometry of the inner disk and the hydrogen bound-free opacity \citep{haubois2014}. Therefore, by fitting the polarization curve with the previous model, we will be able to address the interesting issue whether viscosity varies with radius (since the density and temperature conditions vary in different regions of the disk, it is conceivable that $\alpha$ could vary as well).

In addition to polarimetry, we will also analyze photometry at longer wavelengths ($JHK$ photometry taken sparsely during the last decade and  recent sub-mm observations made at APEX by our group). These data will allow us to check the validity of the VDD over much larger disk volumes. In particular the sub-mm data will be crucial, since we expect it will allow us to determine the actual size of the disk around $\omega$ CMa. 

Another important part of the available data is spectroscopy collected with variety of telescopes and instruments from all over the world. Since different spectral lines originate from different regions of the disk, spectroscopy is very useful to test the VDD model in different physical circumstances.


\section{Organization}
\label{sect:organization}

This PhD thesis is organized as follows. Chap.~\ref{chap:omecma_obs} describes the observational data available for $\omega$ CMa. Chap.~\ref{chap:theories} presents briefly the theoretical concepts that were used in this work. In Chap.~\ref{chap:exp_formula}, the $V$-Band light curve is analyzed without the usage of a physical model. In Chap.~\ref{chap:photometry} I show the results of studying the light curve of $\omega$ CMa in $V$-band using the VDD model. Finally, in Chap.~\ref{chap:other_tech} the results of investigation of the available data in other observables (than $V$-band photometry) are presented. 
\chapter{$\omega$ CMa and Observations}
\label{chap:omecma_obs}

\begin{figure}[!ht]
\begin{center}
\includegraphics[width=1.0 \columnwidth,angle=0]{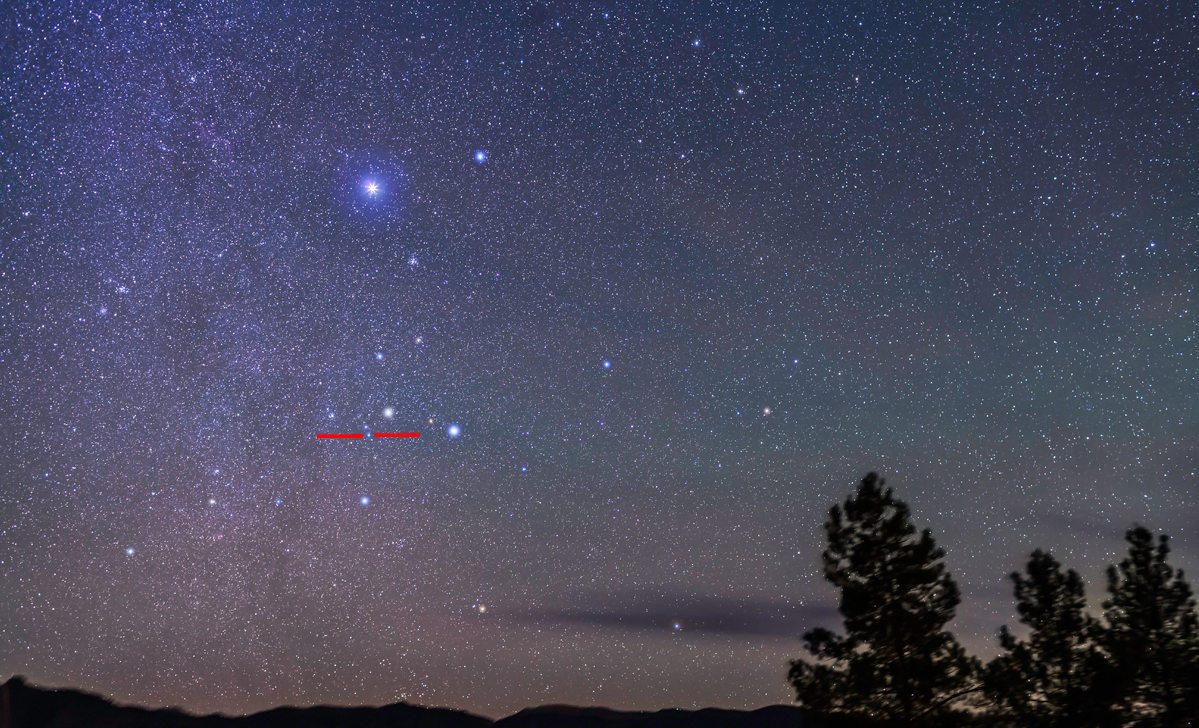}
\caption[$\omega$ CMa's position in the Canis Major constellation]{$\omega$ CMa can be seen as a rather bright star in the Canis Major constellation (Credit: Alan Dyer - \href{www.amazingsky.com}{www.amazingsky.com}).} 
\label{fig:ome_CMa}
\end{center}
\end{figure}

$\omega$ (28) CMa (HD 56139, HR2749; B2 IV-Ve) is one of the brightest Be stars ($m_\mathrm{v} \approx$ 3.6 to 4.2) in the sky (Fig.~\ref{fig:ome_CMa}) and it has been a common target of observers. Therefore, there exists a rich dataset from different observational techniques since 1963. $\omega$ CMa is a blue-white star in the equatorial constellation of Canis Major. Due to its brightness it is visible to the naked eyes even in the urban light pollution. Its measured annual parallax shift as seen from Earth is $\sim$3.58 mas, therefore, it should be located roughly at 280 pc (910 ly) far from the Sun \citep{perryman1997, vanleeuwen2007}. However, the new measurements by GAIA put the star at a closer distance, roughly at 205$^{+19}_{-16}$ pc (670 ly) \citep{gaia2016}, receding from us at the velocity of 23 km/s \citep{gontcharov2006}.

In the period of 34 years (between 1981 and 2015), the star went through four complete cycles of disk formation and dissipation that cause variations in brightness (from the visible to the IR), line profiles, polarization level, and other observables. The brightness and the radial velocity of the star vary with a primary cyclical period of $\sim$1.37 days attributed to a single mode of non-radial pulsation \citep{baade1982a, baade1982b, harmanec1998}. The estimated age of $\omega$ CMa is 22.5 million years \citep{tetzlaff2011}. It is a nearly pole-on star, so the measured projected rotational velocity of 80 km/s \citep{slettebak1975} is only a fraction of the true equatorial velocity, estimated as 350 km/s \citep{maintz2003}. The central star is surrounded by a symmetric circumstellar disk that was formed by decreted material. The disk is being heated by the star, and is the source of emission lines in the observed spectrum. The main object in this research is this disk. The stellar parameters of $\omega$ CMa used in this work are summarized in the Table~\ref{table:stellar_parameter} and were obtained by careful asteroseismological analyses by \cite{maintz2003}.

\begin{table}
\begin{center}
\caption[The stellar parameters of $\omega$ CMa]{The stellar parameters of $\omega$ CMa.}
\begin{tabular}{@{}cccccccccc}
\hline
\hline
Parameter & & & & Value & reference \\
\hline
$L$ & & & & 5224 L$_{\odot}$ & \citealt{maintz2003} \\
$T_\mathrm{pole}$ & & & & 22000 K & \citealt{maintz2003} \\
$R_\mathrm{pole}$ & & & & 6.0 R$_{\odot}$ & \citealt{maintz2003} \\
log g$_\mathrm{pole}$ & & & & 3.84 & \citealt{maintz2003} \\
$M$ & & & & 9.0 M$_{\odot}$ & \citealt{maintz2003} \\
$V_\mathrm{rot}$ & & & & 350 km s$^{-1}$ & \citealt{maintz2003} \\
$V_\mathrm{crit}$ & & & & 436 km s$^{-1}$& \citealt{maintz2003} \\
$R_\mathrm{eq}$ & & & & 7.5 R$_{\odot}$ & \citealt{maintz2003} \\
$i$ & & & & 15$^{\circ}$ & \citealt{maintz2003} \\
distance & & & & 279 $\pm$ 14 pc & \citealt{perryman1997}\\
$m^{*}_\mathrm{v}$ & & & & 4.22 $\pm$ 0.05 & This work \\

\hline
\end{tabular}
\label{table:stellar_parameter}
\end{center}
\end{table}


\section {Photometric data of $\omega$ CMa}
\label{sect:photometry}

\begin{figure}[!t]
\begin{center}
\includegraphics[width=1.0 \columnwidth,angle=0]{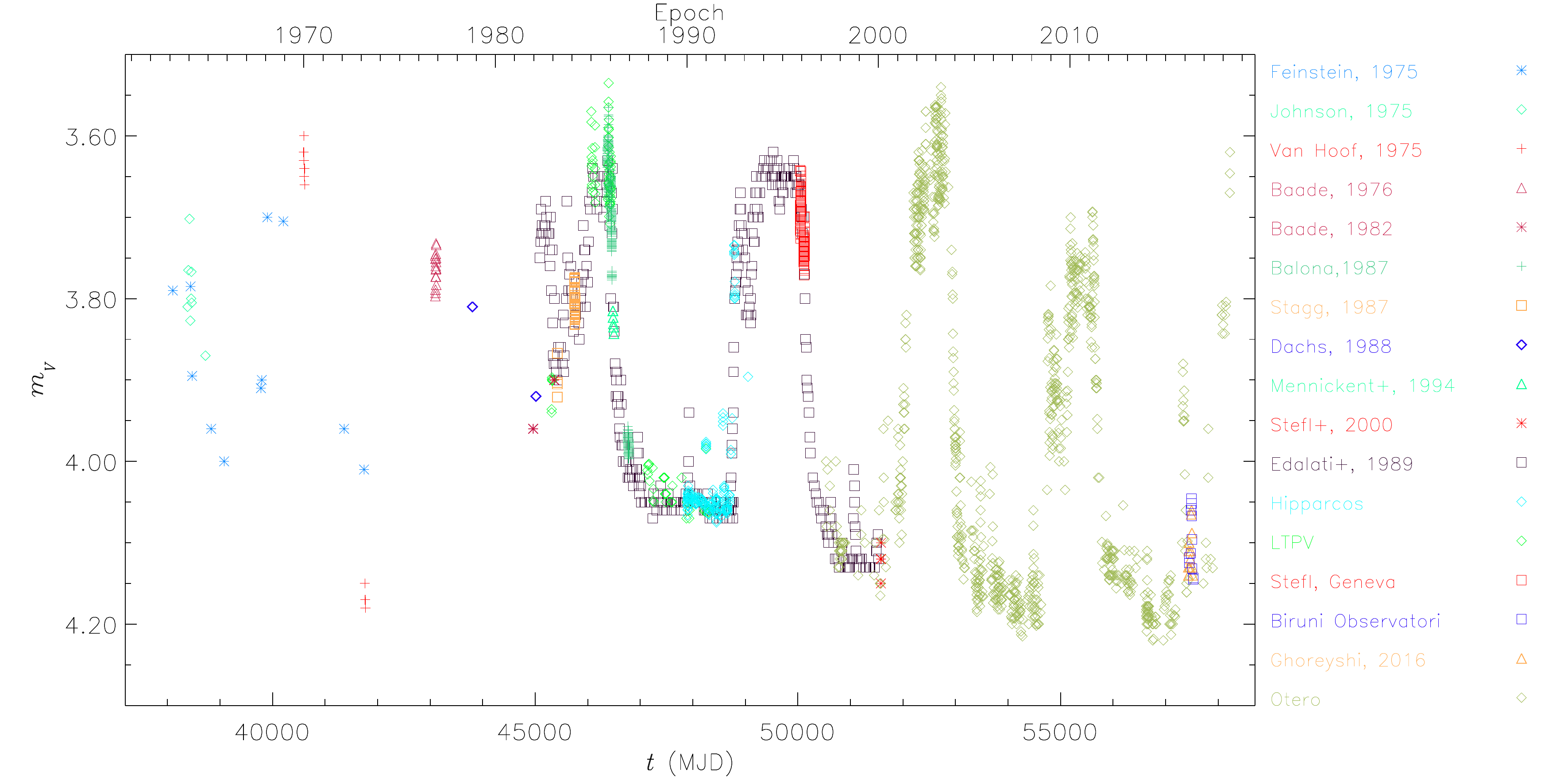}
\caption[Collection of $V$-band data of $\omega$ CMa]{Collection of $V$-band data of $\omega$ CMa with their respective sources indicated. The light curve is a collection of observations from the following sources: \citet{feinstein1975}, \citet{johnson1975}, \citet{vanhoof1975}, \citet{baade1982a}, \citet{balona1987}, \citet{stagg1987}, \citet{dachs1988}, \citet{edalati1989}, photoelectric observations obtained in the Long-Term Photometry of Variables (LTPV, \citealt{manfroid1991}, \citeyear{manfroid1995}; \citealt{sterken1993}), \citet{mennickent1994}, Hipparcos (\citealt{perryman1997}), \citet{stefl2000}, and the visual observations by Otero \citep{stefl2003a}.} 
\label{fig:lightcurve_observer}
\end{center}
\end{figure}

After \cite{stoy1959} reported that $\omega$ CMa is a variable star, it was observed photometrically in a variety of bands, mainly in the $V$-band. We have access to the $V$-band photometric data of $\omega$ CMa which was observed by various observers since 1963, and that is still growing (See Fig.~\ref{fig:lightcurve_observer}). Unfortunately, the data from 1963 to 1982 are very sparse, and do not allow for a sufficient description of the behavior of the star; however, even these sparse data are important because they tell us that the star was active in the sixties and seventies. From now on we focus mostly on the data from 1982 onwards. 

\begin{figure}[!ht]
\begin{center}
\includegraphics[width=0.75 \columnwidth,angle=0]{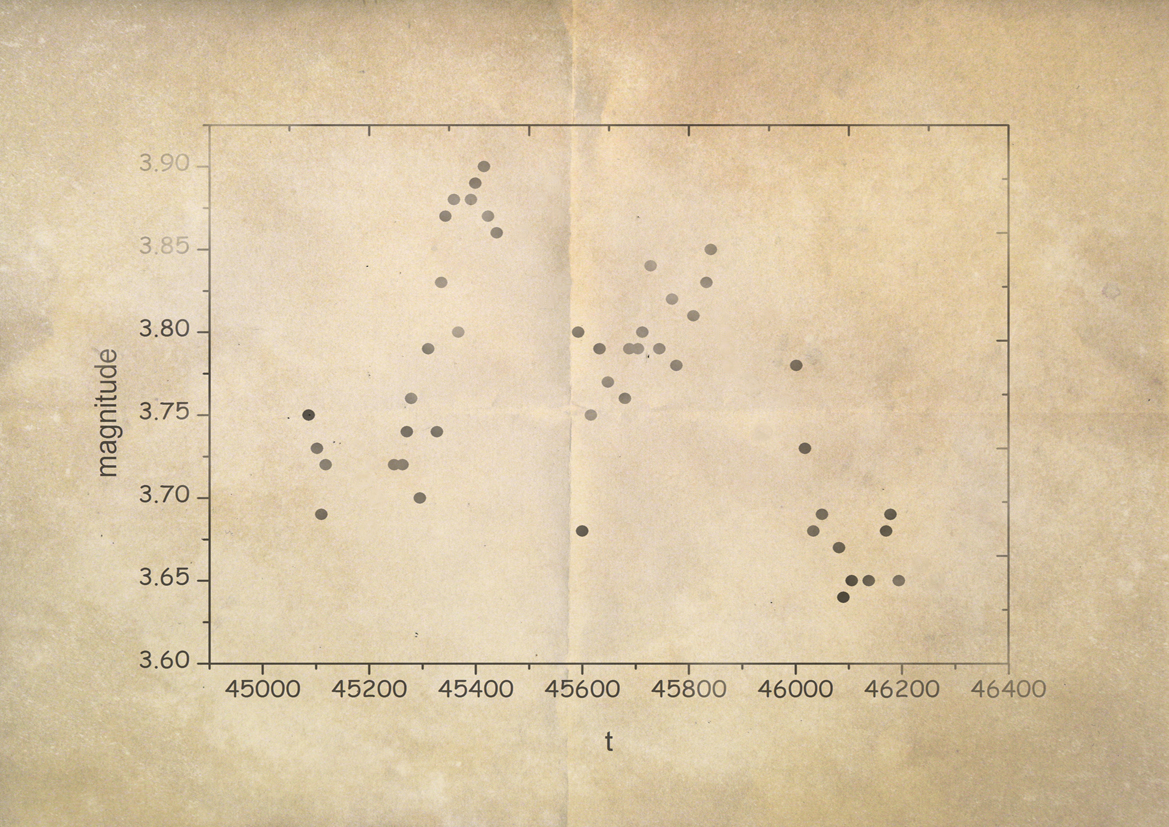}
\caption[$V$-band data of $\omega$ CMa observed by Dr. Edalati.]{One of the figures printed by Dr. Edalati showing the $V$-band data of $\omega$ CMa from 1982 to 1985.} 
\label{fig:edalati}
\end{center}
\end{figure}

\begin{figure}[!ht]
\begin{center}
\includegraphics[width=1.0 \columnwidth,angle=0]{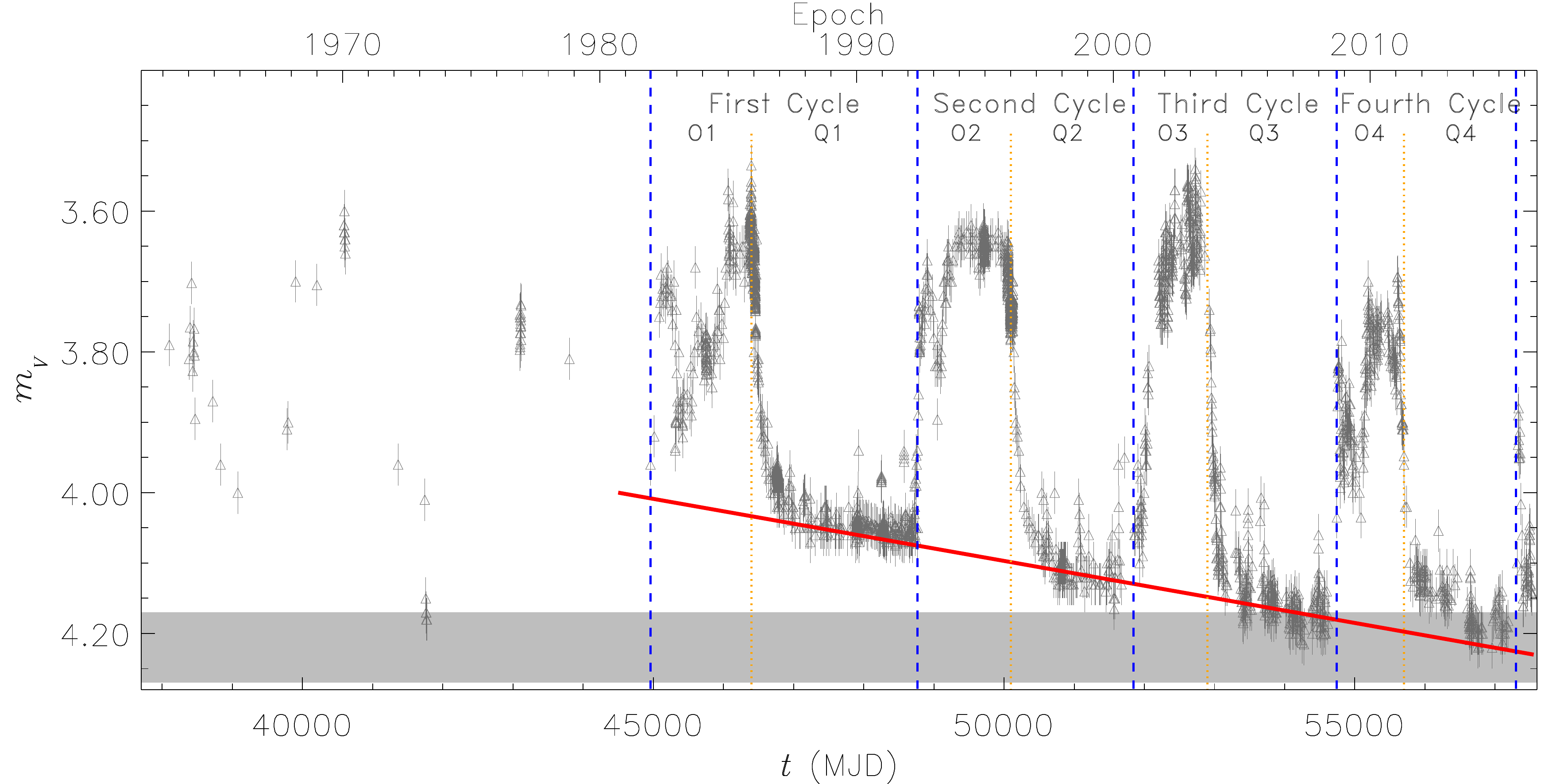}
\caption[Full $V$-band light curve of $\omega$ CMa]{$V$-band light curve of $\omega$ CMa (grey triangles), showing four well monitored cycles, as indicated.  The last 4 outbursts are particularly well monitored. The red line is a linear fit of the last 100\,d of each cycle, indicating that the lowest brightness of each subsequent cycle is fainter than the previous one. O$i$ and Q$i$ stand for outburst and quiescence phases, respectively, where $i$ is the cycle number. The horizontal grey band represents the estimated intrinsic visual magnitude of the central star of $\omega$ CMa (Table \ref{table:Be_stellar_parameter}). The vertical dashed and dotted lines indicate the transitions between quiescence and outburst (and vice-versa).} 
\label{fig:full_lightcurve}
\end{center}
\end{figure}

The most part of the data was observed by Sebasti\'{a}n Otero\footnote{\href{www.aavso.org/sebasti\'{a}n-otero}{www.aavso.org/sebasti\'{a}n-otero}}. The majority of his observed data correspond to visual observations using a modified version of the Argelander method \citep{hirshfeld1985}, carried out from 1997 to the present. To increase the accuracy for estimating the small magnitude variations of $\omega$ CMa, a grid of standard stars (Table 2 in \citealt{stefl2003a}) was used to determine the visual magnitude. Thus, the availability of nearby comparison stars governs the accuracy that is typically better than 0.$^m$05. Another important part of data was observed by Dr. Mohammad Taghi Edalati, a distinguished professor of the Ferdowsi University of Mashhad\footnote{\href{en.um.ac.ir/index.php?module=htmlpages&func=display&pid=10&print=1}{en.um.ac.ir/edalati}}, and his professional and amateur collaborators from Iran, Japan, Australia, Canada and Chile during a survey of observation and investigation of variable stars \citep{edalati1989} which was unfinished because of his death. He was an expert observer who was especifically interested in eclipsing binary stars; therefore, $\omega$ CMa was a strange object for him. No one knows if he could ever realize the true nature of $\omega$ CMa as he did not leave any note about that. I found some of his printed-plotted data under layers of dust in a abondaned box in the Biruni Observatory in Iran. Figure~\ref{fig:edalati} is one of several sheets containing plotted data that I scanned and then digitized. That part of data which will be used in this research is shown in Fig.~\ref{fig:full_lightcurve}, demonstrating that the fortuitous discovery of the Dr. Edalati's data was very important for this project, as it allowed for a nearly continuous coverage of the visual brightness of $\omega$ CMa spanning more than 34 years.

Since 1982, $\omega$ CMa exhibited quasi-regular cycles, each one lasting between 7.0 $\sim$ 10.5 years. Each cycle consists of two main parts: 1) An outburst phase represented by a fast increase in the brightness. This increase is not always smooth, and lasts about 2.5 $\sim$ 4.0 years. 2) A quiescence phase lasting about 4.5 $\sim$ 6.5 that is characterized by a slow (when compared to the outburst phase) decline in brightness. During these phases the brightness of the system in $V$ band changes about 0.$^m$3 $\sim$ 0.$^m$5. Throughout this text we refer to the cycles by C$i$ and to the phases by O$i$ and Q$i$ for outburst and quiescence, respectively, where $i$ is the cycle number. All cycles are marked in Fig.~\ref{fig:full_lightcurve}.

At first glance, the light curve of $\omega$ CMa demonstrates some noticeable features. First of all, one can easily see a decline in the brightness of the system in successive dissipation phases (notice to the red solid line in Fig.~\ref{fig:full_lightcurve}). In other words, there is a 0.$^m$15 difference between the minimum brightness of the first cycle and the fourth one. This secular dimming of the the system (star and disk) has been a long-standing mystery in the literature.

\begin{figure}[!t]
\begin{center}
\includegraphics[width=1.0 \columnwidth,angle=0]{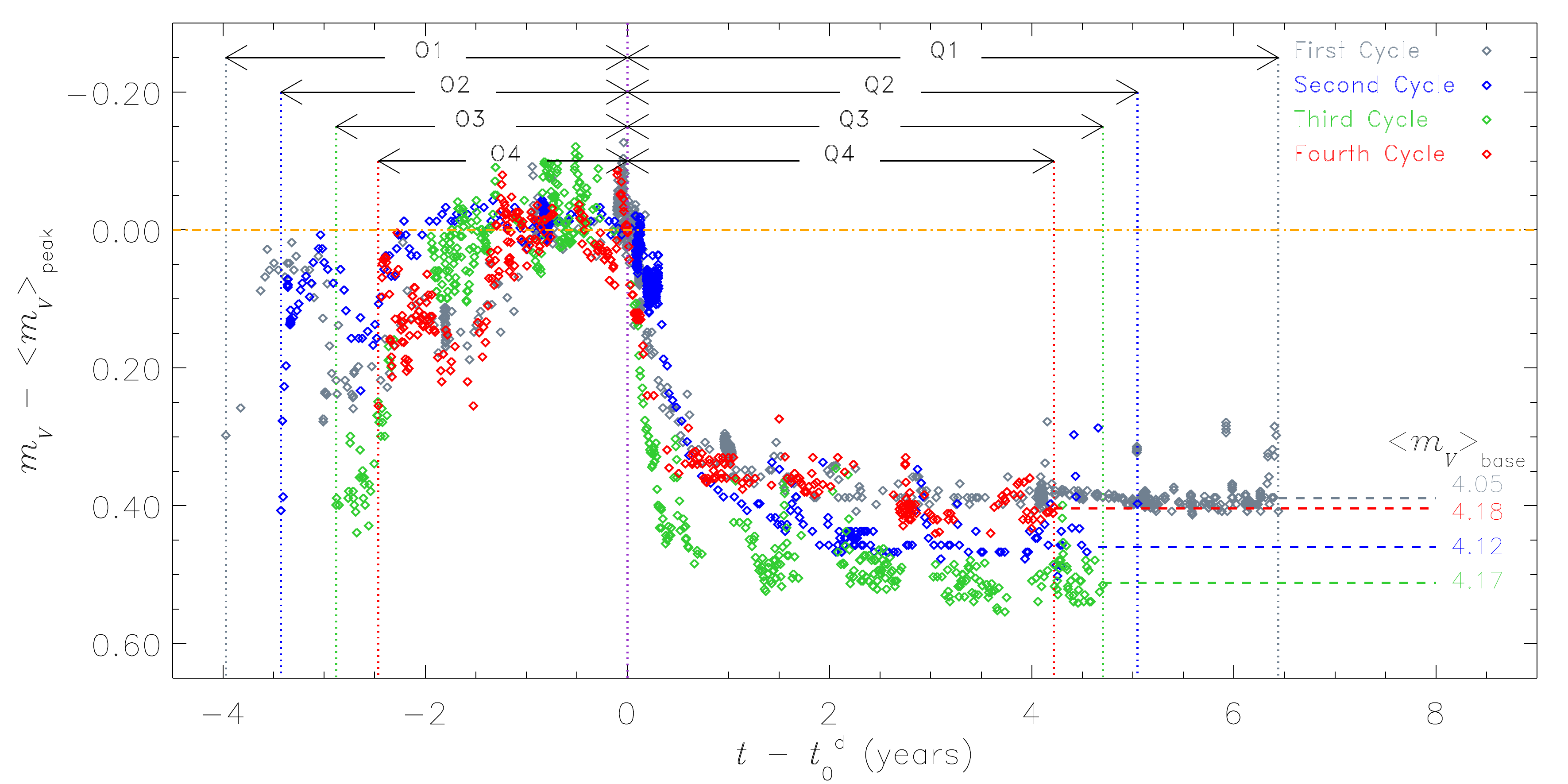}
\caption[Overlaped $V$-band light curve of $\omega$ CMa]{$V$-band photometry of $\omega$ CMa. The data for different cycles are displaced horizontally in time so that the beginnings of all four quiescence phases are roughly aligned (vertical violet dotted line), and vertically so that the average magnitude of the last year of outburst is zero ($\langle m_\mathrm{v}\rangle_\mathrm{peak}$=0). Also, $\langle m_\mathrm{v}\rangle_\mathrm{base}$ is the average magnitude for the last two years of each quiescence phase. Each cycle is shown by an individual color as indicated. The epoch of onset of each quiescence phase ($t_0^{\rm d}$) was determined in Sect.~\ref{chap:exp_formula}, and is listed in Table~\ref{table:results}.} 
\label{fig:overlap}
\end{center}
\end{figure}

Other features of the four well-documented cycles become more apparent when the cycles are superimposed, as in Fig.~\ref{fig:overlap}. This figure was made by aligning all cycles with respect to the onset of the dimming phases Q$i$'s (vertical violet dotted line). Additionally, the data was shifted vertically so that the average magnitude of the last year of outburst is zero. From Fig.~\ref{fig:overlap} some features are visible: 1) The cycle lengths are not equal. Indeed, it seems that during the 34 years the length of the cycles decreased from about 10.5 years for the first cycle to about 7.0 years for the fourth cycle. 2) The rate of variations at Q1 and Q2 are noticably smaller than for Q3 and Q4. 3) It is obvious that the longer the formation phase, the longer the dissipation phase. This observation allowed me to predict (in March, 2015) that a new outburst following Q4 should start in September, 2015. The true outburst started in October, 2015, indicating that the said correlation between the lengths of outburst and quiescence might be valid. 4) The drop in the brightness of the system at the end of each cycle can be seen by noticing the difference between the magnitude level of the tail of Q1 and Q4 at the right side.

\begin{figure}[!ht]
\centering
\includegraphics[width=0.85\linewidth]{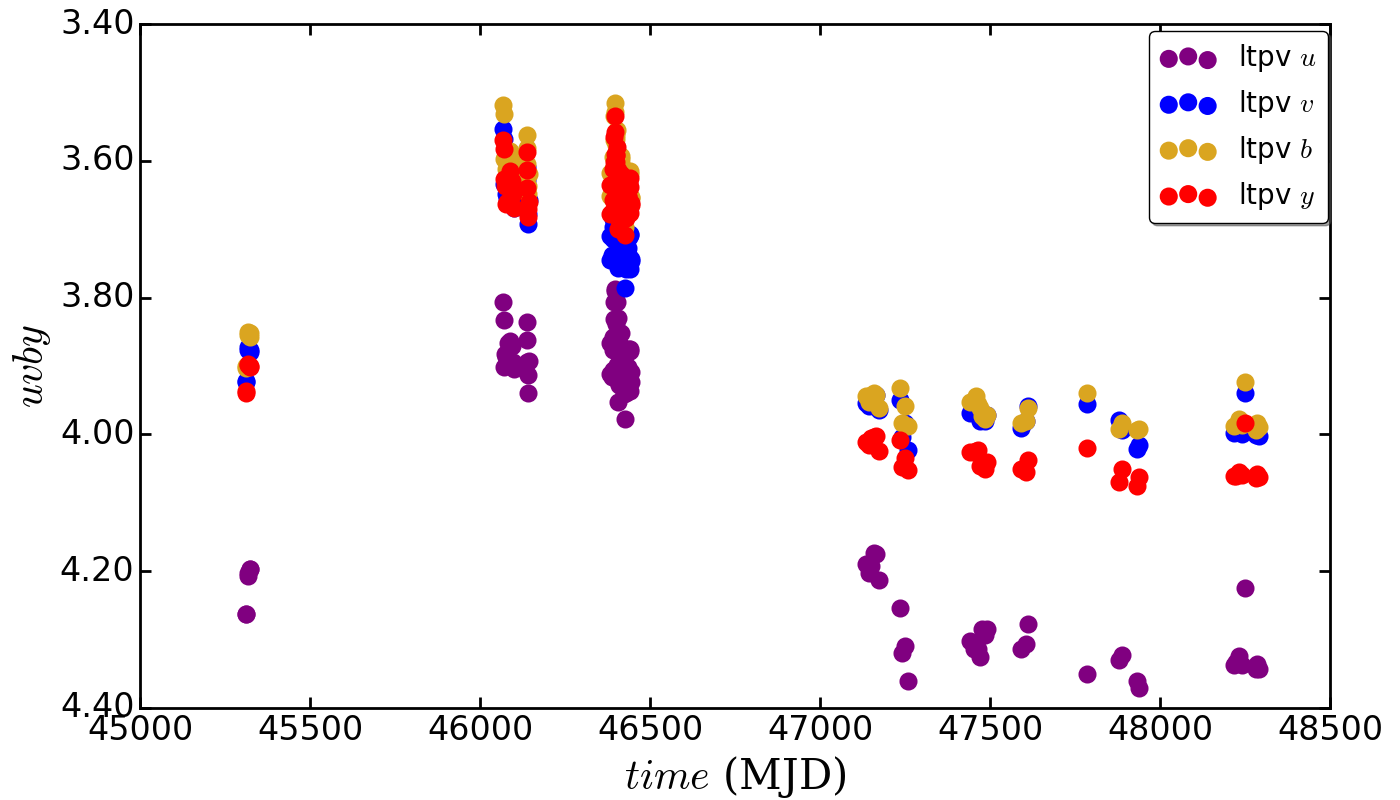}
\caption[$uvby$ band magnitudes of $\omega$ CMa]{$uvby$ band magnitudes of $\omega$ CMa from LTPV.}
\label{fig:uvby}
\end{figure}

Finally, additional photometric data in the $uvby$ Stromgren \citep{stromgren1956, crawford1958} filters were made by \cite{manfroid1995}. This is shown in Fig.~\ref{fig:uvby}.


\section {Other data on $\omega$ CMa}
\label{sect:other_data}

At the end of 2008, when our colleague Sebastian Otero, an Argentinian amateur astronomer member of the American Association of Variable Stars Observers, alerted the start of a new outburst, a broad suite of observations was undertaken. In addition to the dense visual photometry, \textit{JHKL} photometry was obtained with the Mk II photometer of SAAO \citep{glass1973} and CAIN-II Tenerife/TCS camera \citep{cabrera2006}, Q1- and Q3-band measurements were made with VISIR on the VLT/ESO \citep{lagage2004}, and observations at 0.87 mm/345 GHz were secured with LABOCA on APEX \citep{siringo2009}. Optical echelle spectra are available from UVES/VLT \citep{dekker2000} (Oct 2008-Mar 2009), FEROS/La Silla \citep{kaufer1999} and the 1.6m telescope at Observat\'{o}rio Pico dos Dias (OPD/LNA) (Jan 2009 - present) using initially the ECASS spectrograph\footnote{\href{www.lna.br/opd/instrum/cassegr/eficiencia\_cass.html}{www.lna.br/opd/instrum/cassegr/eficiencia\_cass.html}} and more recently, since about 2012, the MUSICOS spectrograph\footnote{\href{www.lna.br/opd/instrum/musicos.html}{www.lna.br/opd/instrum/musicos.html}}. \textit{BVRI} imaging polarimetry was made with the 0.6-m telescope at OPD \citep{magalhaes2006}. Finally, time was obtained with the interferometer AMBER at VLTI/ESO (now decomissioned) in its high-spectral resolution mode for interferometry with three VLTI Auxiliary Telescopes \citep{petrov2007}. 

In addition to the above observational campain for cycle 4, we were able to obtain additional data for some previous cycles from the literature. For the first and second cycles, we obtained spectroscopy from IUE\footnote{\href{archive.stsci.edu/iue/}{archive.stsci.edu/iue/}} and HEROS\footnote{\href{www.lsw.uni-heidelberg.de/projects/instrumentation/Heros/}{www.lsw.uni-heidelberg.de/projects/instrumentation/Heros/}}/FEROS, respectively. For the third cycle, we found spectropolarimetry from FORS \citep{appenzeller1998}, and spectroscopy from BeSS\footnote{\href{basebe.obspm.fr/basebe/}{basebe.obspm.fr/basebe/}}, CES\footnote{\href{www.eso.org/public/teles-instr/lasilla/coude/ces/}{www.eso.org/public/teles-instr/lasilla/coude/ces/}}, FEROS, Lhires spectroscope in Observatoire Paysages du Pilat \footnote{\href{www.parc-naturel-pilat.fr/nos-actions/architecture-urbanisme-paysage/observatoire-du-paysage/
}{www.parc-naturel-pilat.fr/nos-actions/architecture-urbanisme-paysage/observatoire-du-paysage/
}}, and Ondrejov Observatory\footnote{\href{stelweb.asu.cas.cz/web/index.php?pg=2m_telescope}{stelweb.asu.cas.cz/web/index.php?pg=2mtelescope}}. Finally, additional spectroscopic data of the fourth cycle came from BeSS, ESPaDOnS \citep{donati2003}, OPD, PHOENIX \citep{hinkle1998}, Ritter Observatory\footnote{\href{www.utoledo.edu/nsm/rpbo/}{www.utoledo.edu/nsm/rpbo/}}, and UVES.

\begin{figure}[!t]
\begin{minipage}{0.5\linewidth}
\centering
\subfloat[Polarimetry]{\includegraphics[width=1.0\linewidth]{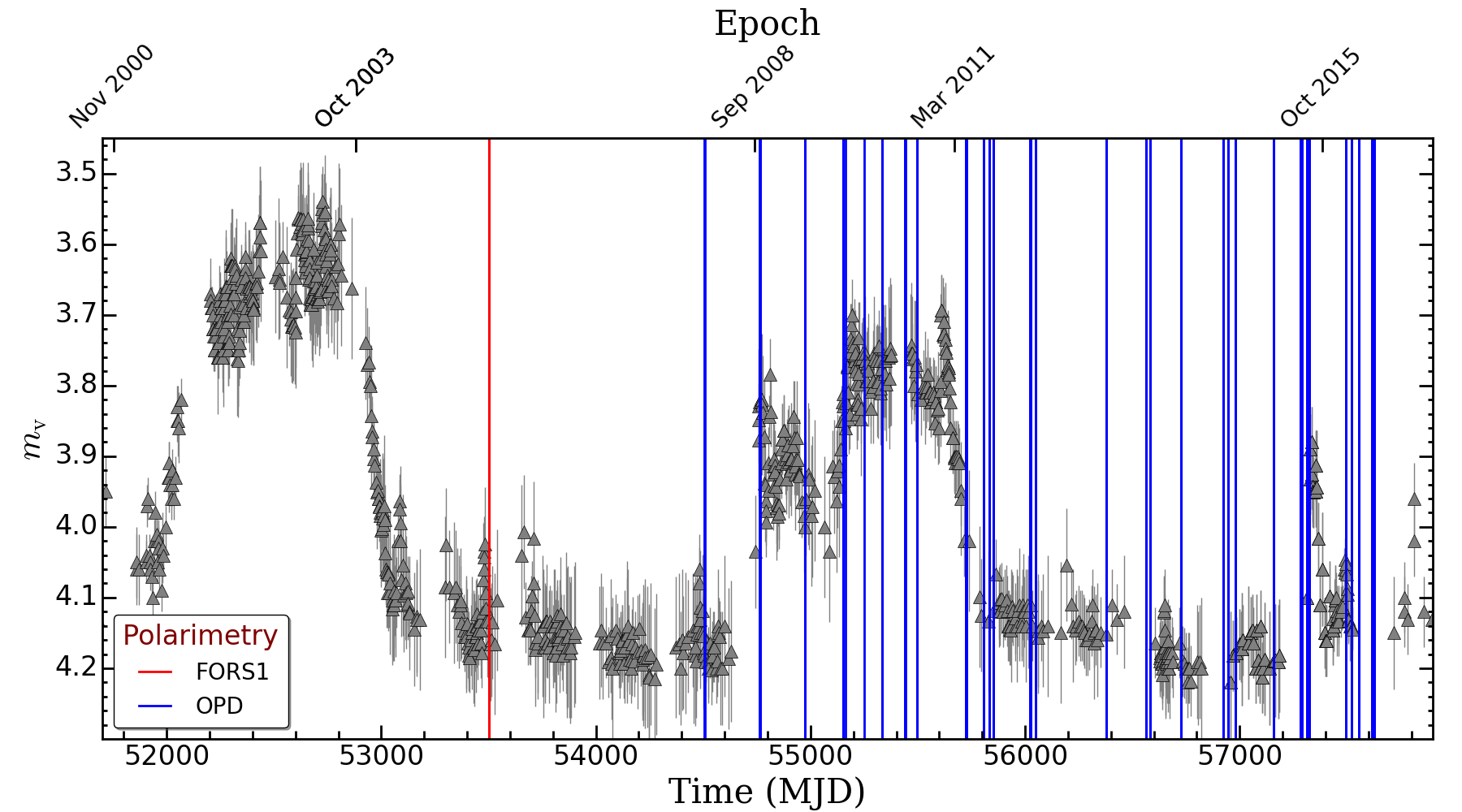}}
\end{minipage}%
\begin{minipage}{0.5\linewidth}
\centering
\subfloat[Spectroscopy]{\includegraphics[width=1.0\linewidth]{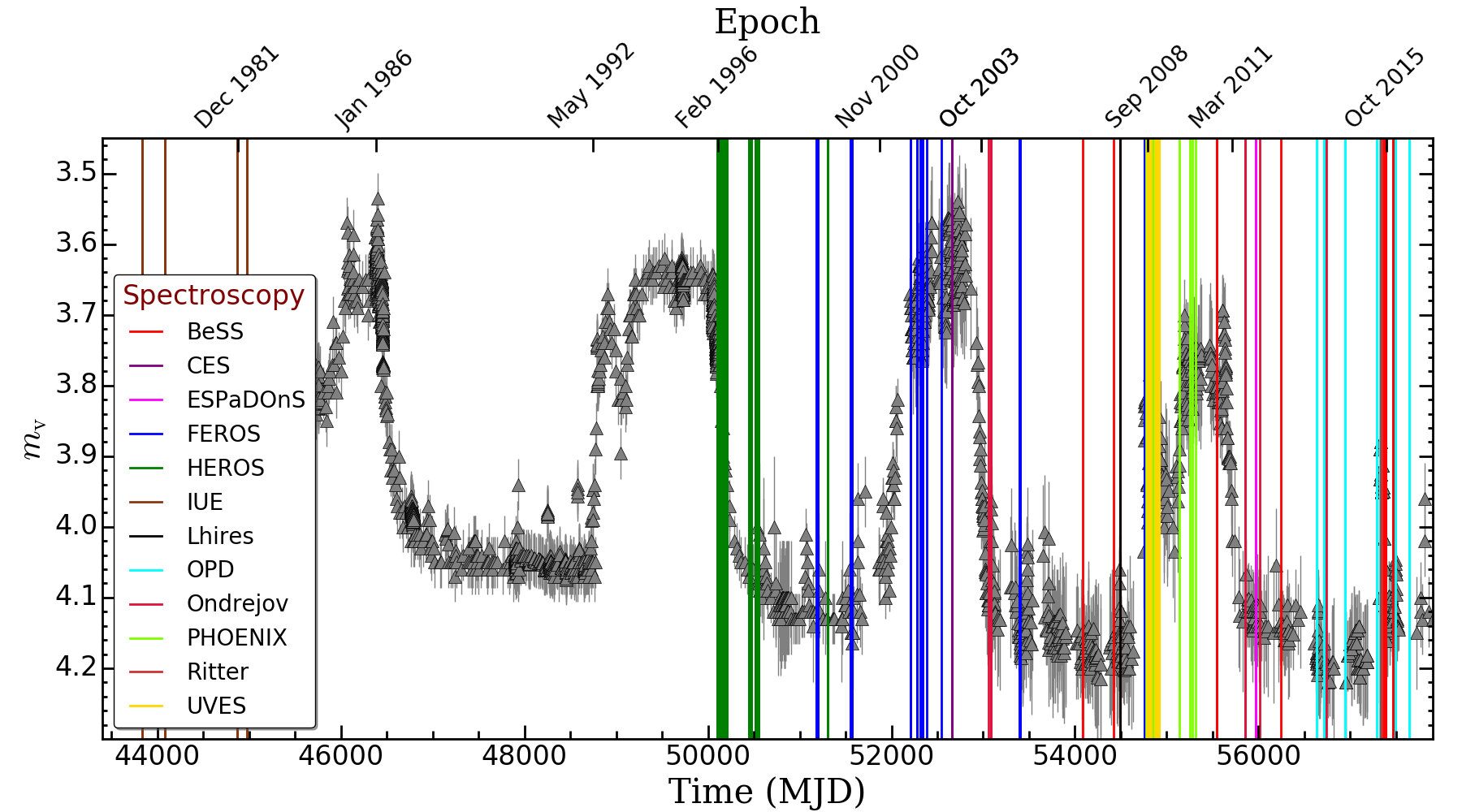}}
\end{minipage}\par\medskip
\begin{minipage}{0.5\linewidth}
\centering
\subfloat[Interferometry]{\includegraphics[width=1.0\linewidth]{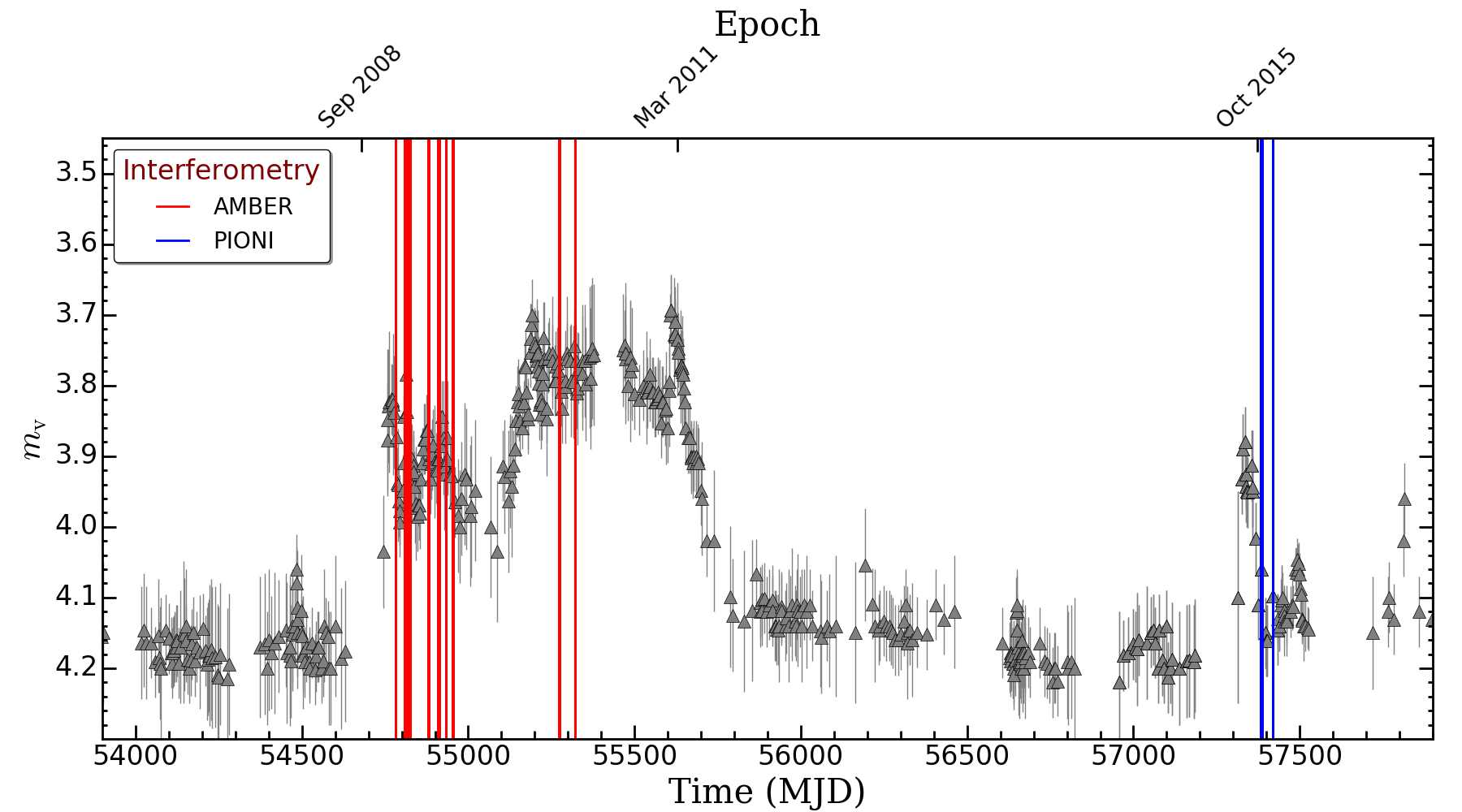}}
\end{minipage}%
\begin{minipage}{0.5\linewidth}
\centering
\subfloat[JHKL and sub-mm photometry by LABOCA/APEX]{\includegraphics[width=1.0\linewidth]{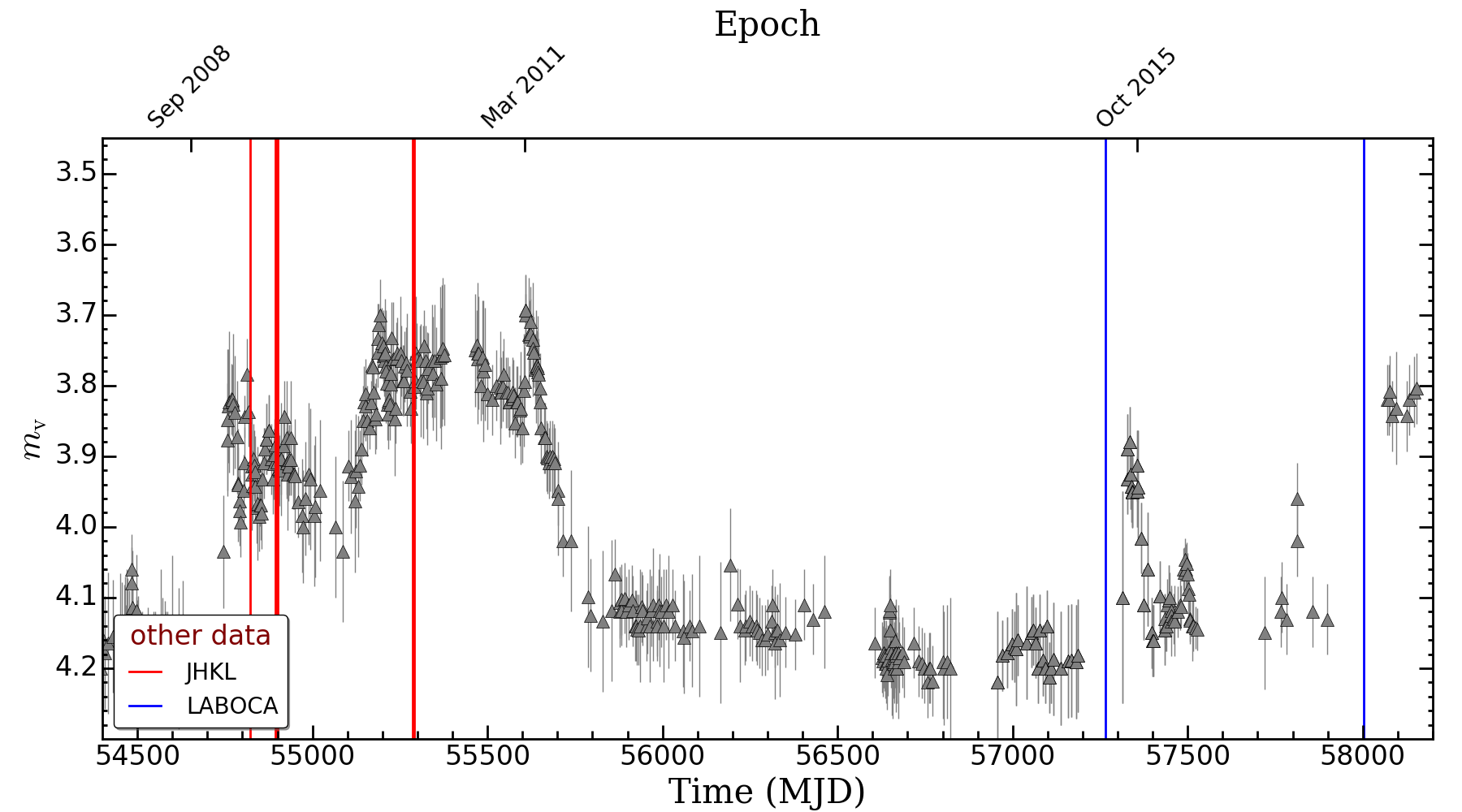}}
\end{minipage}%
\caption[Epochs of all available data of $\omega$ CMa]{Epochs of all available data of $\omega$ CMa in polarimetry (a), spectroscopy (b), interferometry (c), and other techniques (d).}
\label{fig:data_dist}
\end{figure}

All observations other than the optical photometry are listed in Tables~\ref{table:spec_data_log} and~\ref{table:pol_data_log}, and the epochs of all available observations are shown in Fig.~\ref{fig:data_dist}. The rich data set covering the last outburst with several different techniques is one of the most important assets for the current project. 

\begin{figure}[!t]
\begin{minipage}{0.5\linewidth}
\centering
\subfloat[H$\alpha$]{\includegraphics[width=1.0\linewidth]{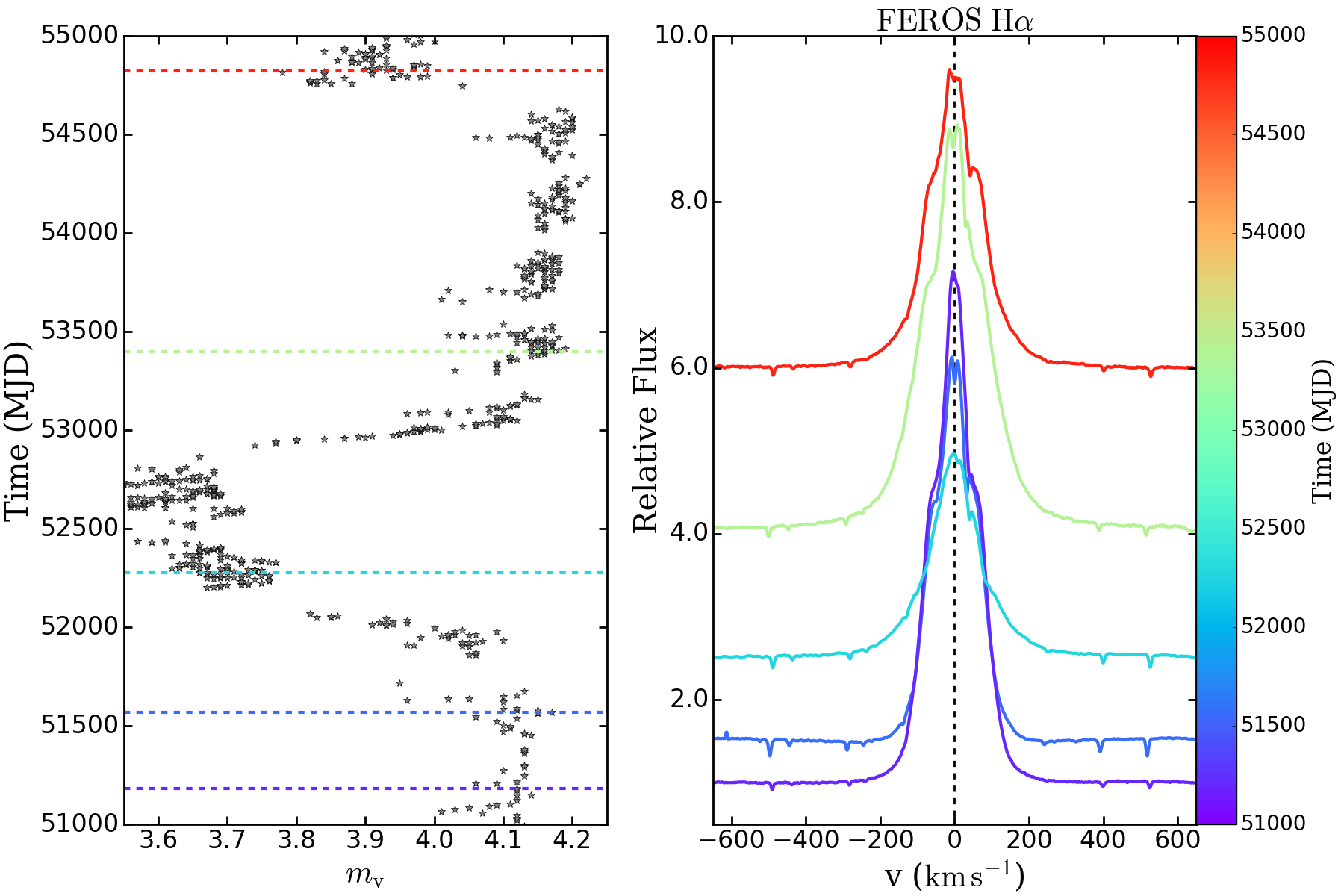}}
\end{minipage}%
\begin{minipage}{0.5\linewidth}
\centering
\subfloat[H$\beta$]{\includegraphics[width=1.0\linewidth]{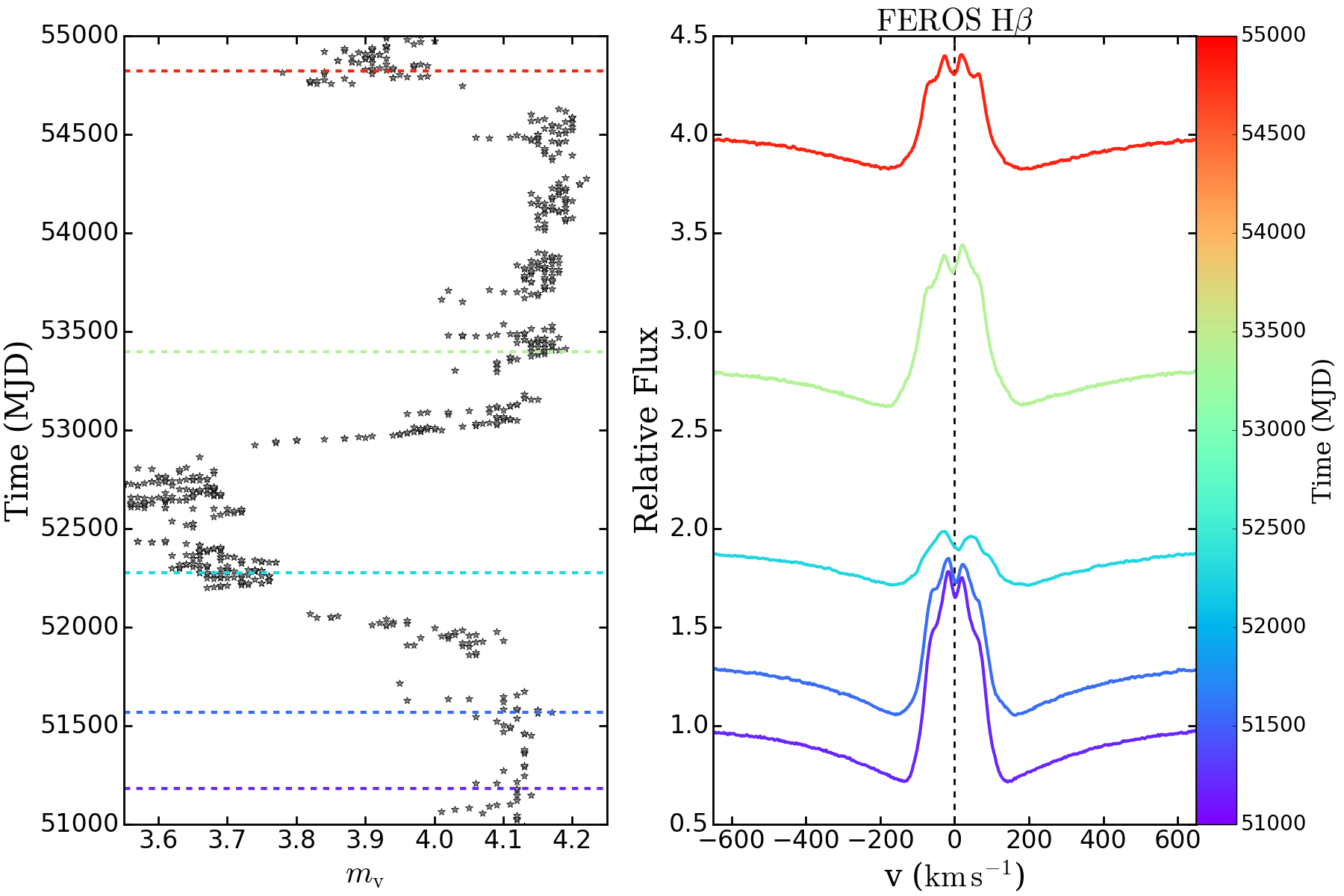}}
\end{minipage}\par\medskip
\begin{minipage}{0.5\linewidth}
\centering
\subfloat[H$\gamma$]{\includegraphics[width=1.0\linewidth]{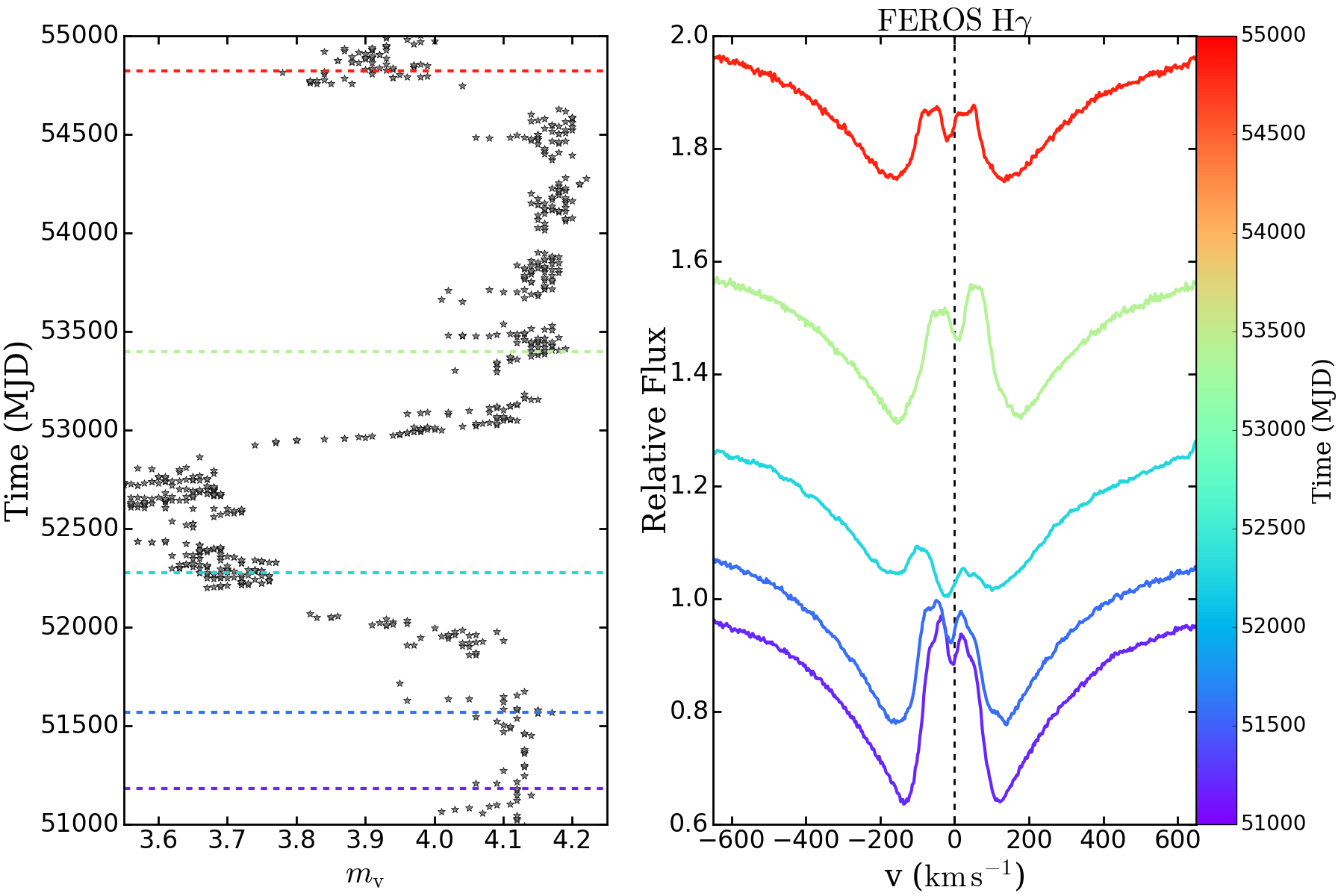}}
\end{minipage}%
\begin{minipage}{0.5\linewidth}
\centering
\subfloat[H$\delta$]{\includegraphics[width=1.0\linewidth]{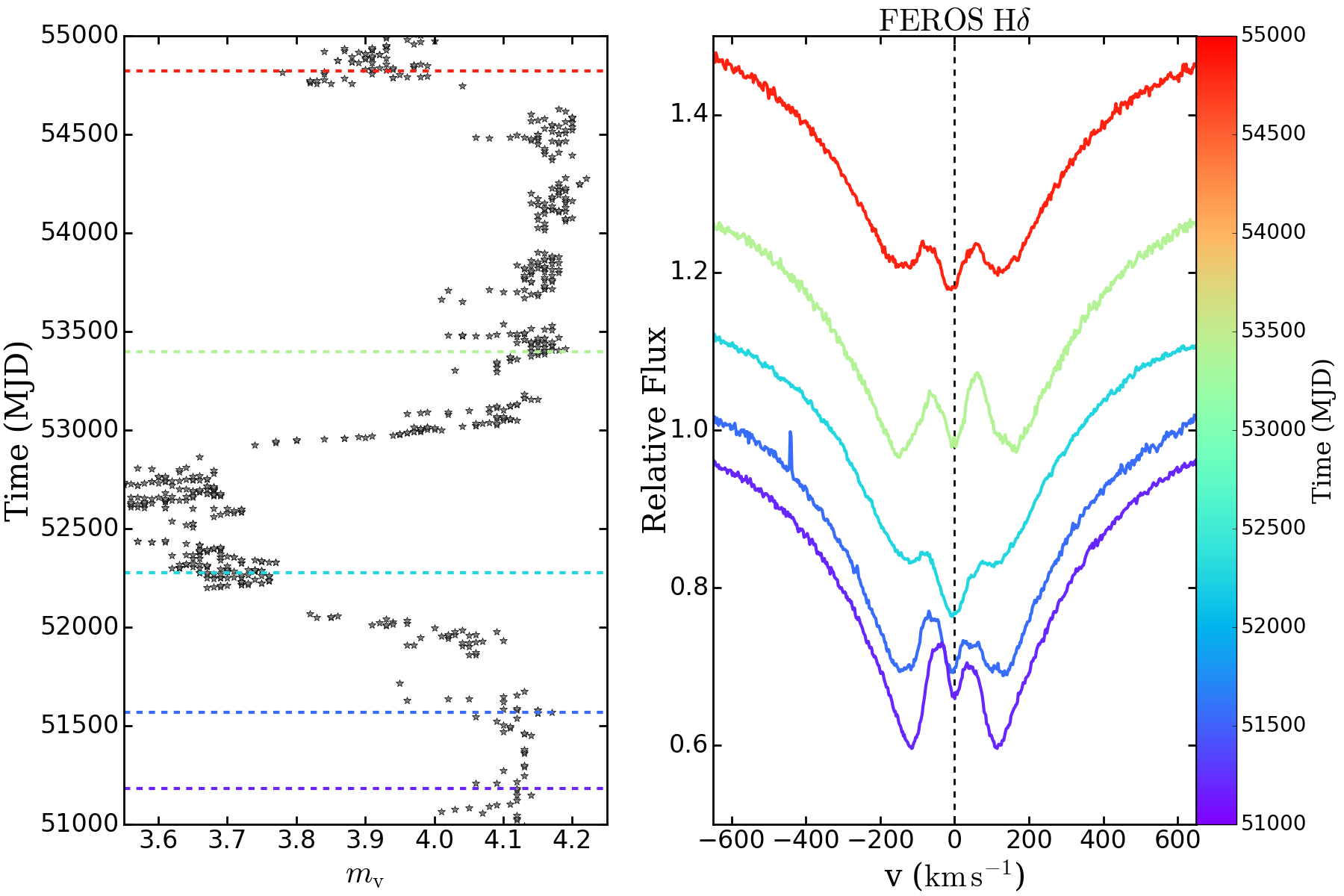}}
\end{minipage}%
\caption[Spectra of $\omega$ CMa observed by FEROS]{Hydrogen lines of $\omega$ CMa observed by FEROS. For all panels, the left plot shows the $V$-band light curve with horizontal lines marking the date the spectrum was taken, and the corresponding spectra (using the same color scheme) are shown in the right plot.}
\label{fig:feros}
\end{figure}

Figure~\ref{fig:feros} stands as an example of observed Hydrogen lines of $\omega$ CMa. Usually the emission over the continuum ratio (E/C) of the H$\alpha$ and H$\beta$ lines is largest at the the end of quiescence, and lower during the outburst. This seemingly contraditory behavior is well explained by the models as will be seen in Chap.~\ref{chap:other_tech}. Also, small peak separations (PS) of the line profiles are expected because of the low inclination angle of $\omega$ CMa \citep[see Fig.~1 in][]{rivinius2013a}

\begin{figure}[!h]
\centering
\includegraphics[height=0.5 \textheight, width=0.5 \columnwidth]{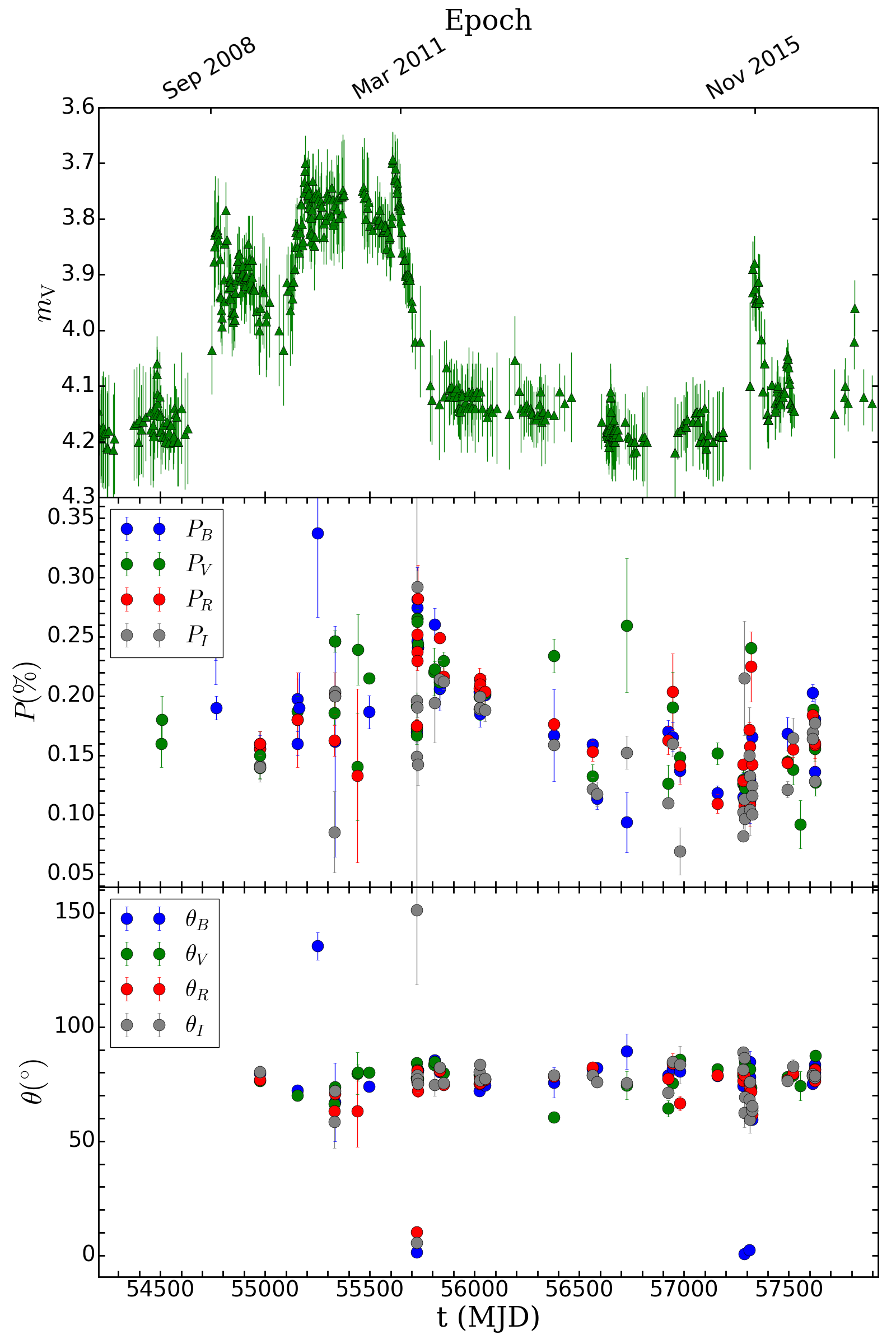}
\caption[Polarization data of the fourth cycle of $\omega$ CMa in $BVRI$ bands]{{\it Top panel}: The $V$-band data of $\omega$ CMa. {\it Middle panel}: Polarization data of the fourth cycle of $\omega$ CMa in $BVRI$ bands. {\it Bottom panel}: The observed position angle of $\omega$ CMa.}
\label{fig:pol_BVRI}
\end{figure}

The group of Prof. Alex Carciofi at IAG/USP has been conducting since 2006 a polarimetric and spectroscopic survey of Be stars at OPD. In the context of this survey, $\omega$ CMa was frequently observed and a good amount of polarimetric data was collected in different {\tt Johnson} filters $BVRI$. Data reduction followed standard procedures outlined by \cite{magalhaes1984, magalhaes1996} and \cite{carciofi2007}. Figure~\ref{fig:pol_BVRI} displays the polarimetry data of the last cycle of $\omega$ CMa in $B$, $V$, $R$, and $I$ filters, respectively, alongside the $V$-band photometric data. The top panel of figure shows the $V$-band light curve of the star and the third panel of figure shows the polarization angle, $\theta$, measured east from celestial north. The polarization level is very small, as expected by the fact that $\omega$ CMa is observed at rather small inclination angles (see Table~\ref{table:stellar_parameter}). This also indicates that the interstellar polarization is likely also very small. Both combined factors makes analysis of the polarization data rather uncertain, as will be discussed in Chap.~\ref{chap:other_tech}.

\begin{figure}[!h]
\centering
\includegraphics[width=0.85\linewidth]{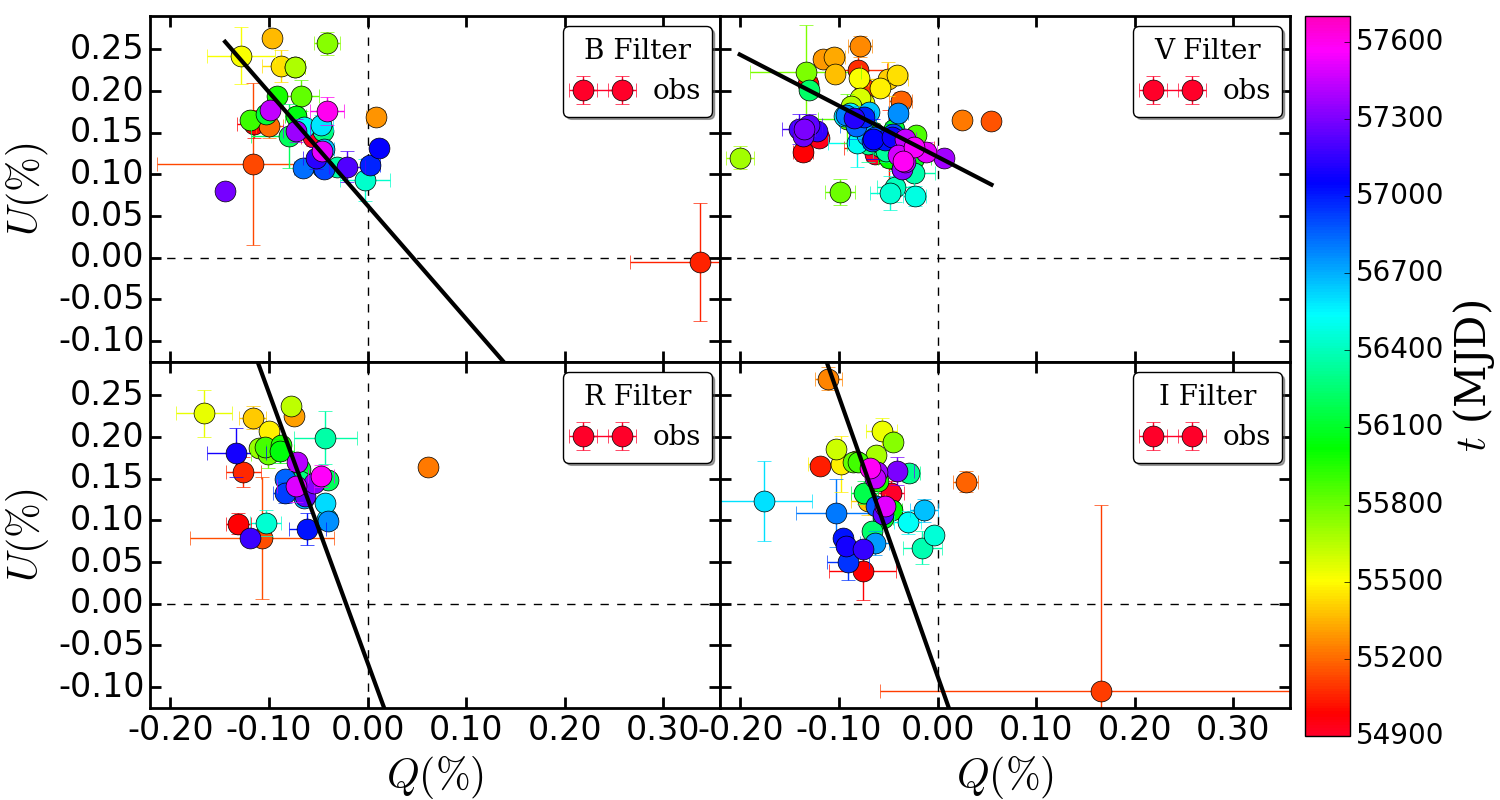}
\caption[$Q-U$ diagram of Polarization data of the fourth cycle of $\omega$ CMa]{$Q-U$ diagram of Polarization data of the fourth cycle of $\omega$ CMa. Different colors indicate the date of observation, according to the legend on the right. The black solid lines are simple linear fits on the data.}
\label{fig:pol_qu}
\end{figure}

Figure~\ref{fig:pol_qu} shows the $Q-U$ diagram of the polarization data of the fourth cycle of $\omega$ CMa. Even though there is lots of scatter in the data, for all four bands the measurements form a straight path in the $Q-U$ diagram. A simple linear fit (solid black lines in the figure) indicates that the angle of this path is 126.4$^\circ$, 148.6$^\circ$, 107.2$^\circ$, and 106.6$^\circ$ for $B$, $V$, $R$, and $I$, respectively. \cite{draper2014} and Bednarksi (2016) indicate that events of disk formation/dissipation of Be stars follow straight lines in the $Q-U$ diagram, which means that the behavior seen for $\omega$ CMa is expected. Furthermore, these authors show that the angle of this path, $\psi$, divided by 2 gives information about orientation of the Be disk in the plane of the sky. More specifically, $\psi/$2 should be parallel to the minor elongation axis of the Be disk. Therefore, the $Q-U$ diagram of Fig.~\ref{fig:pol_qu} gives us the precious information that the orientation of the minimum elongation of the disk of $\omega$ CMa is about 61$^\circ$, East from celestial North. The values derived here were compared with the results of \cite{bednarski2016} in Table.~\ref{table:psi_bednarski} and found to be in general good agreement. The polarization data will be further discussed in Chap.~\ref{chap:other_tech}.

\begin{table}
\begin{center}
\caption[Orientation of the $\omega$ CMa's disk in the plane of the sky]{Orientation of the $\omega$ CMa's disk in the plane of the sky obtained in this work and that of \cite{bednarski2016}.}
\begin{tabular}{@{}cccccc}
\hline
\hline
& $B$ Filter & $V$ Filter & $R$ Filter & $I$ Filter & average \\
\hline
$\psi$($^\circ$)/2 this work & 63.2 & 74.3 & 53.6 & 53.3 & 61.1 \\
$\psi$($^\circ$)/2 \cite{bednarski2016} & 55.5 & 57.7 & 61.5 & 51.0 & 56.4 \\
\hline
\end{tabular}
\label{table:psi_bednarski}
\end{center}
\end{table}


\chapter{Theoretical Considerations}
\label{chap:theories}

As described in Chap.~\ref{chap:intro}, the main goal of this thesis is to employ the VDD model to study the temporal variations seen in $\omega$ CMa's observations. The goal is, on the one hand, to validate the VDD in conditions hitherto not explored in the literature (e.g., to model the disk construction phase), but also to extract from the modeling what are the key parameters governing the temporal evolution of the disk.

The VDD is, therefore, the main theoretical tool employed in this thesis. In this chapter, the main properties of this model are outlined, and the main methods and approximations used to obtain the solutions needed are described. It should be noted that, here, only a brief description of the VDD will be given, to introduce what are the main model assumptions and the main physical quantities. We employ, in this thesis, exactly the same formalism as developed by Leandro R. R\'{i}mulo in his PhD thesis \citep{rimulo2017}. The interested reader is specifically referred to Chap.~2 of his work for more details on the physical implementations of the VDD used in this work. The reader is also referred to the original work by \cite{lee1991}, that laid the foundation of the VDD model, as well as the works of \cite{porter1999, bjorkman2001, okazaki2001, bjorkman2005, krticka2011} that further developed the VDD theory.


\section {Model description}
\label{sect:model_description}

This PhD thesis studies a system composed by a fast-rotating star that irradiates a circumstellar disk. The central star is located at the origin of a cylindrical coordinate system whose vertical axis is parallel to the rotational axis of the star ($z$ direction). Furthermore, the disk is assumed to lie in the equatorial plane of the star. The rotational velocities of the star and the disk are vectors in the azimuthal direction, $\phi$. Also, the disk has radial velocity component ($v_r$) that can be negative (i.e., inflow) or positive (i.e., outflow). 


\subsection {The central star}
\label{subsect:central_star}

As a Be star, $\omega$ CMa is a fast-rotating star, with equariorial radius $R_{\rm eq} = 7.5\,{\rm R}_{\odot}$, polar radius $R_{\rm pole} = 6.0\,{\rm R}_{\odot}$, mass $M = 9.0\,{\rm M}_{\odot}$, and luminosity $L = 5224\,{\rm L}_{\odot}$ (see Table~\ref{table:Be_stellar_parameter}). With a rotational velocity of $v_{\rm rot} = 350\,{\rm km\,s}^{-1}$, the star is rotating at about 80\% of the its critical velocity. 

Because of such a fast-rotational velocity, the star is oblate (notice the difference between $R_{\rm eq}$ and $R_{\rm pole}$). Consequently, the polar regions of the star have a higher effective gravity than the equatorial regions. According to the \cite{vonzeipel1924} theorem, this latitudinal dependence of the effective gravity causes a latitudinal dependence of the flux, in the sense that the poles become brighter (thus hotter) and the equator darker (thus cooler). This so-called gravity darkening effect plays a key role in determining of the surface distribution of the flux, and therefore the latitudinal dependence of the temperature.

In his original formalism, applicable for a purely radiative envelope, the von Zeipel theorem can be written as 
\begin{equation}
    T_{\rm eff}(\theta) \propto g_{\rm eff}^{\beta}(\theta)\,,
    \label{eq:vonzeipel}
\end{equation}
with $\beta = 0.25$. More recent theoretical works \citep{espinosa2011} showed that the value of $\beta$ is a function of the rotational rate of the star. Following these authors, the assumed $\beta$ for $\omega$ CMa was 0.19.


\subsection {The disk}
\label{subsect:disk}

The disk in a Be star system plays a very important role in its observational appearance. Its geometrical shape, density, and inclination to the line of sight cause variations in the observables of the star. According to the VDD model, the disk is fed by the matter ejected from the star. After being injected into the disk at a given point, the matter spreads both inwards and outwards as a result of the viscous diffusion of material. The mechanism that brings matter into, forming the disk, is not known yet. Even though pulsations have gained, recently, a strong support from studies space-based photometry \citep[see, e.g.,][and references therein]{baade2016}. Also, the inner and outer boundary conditions of the disk depends very much on the model considered. In the following the most important concepts and quantities regarding to the Be disk in the paradigm of the VDD model are described.

Relevant disk quantities are the mass flux, $\dot{M}_\mathrm{disk} (r, t)$, and the AM flux, $\dot{J}_\mathrm{disk} (r, t)$. Both quantities will describe the flow of matter and AM in the disk in response to varying inner (e.g., a variable mass loss rate from the star) and outer (e.g., a disk that is truncated by a binary companion) conditions. 
A common assumption made both in VDDs of Be stars and in disks around Young Stellar Objects \citep[e.g.,][]{pringle1981} is that the azimuthal component of the gas velocity, $v_{\phi}$, is Keplerian, i.e., $v_{\phi} \propto r^{0.5}$. As described in details by \cite{rimulo2017} and many other earlier works, under this assumption the mass flux is related to the disk surface density, $\Sigma(r, t)$, by the mass conservation relation (see, e.g., Eqs.~3.1.1 to~3.1.3 of \citealt{rimulo2017})
\begin{equation}
    \label{eq:mdot_disk}
    \dot{M}_\mathrm{disk}(r,t) = 2\pi r \Sigma(r, t) v_r = -4\pi \left(\frac{r^3}{GM_*}\right)^\frac{1}{2}\frac{1}{r}\frac{\partial}{\partial r}\left(\alpha c_s^2 r^2\Sigma(r, t)\right),
\end{equation}
where $v_r$ is the radial speed, $G$ is the universal gravitational constant, $\alpha$ is the viscosity parameter of \cite{shakura1973}, $c_\mathrm{s}$ is the isothermal sound speed, given by $c_s^2=kT_\mathrm{disk}/\mu m_\mathrm{H}$, and $r$ is the distance from the star. The temperature profile of the disk, $T_{\rm disk}$, is well described by a flat blackbody reprocessing disk at the regions very close to the stellar photosphere, as it drops very quickly \citep{adams1987}. The temperature profile follows this curve as long as the disk is vertically optically thick, from which point the temperature rised to a constant temperature of about 60\% of $T_{\rm eff}$ \citep{carciofi2006a}. This value for the $T_{\rm disk}$ is used in this work. The surface density, $\Sigma(r, t)$, is defined as the vertical integral of the mass density, $\rho$, such that
\begin{equation}
    \label{eq:surf_dens}
    \Sigma(r, t) = \int_{-\infty}^{+\infty}\rho(r, t, z)\mathrm{d}z\,.
\end{equation}
Note that in the above expression, an axial symmetry is assumed (i.e., the disk quantities do not depend of $\phi$).

The turbulent viscosity is the source of AM and mass transport in a Keplerian disk \citep{pringle1981}. Originally, \cite{shakura1973} proposed a dimensionless parameter, known as the Shakura--Sunyaev viscosity parameter, $\alpha$ to describe the kinematic viscosity. It is the role of $\alpha$ parameter to link the scale of the turbulence to the (vertical) scale of the disk. The turbulence is composed of eddies (vortices) and $l$ is the size scale of the largest eddies, while $v_\mathrm{tur}$ is the ``turnover'' velocity of the eddies. 
We set $l = H$ since the largest eddies can be at most about the size of the disk scale height. Since the velocity is not known, it is reasonable to assume that it is of the order or smaller than the sound speed ($c_s$). Otherwise, the turbulence would be supersonic and the eddies would fragment into a series of shocks. Again, since the actual velocity is not known, Shakura \& Sunyaev introduced a parameter, $\alpha$, and assumed the viscosity, $\nu$, to be given by $\nu=v_\mathrm{tur}l=\alpha c_s H$. This is the prescription for the viscosity that is widely used in the literature.

The AM flux, according to the AM conservation relation, is given by (see, e.g., Eqs.~3.2.17 to~3.2.24 of \citealt{rimulo2017})
\begin{equation}
\label{eq:jdot_disk}
\dot{J}_\mathrm{disk}(r,t) = 2\pi r \Sigma(r, t) v_r \left(GM_*r\right)^\frac{1}{2} + 2\pi \alpha c_s^2 r^2\Sigma(r, t)\,,
\end{equation}
where the first term is the AM flux that is carried with the radial motion of the gas and the second term is the AM flux due to the torque generated by the viscous force.

The temporal evolution of the surface density is given by the following diffusion-like equation, which holds in the thin disk approximation ($c_\mathrm{s}^2\ll GM_*/r$):

\begin{equation}
\label{eq:sigmadot}
\frac{\partial{\Sigma}}{\partial{t}}=\frac{2}{\bar{r}} \frac{\partial}{\partial{\bar{r}}} \left\{ \bar{r}^\frac{1}{2}\frac{\partial{}}{\partial{\bar{r}}}\left[\frac{\alpha c_\mathrm{s}^2}{(GM_*R_\mathrm{eq})^\frac{1}{2}} \bar{r}^2 \Sigma(\bar{r}, t)\right]\right\}+S_\Sigma,
\end{equation}
where $\bar{r}$ is the normalized radius ($r/R_\mathrm{eq}$) and $S_\Sigma$ represents the mass injection rate per unit area from the star into the disk. 

Following \cite{rimulo2017}, it was assumed that the star injects mass in a Keplerian orbit at a given rate, $\dot{M}_\mathrm{inj}$, at a radius very close to the surface of the star, $R_\mathrm{inj}$ = 1.02 $R_\mathrm{eq}$, at the equatorial plane (see below). This assumption may be too simplistic for some Be stars, because there is observational evidence pointing to asymmetric mass loss (e.g., the short-term V/R variations of $\eta$ Cen, \citealt{rivinius1997}) or matter being ejected at higher latitudes (e.g., \citealt{stefl2003a}). However, this is not an issue for this work, since after a few orbital periods, the matter loses memory of the injection process owing to viscous diffusion and orbital phase mixing. Hence, $S_\Sigma$ is assumed to be given by 
\begin{equation}
\label{eq:s_sigma}
S_\Sigma=\frac{\dot{M}_\mathrm{inj}}{2\pi R_\mathrm{eq}^2}\frac{\delta(\bar{r}-\bar{r}_\mathrm{inj})}{\bar{r}},
\end{equation}
where $\bar{r}_\mathrm{inj}=R_\mathrm{inj}/R_\mathrm{eq}$.

Finally, we assume torque-free boundary conditions at the stellar surface ($r=R_\mathrm{eq}$) and at a very distant outer radius ($r=R_\mathrm{out}$). As the second term in Eq.~(\ref{eq:jdot_disk}) shows, this is accomplished by setting $\Sigma=0$ at those boundaries. The outer boundary, in particular, could represent the limiting radius of the disk due to a binary companion (e.g., \citealt{okazaki2002}) or due to the photoevaporation of the disk (e.g., \citealt{okazaki2001, krticka2011}).

The conservation of AM through the system composed by the star, the disk and the outside medium can be written as follows
\begin{equation}
\label{eq:jdot_star}
\dot{J}_*(t)+\frac{\mathrm{d}\phantom{t}}{\mathrm{d}t}\int_{R_\mathrm{eq}}^{R_\mathrm{out}}(GM_*r)^\frac{1}{2}\Sigma(r,t) 2\pi r\mathrm{d}r+\dot{J}_\mathrm{disk}(R_\mathrm{out},t)=0\,,
\end{equation}
where the first, second and third terms are the variation rates of the AM in the star, the disk and the outside medium, respectively. The second and third terms are obtained using the solution $\Sigma(r,t)$ of Eq.~(\ref{eq:sigmadot}), which allows us to obtain the AM lost by the star (the first term).

In a steady-state mass feeding fashion\footnote{It is very important to distinguish between ``steady-state disk'' and ``steady mass or AM injection rate''. The former refers to a disk whose properties do not vary (or vary very little) with time. The second term refers to a constant mass and AM injection rate.}, AM is injected at a constant rate $(GM_* R_\mathrm{inj})^\frac{1}{2}\dot{M}_\mathrm{inj}$ at the radius $R_\mathrm{inj}$. This AM is divided into a constant AM flux inwards, for $r<R_\mathrm{inj}$, and a constant AM flux outwards, for $r>R_\mathrm{inj}$. Since in steady-state the AM in the disk (second term of Eq.~\ref{eq:jdot_star}) is constant in time, the constant flux outwards in the region $r>R_\mathrm{inj}$ is the AM loss rate of the star. It is given by \cite{rimulo2018}
\begin{equation}
\label{eq:jdot_steady}
-\dot{J}_{*,\mathrm{std}}=\Lambda \left(GM_*R_\mathrm{eq}\right)^\frac{1}{2}\dot{M}_\mathrm{inj}\left(\bar{r}_\mathrm{inj}^\frac{1}{2}-1\right)\,,
\end{equation}
where $\Lambda=1/(1-\bar{r}_\mathrm{out}^{-\frac{1}{2}})$, with $\bar{r}_\mathrm{out}=R_\mathrm{out}/R_\mathrm{eq}$, is a number usually just slightly larger than 1, as $R_\mathrm{out}\gg R_\mathrm{eq}$. 
The steady-state AM loss rate, therefore, depends very little (through the factor $\Lambda$) on the outer radius of the disk, whose value is poorly known for a few Be stars, and completely unknown for most \citep{klement2017}. In the simulations presented in this work, we have set $\bar{r}_\mathrm{out}=1000$.

The steady-state surface density for radii larger than $R_\mathrm{inj}$ is given by 
\begin{equation}
\label{eq:sigma_steady}
\Sigma_\mathrm{std}(r)=\frac{-\dot{J}_{*,\mathrm{std}}}{2\pi\alpha c_s^2}\frac{1}{r^2}\left(1-\frac{r^\frac{1}{2}}{R_\mathrm{out}^\frac{1}{2}}\right)\,,\,R_\mathrm{inj}\leq r \leq R_\mathrm{out}\,,
\end{equation}
from which we see that the disk density, a physical quantity that can be estimated, for instance, from SED modeling, scales with $-\dot{J}_{*,\mathrm{std}}/\alpha$, or, alternatively, with $\dot{M}_{\rm inj}$/$\alpha$. This outlines a very important property to VDDs: the injection rates can only be known provided $\alpha$ is known. This will be a key aspect of the work presented in Chap.~\ref{chap:photometry}. Note that the $\Sigma_\mathrm{std}$ and $\dot{J}_{*,\mathrm{std}}$ are just another way of expressing the rate of mass and AM injection into the disk and their index ``std'' does not have the meaning of steady-state disk but is only to emphasize that these parameters are defined in the steady-state limit. From now on, the mass and AM injection rate will be expressed in terms of $\Sigma_0$, $\dot{J}_{*,\mathrm{std}}$, or $\dot{M}_{\rm inj}$, interchangably, as these three quantities are connected by Eqs.~\ref{eq:jdot_steady} and~\ref{eq:sigma_steady}.

\begin{figure}[!t]
\begin{minipage}{0.5\linewidth}
\centering
\subfloat[middle of formation]{\includegraphics[width=1.0\linewidth]{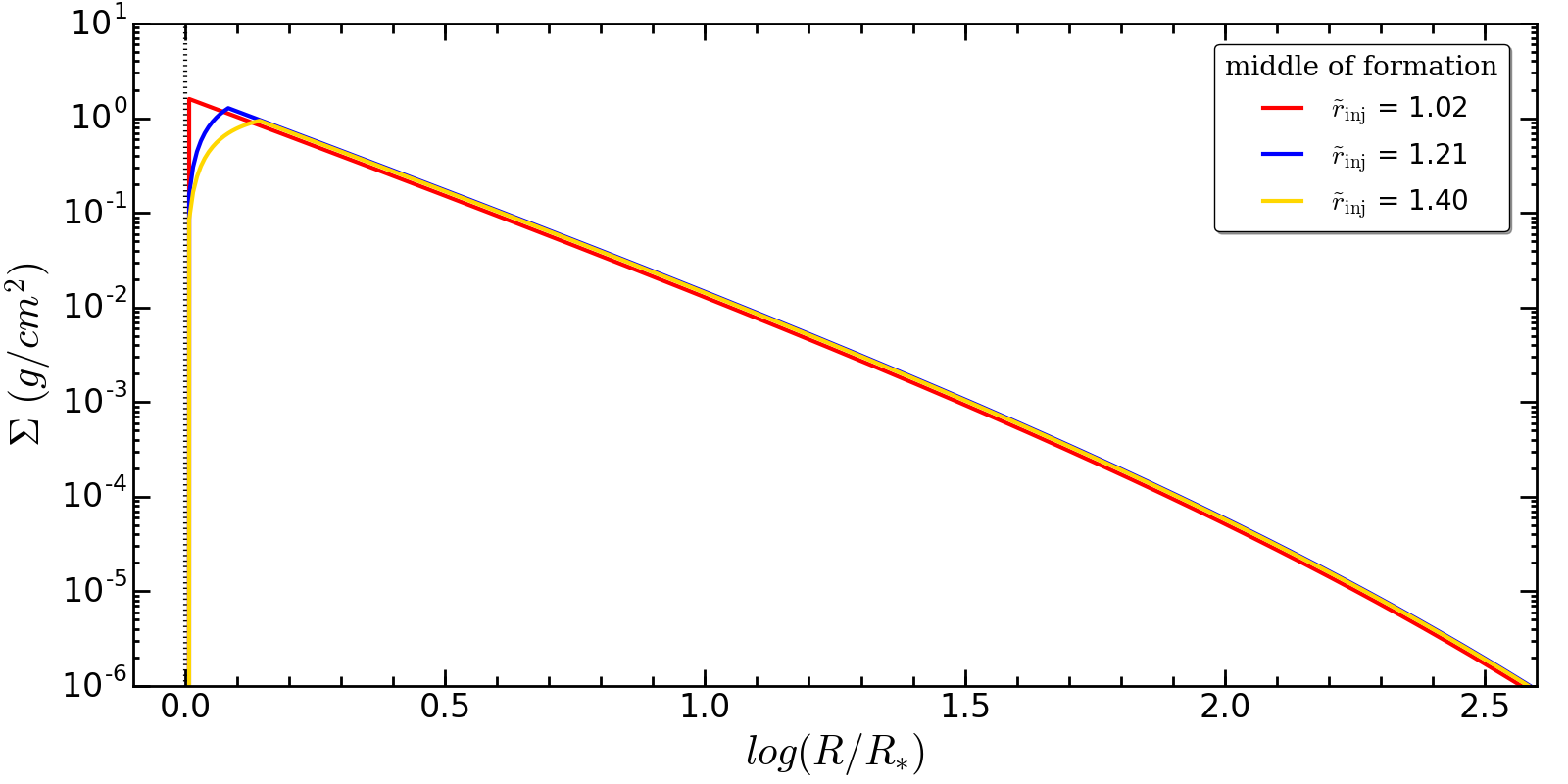}}
\end{minipage}%
\begin{minipage}{0.5\linewidth}
\centering
\subfloat[end of formation]{\includegraphics[width=1.0\linewidth]{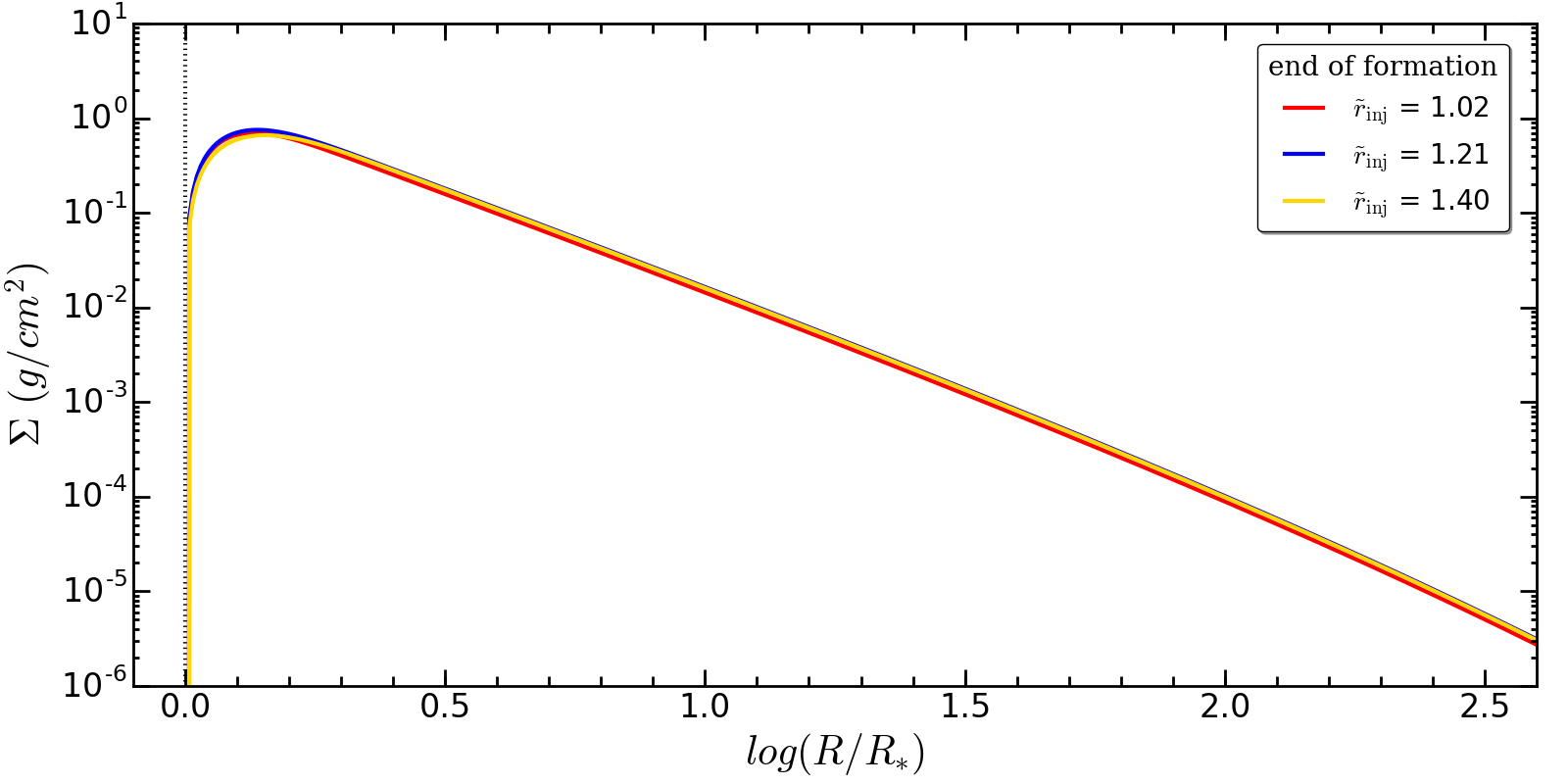}}
\end{minipage}\par\medskip
\begin{minipage}{0.5\linewidth}
\centering
\subfloat[middle of dissipation]{\includegraphics[width=1.0\linewidth]{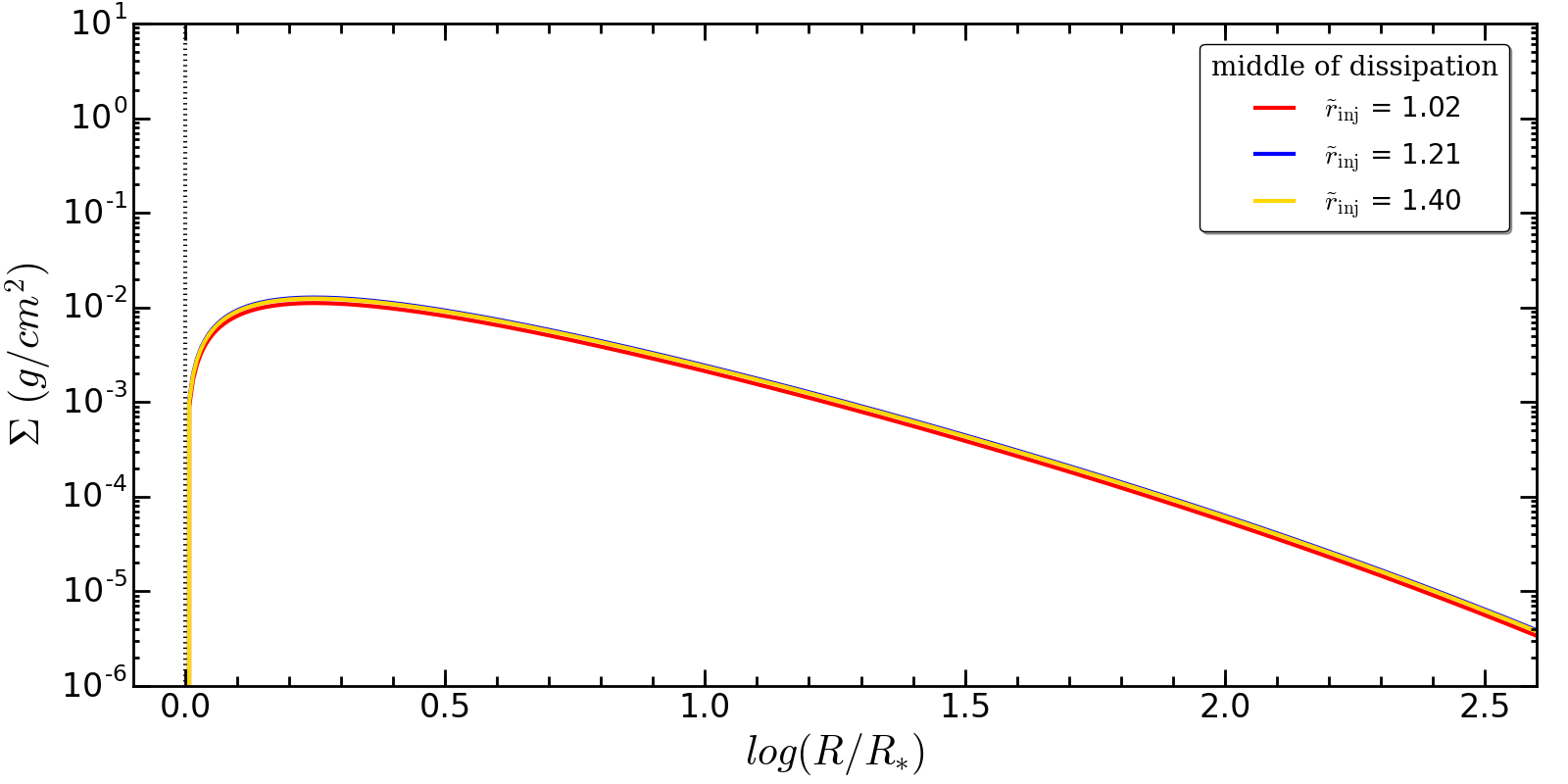}}
\end{minipage}%
\begin{minipage}{0.5\linewidth}
\centering
\subfloat[end of dissipation]{\includegraphics[width=1.0\linewidth]{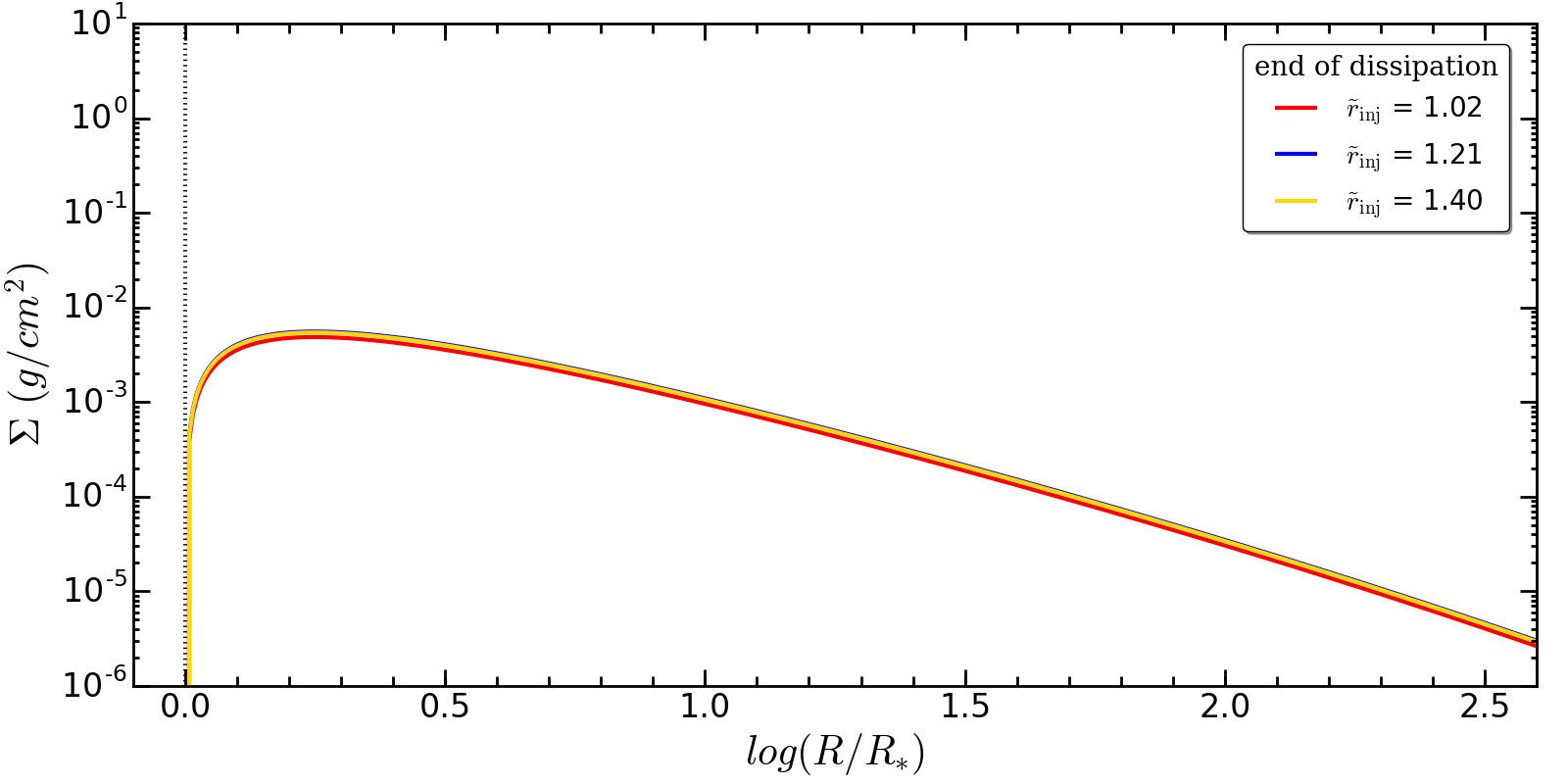}}
\end{minipage}%
\caption[Effect of mass injection point on the surface density profile.]{Surface density as a function of distance from the star for three different injection points, as indicated, at four different epochs: (a) at the middle of disk formation phase, (b) at the end of disk formation phase, (c) at the middle of disk dissipation phase, and (d) at the end of disk dissipation phase.}
\label{fig:injection_point}
\end{figure}

The dynamical evolution of the disk surface density (Eq.~\ref{eq:sigmadot}) is driven by the time-variable source term of Eq.~(\ref{eq:s_sigma}). It has been verified \citep{rimulo2018} that the solution of Eq.~(\ref{eq:sigmadot}) for $r > \bar{r}_\mathrm{inj}$ is negligibly affected by variations of both $\dot{M}_\mathrm{inj}(t)$ and $\bar{r}_\mathrm{inj}$ as long as the product $\dot{M}_\mathrm{inj}(t)(\bar{r}_\mathrm{inj}^\frac{1}{2}-1)$ is kept unchanged. Figure~\ref{fig:injection_point} illustrates this point. There are three curves in this figure: a red, a blue and a gold. They show the solution $\Sigma(r,t)$ at a certain time $t$ (in different stages of disk formation and dissipation) for the case of a disk that is being built-up from a diskless state. The three curves have different values of $\dot{M}_\mathrm{inj}$ and $\tilde{r}_\mathrm{inj}$, but the product $\dot{M}_\mathrm{inj}(\tilde{r}_\mathrm{inj}^{\frac{1}{2}}-1)$ is the same for all of them. Figure~\ref{fig:injection_point}, therefore, shows that $\Sigma(r,t)$ only differs significantly in the small interval $R_\mathrm{eq} < r < R_\mathrm{inj}$ that contributes negligibly to the total emission of the disk as long as $\tilde{r}_\mathrm{inj}$ is close to 1, which is a reasonable assumption; and this small effect is relevant only when the disk is being formed. Since the radius $\bar{r}_\mathrm{inj}$ is a quite unknown quantity, it follows that the mass injection into the disk  is best described by the quantity defined in Eq.~(\ref{eq:jdot_steady}), which roughly represents the net AM injected into the disk. 

At outburst, $-\dot{J}_{*,\mathrm{std}}$ is non-zero, which results in an outflowing ($v_{r}>0$, $\dot{M}_\mathrm{disk} > 0$) decretion disk. During this phase, the disk is slowly built inside-out \citep{haubois2012}. Quiescence is usually understood as a phase where $-\dot{J}_{*,\mathrm{std}}=0$ (however, see Chap.~\ref{chap:photometry}). At this phase, the inner disk passively diffuses inward ($v_{r}<0$, $\dot{M}_\mathrm{disk} < 0$) while the outer disk remains outflowing (see Fig. 10 of \citealt{haubois2012}). 

From Eq.~(\ref{eq:sigmadot}), the rate at which the surface density changes depends not only on the viscosity parameter $\alpha$, but also on the stellar radius (which sets the size of the region over which diffusion happens), and the isothermal sound speed, which was estimated from the approximate relation $T_\mathrm{disk}\approx 0.6T_\mathrm{pole}$ \citep{carciofi2006a}. As $R_\mathrm{eq}$ and $T_\mathrm{pole}$ are well-known for $\omega$ CMa (see Table \ref{table:stellar_parameter}), it follows that the observed changes in the light curve, being the result of variations of the disk density, are controlled only by $-\dot{J}_{*,\mathrm{std}}$ and $\alpha$. This realization forms the basis of the present analysis: $-\dot{J}_{*,\mathrm{std}}$ and $\alpha$ are allowed to vary both in time and magnitude to reproduce the observed light curve. A byproduct of the hydrodynamic simulations is $\dot{M}_\mathrm{disk}$ as a function of time and distance to the star, from which we can determine the amount of mass and AM effectively lost through the disk.


Therefore, another way of writing Eq.~(\ref{eq:jdot_star}), in which $-\dot{J}_{*}$ is the rate of AM being lost by the star, is as follows:
\begin{equation}
-\dot{J}_{*}  = \dot{M}_\mathrm{eq}(GMR_\mathrm{eq})^{\frac{1}{2}} + \dot{M}_\mathrm{inj}(GMR_\mathrm{inj})^{\frac{1}{2}}\,,
\label{eq:jdot_star2} 
\end{equation}
which means that AM appears at the rate $\dot{M}_\mathrm{inj}(GMR_\mathrm{inj})^{\frac{1}{2}}$ in the ring with radius $R_\mathrm{inj}$ and a large fraction of this injected AM disappears (returns to the star) at the rate $\dot{M}_\mathrm{eq}(GMR_\mathrm{eq})^{\frac{1}{2}}$ in the stellar equator. The remaining fraction of the injected AM will be the instantaneous AM loss rate from the star into the disk, $-\dot{J}_{*}$. Simulations, furthermore, show that $-\dot{J}_{*}$ is close to the steady-state AM loss rate $-\dot{J}_\mathrm{*,std}$, given by Eq.~(\ref{eq:jdot_steady}), during most of the time. The physical interpretation of Eq.~\ref{eq:jdot_star2} is as follows: in order that some of the material injected at $R_{\rm inj}$ be driven outwards, thereby gaining wider orbits, some AM must be given to this material; the source of the AM is precisely the material that falls back to the star.

It became therefore clear that, in describing the matter injection from the star into the disk, we should not use the two parameters $M_\mathrm{inj}$ and $\tilde{r}_\mathrm{inj}$, which we cannot determine observationally, but some function proportional to the product $\dot{M}_\mathrm{inj}(\tilde{r}_\mathrm{inj}^{\frac{1}{2}}-1)$, which will generate a certain uniquely defined $\Sigma(r,t)$ profile that can be determined observationally. In the work of \cite{rimulo2018}, the parameter used was $\Sigma_0$ (Eq.~11), which as dubbed ``asymptotic surface density'', and can be calculated from Eq.~\ref{eq:sigma_steady} by making $\bar{r} = 1$. In our work, we chose to parametrize the mass injection rate using $-\dot{J}_\mathrm{*,std}$ (The reason for the different definitions in the two works is that, in the work of \cite{rimulo2018}, the simulations were scaled in time by a parameter that is inversely proportional to the viscosity parameter $\alpha$, in order to generate a grid of synthetic light curves that is independent of $\alpha$. In our work, we used the time $t$ in the true units). The realization of the fact that we should use some function proportional to $\dot{M}_\mathrm{inj}(\tilde{r}_\mathrm{inj}^{\frac{1}{2}}-1)$ came from observing the form of the steady-state solution of Eq.~(\ref{eq:sigmadot}). To find the steady-state solution, we have to set $\partial{\Sigma}/\partial{t} = 0$ in Eq.~(\ref{eq:sigmadot}). It then becomes an ordinary differential equation for $\Sigma_\mathrm{std}(r)$ that is easy to solve analytically. We find that, in steady-state, a fraction $\Upsilon$ of the injected mass flows back to the stellar equator and the remaining fraction, $1-\Upsilon$, flows towards the outer boundary, leaving the system. The fraction is given by 
\begin{equation}
\Upsilon =\frac{\tilde{r}_\mathrm{out}^{\frac{1}{2}}-\tilde{r}_\mathrm{inj}^{\frac{1}{2}}}{\tilde{r}_\mathrm{out}^{\frac{1}{2}}-1}\,,
\label{eq:upsilon} 
\end{equation}
from which we see that, since $\tilde{r}_{\mathrm{inj}}$ is very close to 1, $\Upsilon$ is a number just a little smaller than 1 for every Be disk: almost all the injected mass returns to the star. In steady-state, the rate of mass being absorbed by the stellar equator is $\dot{M}_\mathrm{eq} = -\Upsilon\dot{M}_\mathrm{inj}$ and the rate of mass flowing out of the disk through the outer boundary is $\dot{M}_\mathrm{out} = -(1-\Upsilon)\dot{M}_\mathrm{inj}$. We will see below that the dependence of the steady-state AM loss rate with this quite uncertain parameter is very small. The dependence of the mass loss rate with this parameter, on the other hand, is not small, which then justifies our choice of describing the mass injection by $-\dot{J}_\mathrm{*,std}$ parameter.

Substitution of $\dot{M}_\mathrm{eq} = -\Upsilon\dot{M}_\mathrm{inj}$ into Eq.~(\ref{eq:jdot_star2}) and simple manipulations show that, in steady-state, the AM loss rate from the star is given by
\begin{equation}
    -\dot{J}_\mathrm{*,std} = \Lambda(GMR_\mathrm{eq})^{\frac{1}{2}}\dot{M}_\mathrm{inj}(\tilde{r}_\mathrm{inj}^{\frac{1}{2}}-1)\,,
    \label{eq:jdot_std} 
\end{equation}
that is the same as Eq.~(\ref{eq:jdot_steady}), which is a function proportional to $\dot{M}_\mathrm{inj}(\tilde{r}_\mathrm{inj}^{\frac{1}{2}}-1)$.

With Eq.~(\ref{eq:jdot_std}), we could rewrite the steady-state solution $\Sigma_\mathrm{std}(r)$ as we did in Eq.~(\ref{eq:sigma_steady}). This equation shows that, although we may vary $\dot{M}_\mathrm{inj}$ and $\tilde{r}_\mathrm{inj}$, the function $\Sigma_\mathrm{std}(r)$ would not change as long as $-\dot{J}_\mathrm{*,std}$ was kept constant. 

Now, Eq.~(\ref{eq:jdot_std}) shows that the steady-state AM loss rate has a weak dependence with the outer radius $\tilde{r}_\mathrm{out}$, through the number $\Lambda$. For comparison with that equation, let us write the steady-state mass loss rate from the star,  $-\dot{M}_\mathrm{*,std}$. It is just given by $-\dot{M}_\mathrm{*,std} = -\dot{M}_\mathrm{out} = (1-\Upsilon)\dot{M}_\mathrm{inj}$, or
\begin{equation}
-\dot{M}_\mathrm{*,std} = \frac{\Lambda}{\tilde{r}_\mathrm{out}}\dot{M}_\mathrm{inj}(\tilde{r}_\mathrm{inj}^{\frac{1}{2}}-1)\,,
\label{eq:mdot_std} 
\end{equation}
from which we see that its dependency on $\tilde{r}_\mathrm{out}$ is much greater than the dependency of $-\dot{J}_\mathrm{*,std}$ on the same parameter.

The outer radius  $\tilde{r}_\mathrm{out}$ cannot be determined observationally with visible or IR data, because it is very far away from the regions of the disk where these observables are generated. For single stars, we may assume that the outer radius is given by the formula  $\tilde{r}_\mathrm{out} = 0.3(v_\mathrm{orb}/c_\mathrm{s})^2$ \citep{krticka2011}. For typical Be stars, this formula gives $\tilde{r}_\mathrm{out}$ between 200 and 600. In our dynamical models, $\Lambda = 1.03$, which is very close to 1. If the real outer radius of the Be star system were $\tilde{r}_\mathrm{out} = 100$, meaning that $\Lambda = 1.11$, we would still have a negligible difference between our steady-state AM loss rate estimations and the real AM loss rate. The steady-state mass loss rate estimations on the other hand would differ by more than a factor of 3. Even if our Be star system was a member of a close binary system and the disk was truncated at $\tilde{r}_\mathrm{out} = 8$, then we would have $\Lambda = 1.55$, meaning a good determination of the steady-state AM loss rate, at least in order of magnitude. But, in this case, the steady-state mass loss rate would differ by more than a factor of 10. Interestingly, an infinite steady-state disk ($\tilde{r}_\mathrm{out} \rightarrow \infty$) would have $\Upsilon = \Lambda = 1$. It would transport AM from the star to infinity at a constant rate given by $(GMR_\mathrm{eq})^{\frac{1}{2}}\dot{M}_\mathrm{inj}(\tilde{r}_\mathrm{inj}^{\frac{1}{2}}-1)$, with zero mass
loss.

The relation between the surface density, $\Sigma$, and the volume density, $\rho$, is given by
\begin{equation}
\rho(r,z,t)=\frac{\Sigma(r,t)}{(2\pi)^\frac{1}{2}H}e^{-\frac{z^2}{2H^2}}\,,
\label{eq:rho}
\end{equation}
where $H(r)=c_s/(GM_*/r^3)^\frac{1}{2}$ is the disk hydrostatically supported scaleheight. To obtain Eq.~\ref{eq:rho}, it was assumed that the disk is in hydrostatic equilibrium and has a constant temperature in the vertical direction. See, e.g., \cite{bjorkman2005} for a derivation of Eq.~\ref{eq:rho}.


\section {Mass reservoir effect}
\label{sect:mass_reservoir}

The visible excess emission observed in Be stars is affected mostly by the inner parts of the disk (see Sect.~\ref{sect:vband_importance}). From this it follows that the visible excess is not a good indicator of the total disk mass. This is easily understood if one considers the case of a disk fed at a constant rate for a very long time. Initially the brightness grows very fast, as a result of the increase of the density of the inner disk, but after a few weeks/months the light curve flattens out as the inner disk reaches a near steady-state configuration (See Fig.~\ref{fig:v_band_variation}). The outer parts, however, continue to grow in mass without any effects on the visible light curve. To observationally probe these parts, one would have to follow up the light curve in the IR or radio wavelengths (\citealt{vieira2015}, \citealt{panoglou2016}).

When the disk feeding eventually ceases, the viscous forces that couple all the matter in the disk lead to the reaccretion of part of the disk back onto the star. In that case, disks that are very similar in their inner regions (having, therefore, similar visual excesses) but have different masses (as a result of a different previous evolution) will dissipate at different rates, owing to the fact that the inner region is viscously coupled to the outer parts. More specifically, a more massive disk can supply the inner regions with mass for a longer time than a less massive disk. As a result, the dissipation of more massive disks appears slower than the dissipation of less massive ones.

\begin{figure}[!t]
\centering
{\includegraphics[height=0.55 \textheight, width=0.6 \columnwidth]{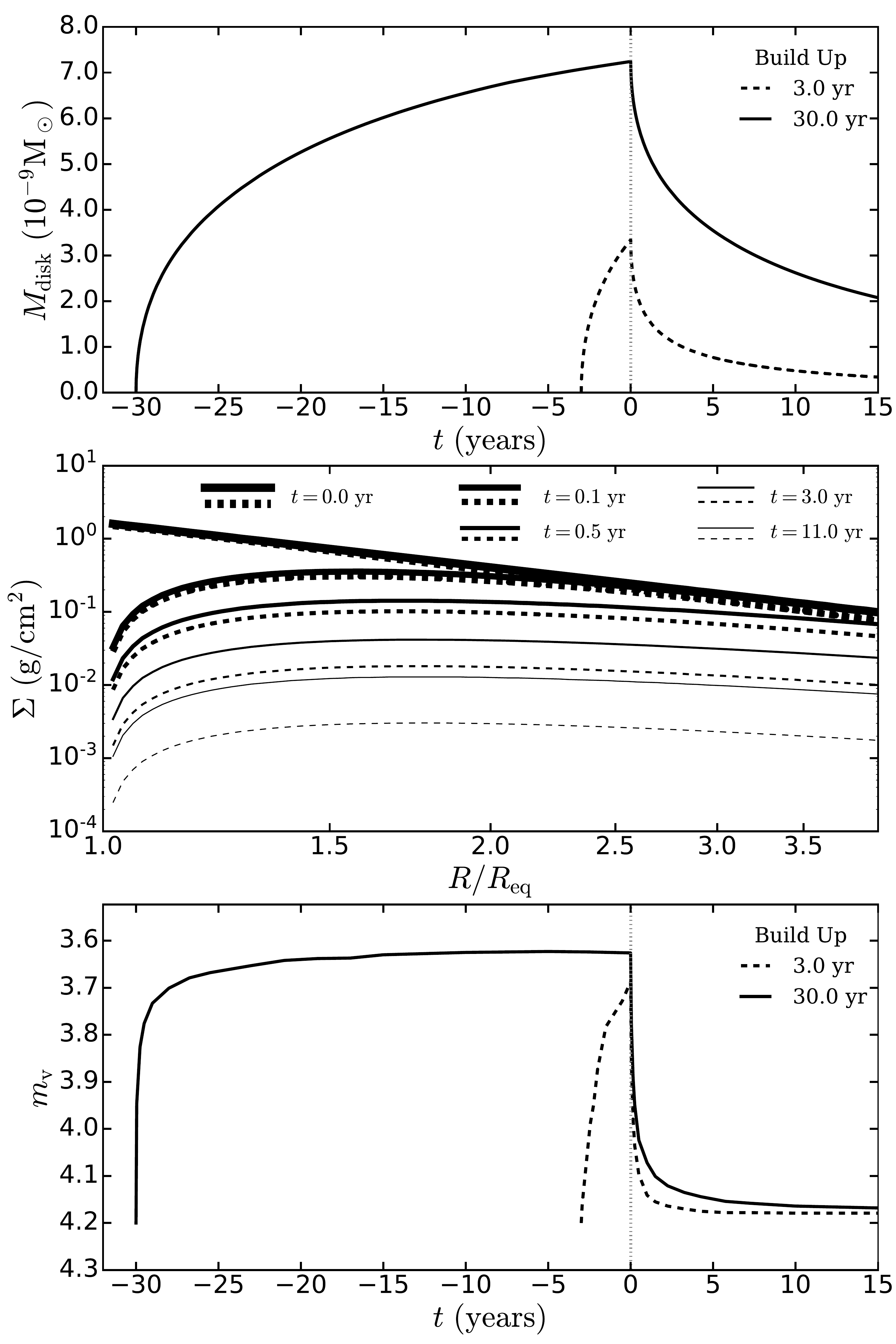}}
\caption[Disk mass and density variation]{
Theoretical comparison between two events of disks formation and dissipation. In one event (solid lines) the disk was fed at a constant rate for 30.0\,yr. The dashed lines show another model for a much shorter build-up time (3.0\,yr). To facilitate comparison, we assumed $t=0$ as the onset of disk dissipation for both models.
 {\it Top}: The total disk mass of the two models as a function of time.
 {\it Middle}: Disk surface density as a function of distance from the star. The two lines trace the disk build-up for different times (as indicated) and the subsequent dissipation.  
 {\it Bottom}: The corresponding light curves for both models.
}
\label{fig:disk_mass}
\end{figure}

In Fig.~\ref{fig:disk_mass}, we compare two disk models that have been fed at the same rate but for different time spans ($3.0~\mathrm{yr}$ and $30.0~\mathrm{yr}$). In this comparison, we arbitrarily fixed the time $t=0$ as the end of the build-up phase. In both models, $-\dot{J}_{*,\mathrm{std}}=2.68\times 10^{36}\,\mathrm{g\,cm^2\,s^{-2}}$ and $\alpha=0.5$. The inner regions of both disk models reach a near-steady surface density by the time of the end of the smallest build-up time ($3.0~\mathrm{yr}$), thus, the surface density for the solid and dashed lines are nearly identical at $t=0$ in the middle panel. The total disk mass is shown in the top panel. Aside from the obvious fact that the disk that was fed longer has a much larger mass, the plot clarifies that the dissipation rate of the more massive disk is much slower than the less massive one. This is more easily seen in the middle panel that displays the temporal evolution of the surface density for the two models. Finally, the third panel exhibits the corresponding light curves, demonstrating that the model that had the longer build-up time (and, hence, is more massive) dissipates at a much slower rate than the other model.

In the remainder of the text, we refer to the effect described above as the \textit{mass reservoir effect}, following \cite{rimulo2018}. This realization is important because it points to a fundamental difference between the light curves during disk build-up and dissipation: for the former, the rate at which the brightness varies depends only on $\alpha$ and 
$\dot{J}_{*,\mathrm{std}}$
 while for the latter the rate also depends on how long the disk was fed. \cite{carciofi2012} did not include the mass reservoir effect in their analysis, as no modeling was done of the build-up phase that preceded the 2004 disk dissipation. As a result, we show in Chap.~\ref{chap:photometry} that the $\alpha$ determination in that study was overestimated by a factor of roughly five.


\section {The computational tools}
\label{sect:tools}

In this PhD project we used mainly two computational codes: the 1D time-dependent hydrodynamics code {\tt SINGLEBE} \citep{okazaki2007} and the {\tt HDUST} code \citep{carciofi2006a, carciofi2008b}. In the following the codes are briefly introduced.


\subsection {The $\alpha$-disk code {\tt SINGLEBE}}
\label{subsect:singlebe}

An important aspect of this research is the ability to deal with the ``dynamical evolution'' of the VDD. Therefore, we need a code to simulate the time-dependent evolution of an axisymmetric $\alpha$-disk. This is what we obtain by the 1D time-dependent hydrodynamics code {\tt SINGLEBE}, that although is a simple code (comparing to other available general 3D hydrodynamical solver codes) it is good enough for our purpose.

The 1D grid used in the {\tt SINGLEBE} code is limitted to the space between $R_\mathrm{eq}$ and $R_\mathrm{out}$. The grid is defined as a logarithmic net with an aribitrary number of cells. One of the cells is selected as the point where the mass from the central star is injected into it.  

{\tt SINGLEBE} solves the isothermal 1D time-dependent fluid equations \citep{pringle1981} in the thin disk approximation (Eq.~\ref{eq:sigmadot}), and provides the disk surface density, $\Sigma (r, t)$, and the mass flux, $\dot{M}_\mathrm{disk} (r, t)$. More details on {\tt SINGLEBE} can be obtained by the original publication \citep{okazaki2007} and in \cite{rimulo2017}.


\subsection {The Monte Carlo radiative transfer code {\tt HDUST}}
\label{subsect:hdust}

The main code used in this work is the Monte Carlo radiative transfer code {\tt HDUST} which incorporates all the physical ingredients need for this work \citep{carciofi2004, carciofi2006a, carciofi2008b}. {\tt HDUST}, a fully three-dimensional (3D) code, solves simultaneously the radiative equilibrium, the radiative transfer, and NLTE statistical equilibrium problems to obtain the ionization fraction, hydrogen level populations, and electron temperature as a function of position in a 3D envelope around the star. Having the above quantities, {\tt HDUST} provides the emergent spectral energy distribution (SED), including emission line profiles, as well as the polarized spectrum and synthetic images. 

To date, {\tt HDUST} was widely used to study a variety of Be stars, presenting both theoretical-only studies \citep[e.g.,][]{carciofi2006a, carciofi2008b, haubois2012, faes2013, haubois2014} and model-observational comparison \citep[e.g.,][]{carciofi2009, carciofi2010, carciofi2012, klement2015, faes2016, klement2017, baade2018, rimulo2018, ghoreyshi2018} investigations.

\chapter{Cycle Lengths, Growth, and Decay Rates}
\label{chap:exp_formula}

In this Chapter, the $V$-band light curve of $\omega$ CMa is analyzed with a simple exponential formula. The goals of this analysis are two-fold: i) identify the main events of the light curve, and ii) measure the lengths of these events. In this way, avoiding the complexities of a physical modeling, obtaining some important parameters will be possible which are necessary for the physical modeling as well. Moreover, analysing the results of the exponential fitting provides independent conclusions about the light curve that can be compared with the results of the physical model described in the next chapter. 

The formula we used was:

\begin{equation} 
\label{eq:magnitude}
m_\mathrm{v}(t) = (m_{0}-m^{\prime})e^{-\frac{(t-t_{0})}{\tau}}+m^{\prime},
\end{equation}
where $m_\mathrm{v}(t)$ is the visual magnitude as a function of time, $m_{0}$ is the visual magnitude at the beginning of each section of the lighcurve ($t=t_0$), $m^{\prime}$ is the asymptotic visual magnitude of each section, and $\tau$ is the associated time-scale of the brightness variation. 
 
 
\begin{figure}[!h]
\begin{minipage}{0.5\linewidth}
\centering
\subfloat{\includegraphics[width=1.0\linewidth]{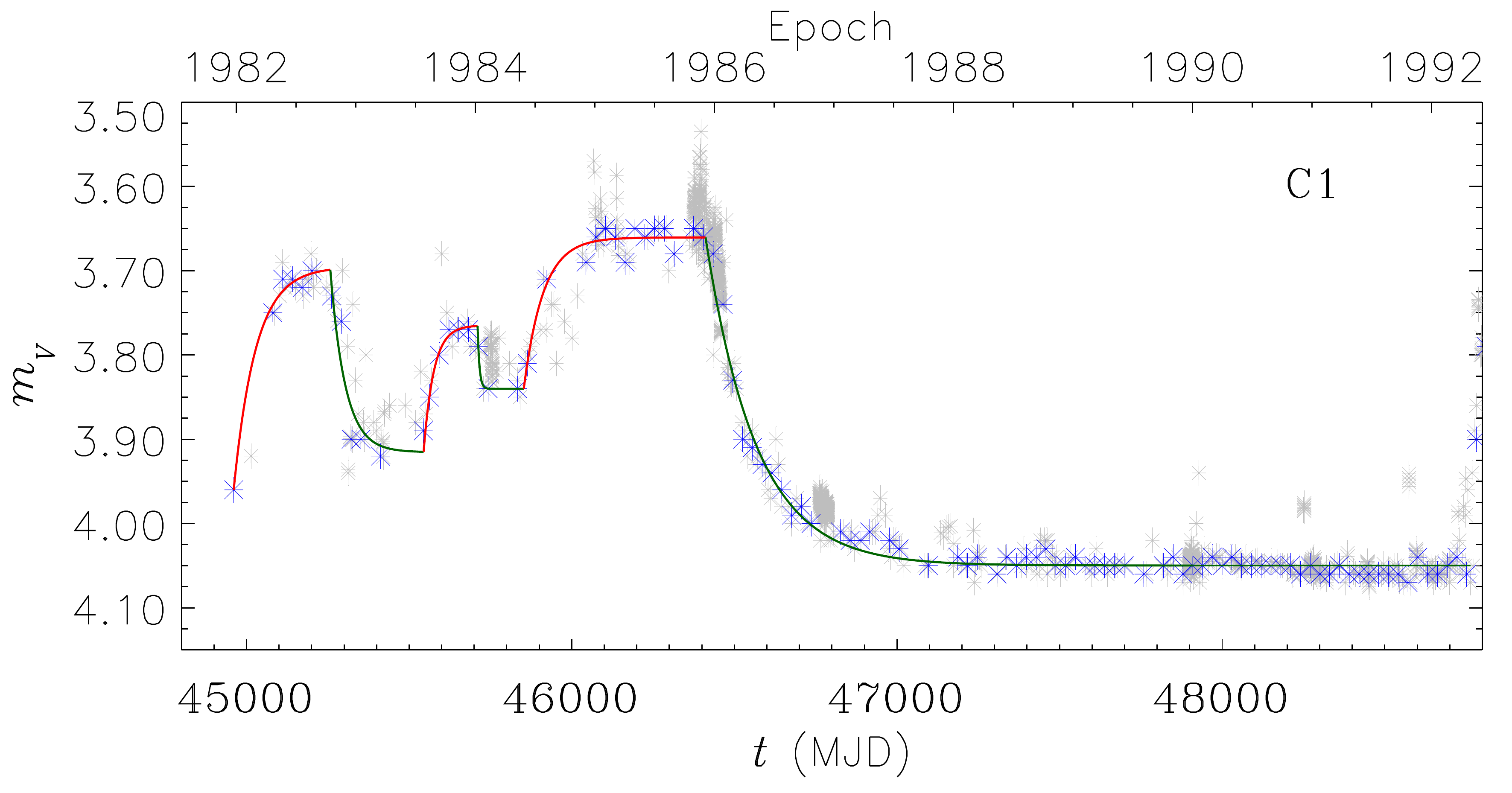}}
\end{minipage}%
\begin{minipage}{0.5\linewidth}
\centering
\subfloat{\includegraphics[width=1.0\linewidth]{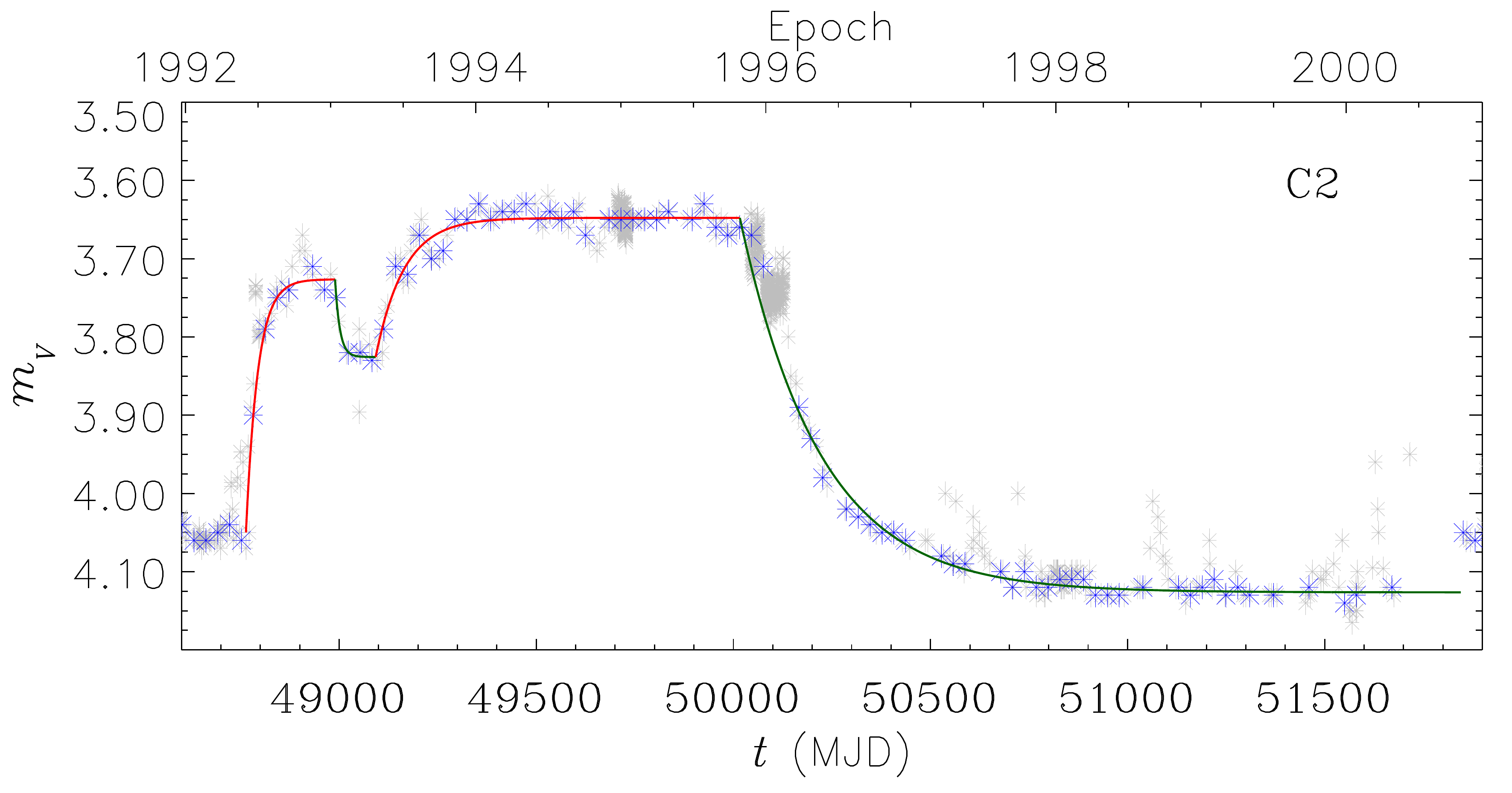}}
\end{minipage}\par\medskip
\begin{minipage}{0.5\linewidth}
\centering
\subfloat{\includegraphics[width=1.0\linewidth]{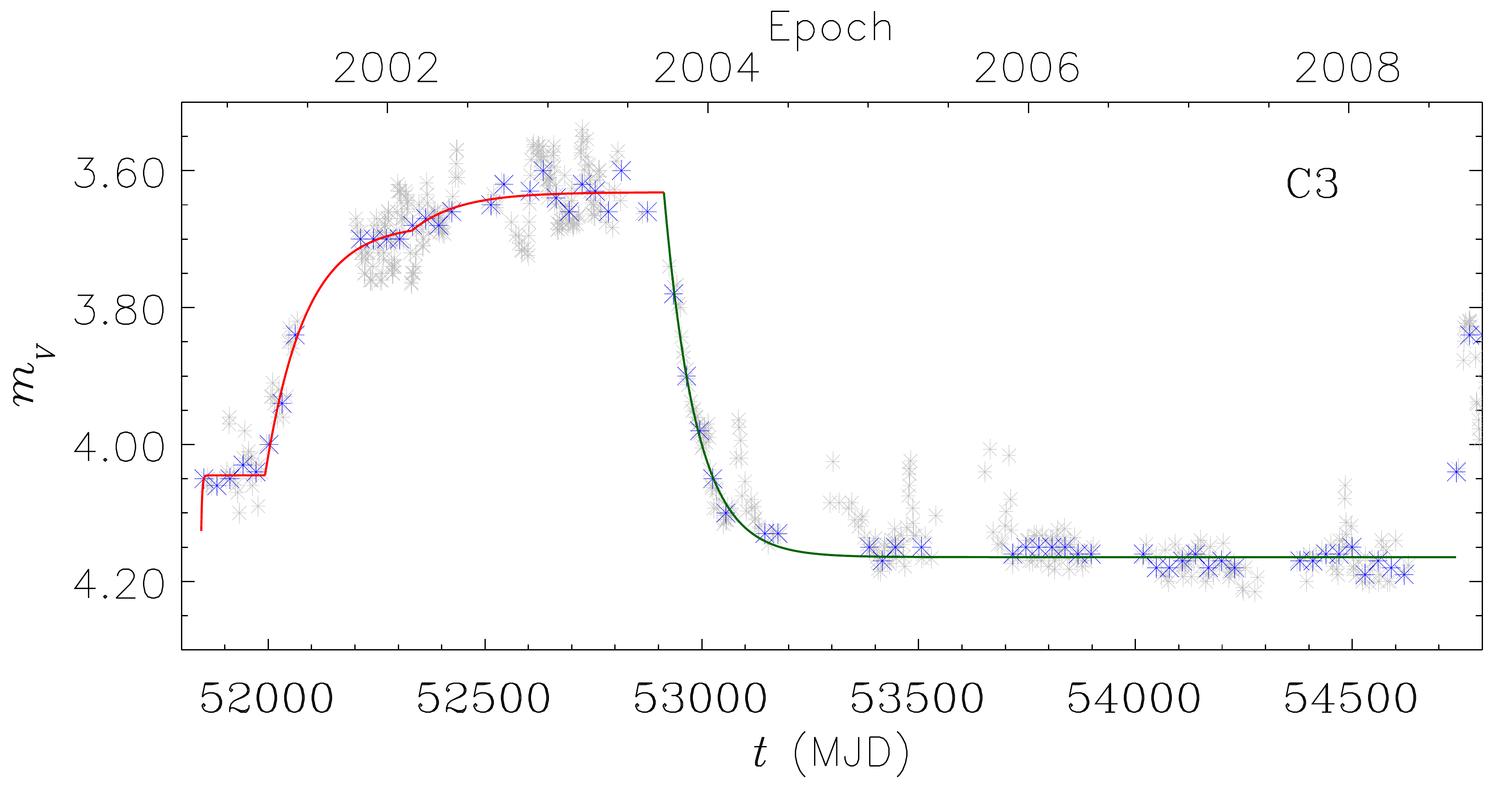}}
\end{minipage}%
\begin{minipage}{0.5\linewidth}
\centering
\subfloat{\includegraphics[width=1.0\linewidth]{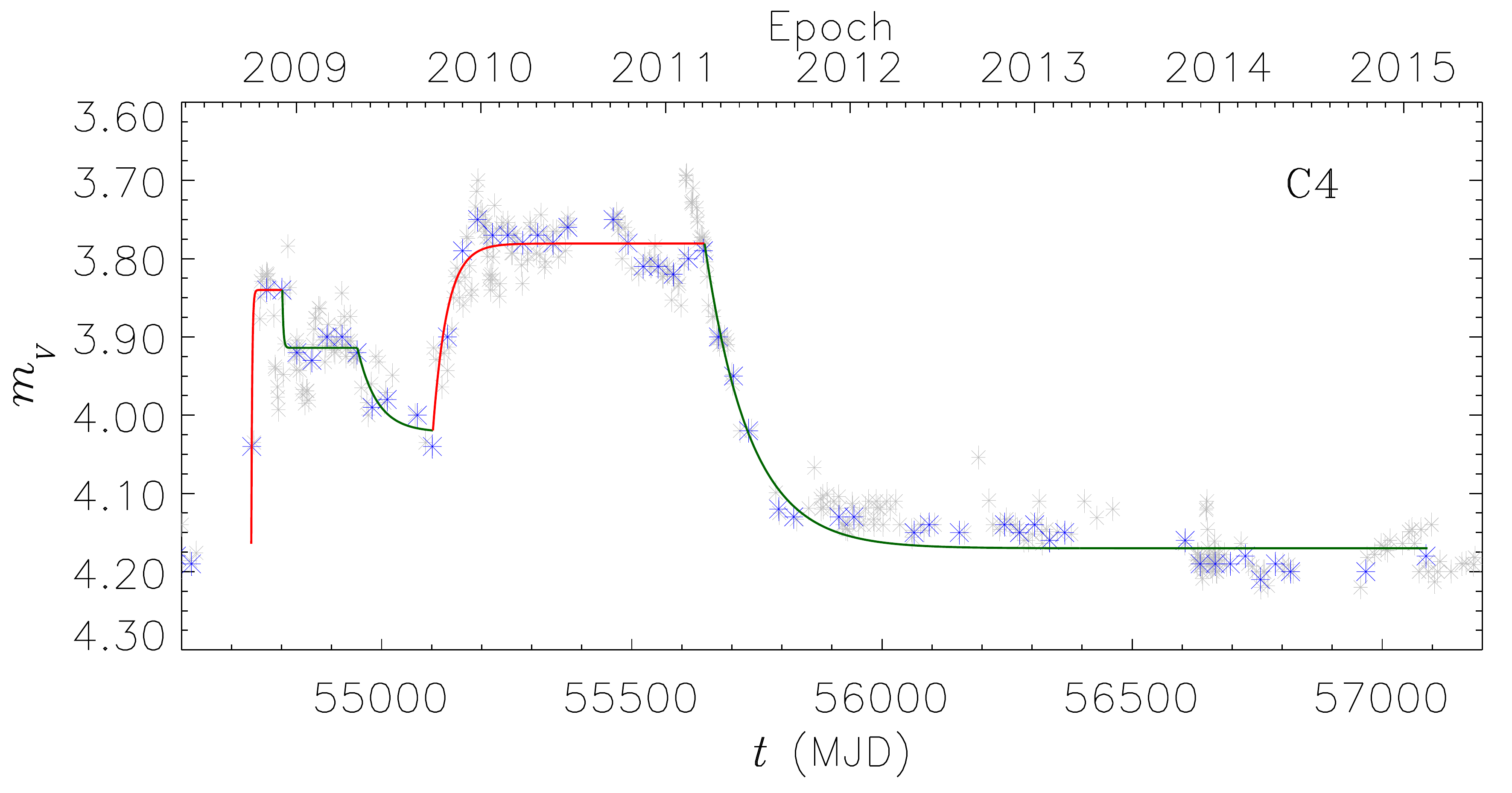}}
\end{minipage}%
\caption[Fit model of the $V$-band data using exponential formula]{Exponential fit on the $V$-band photometric data of $\omega$ CMa for the first (top left), second (top right), third (bottom left), and fourth (bottom right) cycles. The grey stars show the observed data, and the blue stars show the averaged data in 30-day intervals. The red and green solid lines represent the (partial) outburst and quiescence phases, respectively.}
\label{fig:exp_fitting}
\end{figure}


The results of the exponential fitting are shown in Fig.~\ref{fig:exp_fitting}, and the list of sections considered with their respective parameters are presented in Table~\ref{table:results}. Note that all outbursts were divided into several subsections, that are identified either by $\mathrm{O}i_{\mathrm{o}j}$, for an outburst proper, or by $\mathrm{O}i_{\mathrm{q}j}$, for a partial quiescence within an outburst. The observed $V$-band photometric data is shown by grey stars. In blue, we show the data binned to 30-day long bins. This procedure smoothes out the short-term variability of the light curve, and reveals the most important, longer-term trends. Even so, it is clear that none of the outbursts have a simple behavior and, instead, are formed by a succession of smaller outbursts followed by short quiescences. Conversely, after smoothing out the data, all four long-term quiescences are smooth, quiet processes.

A first initial choice of the exponential modeling was to ignore events shorter than 3 months. This arbitrary choice came from a compromise between modeling the curve as best as possible, and having a reasonable number of different events to model. Each cycle was divided into several parts, each part is either a (partial) outburst or a quiescence. For each part the exponential formula was applied to find the best fit by adjusting four free parameters: $m_{0}$, $m^{'}$, $\tau$, and $t_{0}$. 

O1 and O4 have the most complicated structure while O2 and O3 posses simpler ones. Moreover, the main dissipation phase of all cycles show some bumps that increase the brightness as much as $\sim$ 0.$^m$15. These short-term events vanish in the averaged data, and were ignored.

\begin{figure}
\centering
\includegraphics[width=1.0\linewidth]{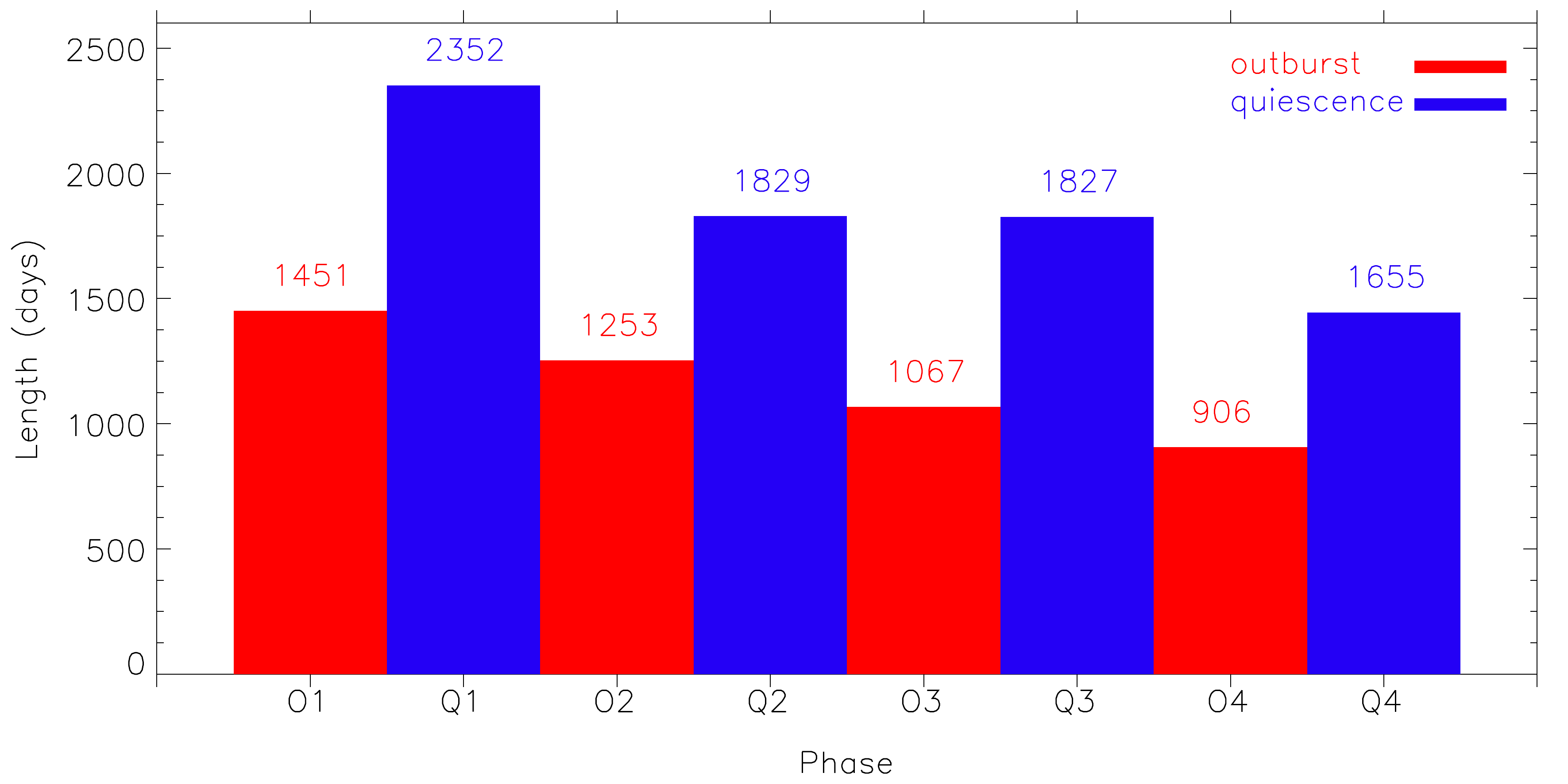}
\caption[Cycle lengths]{The length of each event measured by exponential formula fitting. The red columns represent the length of formation phases and the blue columns represent the length of dissipation phases.}
\label{fig:cycle_length}
\end{figure}

Based on the exponential formula fitting, the intersection point of each two curves (the end of one with the beginning of the next) provides one way of estimating the time of start or end of each event, thus providing a useful way to estimate the total length of each cycle. The results are displayed in Fig.~\ref{fig:cycle_length} which gives a numerical value to the fact that the length of successive formation phases and successive dissipation phases is decreasing. Note that Fig.~\ref{fig:cycle_length} does not display the substructures found in all outbursts, but rather the total length of the outburst. In addition, we find that on average the dissipation is 1.64 times longer than the outburst. Recent works \citep[e.g.][]{rimulo2018, labadie2018} find the same trend for Be star disks, for which the dissipation is usually a (much) longer phase than the formation. The same result was found by \cite{haubois2012} from a theoretical perspective.

\begin{table*}
\begin{center}
\tabcolsep 5.8pt
\footnotesize
\caption[Parameters derived from exponential formula fitting and VDD model]{The parameters derived using exponential formula fitting (columns 2 to 6) and the parameters obtained by the VDD model (columns 7 to 9).}
\begin{tabular}{@{}ccccccccccccccccc}
\hline
\hline
phase & \vline & length & $\vert\Delta m\vert$ & $m_{0}$ & $\tau$ & $t_0^{\rm d}$ & $\alpha$ & $\dot{M}_\mathrm{inj}$ & $-\dot{J}_{*,\mathrm{std}}$ \\

& \vline & (days) & & & (days) & (MJD) & & $(10^{-7}\mathrm{M_\odot yr^{-1}})$ & $(10^{36}\mathrm{g\,cm^2\,s^{-2}})$ \\

\hline  & \vline \\

Cycle 1  & \vline & 3803 & 0.09 & 3.96 & ----- & 44960.5 & ----- & ----- & ----- \\

O1  & \vline & 1451 & 0.30 & 3.96 & ----- & 44960.5 & ----- & ----- & ----- \\

O1$_\mathrm{o1}$  & \vline & 297 & 0.26 & 3.96 & 68.0 & 44960.5 & 1.00 & 3.0 & 4.4 \\

O1$_\mathrm{q1}$  & \vline & 287 & 0.22 & 3.70 & 46.1 & 45257.5 & 1.00 & 1.4 & 2.0 \\

O1$_\mathrm{o2}$  & \vline & 166 & 0.16 & 3.92 & 31.1 & 45544.5 & 1.00 & 2.5 & 3.6 \\

O1$_\mathrm{q2}$  & \vline & 142 & 0.08 & 3.76 & 17.4 & 45710.5 & 1.00 & 1.9 & 2.7 \\

O1$_\mathrm{o3}$  & \vline & 559 & 0.18 & 3.84 & 59.5 & 45852.5 & 1.00 & 3.4 & 4.9 \\

Q1  & \vline & 2352 & 0.39 & 3.66 & 156.3 & 46411.5 & 0.20 $\pm 0.03$ & 0.2 & 0.2 \\

& \vline \\

Cycle 2  & \vline & 3082 & 0.07 & 4.05 & ----- & 48763.5 & ----- & ----- & ----- \\

O2  & \vline & 1253 & 0.40 & 4.05 & ----- & 48763.5 & ----- & ----- & ----- \\

O2$_\mathrm{o1}$  & \vline & 226 & 0.32 & 4.05 & 30.1 & 48763.5 & 1.00 & 3.0 & 4.4 \\

O2$_\mathrm{q1}$  & \vline & 103 & 0.10 & 3.73 & 20.0 & 48989.5 & 1.00 & 1.9 & 2.8 \\

O2$_\mathrm{o2}$  & \vline & 924 & 0.18 & 3.83 & 71.9 & 49092.5 & 1.00 & 3.4 & 4.9 \\

Q2  & \vline & 1829 & 0.48 & 3.65 & 204.1 & 50016.5 & 0.13 $\pm 0.01$ & 4.0$\times$10$^{-3}$ & 5.8$\times$10$^{-3}$ \\

& \vline \\

Cycle 3  & \vline & 2894 & 0.04 & 4.12 & ----- & 51845.5 & ----- & ----- & ----- \\

O3  & \vline & 1067 & 0.49 & 4.12 & ----- & 51845.5 & ----- & ----- & ----- \\

O3$_\mathrm{o1}$  & \vline & 147 & 0.08 & 4.12 & 16.3 & 51845.5 & 1.00 & 3.7 & 5.4 \\

O3$_\mathrm{o2}$  & \vline & 339 & 0.35 & 4.04 & 92.6 & 51992.5 & 0.10 & 0.4 & 0.6 \\

O3$_\mathrm{o3}$  & \vline & 581 & 0.06 & 3.69 & 107.5 & 52331.5 & 0.10 & 0.4 & 0.6 \\

Q3  & \vline & 1827 & 0.53 & 3.63 & 74.6 & 52912.5 & 0.21 $\pm 0.05$ & 4.0$\times$10$^{-3}$ & 5.8$\times$10$^{-3}$ \\

& \vline \\

Cycle 4  & \vline & 2561 & 0.03 & 4.16 & ----- & 54739.5 & ----- & ----- & ----- \\

O4  & \vline & 906 & 0.38 & 4.16 & ----- & 54739.5 & ----- & ----- & ----- \\

O4$_\mathrm{o1}$  & \vline & 62 & 0.32 & 4.16 & 15.2 & 54739.5 & 1.00 & 2.6 & 3.7 \\

O4$_\mathrm{q1}$  & \vline & 150 & 0.07 & 3.84 & 15.8 & 54801.5 & 1.00 & 1.6 & 2.3 \\

O4$_\mathrm{q2}$  & \vline & 151 & 0.11 & 3.91 & 39.4 & 54951.5 & 0.10 & 0.1 & 0.1 \\

O4$_\mathrm{o2}$  & \vline & 388 & 0.26 & 4.02 & 33.8 & 55102.5 & 1.00 & 2.6 & 3.8 \\

O4$_\mathrm{q3}$  & \vline & 155 & 0.05 & 3.76 & 17.1 & 55490.5 & 0.50 & 1.0 & 1.4 \\

Q4  & \vline & 1655 & 0.39 & 3.81 & 85.5 & 55645.5 & 0.11 $\pm 0.03$ & 2.0$\times$10$^{-3}$ & 2.9$\times$10$^{-3}$ \\

\hline uncertainty & \vline & $\pm 1$ & $\pm 0.02$ & $\pm 0.01$ & $\pm 0.5$ & $\pm 0.5$ & ----- & ----- & ----- \\

\hline
\end{tabular}
\label{table:results}
\end{center}
\end{table*}


\begin{figure}
\centering
\includegraphics[width=1.0\linewidth]{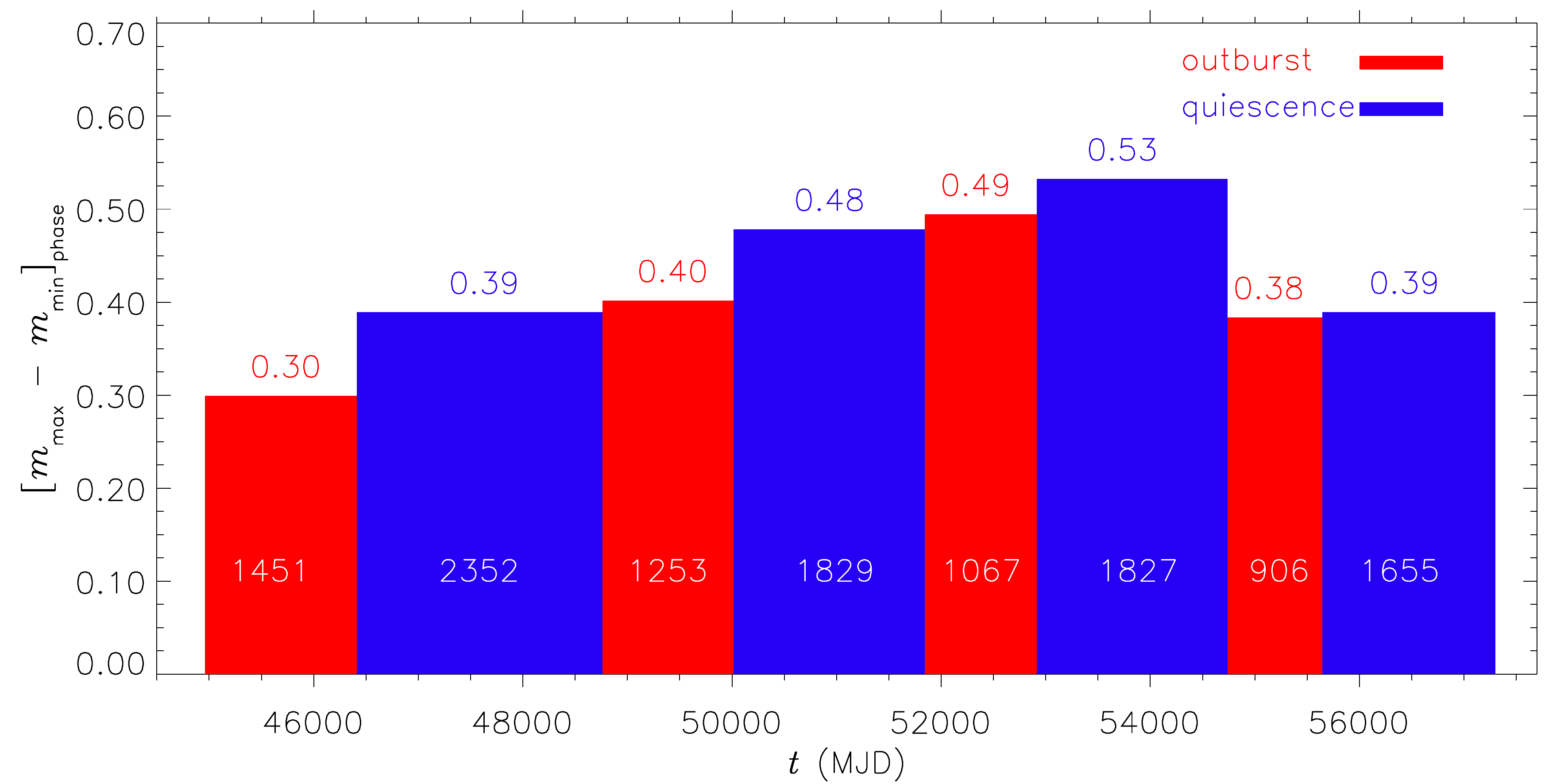}
\caption[Cycle length -- manitude variation relation]{The relation between the sequences of phases, their length and the magnitude changes. The height of each column shows the magnitude change and its width demonstrates the length of that phase. The white numbers written in each column are the length of the phases in days.}
\label{fig:mag_variation}
\end{figure}

Figure~\ref{fig:mag_variation} depicts the relation (if there is any) between event lengths and magnitude change in the event. The first three cycles show a clear correlation between cycle length and magnitude change, but the last cycle is an exception, which may indicate that the former trend is fortuitous.

\begin{figure}
\centering
\includegraphics[width=1.0\linewidth]{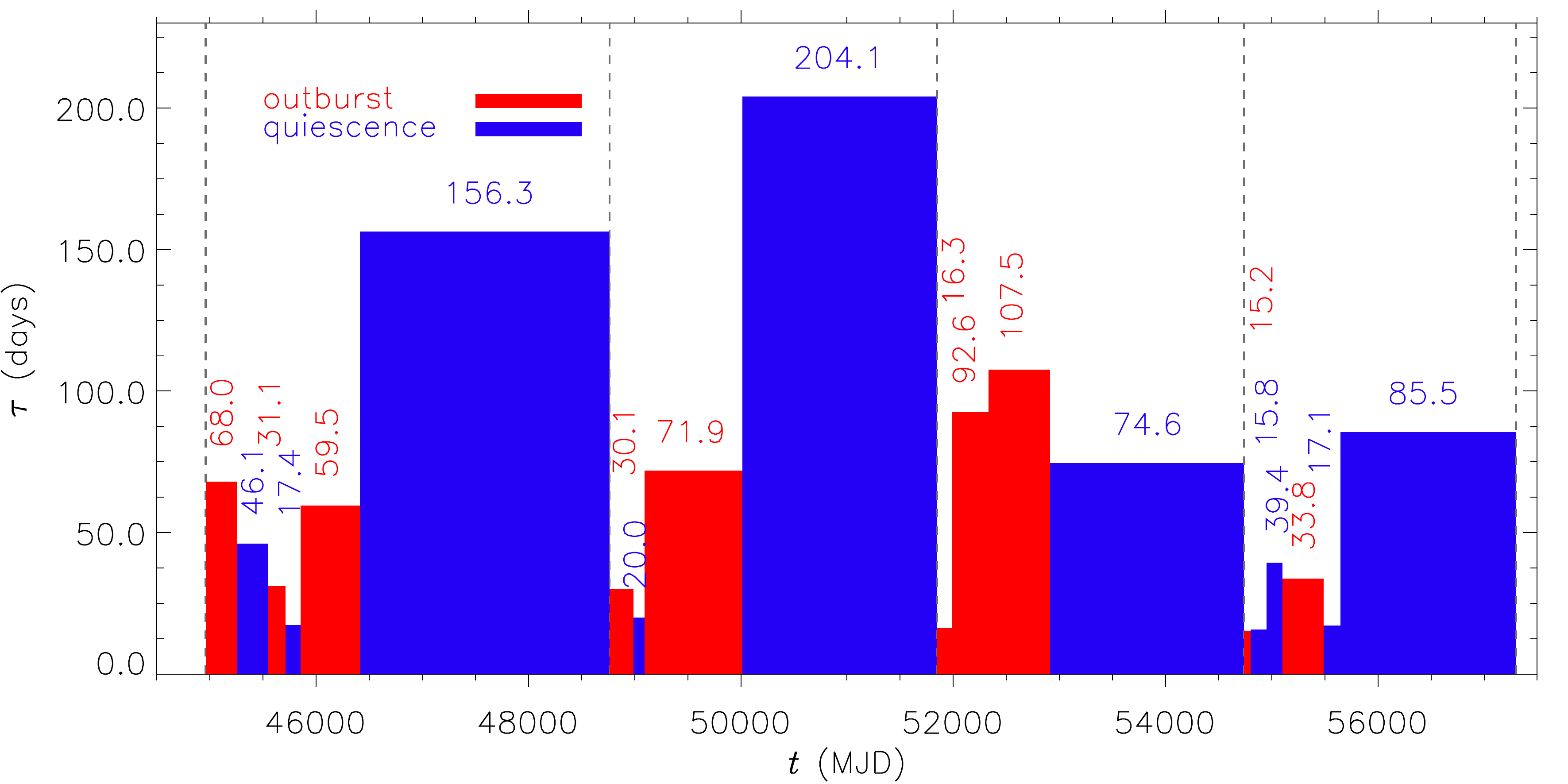}
\caption[Variaion of the light curve slope]{The value of $\tau$ for each section. As we expect the value of $\tau$ should be related to the $\alpha$ parameter, this plot shows that the formation phases (including several partial outbursts and quiescence shown by red and blue columns) should possess higher values of $\alpha$ than the dissipation phases.}
\label{fig:slope_var_rate}
\end{figure}

Finally, we found that the growth and decay time-scales of the light curve, $\tau$, vary from cycle to cycle and also within a given cycle (Table~\ref{table:results} and Fig.~\ref{fig:slope_var_rate}). A large value of $\tau$ is related to a slower rate of magnitude variation, and vice-versa. One intriguing result is the fact that $\tau$ is usually smaller during the outburst phase. Observationally, this means that the rate of variations in this phase are larger; physically, this could mean that the $\alpha$ parameter could also be larger during theses phases. In addition, these results suggest that the viscosity parameter, $\alpha$, is varying, as $\alpha$ is the main parameter controlling the disk evolution time-scales. The selected phases for the exponential modeling, as well as their starting times and duration, will be used in the next chapter as input for the VDD modeling of the light curve of $\omega$ CMa.

\chapter{Analysis of the $V$-band Photometry} 
\label{chap:photometry}

Amongst the available data of $\omega$ CMa, photometric $V$-band data has by very far the best temporal coverage. Moreover, $\omega$ CMa was observed with this technique since the 1960s (\ref{sect:photometry}). Thus, it is expected that analysis of its photometric data provides valuable information. In this chapter we present a detailed physical modeling of the $V$-band light curve using the dynamical VDD model. Part of the contents of this chapter was published in a paper that appeared recently in the Monthly Notices of the Royal Astronomical Society \citep{ghoreyshi2018}.


\section {On the importance of the $V$-band light curve}
\label{sect:vband_importance}

\cite{carciofi2011} discussed the basic aspects of steady-state and dynamical VDD disks, with an emphasis of where in the disk different observables form (e.g., short vs. long wavelength fluxes, line emission, polarization, etc.). This concept of {\it formation loci} was later further developed by \cite{rivinius2013a} and \cite{vieira2015}. The latter authors went further and developed an approximate model for describing the continuum emission processes of gaseous disks, the so-called pseudo-photosphere model. This model was later used to study a large sample of Be stars by \cite{vieira2017}.

In the pseudo-photosphere model, the continuum emission is described as coming from an inner ring of radius $\bar{R}$ (the pseudo-photosphere radius) that is vertically optically thick and an outer, optically thin, diffuse part. $\bar{R}$ grows with wavelength, $\lambda$, approximately as $\bar{R} \propto \lambda^{0.41}$ \citep{carciofi2006b, vieira2015} for an isothermal Be disk with a power-law exponent $n=3.5$, seen with a pole-on orientation. Different inclination angles and power-law exponents will result in a different variation of $\bar{R}$ with $\lambda$ \citep[see Eq. 8][]{vieira2015}.

\begin{figure}[!t]
    \centering
    \includegraphics[width=1.0\linewidth]{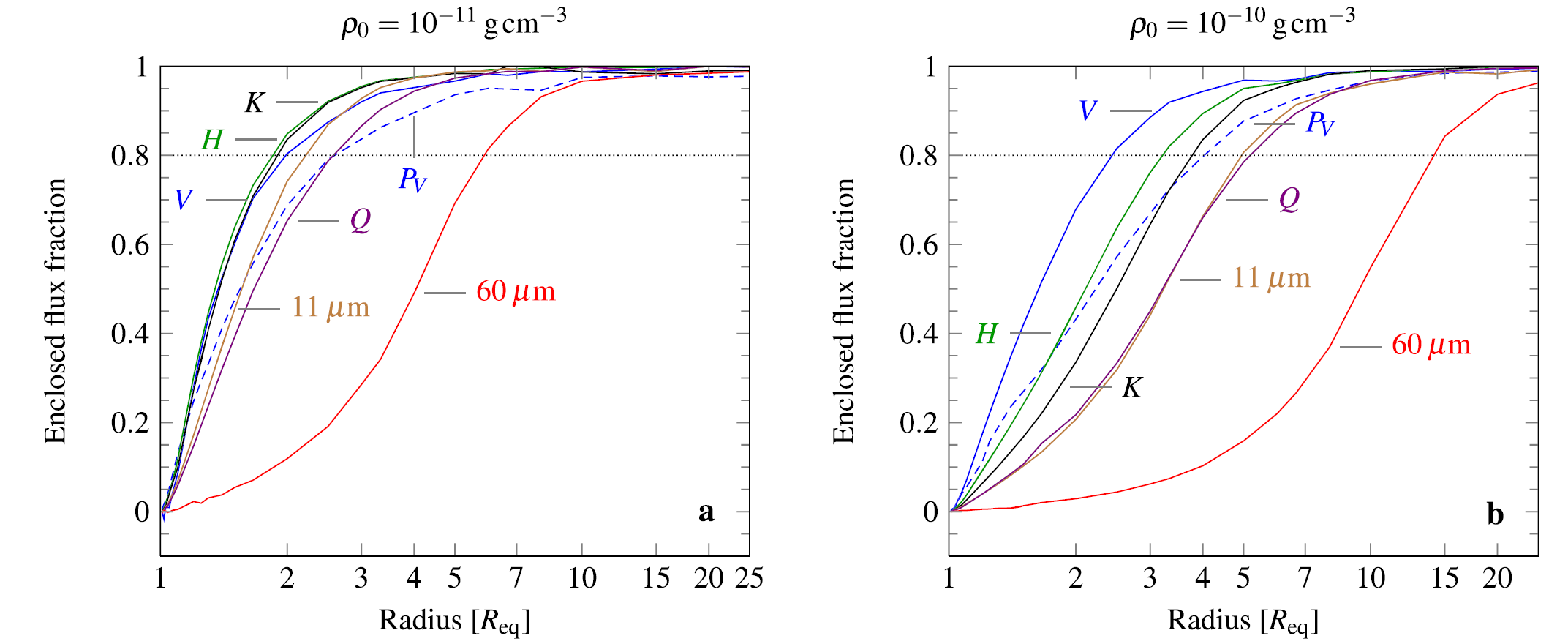}
    \caption[The formation loci of continuum emission]{The formation loci of continuum emission for various wavelength bands and $V$-band polarized flux $P_V$, expressed in accumulated disk contribution as function of increasing radius for two different densities. The stellar flux $F_{\star}$ was subtracted so that each curve starts at zero. The dotted limit marks 80\% of the total flux, which corresponds to the integrated flux inside the FWHM for a Gaussian shaped emission profile. Data were computed with HDUST for the reference model of \cite{faes2013}, for a disk seen at $i=30^\circ$. Reproduced with permission from the authors. (Credit: \citealt{rivinius2013a})}
    \label{fig:flux_loci}
\end{figure}

\cite{carciofi2011} and \cite{rivinius2013a} discuss the formation loci of different observables in terms of their enclosed flux fractions. The formation loci is, thus, the distance from the star from where a given enclosed flux fraction (say, 80\%) arises. One example is shown in Fig.~\ref{fig:flux_loci}. This figure is easy to understand in terms of the pseudo-photosphere model: for each band, the formation loci is closely related to the pseudo-photosphere radius at that band. For intance, for tenuous disks (left plot of Fig.~\ref{fig:flux_loci}) there is no pseudo-photosphere for the $V$, $H$ and $K$, so their formation loci are very similar; only when the density is much larger (right plot of Fig.~\ref{fig:flux_loci}) the pseudo-photosphere of each band becomes relevant, and their formation loci grows with $\lambda$, as expected. In both cases, the pseudo-photosphere is present for the long-wavenlengths bands shown in the plot.

Figure~\ref{fig:flux_loci} reveals immediately the importance of the $V$-band light curve for this project. Because the $V$-band excess emission arises from the very inner parts of the disk, it represents a direct probe of the conditions of this region, which is the first region of the disk affected by mass loss. In other words, the $V$-band excess flux represents a ``snapshot'' of the disk feeding rate and, as such, is well-suited to track how this rate changes over time.


\section {Results of the application of the Viscous Decretion Disk model}
\label{sect:vdd_model}

One of the most striking features of $\omega$ CMa's light curve (Fig.~\ref{fig:full_lightcurve}) is that the star always presents a flux excess, if one assumes that the central star is intrinsically non-variable above the $\sim 0.^m05$ level. This is a reasonable assumption, because known mechanisms that could cause the star to vary in brightness account only for variations of much lower amplitudes. For instance, pulsations could account only for short-period (days or less) variations at the tens of mmag level (e.g. \citealt{baade2016}; \citealt{kurtz2015}; \citealt{balona2011}; \citealt{huat2009}). In the following we therefore assume that a variable $V$-band excess was present in the past 34 years, and this excess is of disk origin.


\subsection {The first scenario for modeling the full light curve of $\omega$ CMa}
\label{subsect:first_model}

\begin{figure}[!t]
    \centering
    \includegraphics[width=0.9\linewidth]{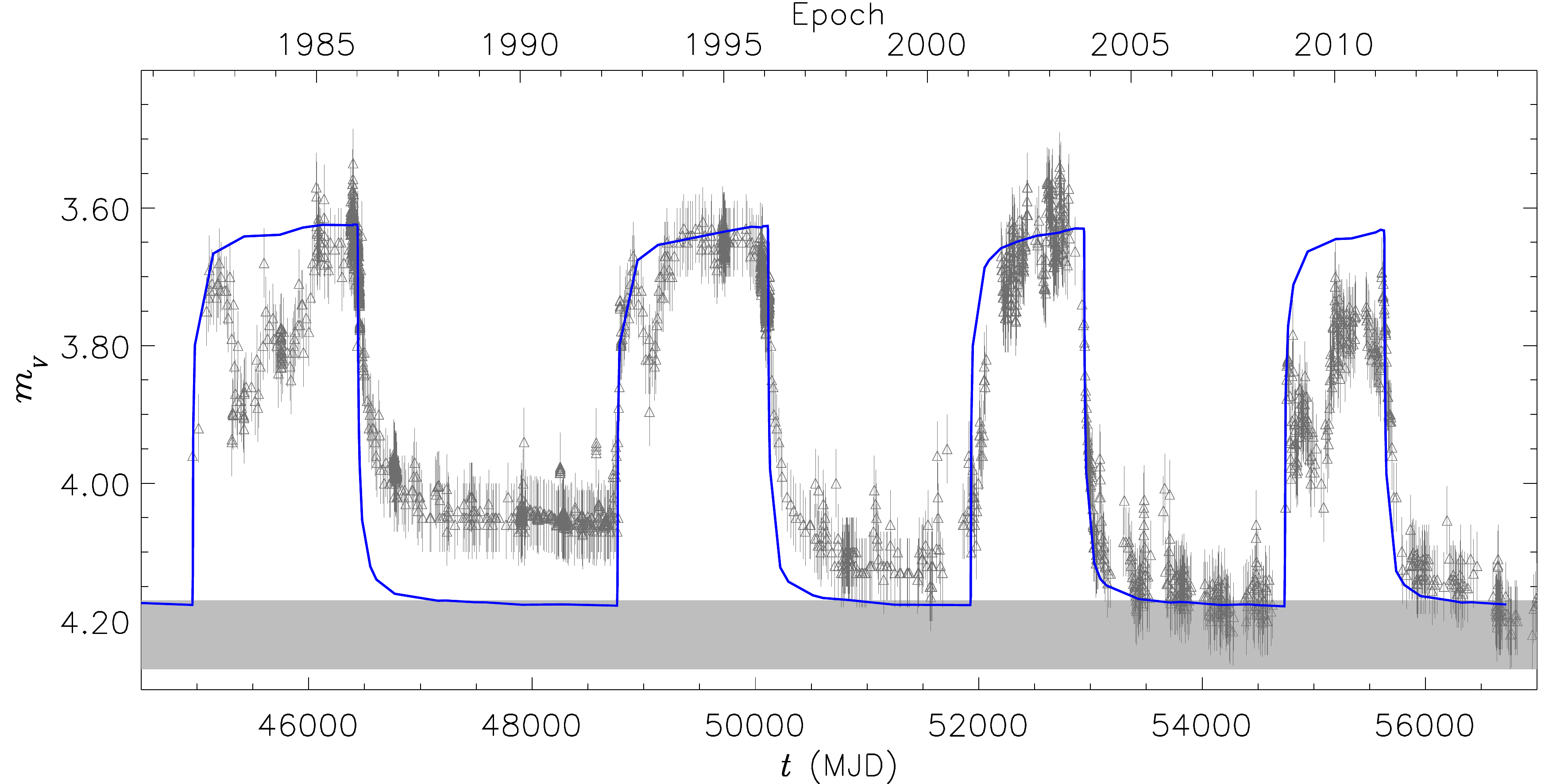}
    \caption[The first scenario for modeling the full light curve of $\omega$ CMa]{The first scenario for modeling the full light curve of $\omega$ CMa. The model employs a viscosity parameter $\alpha$=1.0 and $\dot{M}_\mathrm{inj}=2.6\,\times$ 10$^{-7}$ $\mathrm{M_{\odot}\,yr^{-1}}$ ($-\dot{J}_{*,\mathrm{std}}= 3.6\,\times 10^{36}\,\mathrm{g\,cm^2\,s^{-2}}$). The horizontal grey band represents the estimated intrinsic visual magnitude of the central star of $\omega$ CMa. The data shown with grey triangles are the same as Fig.~\ref{fig:full_lightcurve} (also for the Figs.~\ref{fig:second_model} to \ref{fig:summary}).}
    \label{fig:first_model}
\end{figure}

The light curve of $\omega$ CMa consists of several bumps and dips in brightness. Since $\omega$ CMa is a nearly pole-on star, the bumps and dips correspond to the phases of active disk mass injection and passive disk (partial) dissipation, respectively (for an example of theoretical bump and dips, see Fig.~\ref{fig:v_band_variation}). As a first scenario to fit the data using the hydrodynamic models described in Sect.~\ref{sect:model_description}, each cycle was assumed to consist of a single outburst, during which the disk mass injection had a non-zero value, and a single quiescence, for which $\dot{M}_\mathrm{inj}=0$ ($-\dot{J}_{*,\mathrm{std}}=0$). Moreover, it was assumed that a disk was present before the first observed cycle in the 1980's. The length of outburst and quiescence phases in different cycles have been obtained from the exponential analysis described in the previous chapter. In every outburst the $\alpha$ parameter was assumed to be constant and equal to 1.0, and the mass injection rate to the disk was assumed to be 2.6 $\times$ 10$^{-7}$ $\mathrm{M_{\odot}\,yr^{-1}}$ following \cite{carciofi2012}, which corresponds to $-\dot{J}_{*,\mathrm{std}}= 3.6 \times 10^{36}\,\mathrm{g\,cm^2\,s^{-2}}$. The results are shown in Fig.~\ref{fig:first_model} from which one can see that the model fails to reproduce many aspects of the light curve. For instance, the detailed shape of the outbursts are not reproduced (fault 1), even though the maximum magnitude of the first three agrees with the model, but the last one (fault 2). The first two quiescence phases are not fitted at all (fault 3). Moreover, the overall decline in the brightness of the system in the successive dissipation phases cannot be modeled in this scenario (fault 4), because the model quickly evolves to near-zero disk $V$-band excess owing to the disk dissipation. 


\subsection {The second scenario for modeling the full light curve of $\omega$ CMa}
\label{subsect:second_model}

\begin{figure}[!t]
    \centering
    \includegraphics[width=0.9\linewidth]{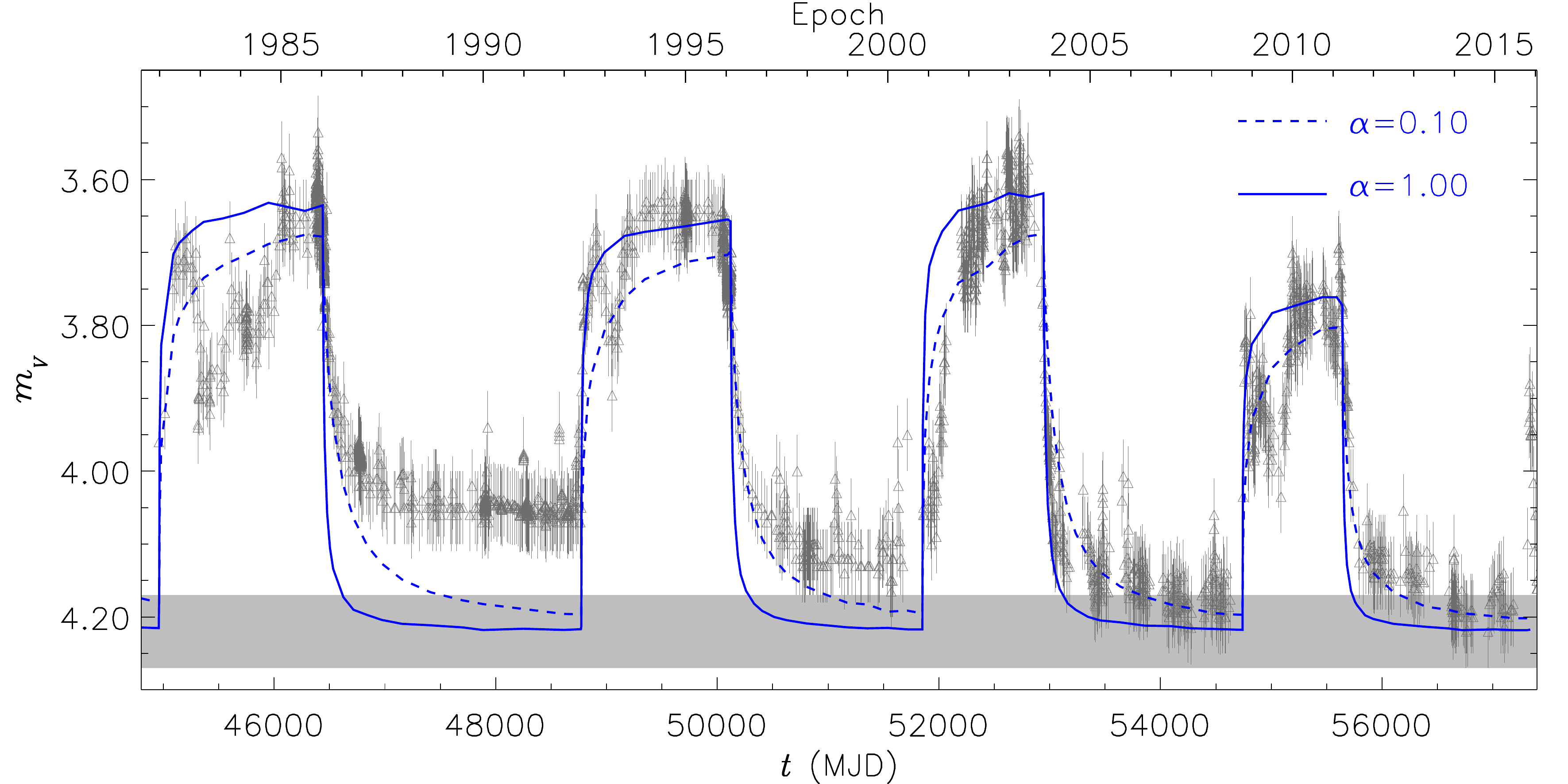}
    \caption[The second scenario for modeling the full light curve of $\omega$ CMa]{The second scenario for modeling the full light curve of $\omega$ CMa. At outburst, different mass injections were assumed in order to fit the brightness level at the end of outburst. The phases of disk dissipation were assumed to have $\dot{M}_\mathrm{inj}=0$ ($-\dot{J}_{*,\mathrm{std}}=0$). Each line type corresponds to different values of viscosity parameter, as indicated.}
    \label{fig:second_model}
\end{figure}

The second scenario follows the same definition of formation and dissipation phases like the last scenario but we allow the $\dot{M}_\mathrm{inj}$ ($-\dot{J}_{*,\mathrm{std}}$) to vary between each outburst. Furthermore, the viscosity was assumed to be constant throughout the four cycles, but contrarily to the previous scenario we also explored different values of $\alpha$. In Fig.~\ref{fig:second_model}, models for two values of $\alpha$ are displayed: 0.1 and 1.0. The values of $\dot{M}_\mathrm{inj}$ and corresponding $-\dot{J}_{*,\mathrm{std}}$ adopted in the models are listed in Table~\ref{table:jdot_steady_value}. 

The employed modifications to the model resulted in some improvements to the fit, but the models are still unable to reproduce the data. The single outburst assumption fails to reproduce the complex observed light curve (fault 1), as expected. In the high-viscosity models the inner disk quickly dissipates, and the flux excess quickly goes to zero, in stark contrast to the general shape of the dissipation phase. Low-viscosity models perform no better: they fail to reproduce the initial phase of disk dissipation (having a too slow flux variation), and also the phase of nearly-constant flux that is reached a few months after dissipation started (fault 3). This simple model also cannot offer any explanation for the secular fading observed at the end of each quiescence phase (fault 4). However, with different mass injection rates for the outbursts the dissimilar height of peaks are reproduced (removing fault 2). Moreover, although using high- and low-viscosities could not completly solve the ``fault 3'', it shows that different $\alpha$s are needed for different dissipation phases, because the initial decline of the dissipation phases are in general better reproduced by the small-$\alpha$ model.


\begin{table}
    \begin{center}
        \caption{Adopted values of $-\dot{J}_{*,\mathrm{std}}$ and $\dot{M}_\mathrm{inj}$ for the models shown in Fig.~\ref{fig:second_model}.}
        \begin{tabular}{@{}ccccc}
            \hline
            \hline
            & \multicolumn{2}{c}{$-\dot{J}_{*,\mathrm{std}}\,(10^{36}\mathrm{g\,cm^2\,s^{-2}})$} & \multicolumn{2}{c}{$\dot{M}_\mathrm{inj}\,(10^{-7}\mathrm{M_{\odot}\,yr^{-1}})$} \\ 
            & $\alpha = 0.1$ & $\alpha = 1.0$ & $\alpha = 0.1$ & $\alpha = 1.0$ \\
            \hline
            $\mathrm{O1}$ & 0.36 & 3.6 & 0.37 & 3.7 \\
            $\mathrm{O2}$ & 0.34 & 3.4 & 0.35 & 3.5 \\
            $\mathrm{O3}$ & 0.39 & 3.9 & 0.40 & 4.0 \\
            $\mathrm{O4}$ & 0.25 & 2.5 & 0.26 & 2.6 \\
            \hline
        \end{tabular}
        \label{table:jdot_steady_value}
    \end{center}
\end{table}


\subsection {The third scenario for modeling the full light curve of $\omega$ CMa}
\label{subsect:third_model}

One of the problems that the previous scenarios could not solve was the overall decline of the brightness of the system in the successive dissipation phases (fault 4). In order to solve this issue, we propose an alternate scenario for the dimming phases of $\omega$ CMa, in which $\dot{M}_\mathrm{inj}$ ($-\dot{J}_{*,\mathrm{std}}$) may be different than zero, as is usually assumed. Therefore, $\dot{M}_\mathrm{inj}$ ($-\dot{J}_{*,\mathrm{std}}$) can have any physically sound value with larger $\dot{M}_\mathrm{inj}$ ($-\dot{J}_{*,\mathrm{std}}$) for outburst phases during which the disk attains a higher density, and smaller $\dot{M}_\mathrm{inj}$ ($-\dot{J}_{*,\mathrm{std}}$) for dissipation phases. In the transition between outburst and dissipation, the disk would therefore switch between a high-density phase to a low-density one. In other words, the dimmings observed in $\omega$ CMa would be partial, rather than full, disk dissipations. 

\begin{figure}[!t]
    \centering
    \includegraphics[width=0.9\linewidth]{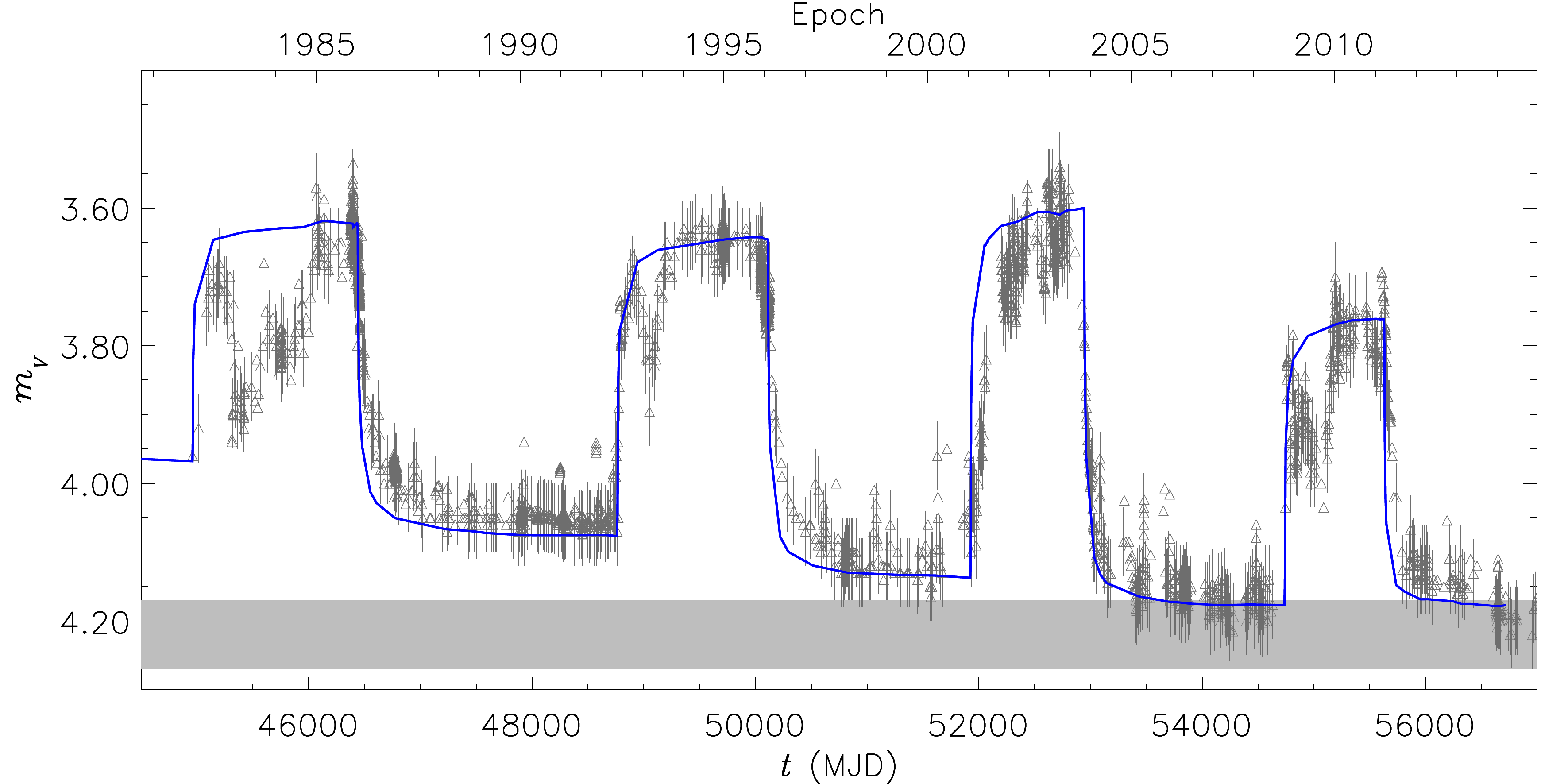}
    \caption[The third scenario for modeling the full light curve of $\omega$ CMa]{Same as Fig.~\ref{fig:first_model} for the third scenario that now assumes a non-zero mass feeding rate even during the quiescent phases.}
    \label{fig:third_model}
\end{figure}

The results of the third scenario are displayed in Fig.~\ref{fig:third_model}. The assumption of a non-zero disk feeding rate even during quiescence has greatly improved the overall fit to the data, as it allowed adjusting the asymptotic magnitude of each quiescence phase. However, two key issues still remain in the model, namely the imperfect reproduction of the outburst structure (fault 1) and the incorrect rate of magnitude variations in several sections of the light curve (see, for instance, the dissipation phases of the first and second cycles; fault 3).


\subsection {The fourth scenario for modeling the full light curve of $\omega$ CMa}
\label{subsect:fourth_model}

\begin{figure}[!t]
    \centering
    \includegraphics[width=0.9\linewidth]{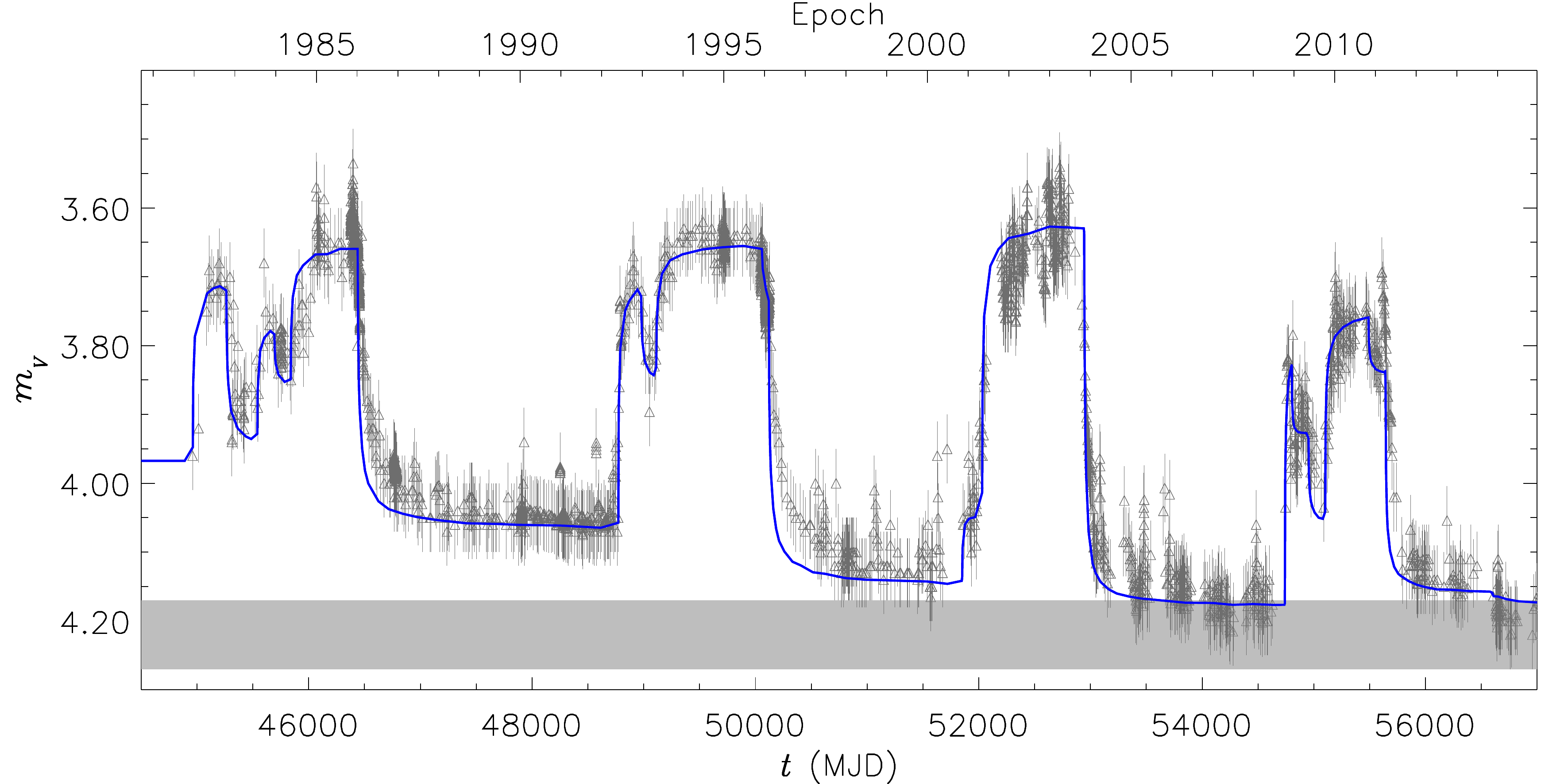}
    \caption[The fourth scenario for modeling the full light curve of $\omega$ CMa]{Same as Fig.~\ref{fig:first_model} for the fourth scenario.}
    \label{fig:fourth_model}
\end{figure}

The analysis of the light curve shown in Chap.~\ref{chap:exp_formula} allowed us to identify substructures within each of the outburst scenarios, composed of several small-scale outbursts and quiescences on top of the large-scale ones. As a refinement of the models shown in Figs.~\ref{fig:second_model} and~\ref{fig:third_model} we included these substructures to our model. The results, shown in Fig.~\ref{fig:fourth_model}, show that this inclusion led to a much better overall fit to the data. However, one issue remains with the model: even though the amplitudes in the light cruve are now well reproduced, the models still fail to reproduce the rate of magnitude variations (fault 3).


\section {Final model to fit the $V$-band photomettric data of $\omega$ CMa}
\label{sect:final_model}

In order to address the issue of the mismatch between the observed rates of magnitude variations and the model, our final ingredient was to assume that the gas viscosity varies with time. This possibility have been hinted already by the results shown in Fig.~\ref{fig:slope_var_rate} that showed that $\tau$ was quite variable for each different subsection.

In the following, we review all our model assumptions made so far, and then provide a detailed comparison between our final model and each observed cycle.

The final model assumptions are:

\begin{itemize} 
    \item The sections defined in Table~\ref{table:results} are modeled individually and in temporal succession. 
    \item For each section of the light curve, the free parameters are the viscosity parameter $\alpha$, the disk mass injection rate, parameterized by $\dot{M}_\mathrm{inj}$ (or alternatively $-\dot{J}_{*,\mathrm{std}}$), and the starting time of the section.
    \item During dissipation, $\dot{M}_\mathrm{inj}$ ($-\dot{J}_{*,\mathrm{std}}$) may have a non-zero value.
    \item Since the observations prior to 1982 are scarce, we did not attempt to model them. However, as observations clearly indicate the presence of a disk prior to 1982 (see Fig.~\ref{fig:full_lightcurve}), the model starts with a steady-state disk followed by a 5-year long dissipation that ends when O1 begins. The choice of 5-year dissipation is justified as this is the typical length of the observed dissipations. The assumed AM injection rate for this previous phase is consistent with the value assumed for the first cycle.
    \item \cite{maintz2003} did not provide estimates for the uncertainties in the stellar parameters and although variations in the stellar parameters might affect our results, we trust their values for modeling simplification. Recall that the uncertainties in the stellar visual brightness (grey bands in Fig.~\ref{fig:first_model} to~\ref{fig:vdd_model_c4}) were estimated from the $1\sigma$ uncertainties in the parallax measurements (see Chap.~\ref{chap:omecma_obs}).
\end{itemize} 

\subsection {Results for the first cycle}
\label{subsect:vdd_c1}

Modeling started by considering a single value of $\alpha = 1.0$ for the first part of O1. 
Other possible values of $\alpha$ were not examined because the data for this part are very sparse. This small-scale outburst (O1$_\mathrm{o1}$) was modeled as the continuation of a 5-year long quiescence that started with a steady-state disk (see above).

\begin{figure}[!t]
    \begin{minipage}{1.0\linewidth}
        \centering
        {\includegraphics[width=0.8\linewidth, height=0.4\linewidth]{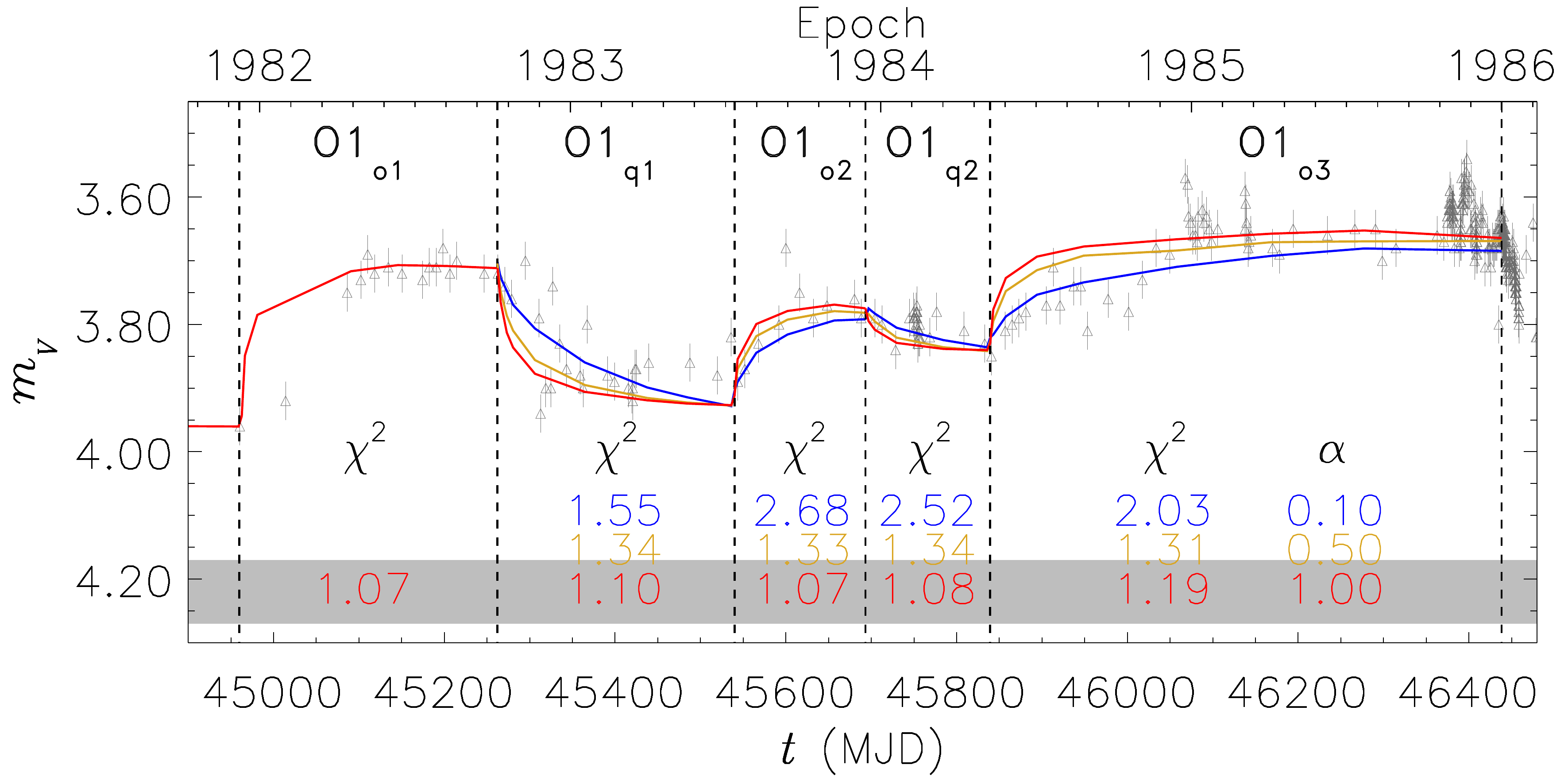}}
    \end{minipage}\par\medskip
    \begin{minipage}{1.0\linewidth}
        \centering
        {\includegraphics[width=0.8\linewidth, height=0.4\linewidth]{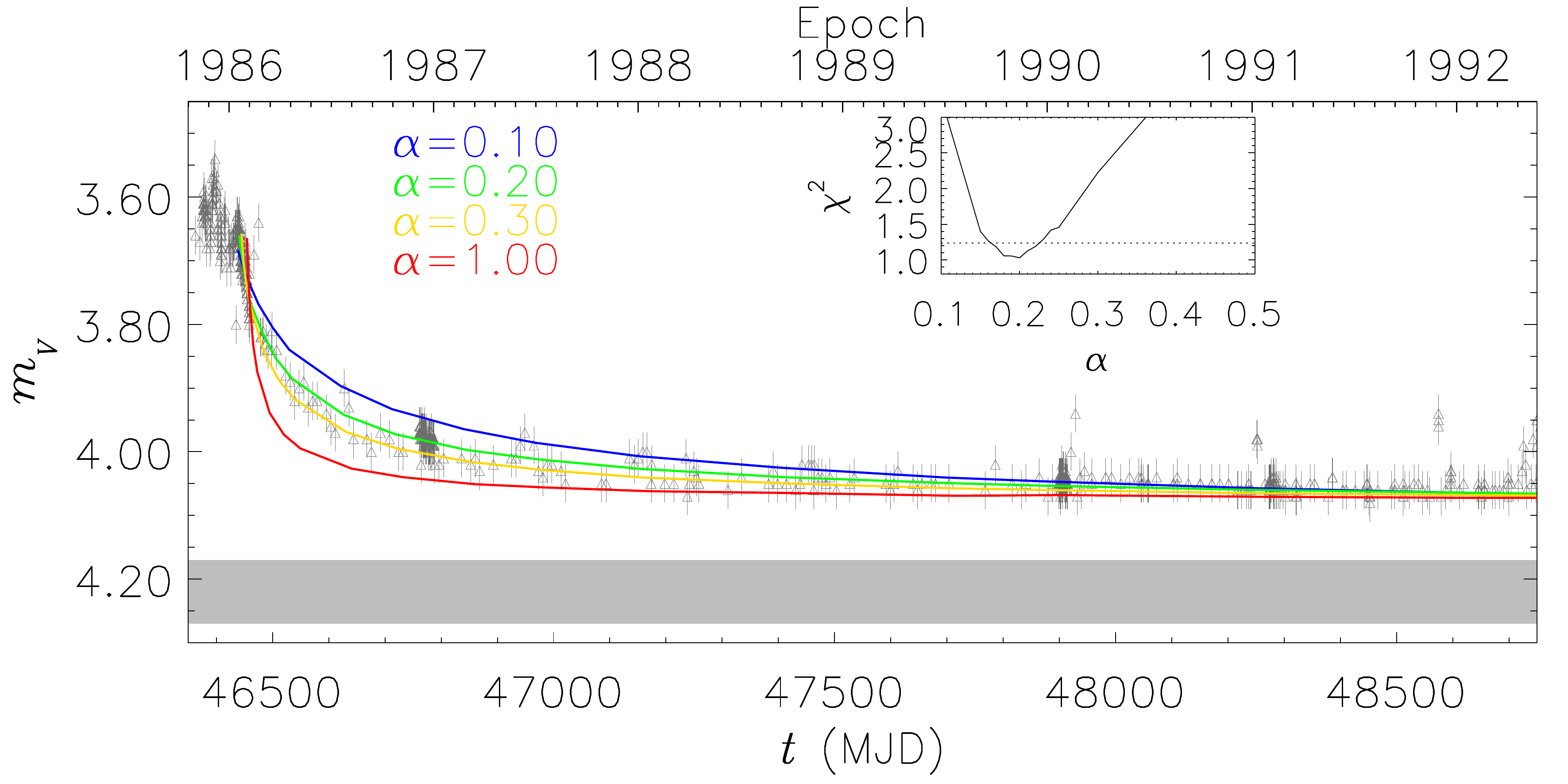}}
    \end{minipage}
    \caption[$V$-band light curve of $\omega$ CMa for the first cycle]{
    $V$-band light curve of $\omega$ CMa for the first cycle (points). 
    {\it Top}: Models for O1 for three different values of $\alpha$: 0.10, 0.50 and 1.00, as indicated. 
    The $\chi^2_\mathrm{red}$ for each model and each section the light curve is indicated.
    {\it Bottom}: Model for Q1 for four values of $\alpha$: 0.10, 0.20, 0.30, and 1.00, as indicated. The estimated best-fit value of $\alpha = 0.20 \pm 0.03$. The $\chi^2_\mathrm{red}$ for different values of $\alpha$ is shown in the inset and the 90\% confidence level is indicated with the horizontal dotted line. The horizontal grey band represents the estimated intrinsic visual magnitude of the central star of $\omega$ CMa.
    }
    \label{fig:vdd_model_c1}
\end{figure}

For the remaining sections of O1, which alternate build-up with dissipation phases, we explored three different values of $\alpha$ (0.10, 0.50, and 1.00 -- Fig.~\ref{fig:vdd_model_c1}, top panel). For the small-scale dissipation phases, the criterium used was to favor models that matched better the lower points in the light curves, as the higher points likely result from small-scale, poorly-sampled outbursts (flickers) on top  of the dissipation. The reduced chi-squared values, $\chi^2_\mathrm{red}$, for each model, shown in the plot, clearly indicate that the $\alpha=1.0$ model better fits all sections of O1. A complex AM injection rate is necessary to reproduce the general behavior of the light curve in O1. A graphical representation of the AM injection rate is shown in the third panel of Fig.~\ref{fig:summary}.

The dissipation phase that followed O1 (Q1) has a much smoother shape, with only a handful of small-scale flickers, which greatly facilitates the modeling (recall that all flickers vanish when the data is smoothed out with a 30-day bin, see Fig.~\ref{fig:exp_fitting}). It is possible that the smoother shape during dissipation is simply a result of the lower mass injection rate: when the star is less active, its mean mass injection rate is also lower, and, as a result, the disk dissipates. We determined that $\dot{M}_\mathrm{inj}$ decreases from  $3.4\,\times\,10^{-7}\,\mathrm{M_{\odot}\,yr^{-1}}$ at the end of O1 to $0.2\,\times\,10^{-7}\,\mathrm{M_{\odot}\,yr^{-1}}$ at the end of Q1 (alternatively, $-\dot{J}_{*,\mathrm{std}}$ decreases from  $4.9\,\times\,10^{36}\,\mathrm{g\,cm^2\,s^{-2}}$ to $2.1\,\times\,10^{35}\,\mathrm{g\,cm^2\,s^{-2}}$), causing an overall density decrement (see Fig.~\ref{fig:summary}) that explains the observed decrease in the disk excess ($\Delta V = 0.39$). We found a best-fit value of $\alpha = 0.20 \pm 0.03$, shown in the bottom panel of Fig.~\ref{fig:vdd_model_c1}, along with models for other values of $\alpha$ for comparison. The inset shows the $\chi^2_\mathrm{red}$ of the fit as a function of $\alpha$. The best fit value was determined from the minimization of the $\chi^2_\mathrm{red}$ value, for which we adopted the following definition:
\begin{equation}
    \label{eq:chisquare}
    \chi^2_\mathrm{red} = \displaystyle\sum_{i=1}^{N} \frac{(m\mathrm{^{obs}_{v,\it{i}}} -m\mathrm{^{mod}_{v,\it{i}}})^2}{\sigma^2_{i}(N-2)},
\end{equation}
where $m\mathrm{^{obs}_{v,\it{i}}}$ and $m\mathrm{^{mod}_{v,\it{i}}}$ are the observed and model magnitude, respectively, $\sigma_{i}$ is the observed magnitude error, and $N$ is the total number of data points that were fitted. The best $\alpha$ is the one that minimizes the $\chi^2_\mathrm{red}$ function, which was computed for a grid with a step size of 0.01. The uncertainty of this value was estimated from the $\Delta\chi^2_\mathrm{red, 0.90} = \chi^2_\mathrm{red} - \chi^2_\mathrm{red, min}$ intersections, where $\Delta\chi^2_\mathrm{red, 0.90}$ is a function of the number of degrees of freedom of the fit for a level of confidence of 90\% \citep[see Chapter~11 of][]{bevington1992}. For all the cases, the derived uncertainty value was found to be larger than the adopted step size, and therefore the $\chi^2_\mathrm{red}$ function was sufficiently well sampled.


\subsection {Results for the second cycle}
\label{subsect:vdd_c2}

\begin{figure}[!t]
    \begin{minipage}{1.0\linewidth}
        \centering
        {\includegraphics[width=0.8\linewidth, height=0.4\linewidth]{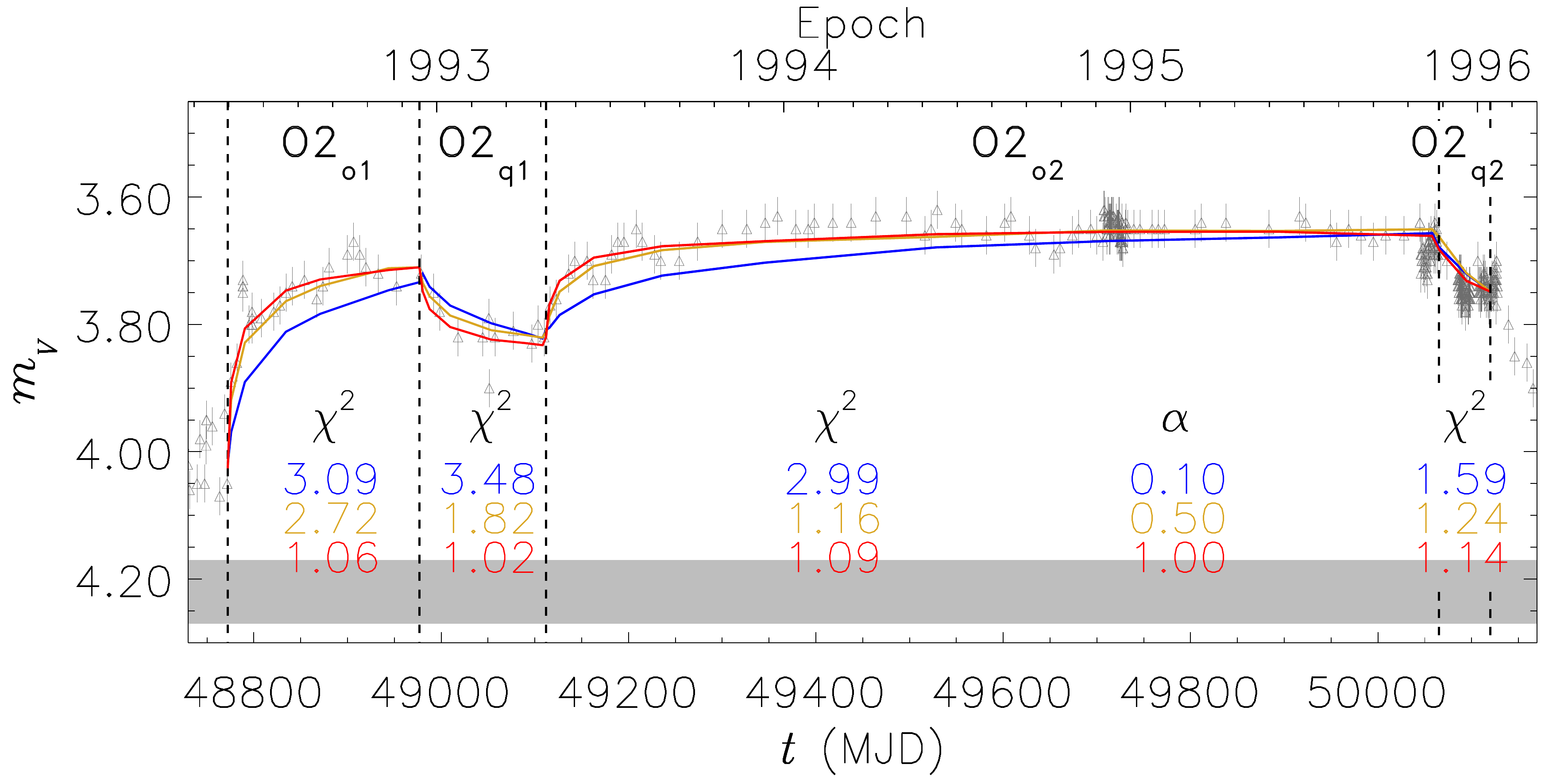}}
    \end{minipage}\par\medskip
    \begin{minipage}{1.0\linewidth}
        \centering
        {\includegraphics[width=0.8\linewidth, height=0.4\linewidth]{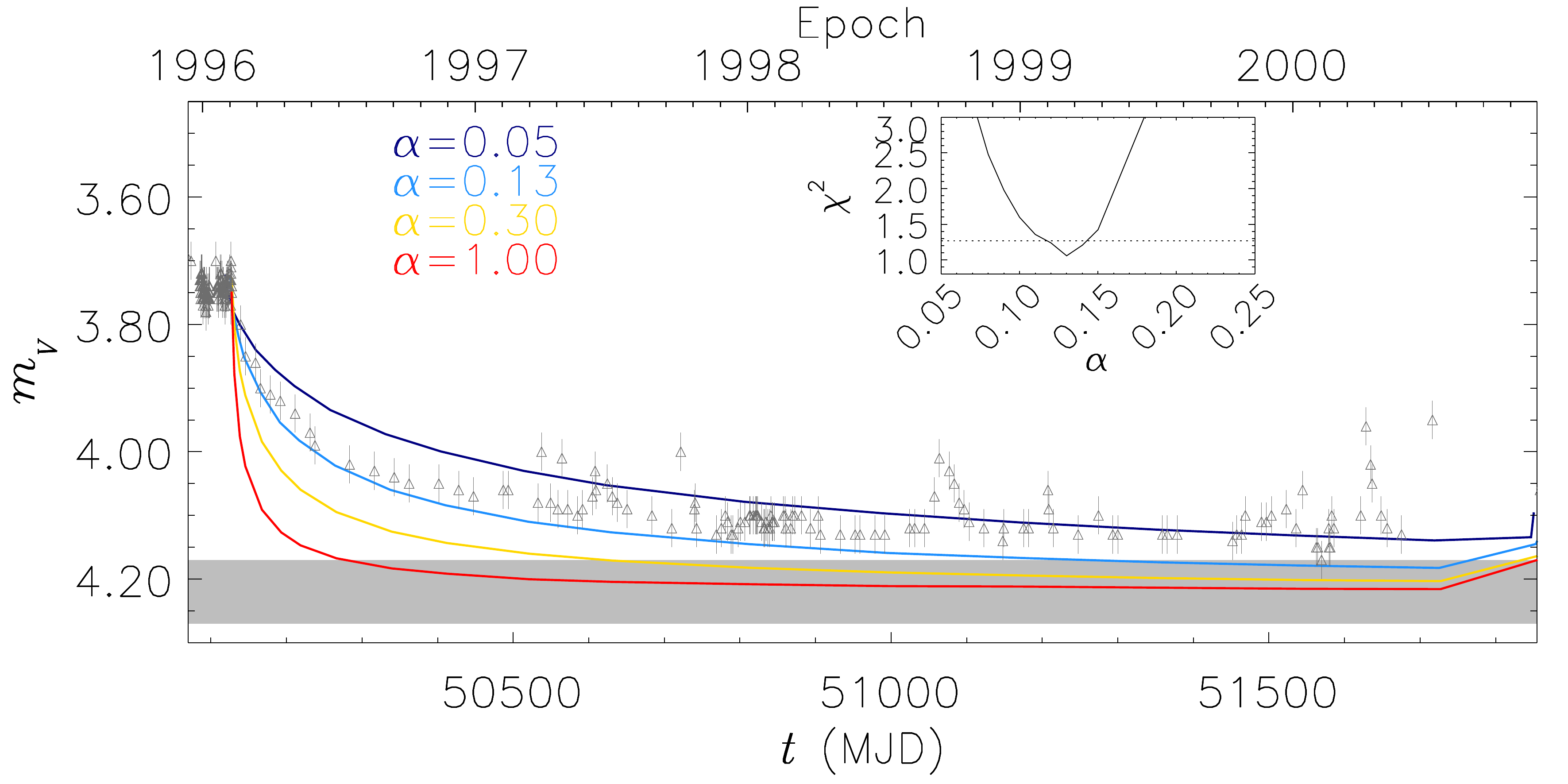}}
    \end{minipage}
    \caption[$V$-band light curve of $\omega$ CMa for the second cycle]{Same as Fig.~\ref{fig:vdd_model_c1} for the second cycle. The best-fit value for the outburst is $\alpha = 1.0$ and for the dissipation, $\alpha = 0.13 \pm 0.01$.}
    \label{fig:vdd_model_c2}
\end{figure}

The model fitting of the second cycle followed the same procedure as presented for the previous cycle. 
In many ways, the second cycle is similar to the first one. As before, O2 is composed of alternating phases of build-up and dissipation, which implies a complex disk feeding history (see Fig.~\ref{fig:vdd_model_c2}). A short dissipation phase (O2$_\mathrm{q2}$) precedes the long dissipation phase Q2.

Again, a value of $\alpha = 1.0$ seems to best represent the entire O2 phase. For Q2, we estimate $\alpha = 0.13 \pm 0.01$ and $\dot{M}_\mathrm{inj} = 4 \times 10^{-10}\,\mathrm{M_{\odot}\,yr^{-1}}$ ($-\dot{J}_{*,\mathrm{std}} = 5.8 \times 10^{33}\,\mathrm{g\,cm^2\,s^{-2}}$).


\subsection {Results for the third cycle}
\label{subsect:vdd_c3}

\begin{figure}[!t]
    \begin{minipage}{1.0\linewidth}
        \centering
        {\includegraphics[width=0.8\linewidth, height=0.4\linewidth]{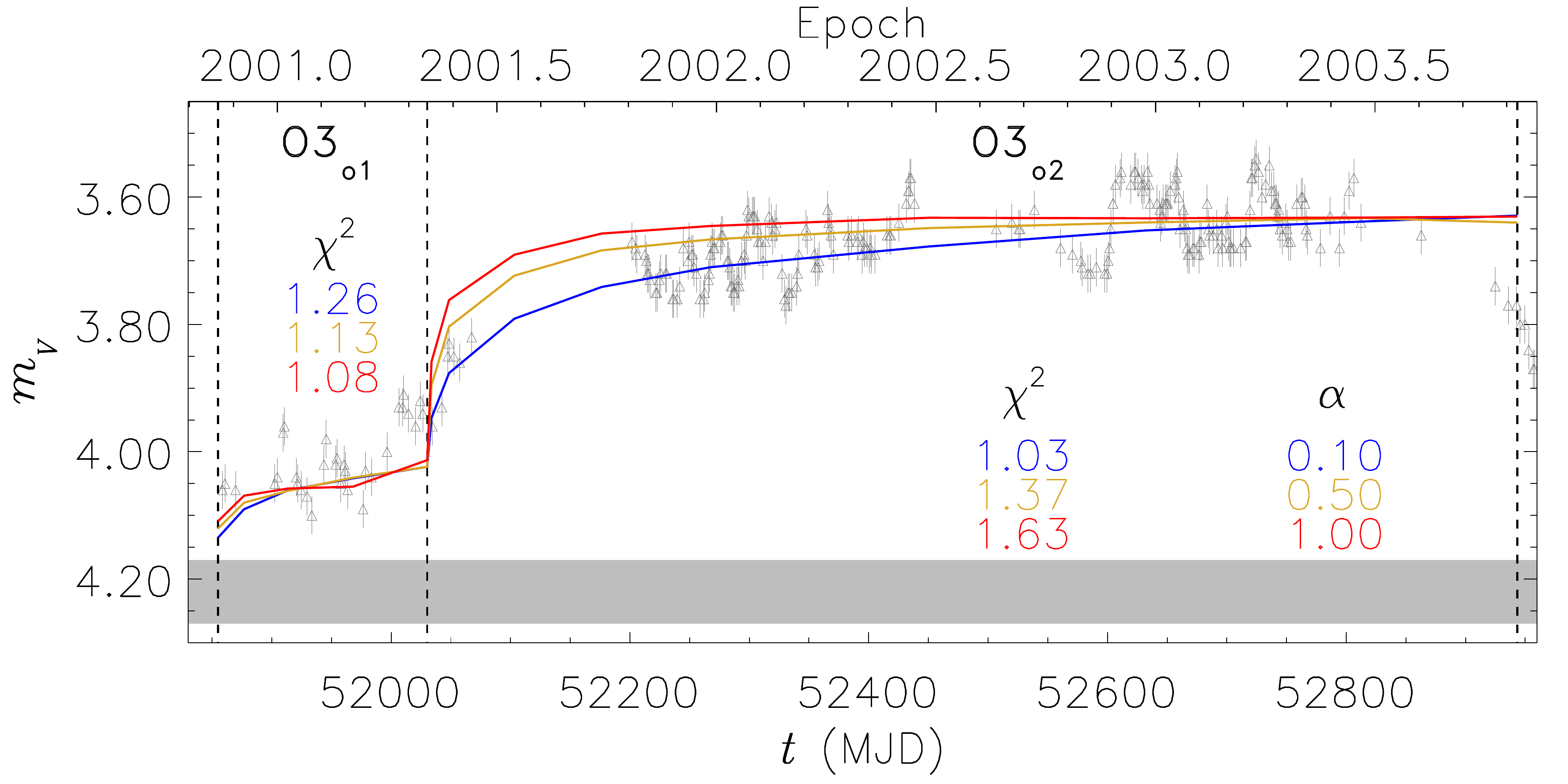}}
    \end{minipage}\par\medskip
    \begin{minipage}{1.0\linewidth}
        \centering
        {\includegraphics[width=0.8\linewidth, height=0.4\linewidth]{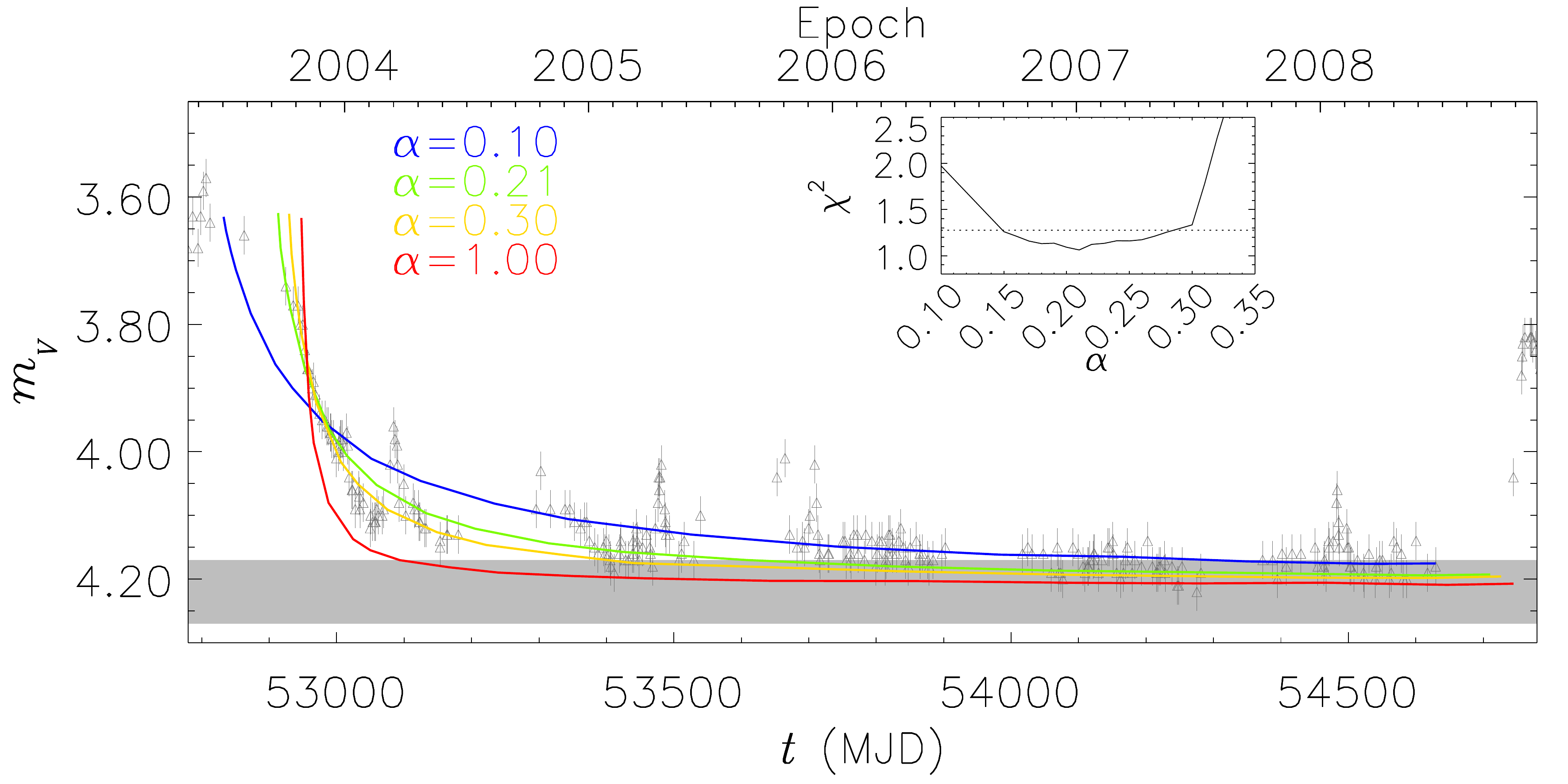}}
    \end{minipage}
    \caption[$V$-band light curve of $\omega$ CMa for the third cycle]{Same as Fig.~\ref{fig:vdd_model_c1} for the third cycle. The best fit values were $\alpha = 1.0$ for O3$_\mathrm{o1}$, $\alpha = 0.1$ for O3$_\mathrm{o2}$, and  $\alpha = 0.21 \pm 0.05$ for Q3.
    }
    \label{fig:vdd_model_c3}
\end{figure}

For the build-up phases of the third cycle (O3), we again explored $\alpha$ values of 0.10, 0.50 and 1.00 (Fig.~\ref{fig:vdd_model_c3}, top panel). The minimum $\chi^2_\mathrm{red}$ values indicated that $\alpha=1.00$ fits better the O3$_\mathrm{o1}$ section, while $\alpha=0.10$ provides a better fit for the O3$_\mathrm{o2}$ section. However, we believe this last value to be of little statistical significance, as the initial phase of O3$_\mathrm{o2}$, when the largest brightness variations happen, is poorly sampled. Furthermore, O3$_\mathrm{o2}$ displays a quite complex behavior, with many small-scale flickers.
As was the case for O1, a complex AM injection rate is needed to explain the photometric behavior of O3, with $\dot{M}_\mathrm{inj}$ ranging between $4.0 \times 10^{-8}\mathrm{M_{\odot}\,yr^{-1}}$ and $3.7 \times 10^{-7}\,\mathrm{M_{\odot}\,yr^{-1}}$ (corresponding to $-\dot{J}_{*,\mathrm{std}}$ ranging between $6.1 \times 10^{35}\,\mathrm{g\,cm^2\,s^{-2}}$ and $5.4 \times 10^{36}\,\mathrm{g\,cm^2\,s^{-2}}$).


Q3 is of particular relevance because it was previously studied by \cite{carciofi2012} who found $\alpha = 1.0 \pm 0.2$, which is about 5 times larger than our result of $\alpha=0.21\pm 0.05$ (Fig.~\ref{fig:vdd_model_c3}, bottom panel). This large discrepancy deserves a careful examination.

\begin{figure}[!t]
    \centering
    \includegraphics[width=0.90\linewidth, height=0.35\linewidth]{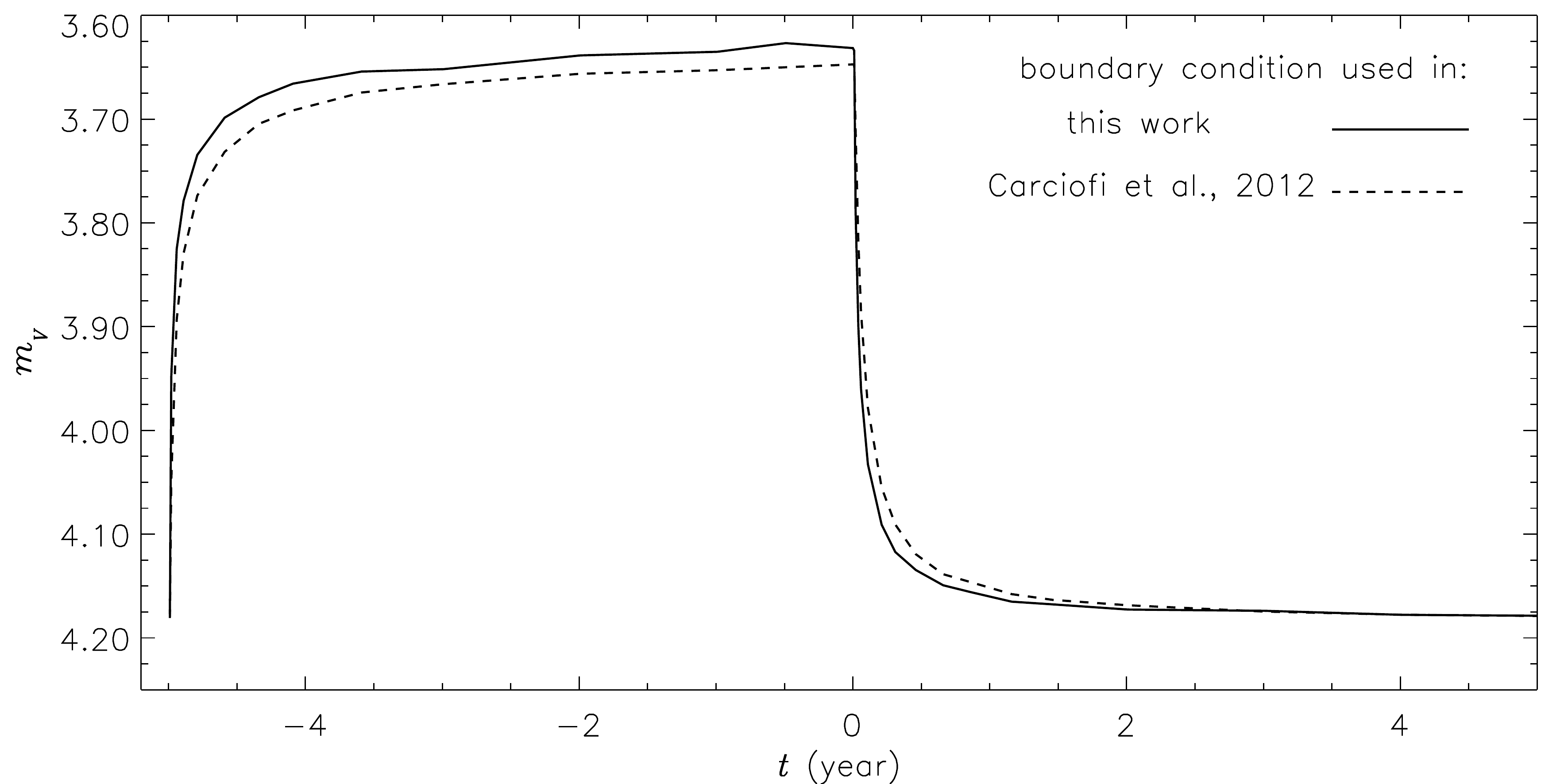}
    \caption[Effect of boundary conditions in the {\tt SINGLEBE} code]{Effect of boundary conditions in the {\tt SINGLEBE} code. The dashed line represents the model in which the inner boundary condition is inside the star (used by \citealt{carciofi2012}) and the solid line shows the model, used in this work, in which the boundary condition is at the stellar equator. The boundary conditions of \cite{carciofi2012} result in faster disk dissipation when compared with the boundary conditions used in this work.}
    \label{fig:boundary}
\end{figure}

One reason behind the discrepancy between our results for Q3 and \cite{carciofi2012}'s is the different boundary conditions used in {\tt SINGLEBE}: \cite{carciofi2012} assumed $\bar{r}_\mathrm{inj} = 1$ and that the inner boundary condition was inside the star (at $r=0.85\,R_\mathrm{eq}$), while we assumed {$\bar{r}_\mathrm{inj} = 1.02$ and that the inner boundary condition is at the stellar equator}. Figure~\ref{fig:boundary} illustrates the effect of this change in the boundary condition, by comparing two equivalent models with the two different boundary conditions. The boundary conditions used in this work result in a faster disk dissipation, which in turn means that the  $\alpha$ required to match the observed rate of dissipation is smaller. However, this is a small effect, and can explain differences of only $\sim$30$\%$ between the results.
 
The main reason for the discrepancy between the results of \cite{carciofi2012} and ours is the mass reservoir effect (Sect.~\ref{sect:mass_reservoir}). \cite{carciofi2012} modeled Q3 by assuming a long previous build-up phase. Therefore, their model overestimated the disk mass at the onset of Q3, which, in turn, caused $\alpha$ to be overestimated. Our results do not suffer from this issue, as we properly took the previous history into account.


\subsection {Results for the fourth cycle}
\label{subsect:vdd_c4}

\begin{figure}[!t]
    \begin{minipage}{1.0\linewidth}
        \centering
        {\includegraphics[width=0.8\linewidth, height=0.4\linewidth]{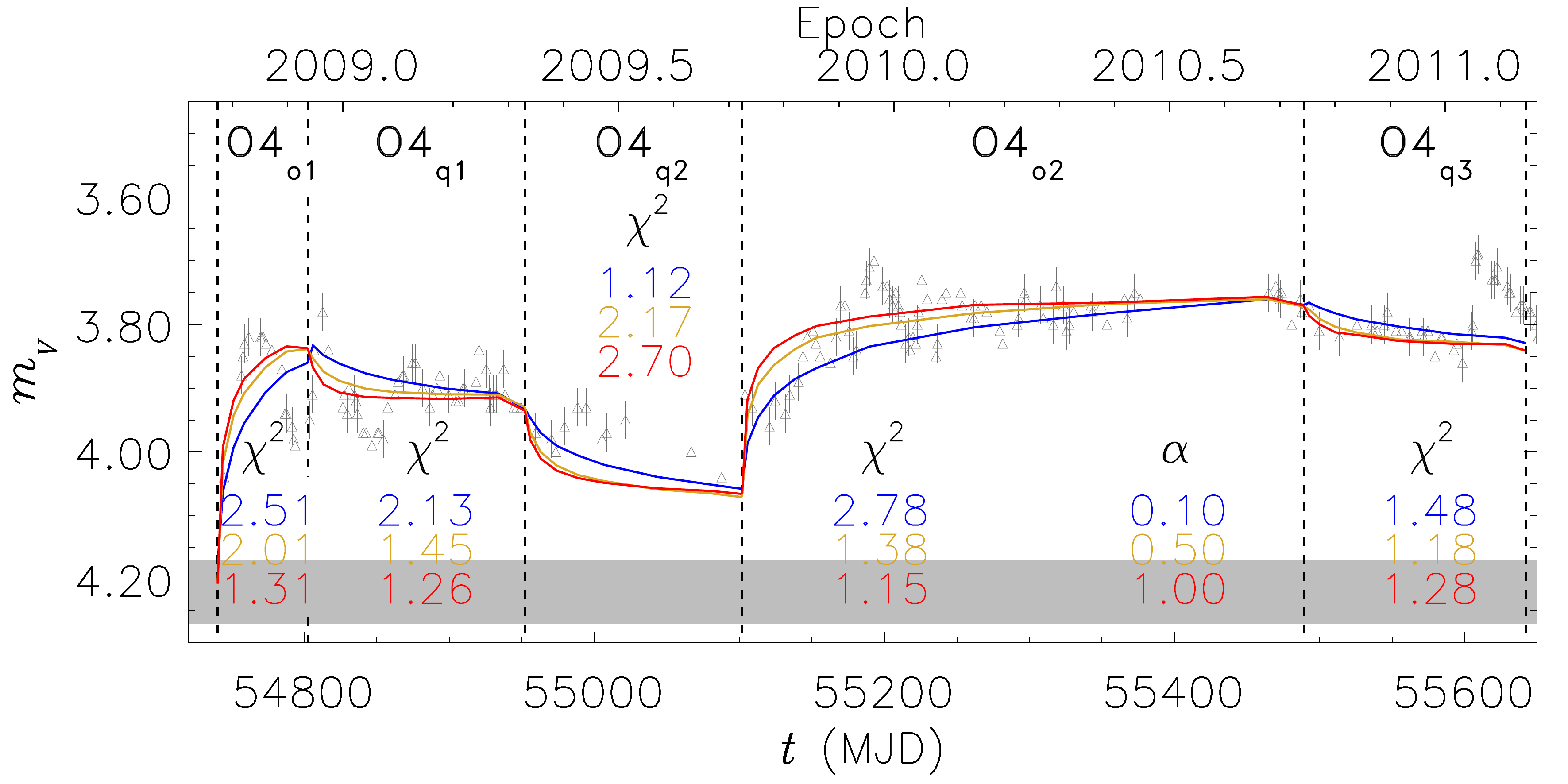}}
    \end{minipage}\par\medskip
    \begin{minipage}{1.0\linewidth}
        \centering
        {\includegraphics[width=0.8\linewidth, height=0.4\linewidth]{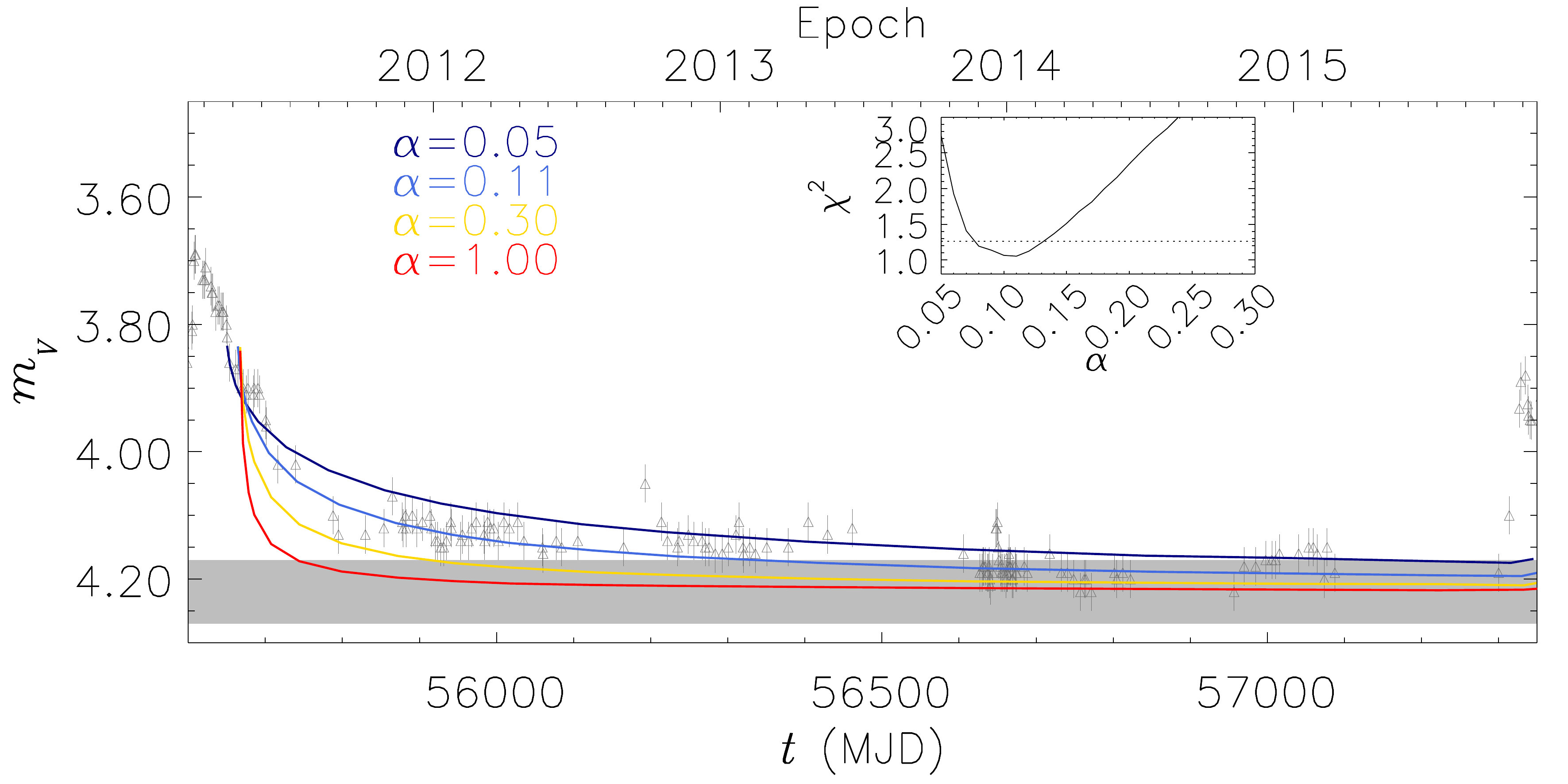}}
    \end{minipage}
    \caption[$V$-band light curve of $\omega$ CMa for the fourth cycle]{Same as Fig.~\ref{fig:vdd_model_c1} for the fourth cycle. The best fit for O4 was obtained with $\alpha$ = 1.00 for the first, second, and fourth sections and $\alpha$ = 0.10 and 0.50 for third and fifth sections. For the dissipation phase, $\alpha = 0.11 \pm 0.03$.}
    \label{fig:vdd_model_c4}
\end{figure}

As before, to study O4 we explored three values of $\alpha$ for each section. O4 displays the most complex behavior of all the outbursts, with rapid switches between brightenings and fadings. As a result of our modeling assumptions, according to which we do not attempt to model phases shorter than 60 days, the model fails at reproducing the detailed behavior of the light curve. Figure~\ref{fig:vdd_model_c4}, top panel, shows how each model compares with the data. Formally, the $\chi^2_\mathrm{red}$ values of each section suggest that $\alpha$ may be varying during the outburst from 0.1 to 1.0.

However, we believe these results should be viewed with some caution, as they may depend on the particular choice for the beginning and end of each section. The important point to emphasize is that the modeling, irrespective of the particular choice of $\alpha$ reproduces the general behavior of the light curve, which is enough to estimate the total disk mass at the end of O4. As previously discussed, knowing the total disk mass at the end of the outburst is necessary to properly model the subsequent dissipation phase. The complex behavior of O4 is suggestive of a quite complex AM injection history, as shown in Fig.~\ref{fig:summary}.

The model for Q4 results in $\alpha = 0.11 \pm 0.01$ and $\dot{M}_\mathrm{inj} = 2.0 \times 10^{-10}\,\mathrm{M_{\odot}\,yr^{-1}}$ (equivalent with $-\dot{J}_{*,\mathrm{std}} = 2.8 \times 10^{33}\,\mathrm{g\,cm^2\,s^{-2}}$). It is interesting to note that even though small, this mass injection rate (or $-\dot{J}_{*,\mathrm{std}}$) is required to reproduce the slight excess still present at the end of Q4 under the hypothesis of reliable stellar parameters.


\section {An alternate scenario to explain the secular decrease in brightness}
\label{sect:other_models}

As mentioned in Chap.~\ref{chap:omecma_obs}, the secular decrease in brightness that $\omega$ CMa displayed in the last 4 cycles has remained a mystery. Our model for the light curve proposes one explanation for this behavier where it is assumed that instead of entering a true quiescence (i.e., zero mass loss rate), the star shifts between higher (outburst) and lower (quiescence) mass loss rates.

Because such scenario has never been reported in the literature before, we attempted to find an alternate explanation for the secular fading. One model we explored, in particular, involved a radially decreasing $\alpha$. The idea behind this model is that it would form, with time, a huge mass reservoir that would be able to feed the disk for longer time once quiescence started, thus giving the appearance of a partial dissipation. 

\begin{figure}[!t]
    \begin{minipage}{1.0\linewidth}
        \centering
        \subfloat{\includegraphics[width=0.8\linewidth, height=0.4\linewidth]{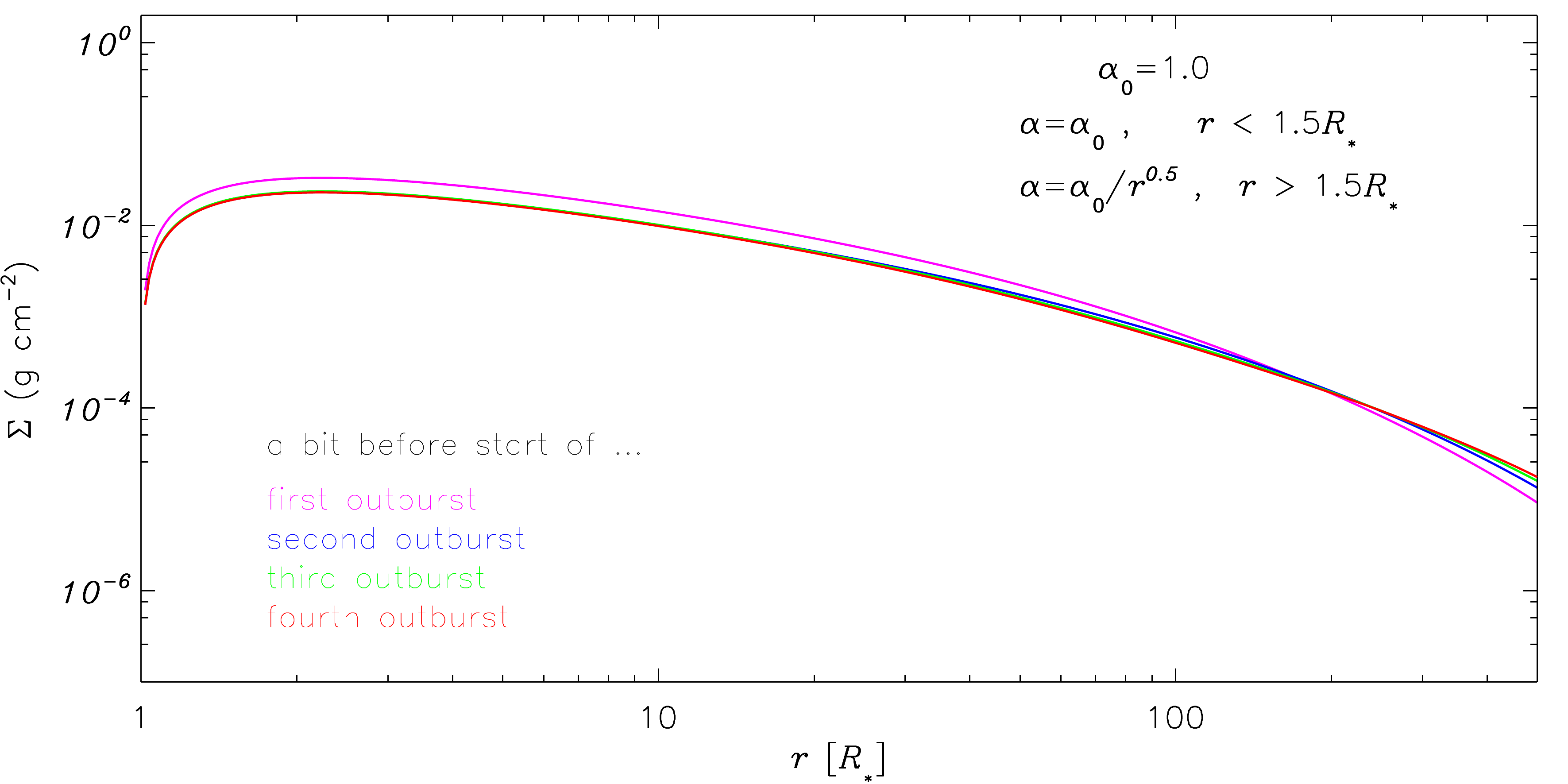}}
    \end{minipage}\par\medskip
    \begin{minipage}{1.0\linewidth}
        \centering
        \subfloat{\includegraphics[width=0.8\linewidth, height=0.4\linewidth]{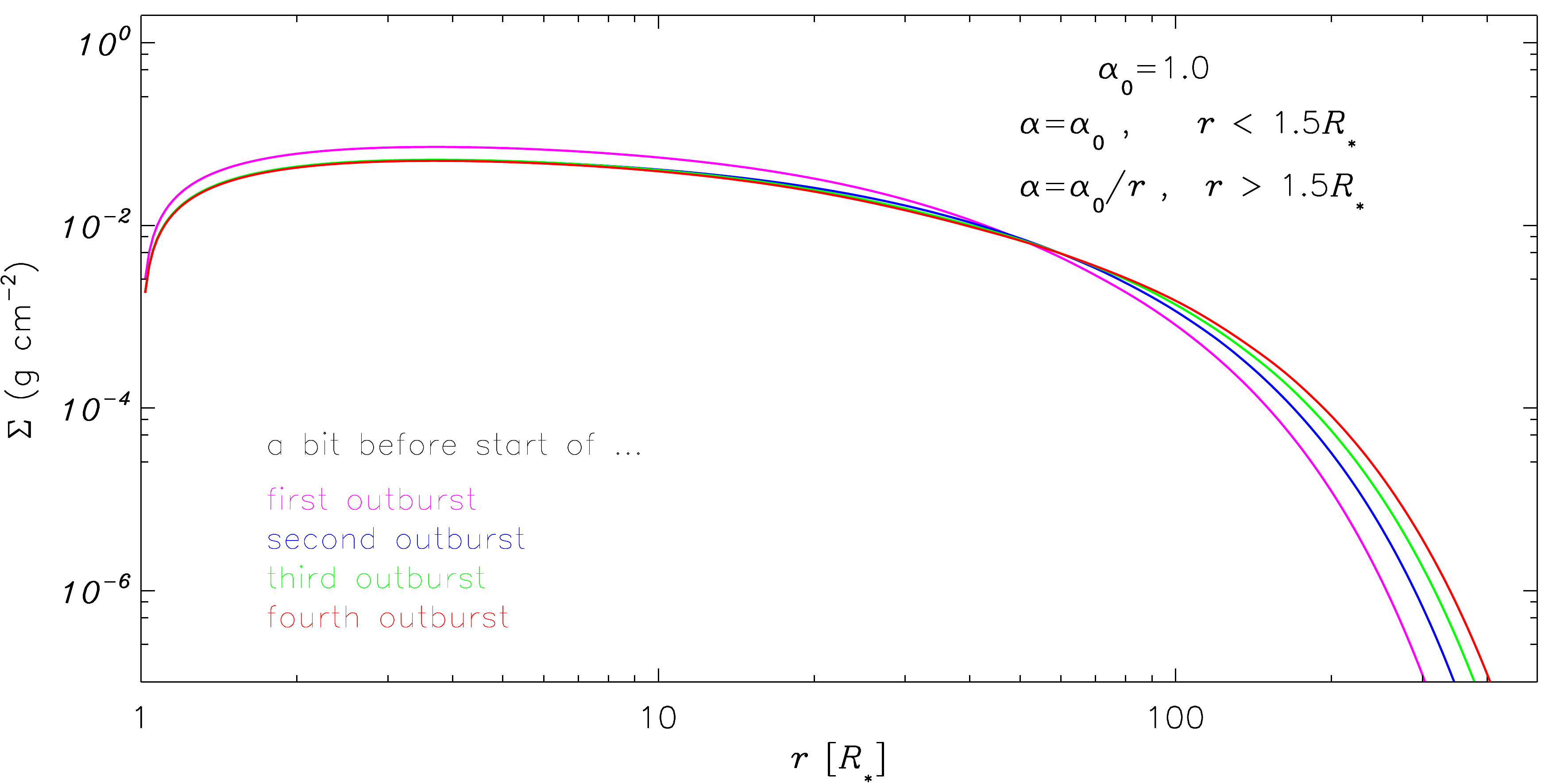}}
    \end{minipage}
    \caption[Seeking to the other models]{Surface density versus radius at the beginning of each cycle for two different scenarios in which $\alpha$ varies with radius. Top: $\alpha \propto r^{-0.5}$, and bottom: $\alpha \propto r^{-1.0}$. None of these scenarios could explain the decline in the brightness of the system at the end of each successive dissipation phase.}
    \label{fig:other_models}
\end{figure}

We explored several such scenarios, for instance one scenario in which $\alpha \propto r^{-0.5}$ and another for which $\alpha \propto r^{-1.0}$. Detailed hydrodynamical calculations (Fig.~\ref{fig:other_models}) shows that while a radially decreasing $\alpha$ indeed helps forming a higher mass reservoir in the outer disk, the impact on the surface density of the inner part is rather low. As a result, and because the $V$-band is formed in the inner part of the disk (Fig.~\ref{fig:flux_loci}), a radially varying $\alpha$ had no relevant impact on the light curve and, therefore, cannot explain the secular fading seen in $\omega$ CMa.


\section{The life cycles of $\omega$ CMa's disk}
\label{sect:life_cycle}

\begin{figure}[!t]
    \centering
    {\includegraphics[width=1.0\linewidth]{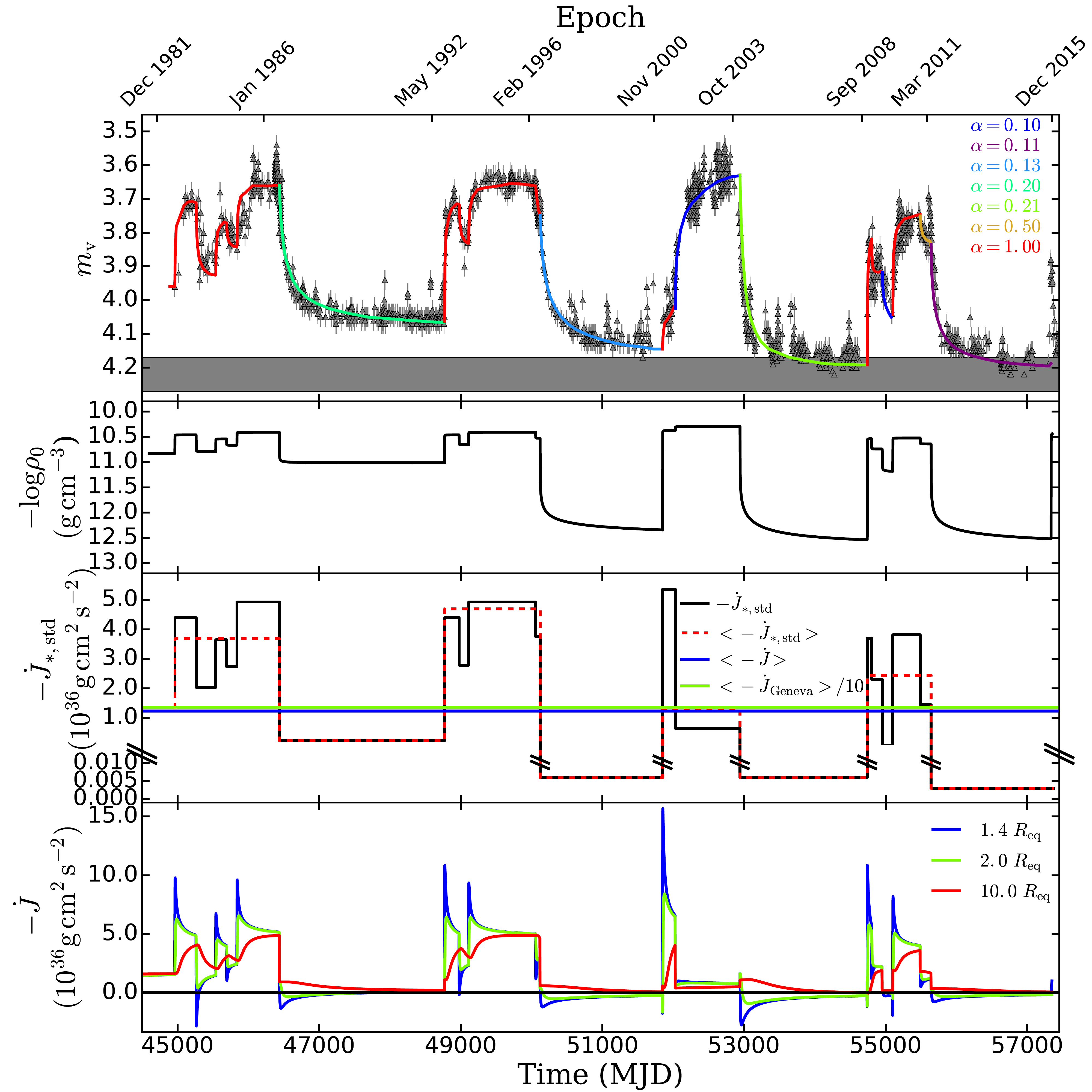}}
    \caption[The summary of the first part of this work]{
    Upper panel: Model fit of the full light curve of $\omega$ CMa. Each colored solid line represents an individual value for the $\alpha$ parameter, as indicated.
    Second panel: Time evolution of the mass density at the base of the disk ($\rho_0$).
    Third panel: The history of $-\dot{J}_{*,\mathrm{std}}$ of the best fit model (the black lines). The dashed red lines display the mean AM loss rate of the star at each phase. The blue line shows the mean AM loss rate of the star during 34 years. The green line shows ten percent of the mean AM loss during 34 years according to \citet{granada2013}. The ordinate is broken for better presentation of the low values against the high ones. 
    Fourth panel: AM flux as a function of time, at different positions in the disk. The blue, green and red lines correspond to $r = 1.4\,R_\mathrm{eq}$, $r = 2.0\,R_\mathrm{eq}$, $r = 10.0\,R_\mathrm{eq}$, respectively.
    }
    \label{fig:summary}
\end{figure}

In Sect.~\ref{sect:final_model} we presented the first physical model (VDD model) to fit the light curve of a Be star including several formation and dissipation phases. It was necessary to consider varying the $\alpha$ parameter to achieve a satisfactory result. The top panel of Fig.~\ref{fig:summary} summarizes our model results. The light curve was fitted using higher values of $\alpha$ for formation phases while the dissipation phases needed lower values (see Sect.~\ref{sect:temperature_evolution}). Moreover, we found that $\alpha$ is not related to the cycle to cycle variations, and is not increasing nor decreasing continuously. The separation between the horizontal grey band (expressing the intrinsic magnitude of the star) and the model curves suggests that $\omega$ CMa never experiences a true quiescence, but instead switches between a high-density phase (outburst) and a low-density one (dissipation). A true quiescence may only have been reached at the end of the last cycle.

The estimated values of $\alpha$ are in rough agreement with the values derived for dwarf novae \citep{king2007, kotko2012}, as well as with the values measured for the SMC by \citet{rimulo2018}. They are, however, an order of magnitude or more above the usual values obtained in magnetohydrodynamic simulations that employ the 
magnetorotational instability \citep{balbus1991} as a possible mechanism to explain the viscosity.

Some studies in the literature \citep[e.g.,][]{touhami2011, vieira2017} use the following power-law approximation for the disk density:
\begin{equation}
    \label{eq:volume_density}
    \rho = \rho_\mathrm{0}\left(\frac{r}{R_\mathrm{eq}}\right)^{-n}\,,
\end{equation}
where $\rho_\mathrm{0}$ is the volume density at the inner rim of the disk (base density). The density slope $n$ varies between 1.5 -- 5.0 with a statistical peak probability around 2.4 \citep{vieira2017}. Studies of individual stars reported values between 1.5 to 4.2  for $n$ and $7 \times 10^{-13}$ to $4.5 \times 10^{-10}$ $\mathrm{g\,cm^{-3}}$ for $\rho_0$ (e.g., \citealt{carciofi2006b}, \citeyear{carciofi2007}, \citeyear{carciofi2009}, \citealt{gies2007}, \citealt{jones2008}, \citealt{tycner2008}, \citealt{klement2015}, and \citealt{vieira2015}). The values of $\rho_0$ reported in the literature can be compared to the ones obtained for $\omega$ CMa in the second panel of Fig.~\ref{fig:summary}. We found that $\rho_{0}$ is varying between $2.9 \times 10^{-13}$ to $5.0 \times 10^{-11}\ \mathrm{g\,cm^{-3}}$ which is almost two orders of magnitude in range. 


The life cycles of Be star disks were discussed by \cite{vieira2017}, who modeled the infrared emission of $80$ objects, at up to three different epochs (\textit{IRAS}, \textit{WISE} and \textit{AKARI} data). By adopting a simple power-law prescription for the disk density profile, they estimated the $\rho_0$ and $n$ values for this large sample. Their results suggest a strong correlation between these parameters that were interpreted in terms of the time-dependent VDD properties. For that purpose, they computed representative evolutionary tracks over the $n-\log\rho_0$ diagram, representing scenarios such as a long disk build-up followed by a full disk dissipation, and cyclic cases where build-up and dissipation alternate on a regular basis. Their results suggested an evolutionary interpretation of the disk as forming, steady-state or dissipating in different parts of the $n-\log\rho_0$ diagram.

\begin{figure}[!t]
    \centering
    \includegraphics[width=0.75\linewidth]{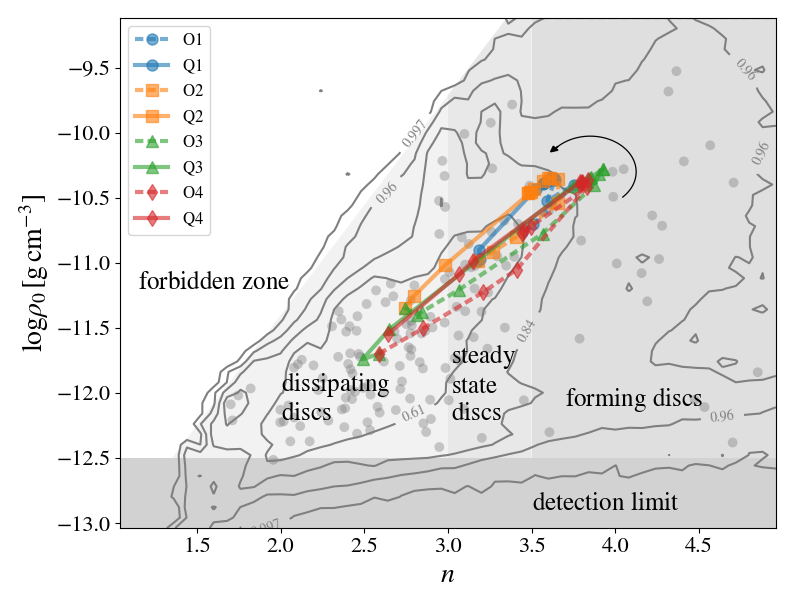}
    \caption[Evolutionary tracks computed for each of $\omega$ CMa's cycles over the $n-\log\rho_0$ diagram]{Evolutionary tracks computed for each of $\omega$ CMa's cycles over the $n-\log\rho_0$ diagram. The epochs were selected to cover the beginning, middle and the end of each phase. These tracks were superimposed to the parameter distribution of the sample of \citet[][gray contours, integral normalized to unity]{vieira2017}. The evolutionary regions proposed by these same authors are also indicated (shaded regions). The scatter distribution of sample Galactic stars is represented by the grey circles.}
    \label{fig:evolution_track}
\end{figure}

Figure~\ref{fig:evolution_track} shows the tracks of $\omega$ CMa's photometric cycles over the $n-\log\rho_0$ diagram, calculated according to the procedure used by \cite{vieira2017}. Each  of $\omega$ CMa's cycles corresponds to a loop in this diagram. Major outbursts are accompanied by a rapid excursion to the forming disks region, where the disk has both a large base density and density slope. Once the outburst developed for a sufficiently long time, the track reaches the steady-state region, where the disk inner region is fully built-up (large $\rho_0$) and $n\simeq 3.5$. This situation is especially clear for the first two cycles, for which the build-up phases are longer.

Once the outburst ends, the mass injection rate drops abruptly. As a result, the base density decreases, and the density profile becomes flatter (smaller $n$). The subsequent loops grow wider with time, reaching smaller $n-\rho_0$ values for the last cycles. This happens because the quiescent level of $-\dot{J}_{\mathrm{*,std}}$ also decreases with time (Fig.~\ref{fig:summary}), which causes the disk to reach smaller densities. 

\cite{vieira2017} concluded that there is a correlation between the spectral type and the disk density with the specific sub-population B2 type stars being more concentrated in the central region. The results for the case of $\omega$ CMa (B2 type star) shown in Fig.~\ref{fig:evolution_track} are in very good agreement with those found for similar stars in the Galaxy. 


\section{Angular momentum loss}
\label{sect:angular_momentum_loss}

\begin{figure}[!t]
    \centering
    \includegraphics[width=0.75\linewidth]{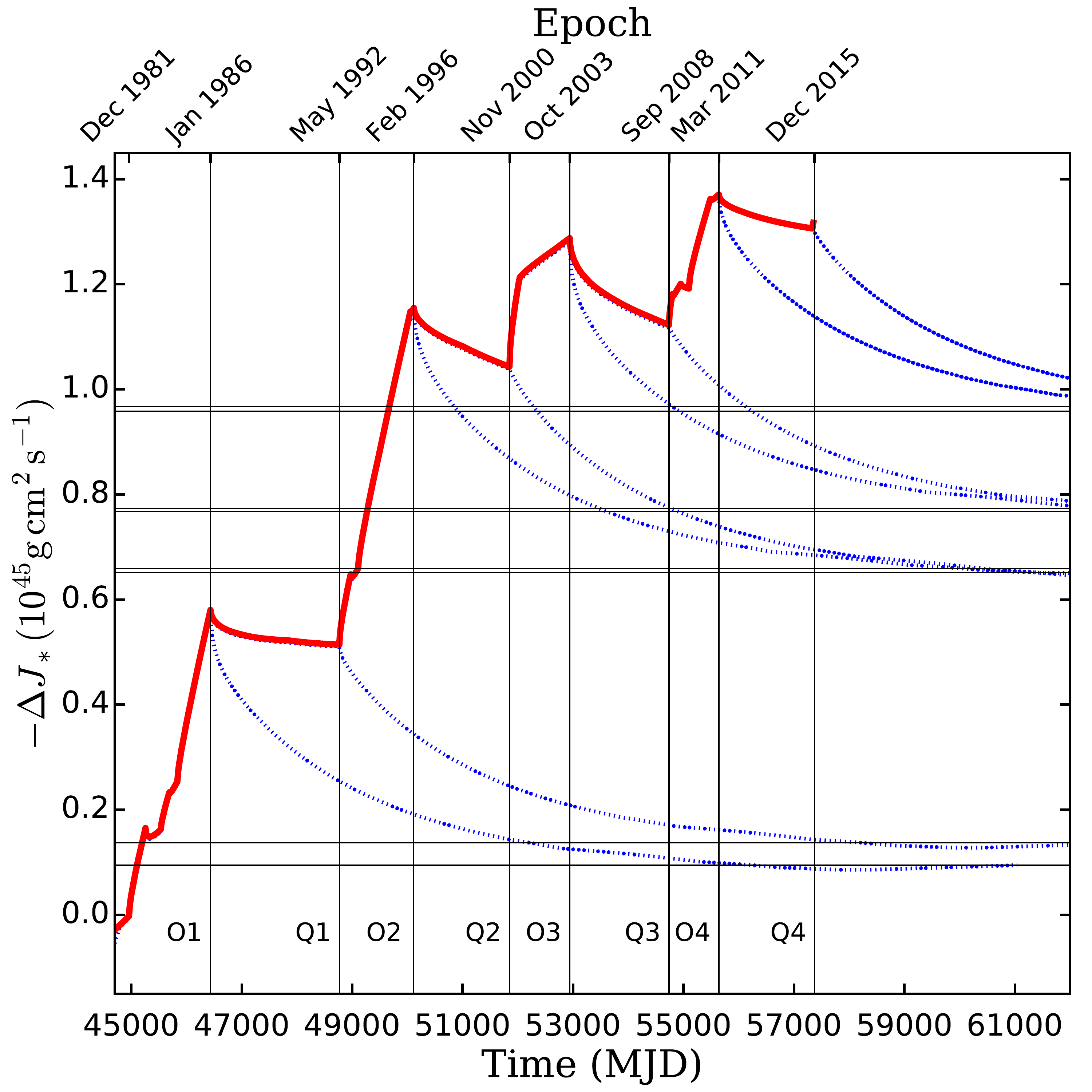}
    \caption[AM lost by the star]{Thick solid red curve: AM lost by the star, $-\Delta J_*(t)$, from $\mathrm{MJD}_0=44960.5$ on. 
    Dotted blue curves: Continuations of the AM lost by the star, if the injection of matter were to cease at that time with $\alpha=1$. 
    The black horizontal lines are showing the integrated amount of AM lost at the end of each phase. 
    }
    \label{fig:angular_momentum}
\end{figure}

The modeling of four complete cycles of $\omega$ CMa allows us to calculate the total AM lost by the star in the past 34 years. This quantity is given by
\begin{equation}
    -\Delta J_*(t)=-\int_{t_0}^{t}\dot{J}_*(t')\mathrm{d}t'\,,
    \label{eq:deltaj}
\end{equation}
where $\dot{J}_*(t)$ comes from Eq.~(\ref{eq:jdot_star}). In Fig.~\ref{fig:angular_momentum}, the thick solid red curve shows $-\Delta J_*(t)$ for $t_0=44960.5$. 
The outburst phases O1, O2, O3 and O4 correspond to intense AM losses, while during quiescence (Q1, Q2, Q3 and Q4) some of the AM comes back to the star with the part of the disk matter that was reaccreted during these phases. At each inflection of the curve, representing the transition between outburst and quiescence or vice-versa, the dotted blue curves show the AM that would have been lost by the star had the AM injection been completely ceased from that instant on. The horizontal lines indicate the asymptotic value of these curves. Because in $\omega$ CMa the quiescence phases are not true quiescences, a net AM is still lost in these phases. 

\begin{figure}[!t]
    \centering
    \includegraphics[width=0.75\linewidth]{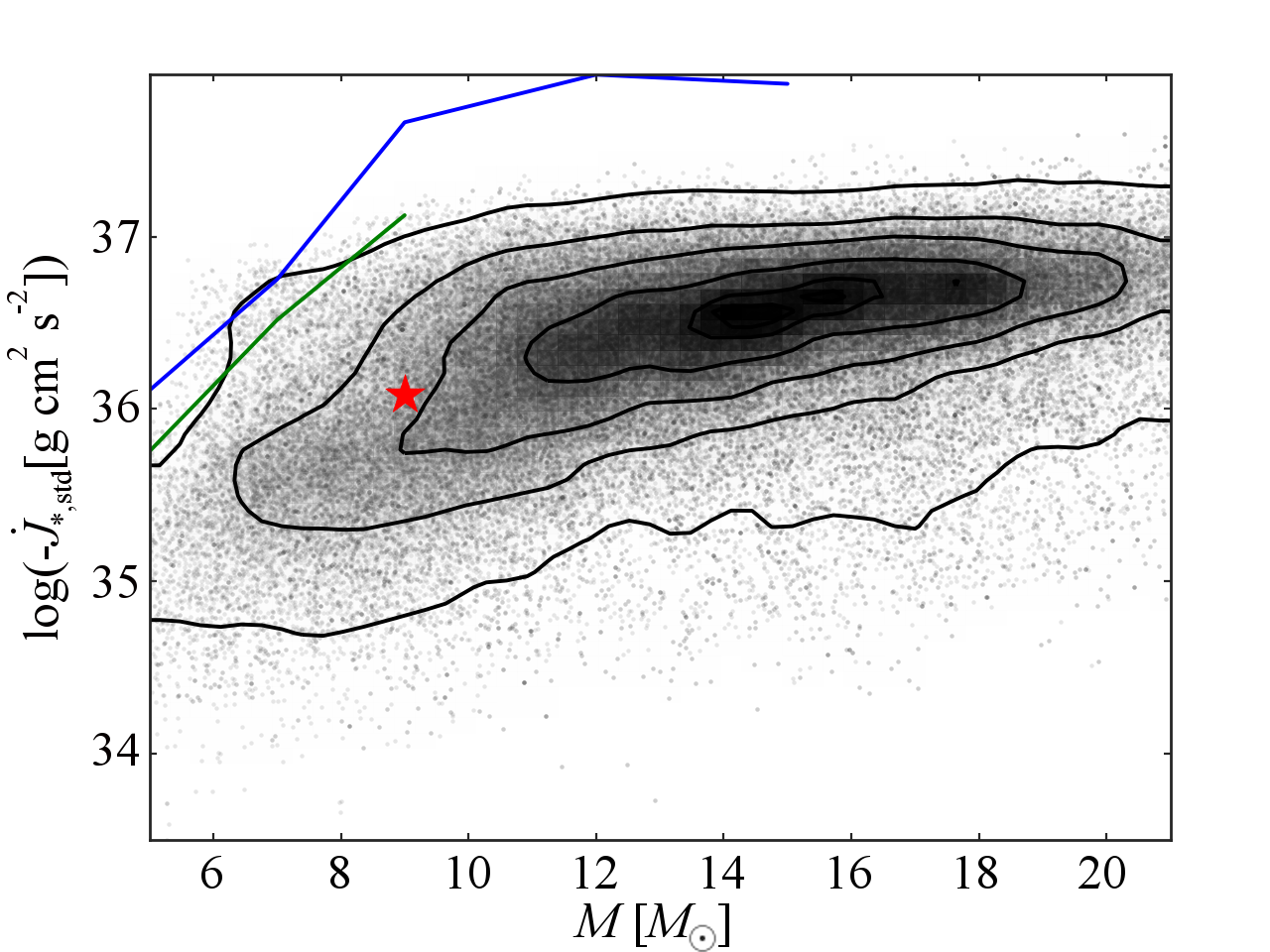}
    \caption[The position of $\omega$ CMa in $-\dot{J}_{*,\mathrm{std}}$ vs $M$ diagram]{The position of $\omega$ CMa in $-\dot{J}_{*,\mathrm{std}}$ vs $M$ diagram calculated by \citet{rimulo2018} for the Be stars in the SMC, in comparison to the estimated values by \citet{granada2013} for the Galaxy (green curve) and the SMC (blue curve).}
    \label{fig:logjdot}
\end{figure}

It has been proposed (e.g., \citealt{krticka2011}) that, as the star evolves through the main sequence, the formation of the VDD would be a natural mechanism to extract AM from the outer layers of the star, preventing it from exceeding the break up velocity. \cite{granada2013}, using the Geneva stellar evolution code, 
estimated the theoretical rate of AM loss expected during the main sequence evolution of stars with different masses and metallicities. To do so, they assumed the appearance of a steady-state VDD every time the outer layers of the star reached a given rotation rate limit. \cite{rimulo2018} found that the predictions of the Geneva code for AM loss rates are much larger than the actual rates observed in Be stars from the SMC (compare the blue line with the shaded areas in Fig.~\ref{fig:logjdot}). Our results suggest a similar disagreement for the Galactic Be star $\omega$ CMa. For a star of Galactic metallicity and 9.0\,M$_{\odot}$, the AM loss rate computed by \cite{granada2013} was $1.3\times 10^{37}\,\mathrm{g\,cm^2\,s^{-2}}$ (green line in the third panel of Fig.~\ref{fig:summary}), which is almost eleven times larger than the mean value estimated from this work. Incidentally, it is interesting to point out that $\omega$ CMa loses AM at a rate similar to that of a similar 9.0\,M$_{\odot}$ Be star in the SMC. Therefore, at this point our result does not indicate any difference in the AM loss rate as a function of metallicity. Clearly, to properly address this issue a large sample of Galactic Be stars must be investigated.

The third panel of Fig.~\ref{fig:summary} shows the evolution of $-\dot{J}_{*,\mathrm{std}}$ during 34 years (black line) that varies between $3.0 \times 10^{33}$ to $5.4 \times 10^{36} \mathrm{g\,cm^2\,s^{-2}}$. During the formation phases the value of $-\dot{J}_{*,\mathrm{std}}$ increased noticeably, except for the O3 section that corresponds to a low value of the $\alpha$ parameter. The dissipation phases match the decrease in $-\dot{J}_{*,\mathrm{std}}$ but may never go to zero. The red lines represent the mean AM loss rates of the star in each phase. The blue horizontal line represents the mean AM loss rate in the period of 34 years ($<-\dot{J}>$) covered by the light curve of $\omega$ CMa. This rate is $1.2\times 10^{36}\,\mathrm{g\,cm^2\,s^{-2}}$, and the total AM lost in the whole studied period was $1.3\times 10^{45}\,\mathrm{g\,cm^2\,s^{-1}}$. From the Geneva evolutionary models, a B2 star with W=0.73 has a total AM content of $2.1\times 10^{53}\,\mathrm{g\,cm^2\,s^{-1}}$. Thus, the AM lost over 34 years, is just $6\times 10^{-9}$ of that. Even if we extrapolate the measure of AM loss rate for the entire main sequence lifetime, which is about $3.25\times 10^7$ yr, this would correspond to 0.006 of the total AM of the star. 

The fourth panel of Fig.~\ref{fig:summary} demonstrates the variation of the AM flux ($-\dot{J}$) for the four cycles of $\omega$ CMa at different radii (1.4, 2.0 and $10.0\,R_\mathrm{eq}$). While $-\dot{J}$ has both positive (losing) and negative (gaining) values at 1.4 and $2.0\,R_\mathrm{eq}$ it never has a negative value at the outer radius $10.0\,R_\mathrm{eq}$ meaning that, from this distance outward, the disk remains completely coupled with viscosity, while the inner disk falls in the star so quickly that it is no longer viscosity-supported.

\cite{rivinius1999} provided strong evidence for the quasi-Keplerian nature of Be disks even before they were resolved by interferometry. The values of AM presented above correspond to the radial velocity ($v_{r}$) between $\sim$ -3 to $\sim$4 km/s in the $\omega$ CMa's disk which is in good agreement with the typical values presented in \cite{rivinius1999}. Moreover, comparing these values to the orbital velocity ($v_\mathrm{orb}$) of $\sim$500 km/s in the inner rim of the disk, we find that our results match the detailed hydrodynamic simulations which show that for most of the disk (within at least several tens of stellar radii), the azimuthal velocity is much larger than the radial velocity \citep{krticka2011}.


\section{Evolution of the disk temperature}
\label{sect:temperature_evolution}


\begin{figure}[!t]
    \centering
    \includegraphics[width=1.0\linewidth]{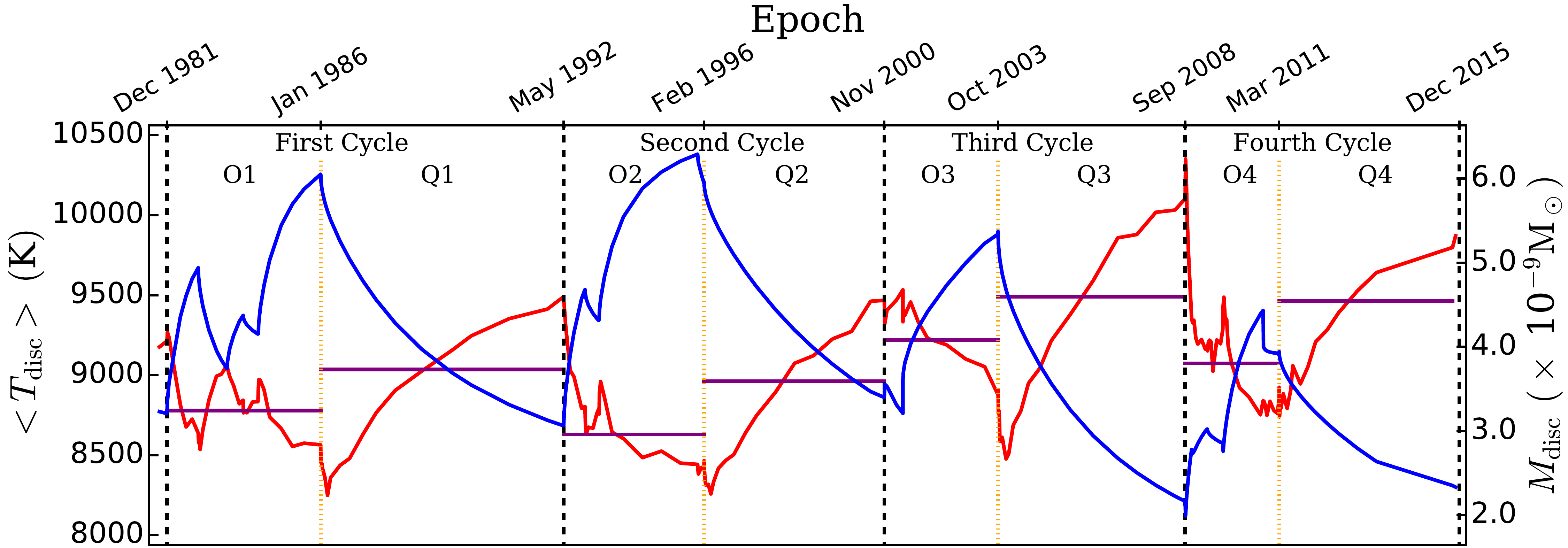}
    \par\medskip
    \centering
    {\includegraphics[width=1.0\linewidth]{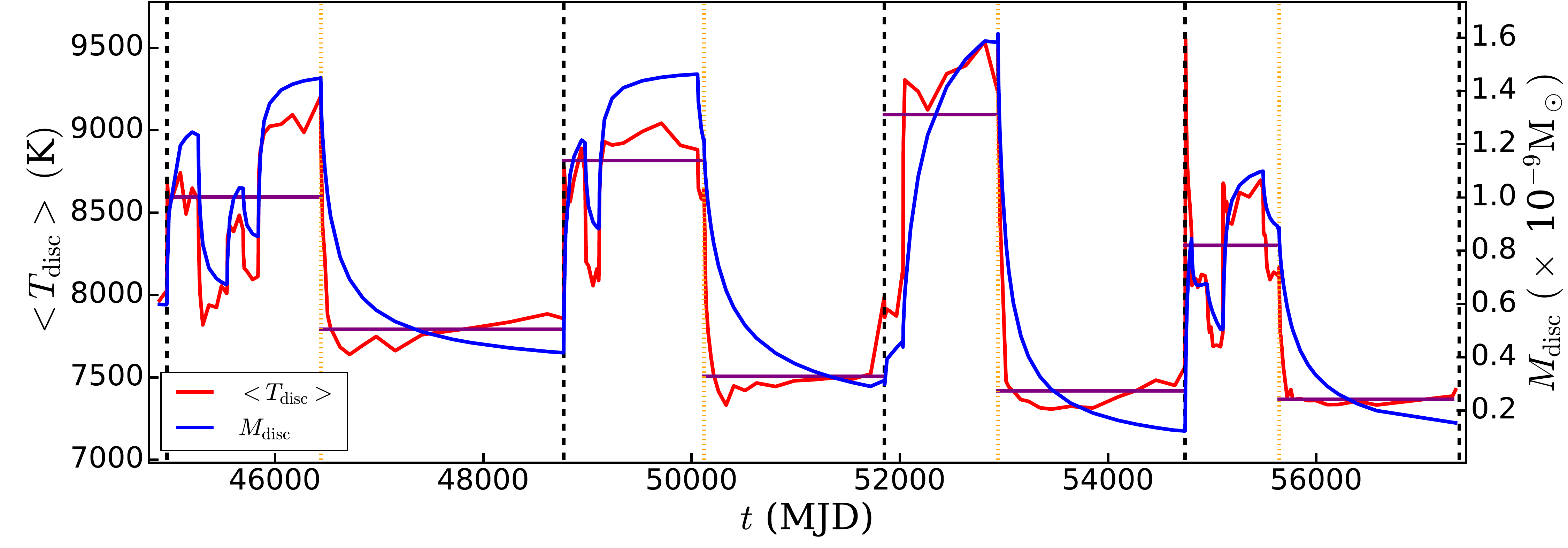}}
    \caption[Evolution of mass-averaged temperature and total mass of $\omega$ CMa's disk]{Evolution of mass-averaged temperature (red line, Eq.~\ref{eq:mean_disk_temp}) and total mass (blue line) of $\omega$ CMa's disk for the whole (top panel) and inner part ($r < 3.0\,R_{\rm eq}$) of the disk (bottom panel). The horizontal purple lines represent the mean temperature during each phase.}
    \label{fig:disk_temperature}
\end{figure}


\begin{figure}[!ht]
    \centering
    \includegraphics[width=0.8\linewidth, height=0.4\linewidth]{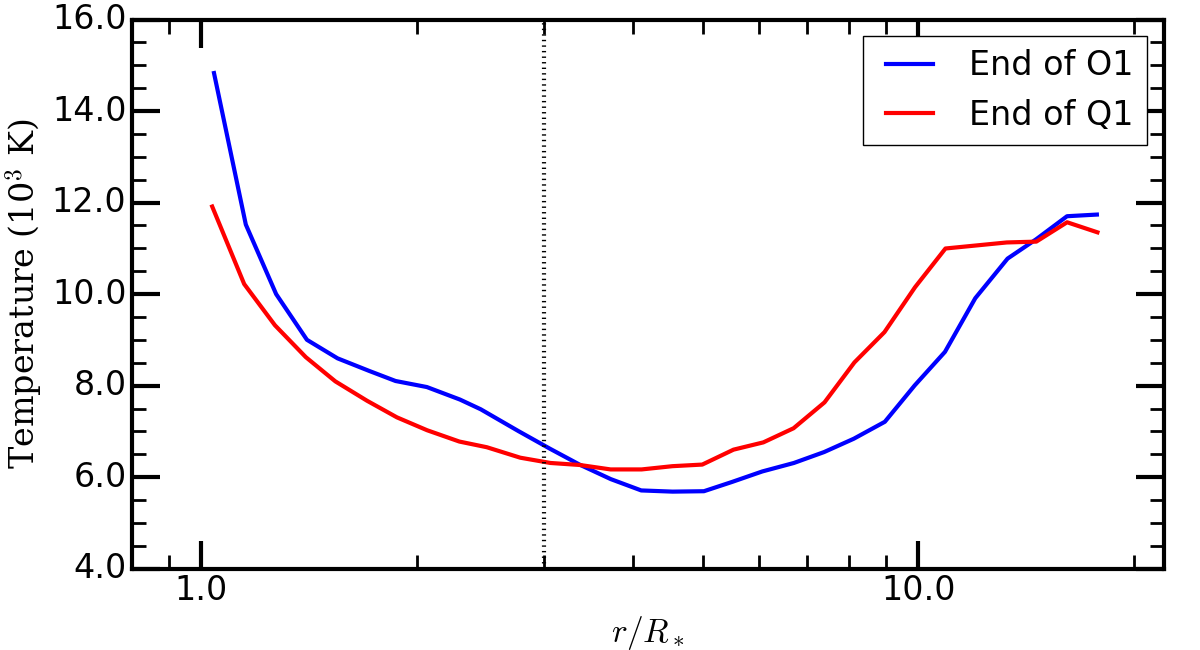}
    \caption[Temperature of the disk as a function of radius]{Temperature of the disk as a function of radius at the end of O1 (blue line) and Q1 (red line) representing the fact that the inner disk is hotter in the high-mass state while the bulk of the disk is hotter in the low-mass state.}
    \label{fig:temperature_vs_r}
\end{figure}


One of the significant results of this work is that, with the exception of O3, the values of $\alpha$ are larger during the formation phase than during dissipaiton. As described in Sect.~\ref{subsect:vdd_c3}, however, the result of O3 may be of little statistical significance as the early stages of the outburst are poorly sampled. This result is consistent with the findings of \cite{rimulo2018} who studied 81 disk formation and dissipation phases of 54 Be stars in the SMC. As Eq.~(\ref{eq:sigmadot}) shows, the timescale of disk evolution is set by the factor $\alpha T$. In the {\tt SINGLEBE} hydrodynamic calculations the disk was set to a constant isothermal temperature of $T_\mathrm{isot}=13200\,\mathrm{K}$, following \cite{carciofi2006a}. However, the {\tt HDUST} simulations that are run for selected times of the disk evolution calculate the coupled problem of radiative and statistical equilibrium, thus determining the gas kinematical temperature as a function of position and time. 
In this section we investigate whether this inconsistency in the modeling might be responsible for $\alpha$ being larger at build-up than dissipation.

We begin by defining the mass-averaged temperature of the disk inside a spherical shell of inner radius $R_\mathrm{eq}$ and outer radius $r$
\begin{equation}
    \left\langle T_\mathrm{disk}\right\rangle_r(t)=\frac{1}{M_{\mathrm{disk},r}}\int_{R_\mathrm{eq}}^r\oint_{4\pi} T_\mathrm{disk}(\boldsymbol{s},\Omega,t)\rho(\boldsymbol{s},\Omega,t) \mathrm{d}\Omega s^2\mathrm{d}s\,,
    \label{eq:mean_disk_temp}
\end{equation}
where $M_{\mathrm{disk},r}$ is the mass of the disk inside the same spherical shell, given by 
\begin{equation}
    M_{\mathrm{disk},r}(t)=\int_{R_\mathrm{eq}}^r\oint_{4\pi}\rho(\boldsymbol{s},\Omega,t) \mathrm{d}\Omega s^2\mathrm{d}s\,,
    \label{eq:mass_disk_r}
\end{equation}
and $T_\mathrm{disk}$ is the disk temperature as a function of position as calculated by {\tt HDUST}.

Figure~\ref{fig:disk_temperature} shows the evolution of $\left\langle T_\mathrm{disk}\right\rangle_r$ and $M_{\mathrm{disk},r}$ for two values of $r$: $r=3R_\mathrm{eq}$, corresponding to the inner disk, where most of the visual continuum flux comes from (lower panel), and $r=18R_\mathrm{eq}$, corresponding to the whole simulated disk by {\tt HDUST} (upper panel).

The behavior of $M_{\mathrm{disk},3}(t)$ and $M_{\mathrm{disk},18}(t)$ qualitatively follows the behavior of the light curve, as expected, since the formation and dissipation phases are associated with increasing and decreasing disk mass. 
The rate of evolution of $M_{\mathrm{disk},3}(t)$ is much faster than $M_{\mathrm{disk},18}(t)$ because the viscous timescale scales as $r^{0.5}$. In other words, the inner disk reacts much faster to changes in the disk feeding rate than the whole disk does. The mass-averaged temperature, however, presents a different behavior: in the inner disk, the average temperature is positively correlated with the disk mass, but the opposite happens for the whole disk. To understand this, we plotted the temperature profile of $\omega$ CMa at the end of O1 (high-mass state) and at the end of Q1 (low-mass state) in Fig.~\ref{fig:temperature_vs_r}, that shows that the inner disk is hotter in the high-mass state while the bulk of the disk is hotter in the low-mass state which naturally explains the above correlations.

From Fig.~\ref{fig:disk_temperature} we conclude that the average temperature at the inner disk during build-up is larger than during dissipation by $\sim 800$--$1700\,\rm K$, and the reverse is true for the whole disk, but with smaller temperature differences ($\sim 250$--$350\,\rm K$). There is spectroscopic evidence for such variations for the stars $\delta$ Sco and $\eta$ Cen, notably the coexistence/balance between FeIII and FeII emission/absorption, but also HeI/HeII (unpublished data, Thomas Rivinus, priv. comm.).

Let us now examine the effects that the temperature differences in both phases might have on the $\alpha$ determination. Any combination of $\alpha$ and $T$ that keeps $\alpha T$ unchanged does not affect the surface density solution $\Sigma(r,t)$. Therefore, assuming that the \emph{real} value of the viscosity, $\alpha_{\rm real}$, is associated with the temperature calculations shown in Fig.~\ref{fig:disk_temperature}, by using an isothermal disk temperature in the hydrodynamic calculations we introduce a bias in the $\alpha$ determination given roughly by
\begin{equation}
    \frac{\alpha_{\rm biased}}{\alpha_{\rm real}} = \frac{\left\langle T_\mathrm{disk}\right\rangle_r}{T_\mathrm{isot}}\,.
    \label{eq:mean_disk_temp2}
\end{equation}
Here, we ignore possible temporal and radial variations of $\alpha$.
Given that the hydrodynamic calculations are isothermal, this bias is certainly present in our $\alpha$ determinations. However, the effect is likely small, as the temperature variations between the different phases amount to only $~5$--$10\%$ (Fig.~\ref{fig:disk_temperature}), which means that the bias in $\alpha$ would be at most around 10\%. To properly address this issue, non-isothermal hydrodynamic calculations, coupled with radiative transfer, should be performed, which is beyond the scope of this work. However,  because the calculated differences in $\alpha$ for the build-up and dissipation phases are much larger than the above, we conclude that these differences are real. 

On the other hand, in a recent paper, \cite{kee2016} discuss the role of radiative ablation in the mass budget of Be disks. They show that radiation forces can play an important role in dissipating the disk, and some of their model calculations show that a disk wind can ablate the entire disk in timescales of order months to years comparable to what is observed. However, their model calculations were performed in the optically thin approximation, which clearly do not apply in our case. 
Therefore, it is possible that radiative ablation plays a role in Be disk dynamics, but the extent of this role remains to be determined. In principle, if radiative ablation is important in $\omega$ CMa's case, the values of $\alpha$ quoted in this paper are likely upper limits. Future work is necessary to answer this question.


		%
\chapter{Multi-technique Modeling}
\label{chap:other_tech}

In Chap.~\ref{chap:photometry} we presented, for the first time, a complete model of the dynamical evolution of a Galactic Be star, $\omega$ CMa. However, the model was limited only to the $V$-band which in turn is sensitive only to variations in the disk regions very close to the star (see Fig.~\ref{fig:flux_loci}). A very important next step consists of extending the analysis to other observables, which probes different disk regions. This represents a crucial test for the model presented in the last chapter, with important consequences for the VDD scenario. For instance, if the current model fits the data, it means that the model also reproduces well the disk structure as a whole. On the other hand, failure to reproduce the data will likely require a revision of the model. 

\begin{figure}[!h]
    \centering
    \includegraphics[width=1.0\columnwidth]{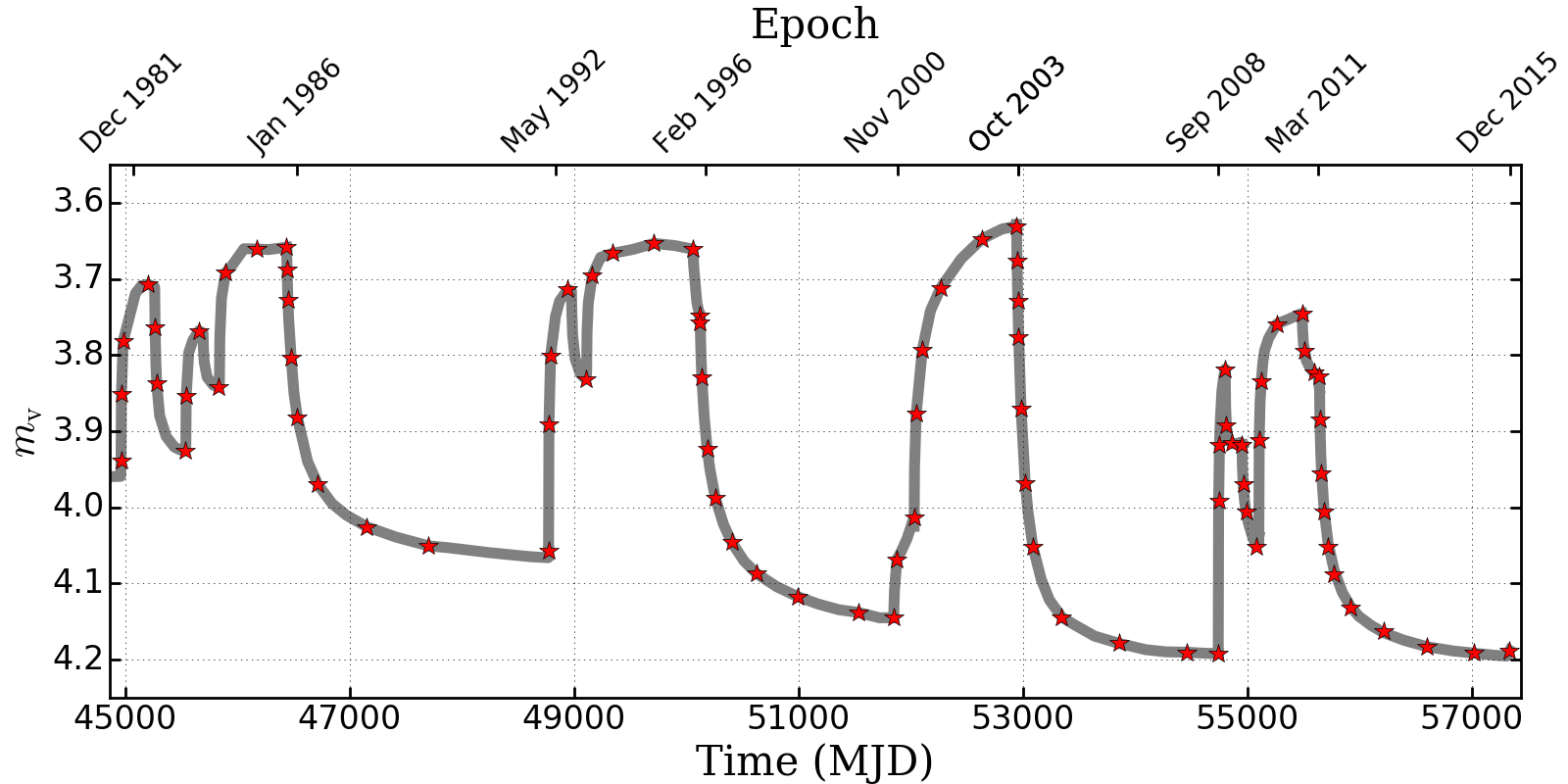}
    \caption[Selected epochs for modeling the multi-technique observations of $\omega$ CMa]{Selected epochs for modeling the multi-technique observations of $\omega$ CMa. The final synthetic light curve of $\omega$ CMa is shown as the grey line and the selected epochs are marked with red stars.}
    \label{fig:selected_epochs}
\end{figure}

As a first step towards a complete, multi-technique model of $\omega$ CMa, we used the same scenario presented in Sect.~\ref{sect:final_model} and computed, with HDUST, various line profiles and the entire SED from the UV to the radio for about 80 selected epochs covering the most important phases of the disk evolution. The selected epochs are shown with red stars in Fig~\ref{fig:selected_epochs}.


It is important to mention that some of the results shown in this chapter are still preliminary.


\section{Polarimetry}
\label{sect:polarimetry}

According to Fig.~\ref{fig:data_dist}, our available data in polarimetry are limited to the last cycle, mostly observed at OPD. 

\begin{figure}
    \begin{minipage}{0.5\linewidth}
        \centering
        \subfloat{\includegraphics[height=0.25 \textheight, width=0.9\linewidth]{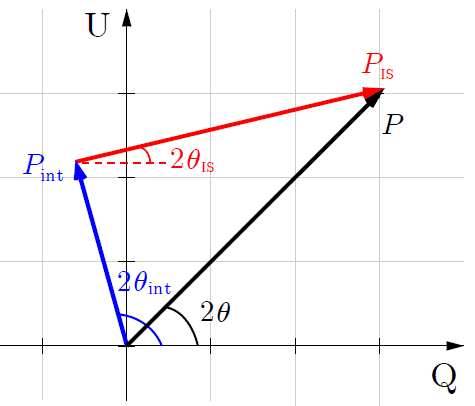}}
    \end{minipage}
    \begin{minipage}{0.5\linewidth}
        \centering
        \subfloat{\includegraphics[height=0.25 \textheight, width=0.9\linewidth]{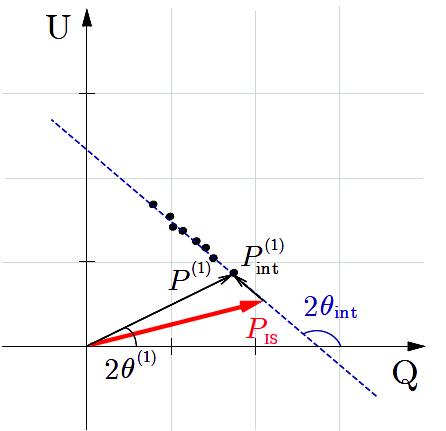}}
    \end{minipage}%
    \caption[Components of polarization vectors, and their evolution]{{\it Left:} The components of polarization vector: interstellar polarization and disk polarization. {\it right:} the variations of polarization vector and its components with the disk growth and decay (Adapted from \citealt{bednarski2016}).}
    \label{fig:bedna_pol}
\end{figure}

One common issue regarding interpretation of polarimetric data is the removal of the interstellar contribution to the data. The observed polarization, decomposed in its Stokes $Q$ and $U$ parameters, can be written as:
\begin{equation}
    Q_\mathrm{obs} = Q_\mathrm{IS} + Q_\mathrm{disk}\,,
    \label{eq:q_obs}
\end{equation}
and
\begin{equation}
    U_\mathrm{obs} = U_\mathrm{IS} + U_\mathrm{disk}\,,
    \label{eq:u_obs}
\end{equation}
i.e., without knowing the interestellar components ($Q_\mathrm{IS}$ and $U_\mathrm{IS}$) of the observed polarization, the intrinsic polarization ($Q_\mathrm{disk}$ and $U_\mathrm{disk}$) cannot be known. This is shown schematically in the left panel of Fig.~\ref{fig:bedna_pol}.


Fortunatelly, $\omega$ CMa's case gives us an almost direct way of measuring $Q_\mathrm{IS}$ and $U_\mathrm{IS}$. We begin examining the right panel of Fig.~\ref{fig:bedna_pol}, that shows, in an schematic way, how the process of disk formation and dissipation of a Be disk appears in the $Q-U$ diagram. The disk has a polarization angle on the Sky given by $\theta_{\rm disk}$. When the intrinsic polarization is zero (no disk), the observed polarization will be the IS one. As the disk grows (and dissipates), the magnitude of the intrinsic polarization changes, but not the angle (assuming that the disk is axisymmetric). This is shown in the right panel of Fig.~\ref{fig:bedna_pol}, as the track of points along the $\theta_{\rm disk}$ direction. Therefore, measuring the angle of this track gives one first piece of information, $\theta_{\rm disk}$. This has been done in Chap.~\ref{chap:omecma_obs} (Fig.~\ref{fig:pol_qu}), and the results are summarized in Table~\ref{table:pol_angle_value} ($\psi$/2\footnote{$\psi$/2 is equivalent of $\theta_\mathrm{disk}$ but derived from $Q-U$ diagram.} row).



\begin{figure}
    \begin{minipage}{0.5\linewidth}
        \centering
        \subfloat{\includegraphics[height=0.40 \textheight, width=1.0\linewidth]{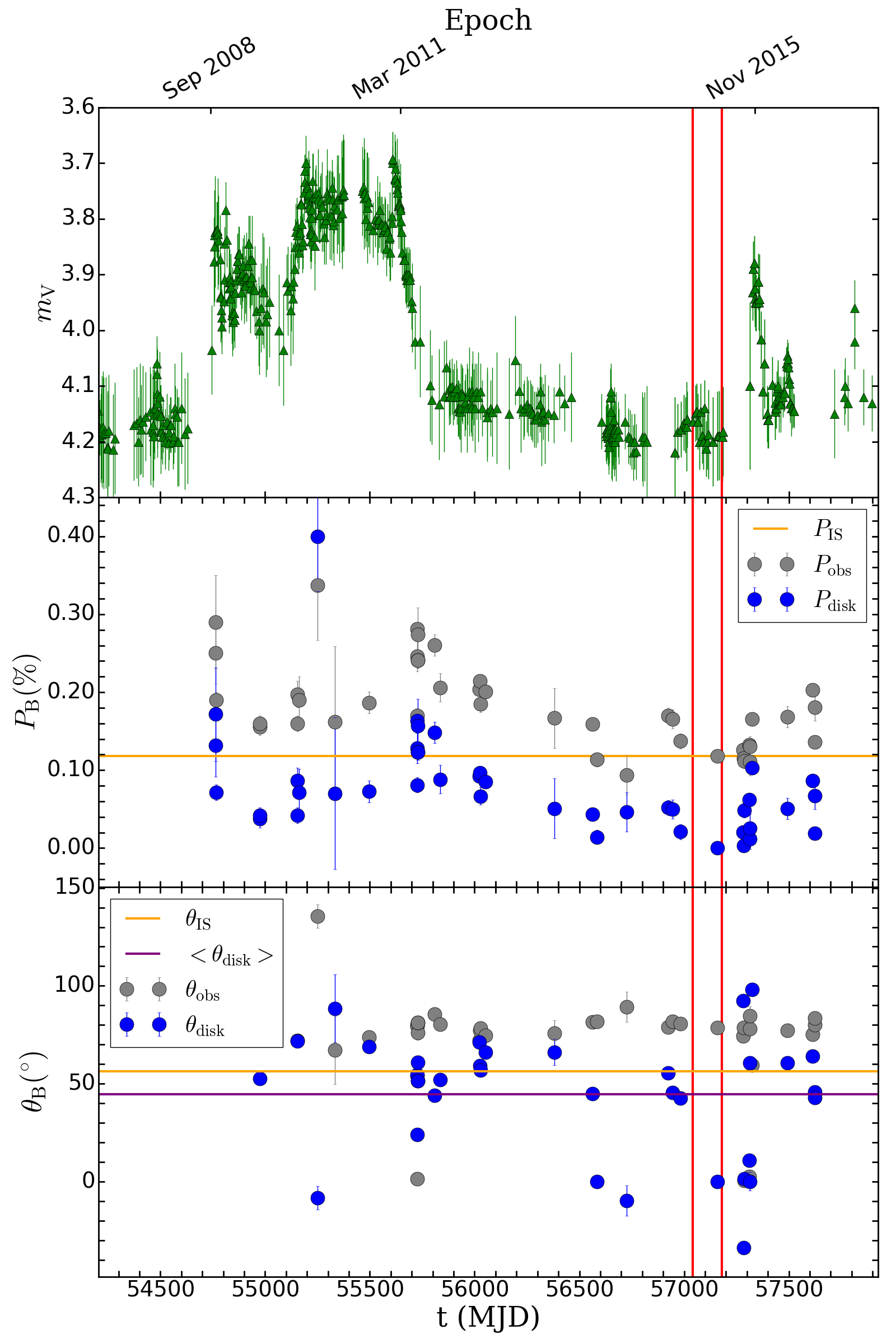}}
    \end{minipage}%
    \begin{minipage}{0.5\linewidth}
        \centering
        \subfloat{\includegraphics[height=0.40 \textheight, width=1.0\linewidth]{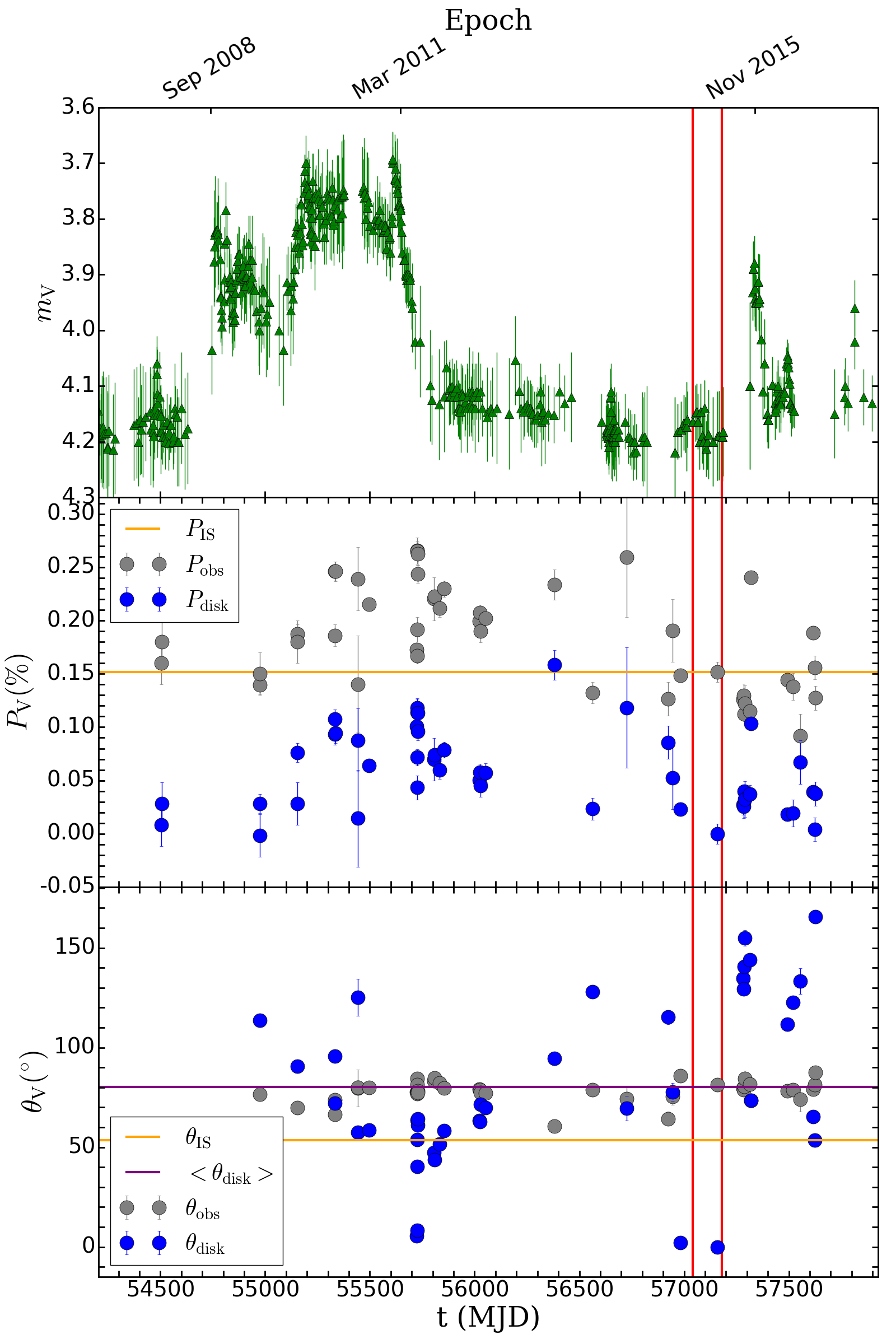}}
    \end{minipage}\par\medskip
    \begin{minipage}{0.5\linewidth}
        \centering
        \subfloat{\includegraphics[height=0.40 \textheight, width=1.0\linewidth]{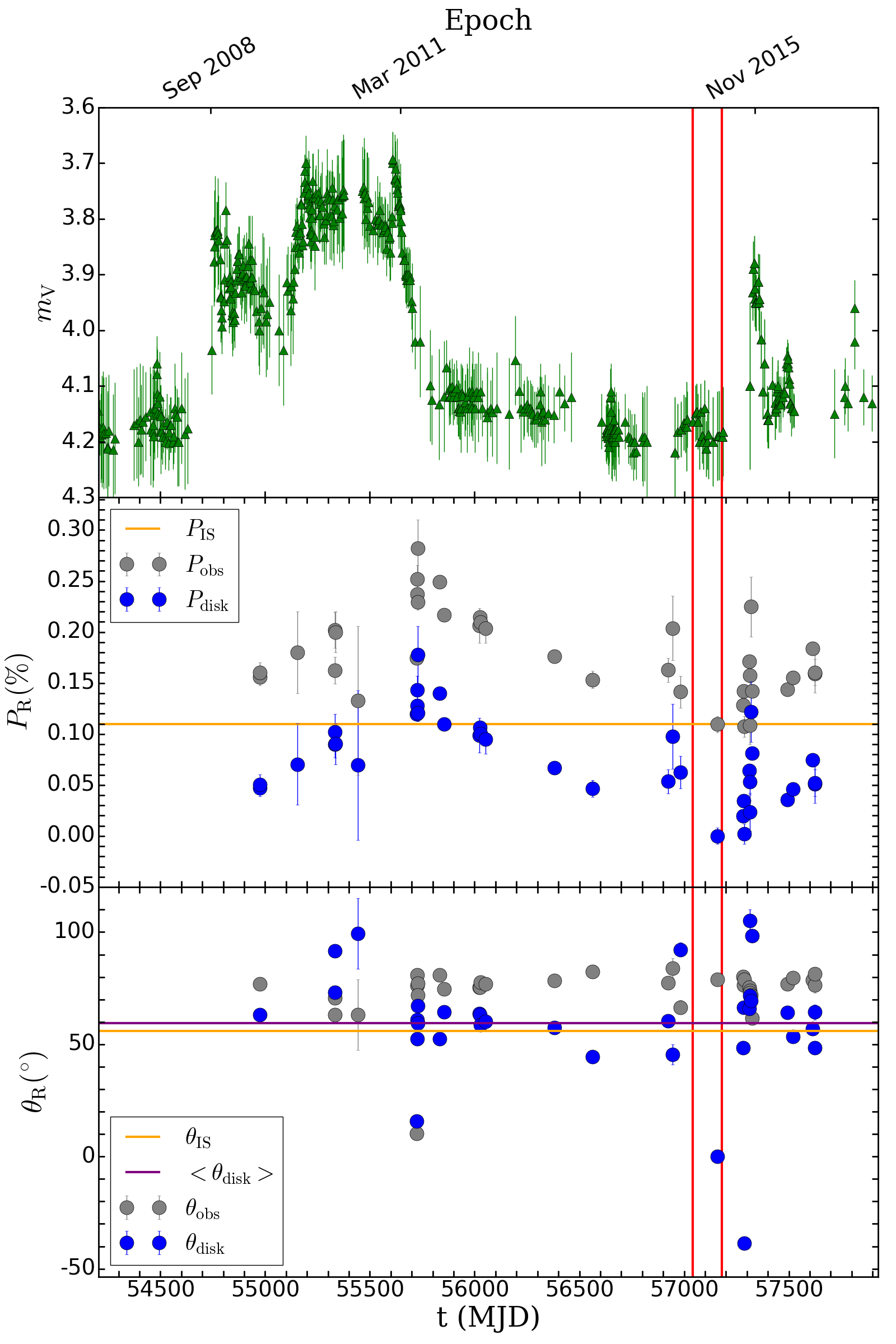}}
    \end{minipage}%
    \begin{minipage}{0.5\linewidth}
        \centering
        \subfloat{\includegraphics[height=0.40 \textheight, width=1.0\linewidth]{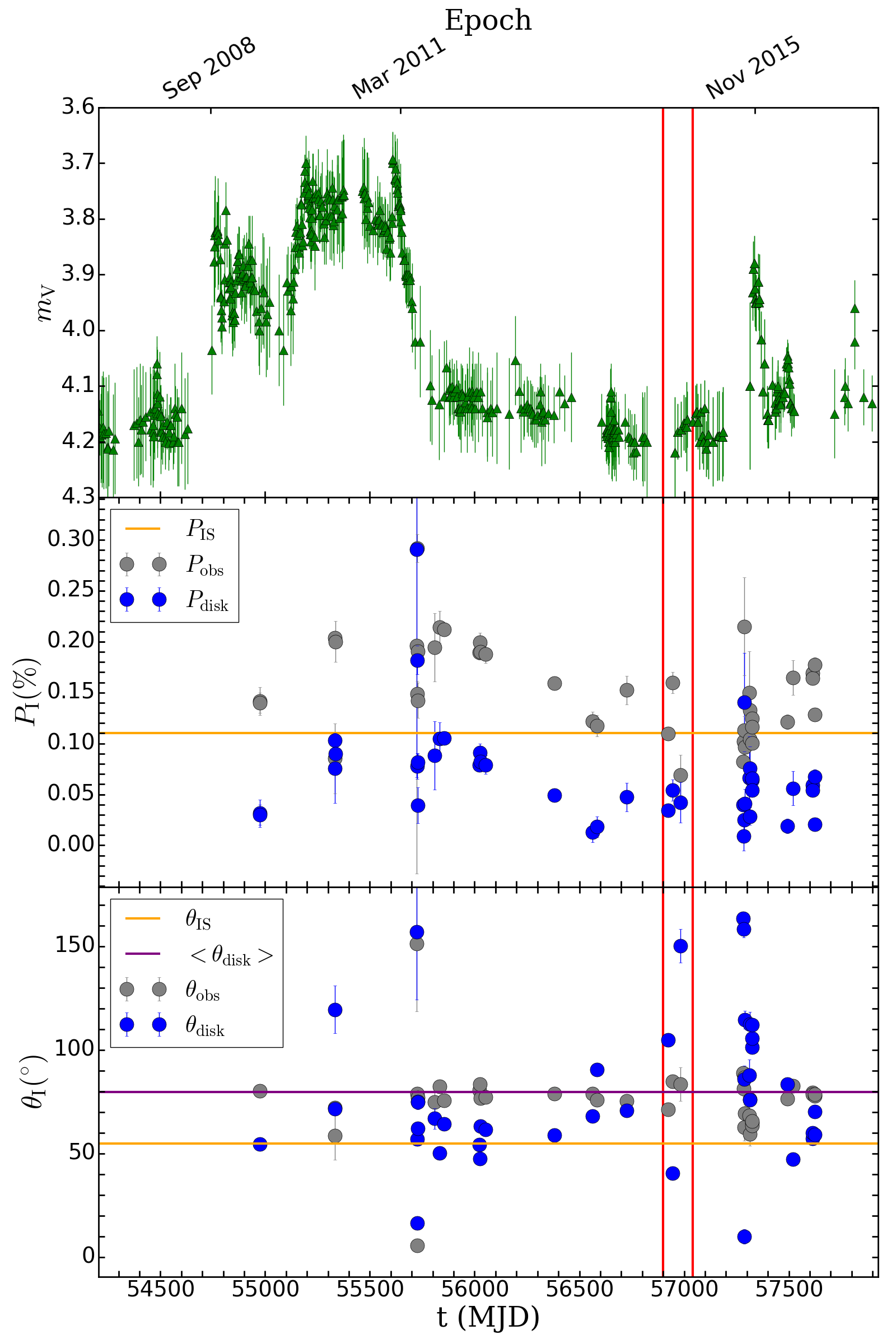}}
    \end{minipage}%
    \caption[Observed and intrinsic polarization of the fourth cycle of $\omega$ CMa in $BVRI$ band]{{\it Top plots}: The $V$-band observed data of $\omega$ CMa. Two vertical red lines indicate the epoch during which the star was assumed to be diskless. {\it Middle plots}: The observed (grey circles) and intrinsic polarization (blue circles) of the fourth cycle of $\omega$ CMa. The estimate of $P_\mathrm{IS}$ is shown as the horizontal orange line. {\it Bottom plots}: The observed (grey circles) and intrinsic polarization angles (blue circles) of $\omega$ CMa. The $\theta_\mathrm{IS}$ and the average value of the intrinsic polarization angle are shown with orange and purple horizontal lines, respectively. Each panel shows the data for a different filter: $B$ (top left), $V$ (top right), $R$, bottom left, and $I$ (bottom right).}
    \label{fig:pol_bvri_corrected}
\end{figure}


The second piece of information comes from the light curve itself. The results of Sect.~\ref{sect:final_model} indicate that at the end of C4, the $V$ band excess is very small. Acording to the best-fit models, the inner disk at that phase is very tenuous (Fig,~\ref{fig:summary}). Therefore, if one assumes that the intrinsic polarization at that phase is very small, the observed polarization will be very close to the IS one. 

The top and middle of each panel in Fig.~\ref{fig:pol_bvri_corrected} show the $V$-band photometric data and polarimetric data of $\omega$ CMa in $B$, $V$, $R$, $I$ bands, respectively. The observed polarization is shown as the grey symbols. Two vertical red lines in the figures indicate the boundaries of the phase during which we assumed the star to be almost diskless. An average of the data at that phase, therefore, gives us $Q_\mathrm{IS}$ and $U_\mathrm{IS}$, which in turn provide
\begin{equation}
    P_{\rm IS} = \sqrt{Q_{\rm IS}^2 + U_{\rm IS}^2}\,,
    \label{eq:p_vector}
\end{equation}
and 
\begin{equation}
    \theta_{\rm IS} = \frac{1}{2} arctan(\frac{U_{\rm IS}}{Q_{\rm IS}}).
    \label{eq:theta_formula}
\end{equation}

Having an estimate of $P_{\rm IS}$, the intrinsic polarization of $\omega$ CMa, $P_{\rm disk}$ can be calculated from Eqs.~\ref{eq:q_obs} and~\ref{eq:u_obs}. The results are shown in the bottom panels of Fig.~\ref{fig:pol_bvri_corrected}. The estimated value of $P_\mathrm{IS}$ is shown as the horizontal orange line in the middle frame of each panel. The conformance of the polarization level to the increments and decrements of the star brightness can be seen for all the filters. An eye inspection reveals that polarization level follows the variation of the $V$-band photometric data with a lag, i.e., in the dissipation phase the drop of polarization is slower than the drop in brightness. This agrees with the theoretical studies of \cite{haubois2014}. On top of the variation, there are some intrinsic scatter in the data that can have two origins. One may be of intrinsic origin, as the polarization is rather sensitive to the conditions of the inner disk \citep{carciofi2007}. The other is observational, as the measured polarization is rather small. We believe a re-reduction of the data is necessary before final publication; in the re-reduction process we will investigate closely the more deviant points, in order to investigate whether the deviations are real or of observational origin.  

The bottom of each panel shows the polarization angle, $\theta$. As before, the $\theta_\mathrm{IS}$ and $\theta_\mathrm{disk}$ are shown as the orange and purple horizontal lines, respectively, and their average are given in Table~\ref{table:pol_angle_value}. A very important result is that the estimated $\theta_\mathrm{disk}$ from the $P_{\rm IS}$-corrected data is on average 66.1$^\circ$. This agrees well with the value estimated from the $Q-U$-diagram method described above (61.1$^\circ$). This is an indirect evidence that the estimated $P_{\rm IS}$ and $\theta_\mathrm{IS}$ are correct.


\begin{table}
    \begin{center}
        \caption[$\theta_\mathrm{disk}$ and $\theta_\mathrm{IS}$ for different filters]{$\theta_\mathrm{disk}$ and $\theta_\mathrm{IS}$ for different filters.}
        \begin{tabular}{@{}cccccc}
            \hline
            \hline
            & $B$ Filter & $V$ Filter & $R$ Filter & $I$ Filter & average \\
            \hline
            $\theta_\mathrm{IS}$($^\circ$) & 56.4 & 53.5 & 56.0 & 54.8 & 55.2 \\
            $\theta_\mathrm{disk}$($^\circ$) & 44.7 & 80.4 & 59.4 & 79.7 & 66.1 \\
            $\psi$($^\circ$)/2 & 63.2 & 74.3 & 53.6 & 53.3 & 61.1 \\
            \hline
        \end{tabular}
        \label{table:pol_angle_value}
    \end{center}
\end{table}


\begin{figure}[!ht]
    \begin{minipage}{1.0\linewidth}
        \centering
        \subfloat{\includegraphics[height=0.25 \textheight, width=0.8\linewidth]{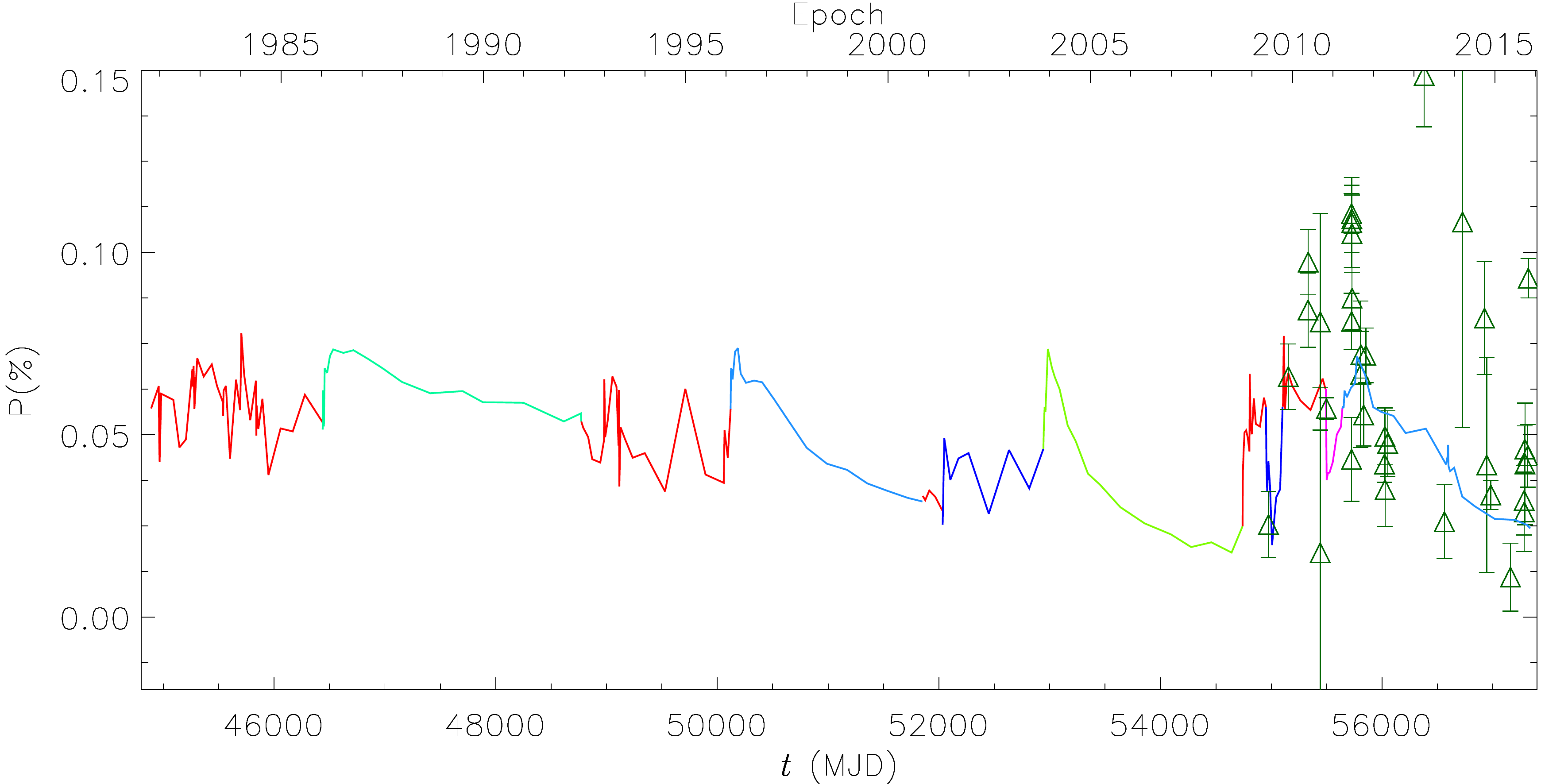}}
    \end{minipage}\par\medskip
    \begin{minipage}{1.0\linewidth}
        \centering
        \subfloat{\includegraphics[height=0.25 \textheight, width=0.8\linewidth]{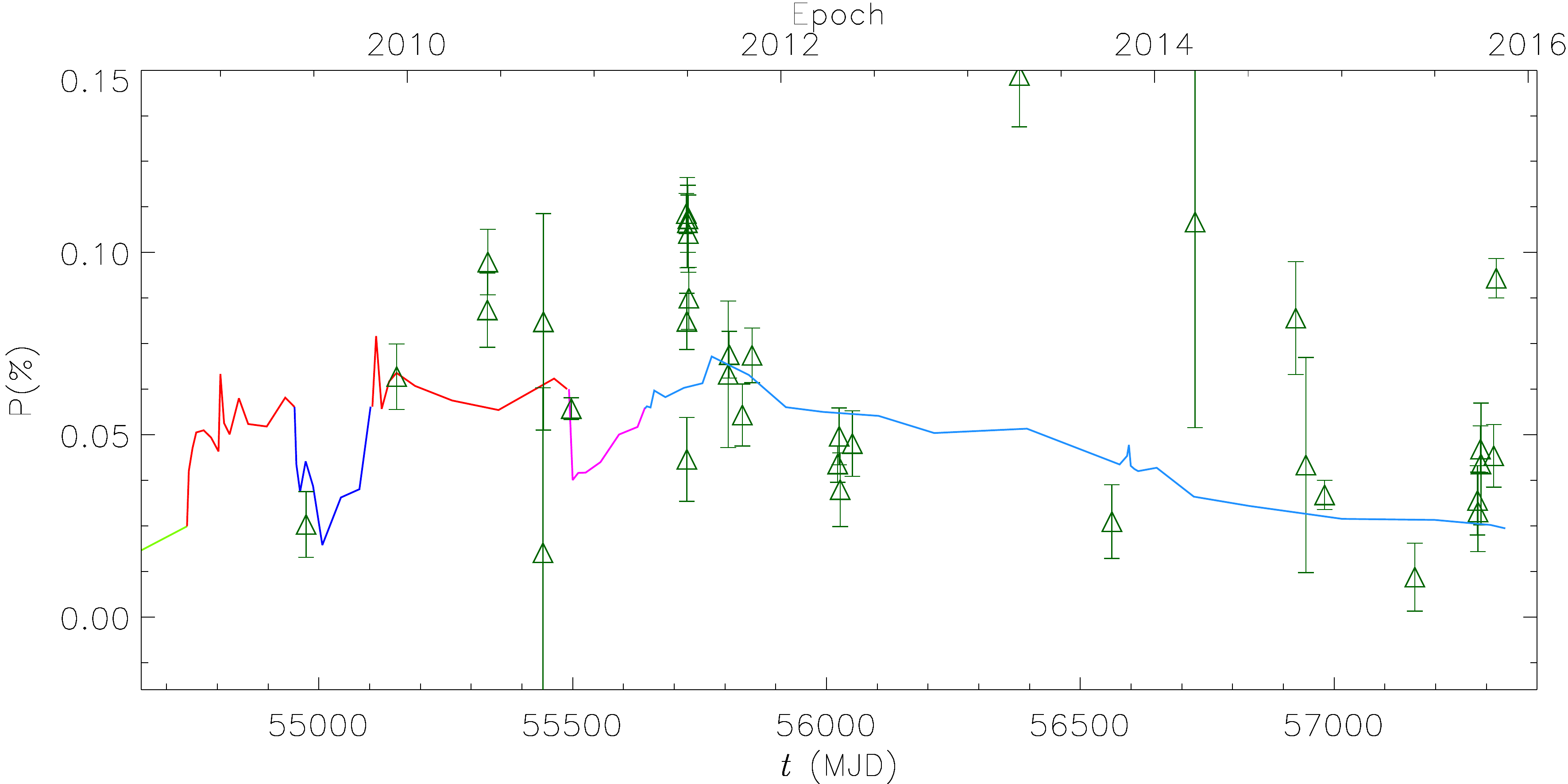}}
    \end{minipage}
    \caption[Polarimetric data of $\omega$ CMa vs. model]{$V$-band polarimetric data of $\omega$ CMa vs. model for the four cycles (top), and the last cycle (bottom). The data are shown with green triangles. The colored lines show the synthetic polarization, with the colors representing the value of the $\alpha$ parameter matching Fig.~\ref{fig:summary}.}
    \label{fig:polarimetry_model}
\end{figure}


Figure~\ref{fig:polarimetry_model} shows the fit of the VDD model presented in Sect.~\ref{sect:final_model} to the observed data. The agreement between the model and data is reasonable. It follows the short (partial) formations and dissipations during the main formation phase (O$_4$), and roughly matches the dissipation phase. The results shown in Fig.~\ref{fig:polarimetry_model} are based on the inclination angle 12$^\circ$. Thus, being an almost face-on system, it was expected to have low level of polarization. During the last dissipation, there are some jumps in the polarization level that can be related to the flickering events. As we did not consider very short flickers, our model does not match these ponits (e.g., epochs 56300 to 56800).

\section{Magnitudes and colors}
\label{sect:mag_color}

The $V$-band photometric data of $\omega$ CMa was fitted well by the VDD model, covering all the disk formation and dissipation phases. In this section we show how the model compares with other photometric data.

\begin{figure}[!ht]
    \centering
    {\includegraphics[height=0.45 \textheight, width=0.8 \columnwidth]{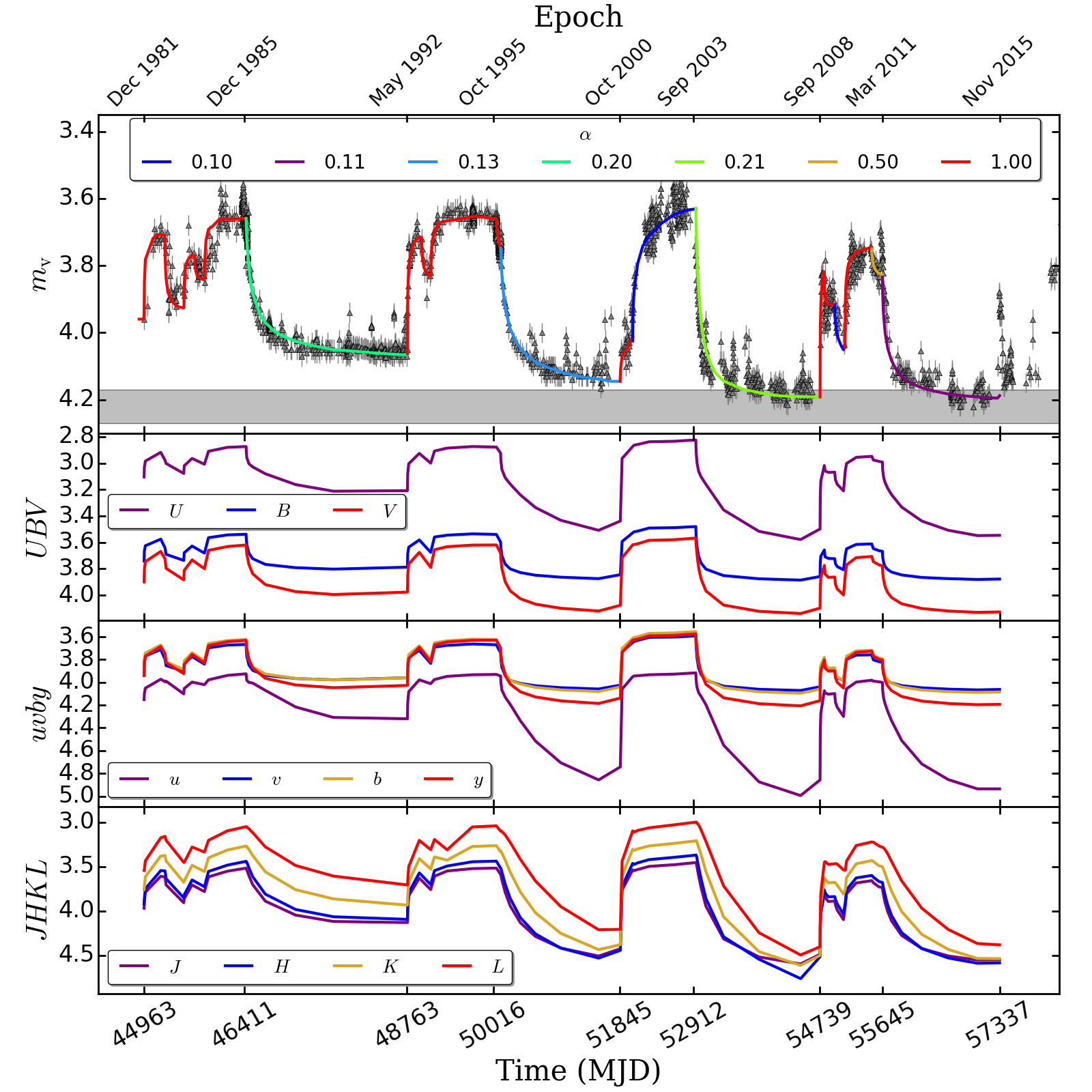}}
    \caption[Synthetic light curves of $\omega$ CMa at different bands from the UV to the far IR.]{Synthetic light curves of $\omega$ CMa at different bands from the UV to the far IR. The upper plot of each panel shows the fit of the VDD model to the $V$-band photometric data (see Fig.~\ref{fig:summary} for more details). The other plots shows the model light curvers for $UBVJHKL$ Johnson bandpasses \citep{johnson1975} and Str\"{o}mgren $uvby$ \citep{stromgren1956}}
    \label{fig:mag_color}
\end{figure}

Figure~\ref{fig:mag_color} displays the synthetic light curves of $\omega$ CMa in different magnitudes from UV to far infrared. This figure only presents the models. The top panel of the figure shows the $V$-band modeling presented in Sect.~\ref{sect:final_model}. The second plot of the top panel shows the synthetic $UBV$ bands together. Interestingly, the $U$-band light curve displays the largest variations, which is expected as this band should be more sensitive to the inner disk conditions, whose density values vary widely (Fig.~\ref{fig:summary}). The model predicts a complex behavior as different bandpasses are considered. In general, we see that the longer the wavelength, the slower the rate of magnitude variations in the model light curve. This is explained by the fact that larger disk volumes responds (from where the long wavelength bandpasses come) respond slower to variations in the inner disk.


\begin{figure}
    \begin{minipage}{\linewidth}
        \centering
        {\includegraphics[height=0.45 \textheight, width=1.0 \columnwidth]{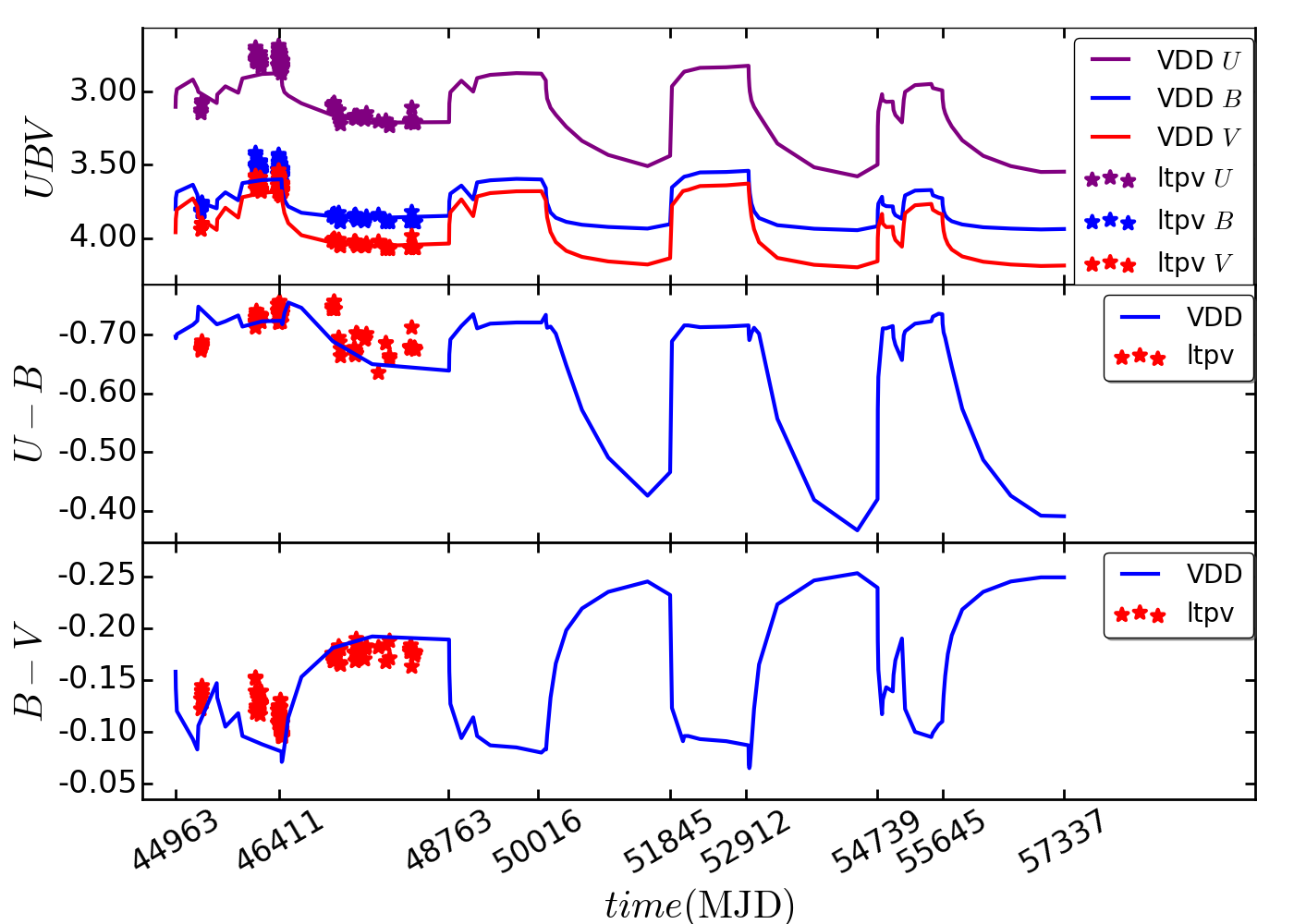}}
    \end{minipage}\par\medskip
    \begin{minipage}{\linewidth}
        \centering
        {\includegraphics[height=0.45 \textheight, width=1.0 \columnwidth]{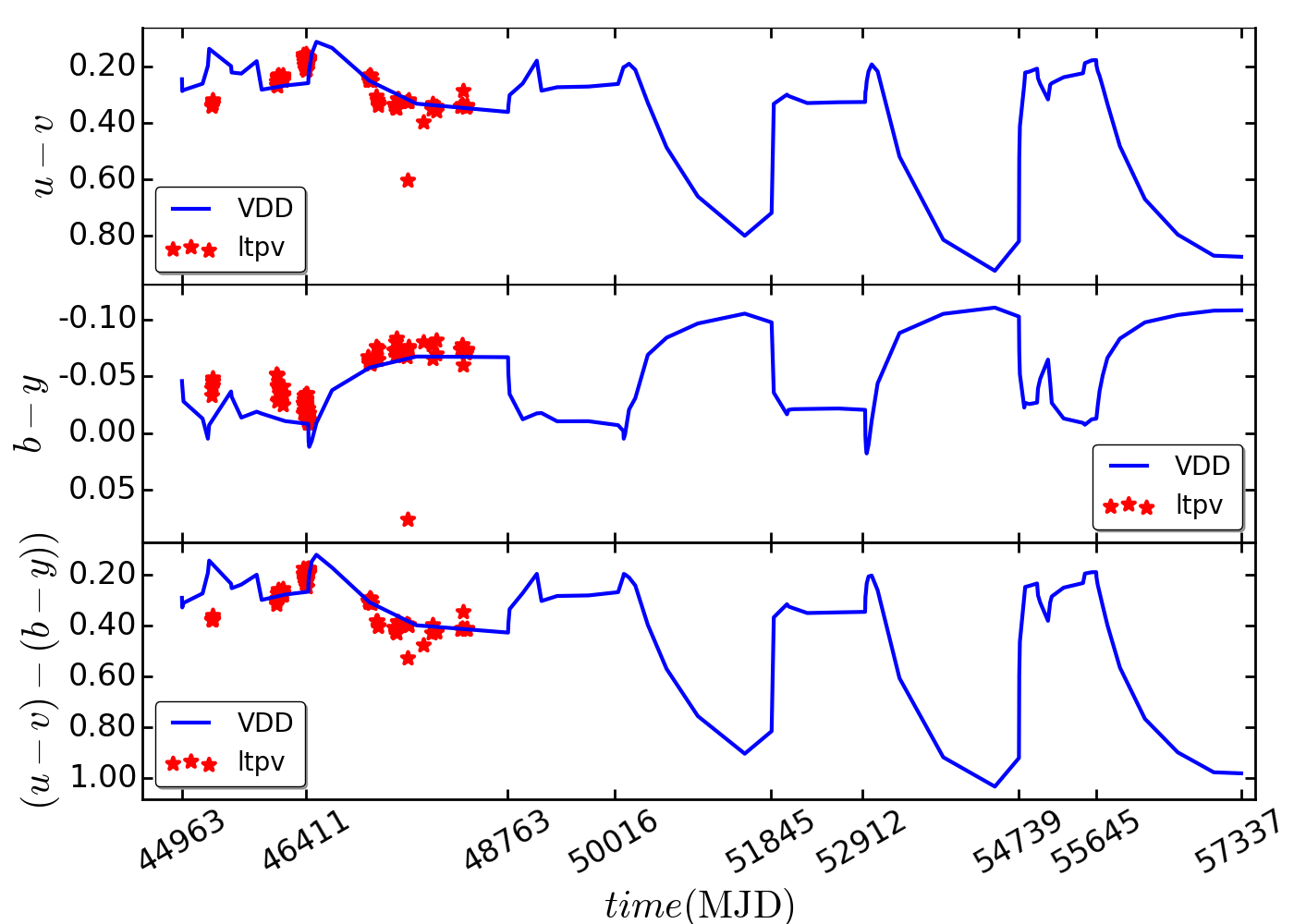}}
    \end{minipage}
    \caption[Comparison between magnitudes and color indices of the VDD model and observed data in Johnson $UBV$ and Str\"{o}mgren $uvby$ filters]{Comparison between magnitudes and color-indices (points) and the VDD model (blue lines). Top: $UBV$. Bottom: Str\"{o}mgren $uvby$, both from LTPV.}
    \label{fig:mag_time}
\end{figure}

The top panel of Fig.~\ref{fig:mag_time} shows the comparison between the observed $UBV$ data and the model. The color-indices are compared as well. The data cover only the C1 and were fitted well by the $UBV$ magnitudes models during Q1. The fit during O1 is not as good as the Q1. We have multi-band data only for C1, that was observed with LTPV (see Fig.~\ref{fig:lightcurve_observer}). There is a slight mismatch between the LTPV magnitudes and the one of other authors, which explains why the LTPV data always lie a little above the models. The general shape of the curve, as well as the colors, are however quite well reproduced by the model.

\begin{figure}
    \begin{minipage}{\linewidth}
        \centering
        {\includegraphics[width=0.9 \columnwidth]{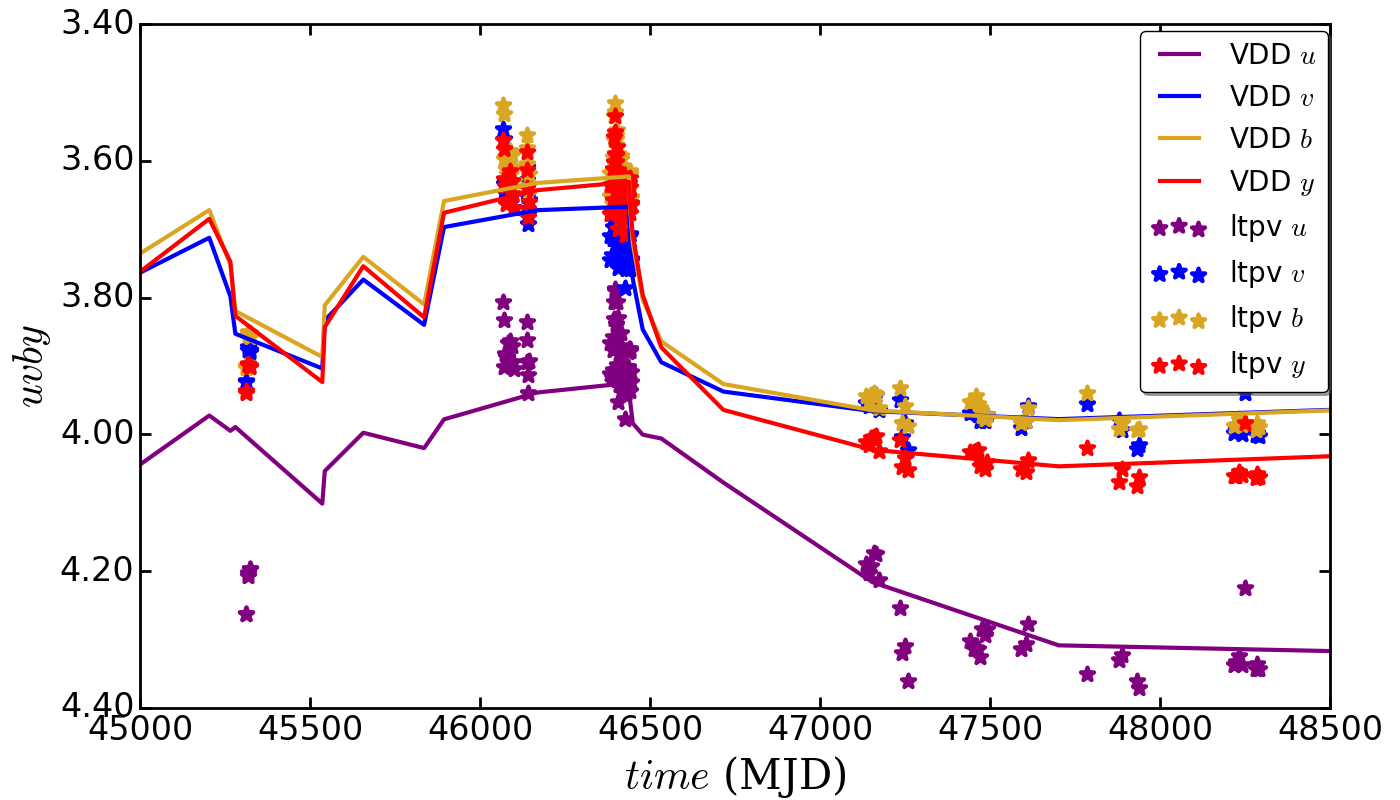}}
    \end{minipage}
    \caption[Comparison between the modeled magnitudes and observed data in $uvby$ filters]{Comparison between the model (solid lines) and the $uvby$ magnitudes (symbols, LTPV)}
    \label{fig:uvby_time}
\end{figure}

The bottom panel of Fig.~\ref{fig:mag_time} shows the comparison with the $uvby$ color-indices. The quality of fit is the same as $UBV$ colors. However, $u-v$ and $(u-v)-(b-y)$ color indices do better than $b-y$ color index. In Fig.~\ref{fig:uvby_time} a comparison is made with the $uvby$ magnitudes, with generally good results.

The ability of the VDD model to match the $UBV$ and $uvby$ colors and magnitudes is a very significant result. As these band-passes probe slightly different disk regions, this good agreement means that both the density levels and the density radial gradients in the inner disk are well reproduced by the models.

\begin{figure}
    \begin{minipage}{\linewidth}
        \centering
        {\includegraphics[width=0.9 \columnwidth]{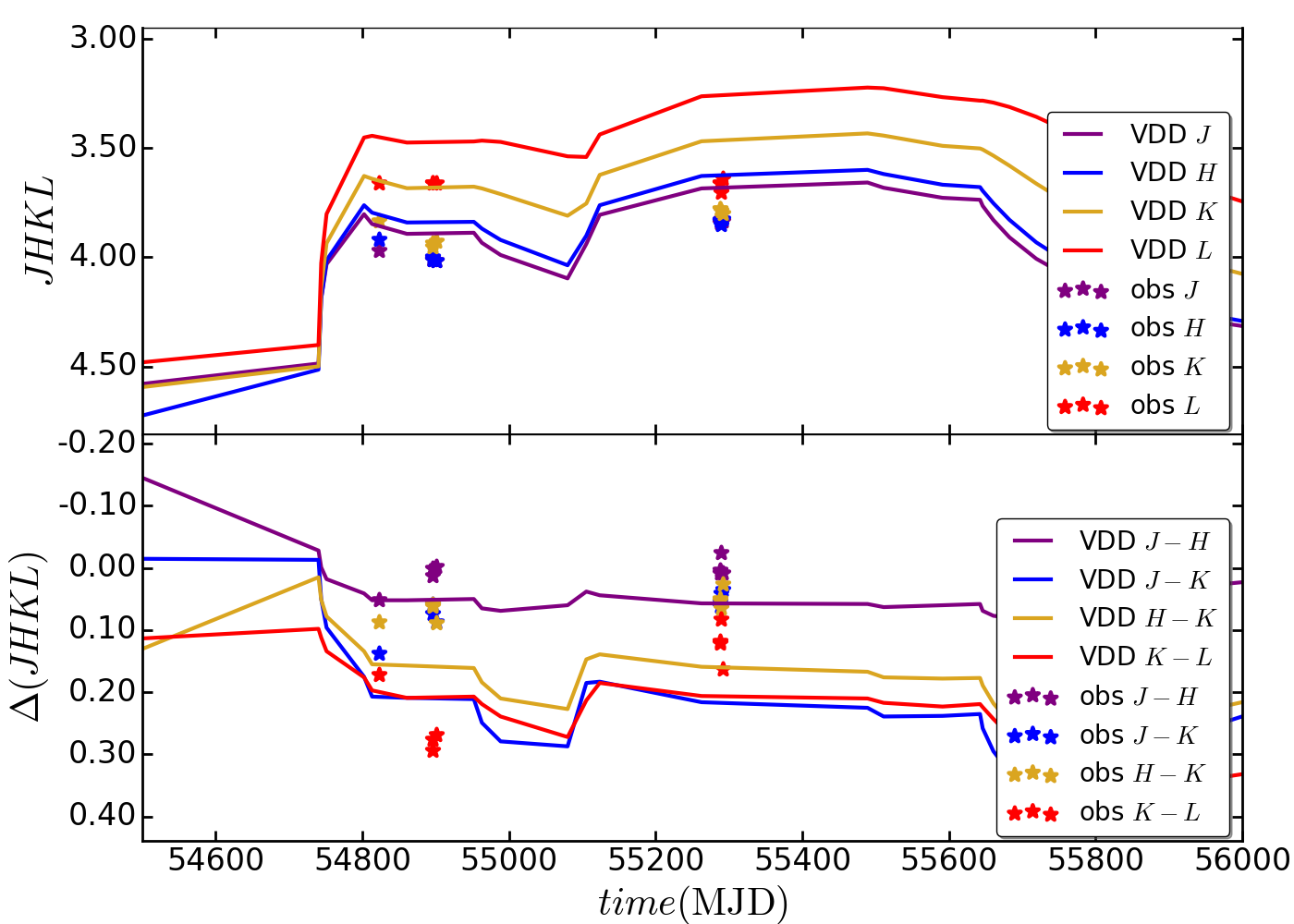}}
    \end{minipage}
    \caption[Comparison between magnitudes and color indices of the VDD model and observed data in $JHKL$ filters]{Comparison between magnitudes and color indices of the VDD model and observed data in $JHKL$ filters. The VDD models are shown with solid lines and the data are shown as colored stars. Each color represents one specific filter as indicated in the legend of the plot.}
    \label{fig:jhkl_time}
\end{figure}

Figure~\ref{fig:jhkl_time} shows the comparison between the observed $JHKL$ magnitudes and colors and the model. The top panel reveals that the model is always a bit brighter than the data. The discrepancy is maximum for the $L$ band and minimum for the $J$ band. Thus, the longer the wavelength, the larger the discrepancy. The bottom panel shows that the color indices have the same issue like magnitudes. 



Although the scarcity of IR data makes it difficult to have a firm conclusion, it still appears that the $JHKL$ color indices are more or less well reproduced by the VDD model, while the actual magnitudes are not. The modeled disk is always systematically brighter in the $JHKL$ than the observations, which means that the model may be too dense. This might argue for a higher $\alpha$ in the outer parts, that would drain the disk faster, making it less dense. See next section for a more detailed discussion on the hypothesis of a variable $\alpha$.


\section{Spectroscopy}
\label{sect:spectroscopy}

\begin{figure}[!ht]
    \begin{center}
        \includegraphics[height=0.38 \textheight, width=0.9\linewidth]{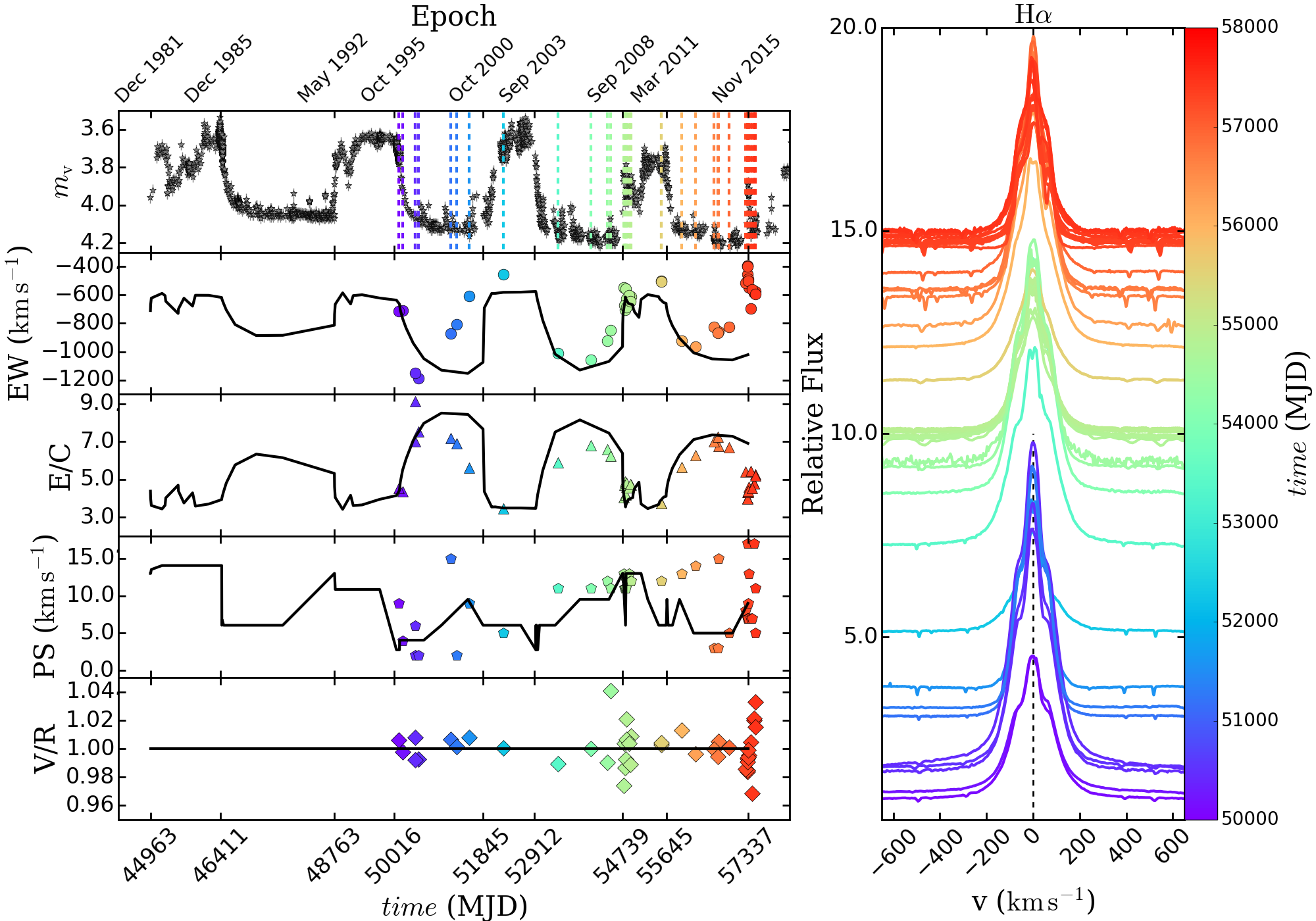}
        \caption[Spectroscopic data (H$\alpha$) of $\omega$ CMa vs. model]{Comparison between the models (lines) and the H$\alpha$ EW, E/C, PS, and V/R. The top plot at the left panel shows the $V$-band light curve with horizontal lines marking the date the spectrum was taken, and the corresponding spectra (using the same color scheme) are shown in the right panel. Different colors indicate the date of observation, according to the legend on the right. Observed EW, E/C, PS, V/R are shown with colored circles, triangles, pentagons, and diamonds, respectively, in the second to fifth panels.} 
        \label{fig:spectroscopy_model_Ha}
    \end{center}
\end{figure}

\begin{figure}[!ht]
    \begin{center}
        \includegraphics[height=0.38 \textheight, width=0.9\linewidth]{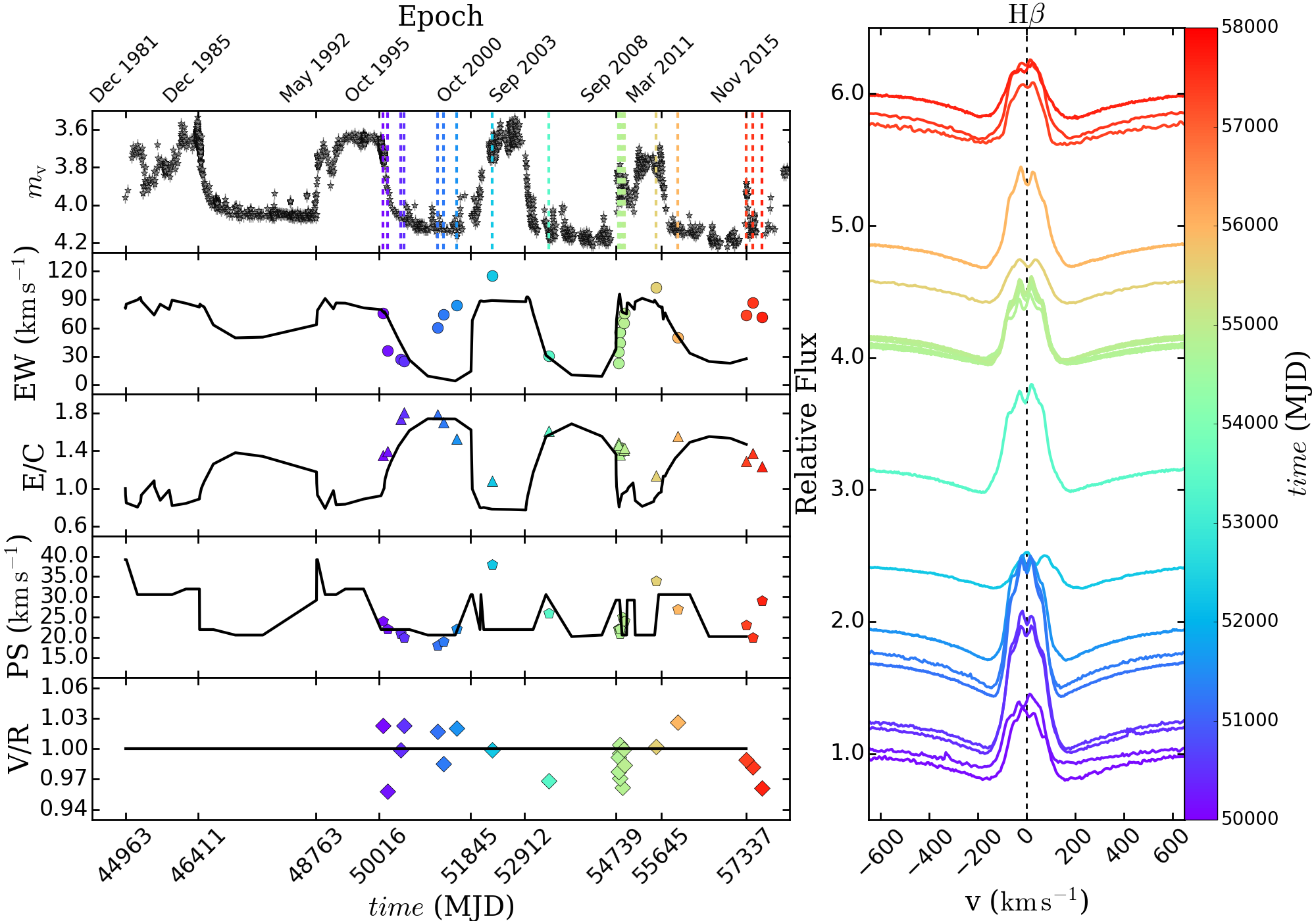}
        \caption[Spectroscopic data (H$\beta$) of $\omega$ CMa vs. model]{Same as Fig.~\ref{fig:spectroscopy_model_Ha} for H$\beta$.} 
        \label{fig:spectroscopy_model_Hb}
    \end{center}
\end{figure}

\begin{figure}[!ht]
    \begin{center}
        \includegraphics[height=0.40 \textheight, width=0.9\linewidth]{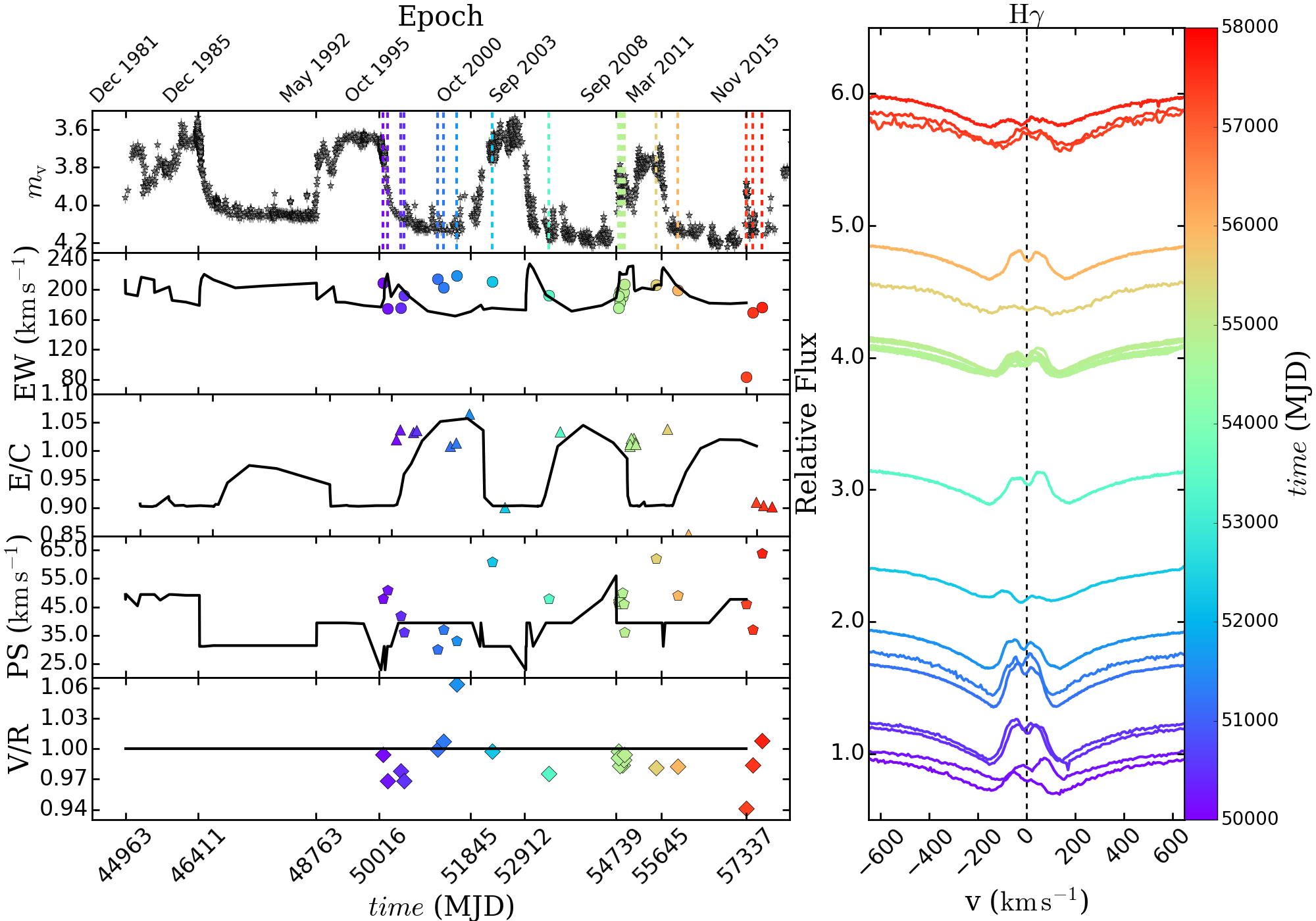}
        \caption[Spectroscopic data (H$\gamma$) of $\omega$ CMa vs. model]{Same as Fig.~\ref{fig:spectroscopy_model_Ha} for H$\gamma$.} 
        \label{fig:spectroscopy_model_Hg}
    \end{center}
\end{figure}

\begin{figure}[!ht]
    \begin{center}
        \includegraphics[height=0.40 \textheight, width=0.9\linewidth]{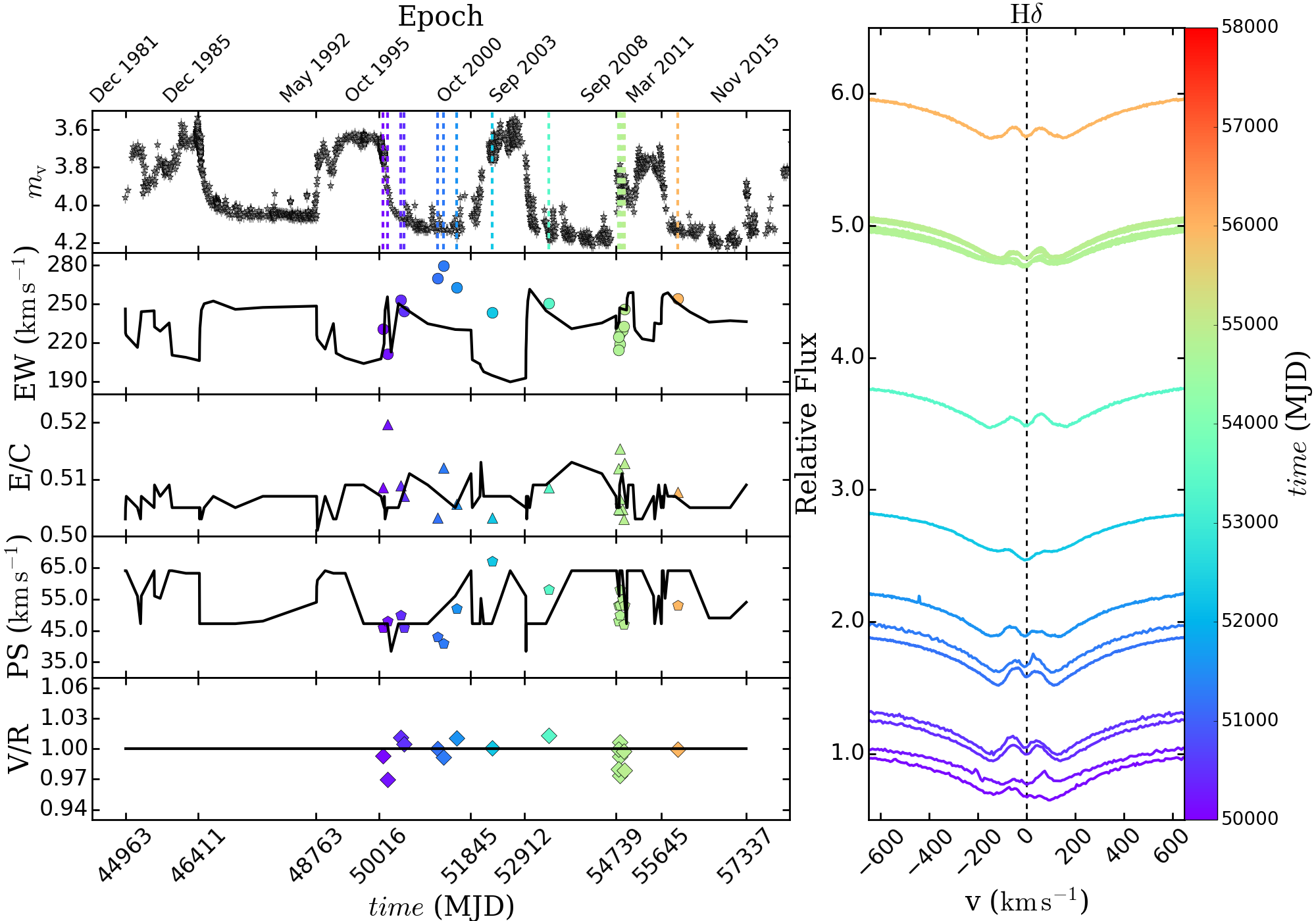}
        \caption[Spectroscopic data (H$\delta$) of $\omega$ CMa vs. model]{Same as Fig.~\ref{fig:spectroscopy_model_Ha} for H$\delta$.} 
        \label{fig:spectroscopy_model_Hd}
    \end{center}
\end{figure}

Spectra of $\omega$ CMa were extensively taken during the last decades with different instruments covering many of the most important events in $\omega$ CMa's history (Sect.~\ref{sect:other_data}). In particular, we have a complete coverage of the cycles 3 and 4 (Fig.~\ref{fig:data_dist}b), as well as of Q2, which provides us with a unique opportunity for testing the model. Figures.~\ref{fig:spectroscopy_model_Ha} to \ref{fig:spectroscopy_model_Hd} show the comparison between the VDD model and observed equivalent width (EW), E/C, PS, and V/R for H$\alpha$, H$\beta$, H$\gamma$, and H$\delta$. The right panel in each figure shows the spectrum and its evolution according to the evolution of the disk. The V/R and the PS were computed by fitting a gaussian curve to each emission peak, in order to determine its height against the neighboring continuum (which was normalized to one), as well as its velocity. These procedures made usage of the ``pyhdust'' package\footnote{\href{https://pyhdust.readthedocs.io/en/latest/}{https://pyhdust.readthedocs.io/en/latest/}}.

It is important to mention that we did not adjust the model to obtain an optimum fit. We only and directly used the same model and scenario described in Sect.~\ref{sect:final_model} to calculate these profiles, in order to evalutate how this model, which was developed using only the $V$-band as a constraint, performs when compared to multi-technique data. Obtaining the optimum model for these observables is not the aim of this work.


The second and third panels of each figure displays the EW and E/C of the line. The EW and the E/C ratio of a line reveals a complex interplay between the line emission proper and that of the adjacent continuum. The line emission comes from a very large volume of the disk \citep{carciofi2011}, that responds very slowly to changes in the disk feeding rate. Conversely, the adjacent continuum responds very quickly to these changes, as discussed in Chap.~\ref{chap:photometry}. Therefore, when the continuum emission rises (e.g., during an outburst), the EW initially drops in magnitude and the E/C ratio will fall. Conversely, when the continuum emission drops (e.g., during quiescence), the EW will increase in magnitude and the E/C ratio will increase.

Figure~\ref{fig:spectroscopy_model_Ha} can be interpreted with the above scenarios in mind. In all quiescence phases, the EW increases in magnitude (becoming more negative) and the E/C increases, as a result of the quick suppression of the inner disk, that causes the emission in the adjacent continuum to drop quickly. This initial dissipation of the inner disk hardly affects the line emission though. Only much later in the dissipation, the disk as a whole start to decline, and the line emission drops. Consequently, the EW will decrease in magnitude and E/C will also drop. At outburst, the converse happens: the inner disk fills up quickly, giving rise to a sudden increase in the continuum. As a result, the EW decreases in magnitude and E/C decreases. 

The model can reproduce these variations in a qualitative way, but not in a quantitative way. After the quick increase in magnitude of the EW at the onset of dissipation, which both the model and observations agree on, the observed EW decrease in magnitude much before the model does (the same is observed with the E/C). 

In the following we provide some speculative explanation for this mismatch between the model and the data. One possibility is to consider that, in the models, the outer disk is not being drained of material fast enough. To achieve that, either the temperature of the disk rises with radius (which is not physically justified) or $\alpha$ grows with distance from the star (see Sect.~\ref{sect:temperature_evolution}). Therefore, this might be the first evidence of a radially varying $\alpha$. Another possibility is to consider that there is an unknown binary companion truncating the disk of $\omega$ CMa at radii smaller than the 50$R_{\rm eq}$ assumed here (recall that this was an arbitrary assumption). If that were the case, the mass reservoir of the outer disk would be smaller and it would dissipate faster, as suggested by the observations. Clearly all this is still very speculative, and further work is necessary for harmonizing the model with the observations.

The low inclination angle of $\omega$ CMa causes an almost single-peak profile in H$\alpha$ whose flux comes mainly from the outer part of the disk (in comparison to the other Hydrogen lines). However, it is still possible to recognize some double-peak structure in the spectra \citep{rivinius2013a}. This creates some difficulties to calculate the PS, which has no meaning in the case of a single-peak spectrum. Therefore, the data shown in the fourth panel of Fig.~\ref{fig:spectroscopy_model_Ha} is somewhat uncertain. Although the observed PS for H$\alpha$ was relatively well fitted by the model during O3 to O4, there is not a good match for Q2 and Q4. 

The fifth panel of Fig.~\ref{fig:spectroscopy_model_Ha} displays the V/R variations measured in the H$\alpha$ line. All measurements indicate that the line remained very symmetric, with a very low V/R values. Of course V/R equals 1 for the model, which assumes axial symmetry for the disk.

In general, the results for H$\beta$ are similar to those for H$\alpha$. The EW curve is qualitatively reproduced, but a qualitative comparison fails mainly during the quiescence phases. Of particular significance is the close match between data and the model for the fast decline in magnitude of O4. The E/C ratio is much better reproduced, however. It is is important to recall that, since the H$\beta$ opacity is smaller than H$\alpha$, the formation volume of this line is smaller \citep{carciofi2011}. This can be seen by the PS values, which for H$\beta$ lies around 25 $\rm{km\,s^{-1}}$, while for H$\alpha$ it is in general smaller than 10 $\rm{km\,s^{-1}}$. The larger PS indicate that H$\beta$ is indeed formed closer to the star, where the rotational velocities are larger. The fact that the model reproduces well this behavior is a very significant result.

The results for H$\gamma$ (Fig.~\ref{fig:spectroscopy_model_Hg}) and H$\delta$ (Fig.~\ref{fig:spectroscopy_model_Hd}) are ambiguous. The model reproduces the data quite well for the whole of forth cycle, but in both cases Q3 is not well reproduced. At this stage we have no explanation for that, and further studies are necessary to explain this behavior. H$\gamma$ and H$\delta$ are formed progressively closer to the star, as revealed by their increasing PS ($\approx 40$ and $\approx  55$ $\rm{km\,s^{-1}}$, respectively). This is well-reproduced by the models.

%
\chapter{Conclusions and Perspectives}
\label{conc}

We investigated a large suite of observations of $\omega$ CMa that, in the past 34 years, underwent four complete cycles of disk formation followed by a partial dissipation. Typically, formation phases lasted between $\sim$ 2.5--4.0 years and the dissipation phases $\sim$ 4.5--6.5 years. The results of a detailed VDD hydrodynamic modeling coupled with 3-D radiative transfer calculations suggest six main conclusions:

\begin{itemize}

\item[1)] We demonstrate that the VDD model is capable of reproducing the disk variability during both build-up and dissipation phases. This result has an important theoretical consequence, as it supports the fact that viscosity is the main driver of the Be disks.
\item[2)] Different values of $\alpha$, in the range of $0.1-1.0$, were required to model the data. Typically, the values of $\alpha$ during the formation phases are higher than during dissipation. A similar trend was recently found for a sample of Be stars in the SMC \citep{rimulo2018}. Possible causes for this phenomenon are still being investigated.
\item[3)] Contrarily to what is widely accepted in the literature, we provide strong evidence that the quiescence phases of $\omega$ CMa are not true quiescences, because a non-zero disk feeding rate is always necessary to explain the data. This finding may have important consequences for our understanding of the life cycles of Be disks. It also provides a further puzzle to the Be phenomenon, as future models must not only explain how stellar mass loss is turned on and off, as observed in many Be stars, but rather switches from a high mass loss phase to a low mass loss one, as seems to be the case for $\omega$ CMa.
\item[4)] Depending on the distance from the star, the AM flux may be positive (decretion) or negative (accretion). The average AM lost by the star is always positive, even during apparent quiescence.
\item[5)] The total AM lost by the star during the 34 years of observations is $1.3\times 10^{45}\,\mathrm{g\,cm^2\,s^{-1}}$ which is $6\times 10^{-9}$ of the total AM of the central star. If extrapolated for the whole main sequence, this would suggest that about 0.006 of the initial AM content of the star would be lost by disk events during the Main Sequence lifetime. This value is about eleven times smaller than the predictions of the Geneva evolutionary models of fast rotating stars. This result agrees with the study of 54 Be stars in the SMC done by \cite{rimulo2018} and is, perhaps, the result from this PhD thesis that will have a broader impact in stellar astrophysics. The disagreement between our measurements and the Geneva models fostered a collaboration between S\~{a}o Paulo and Geneva, whose outcome may very well be, at the end, that the efficiency of transport mechanisms operating inside massive stars must be revised, with important impacts in the theoretical predictions, such as mixing of elements, main sequence lifetimes, etc.
\item[6)] When compared to Galactic Be stars of similar spectral type, $\omega$ CMa displays a similar disk density scale.

\end{itemize}

Another novel result of this PhD thesis is the application of the VDD theory to model the temporal evolution of other observables, such as polarimetry and spectroscopy. Probing the disk with different observables is very important, as it provides a window to study the fluid dynamics of different parts of the disk.

Overall, the results of this multi-technique study were very positive, with a good match for multi-band photometry, polarization, and some spectroscopic characteristics. This is a very relevant result, as it proved that a model that was constructed from constraints only from the very inner part of the disk (the $V$-band light curve), could be extended to the whole disk and to other physical processes, such as scattering and recombination.

This work engenders many future prospects:

\begin{itemize}
\item[1)] To extend the analysis to the available interferometric data on $\omega$ CMa. Since interferometry is very sensitive to the size of the emitting region, this will represent a crucial test for the model.

\item[2)] To investigate the mismatch between the line characteristics at some of the phases of $\omega$ CMa's history. Some physical causes for the mismatch have been speculated (e.g., radially varying $\alpha$ or a disk truncated by an unseen binary companion), but more work must be done to decide whether and how the current model must be changed to fit better the observations. Extending the analysis to Balmer decrements, that are rather sensitive density diagnostics, may help to settle this issue.

\item[3)] In this work, the hydrodynamic calculations were done assuming isothermal disks, while the radiative transfer calculations show that the disks are not isothermal (see Sect.~\ref{sect:temperature_evolution}). Future non-isothermal hydrodynamic models will address the issue of how these approximations affect our results.

\item[4)] We discussed in Sect.~\ref{sect:temperature_evolution} that the derived values of $\alpha$ in this work are the upper limit if the ablation effect is significantly important for the case of $\omega$ CMa. Therefore, a complete model for the disk dynamics must include, in the future, ablation as one of its ingredients.

\item[5)] Some line ratios are sensitive to the temperature. The available spectroscopic data can therefore be used to search for temperature variations in the disk. This would be very valuable as it could provide some first-hand evidence of non-isothermality in a Be star disk.

\end{itemize}
		%
       						%
						%
						%
\bibliography{tex/bibliografia}	
						%
\begin{apendice}			
\chapter{Observational Log and Plots}\label{ap_a}


\section{Spectroscopic plots}\label{ap_a_1}


\begin{figure}
\begin{minipage}{1.0\linewidth}
\centering
\subfloat[H$\alpha$]{\includegraphics[width=1.0\linewidth]{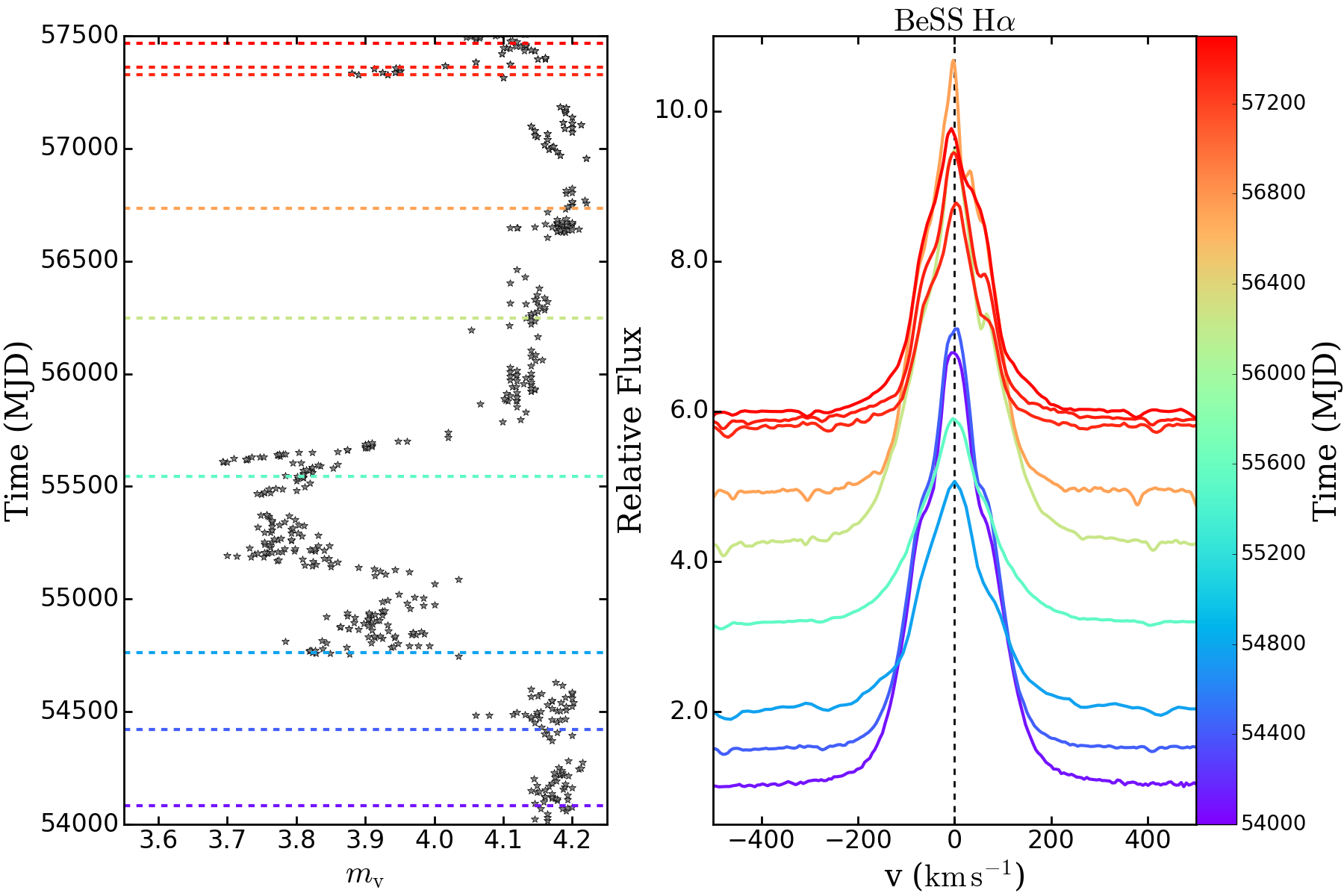}}
\end{minipage}\par\medskip
\begin{minipage}{0.5\linewidth}
\centering
\subfloat[H$\beta$]{\includegraphics[width=1.0\linewidth]{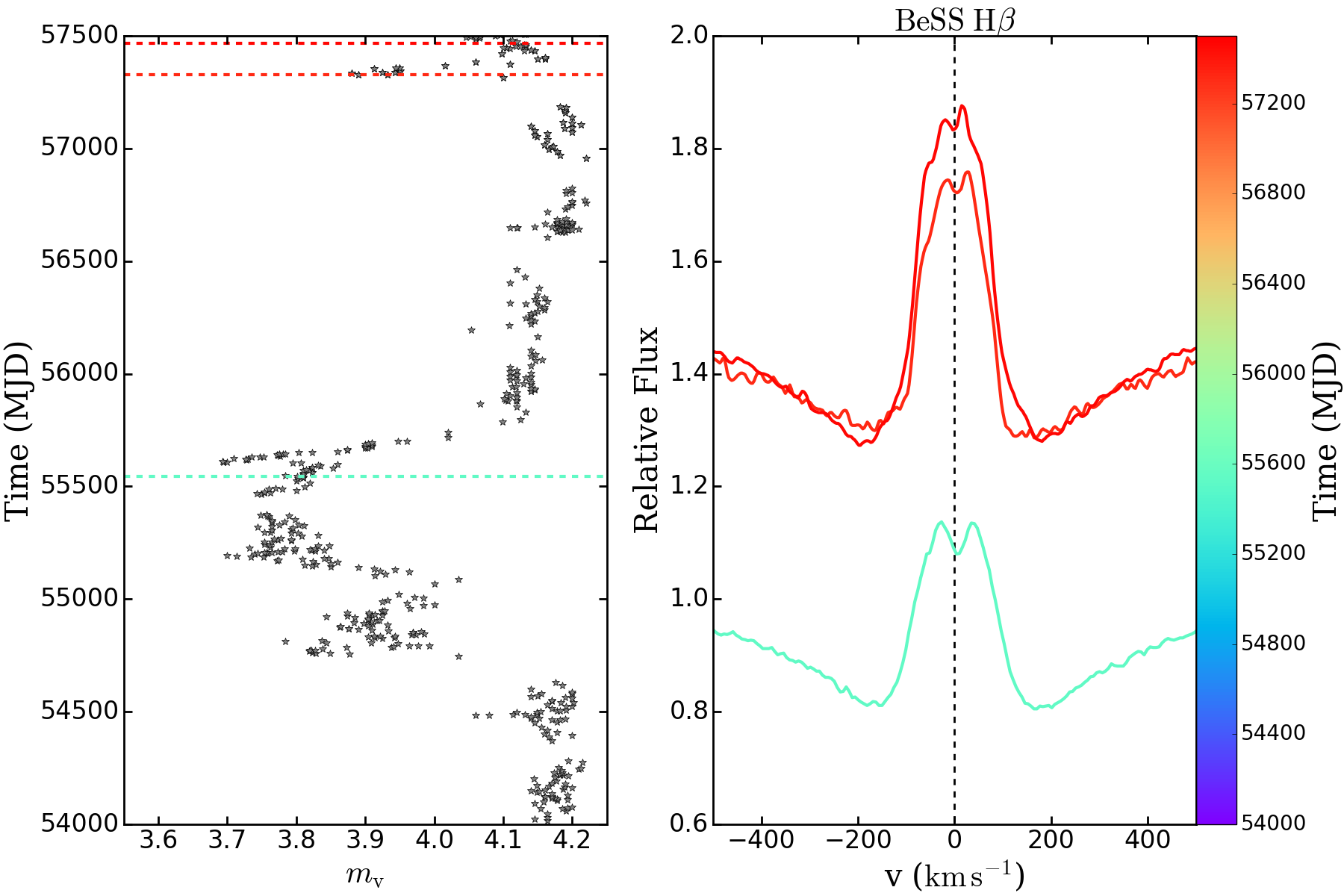}}
\end{minipage}%
\begin{minipage}{0.5\linewidth}
\centering
\subfloat[H$\gamma$]{\includegraphics[width=1.0\linewidth]{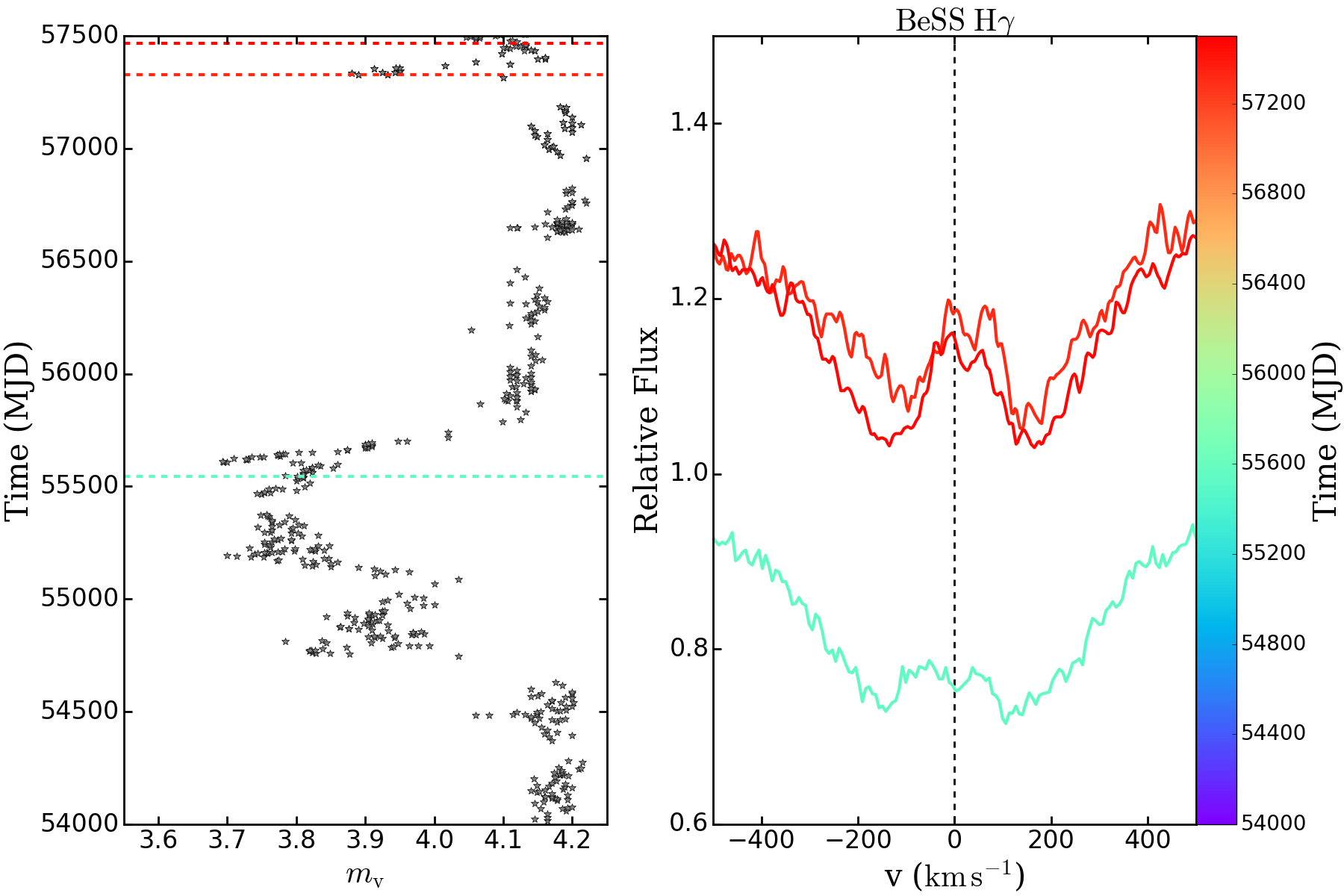}}
\end{minipage}%
\caption[Spectra of $\omega$ CMa observed by BeSS]{Hydrogen lines of $\omega$ CMa observed by BeSS.}
\label{fig:bess}
\end{figure}


\begin{figure}
\begin{minipage}{0.5\linewidth}
\centering
\subfloat[H$\alpha$]{\includegraphics[width=1.0\linewidth]{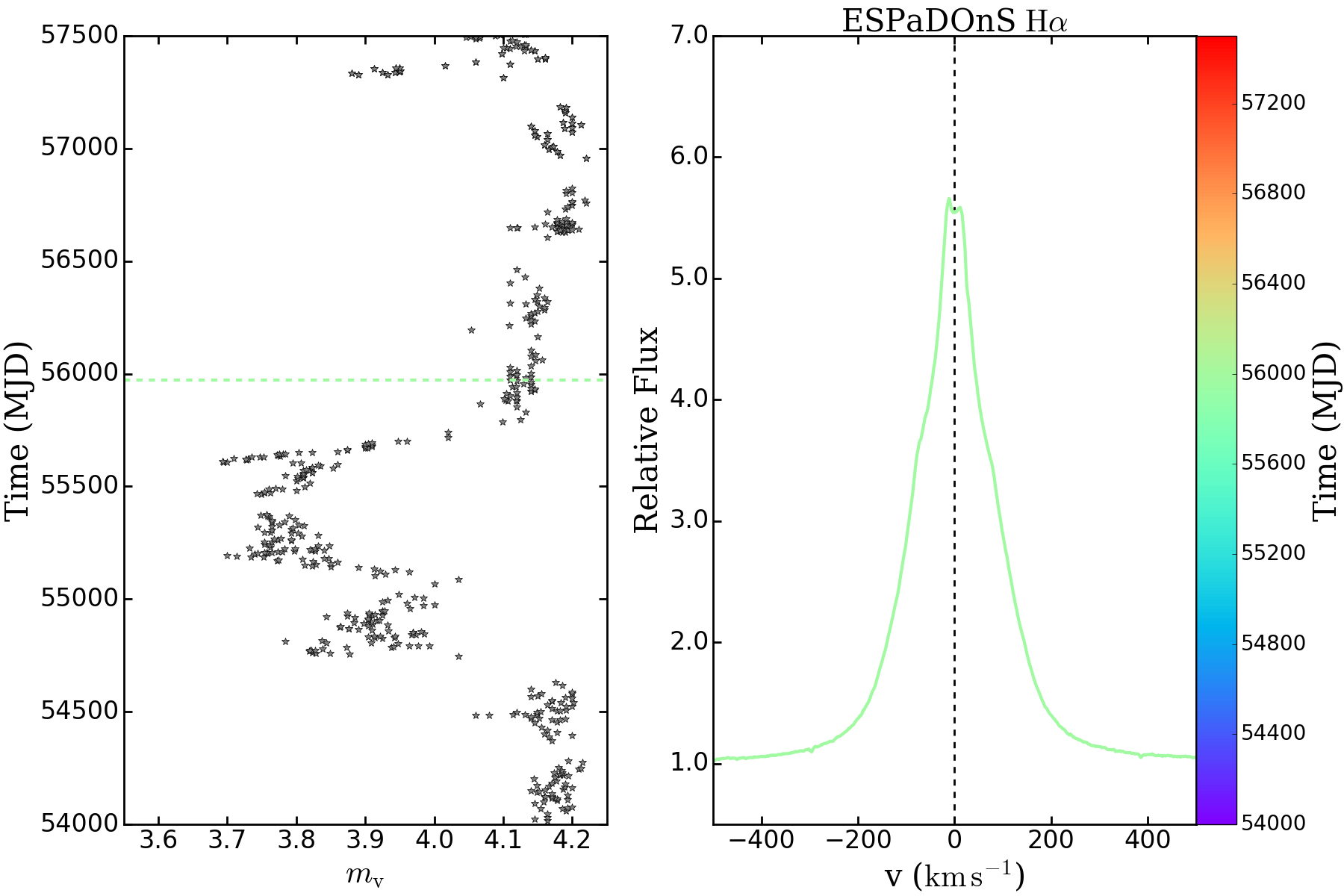}}
\end{minipage}%
\begin{minipage}{0.5\linewidth}
\centering
\subfloat[H$\beta$]{\includegraphics[width=1.0\linewidth]{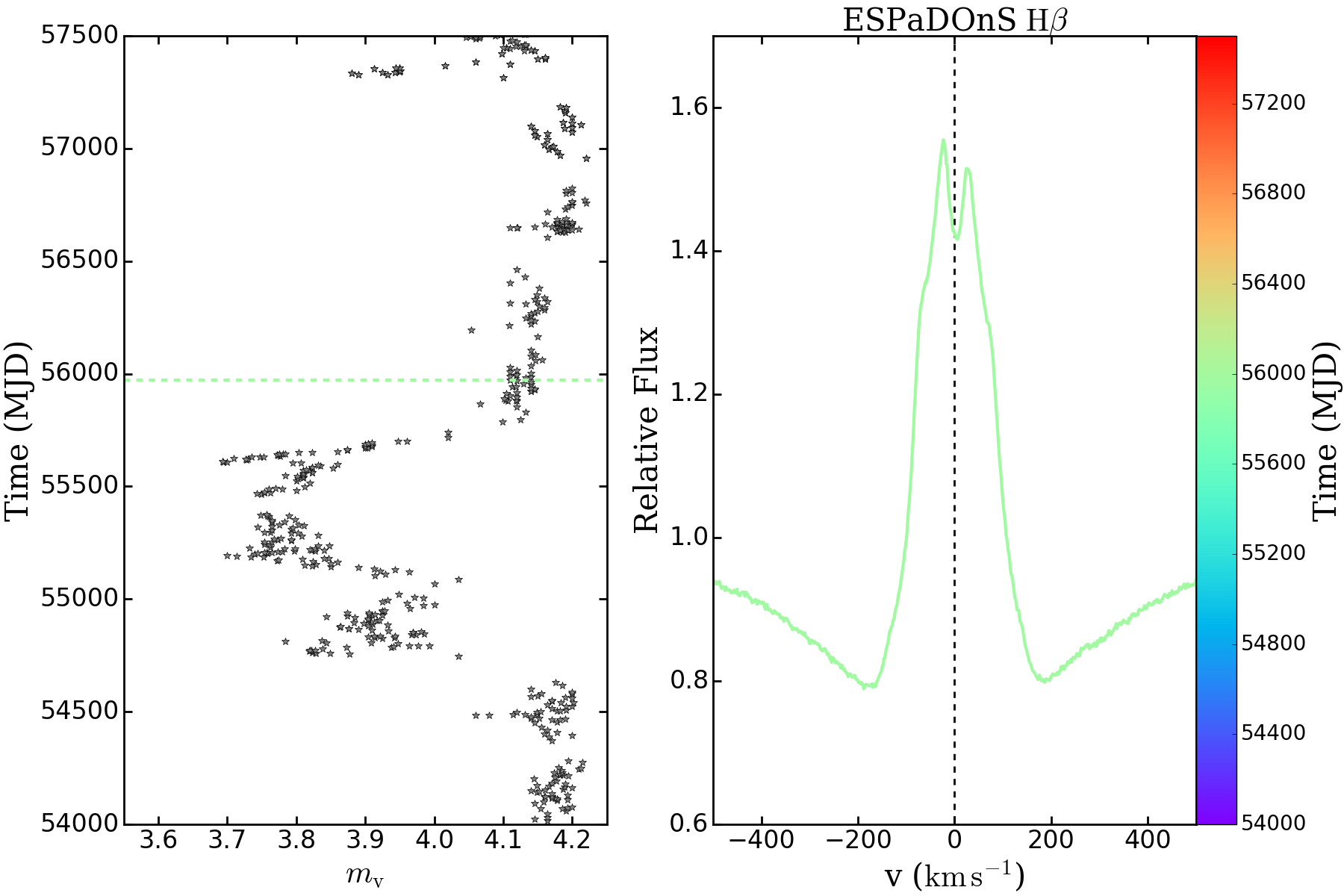}}
\end{minipage}\par\medskip
\begin{minipage}{0.5\linewidth}
\centering
\subfloat[H$\gamma$]{\includegraphics[width=1.0\linewidth]{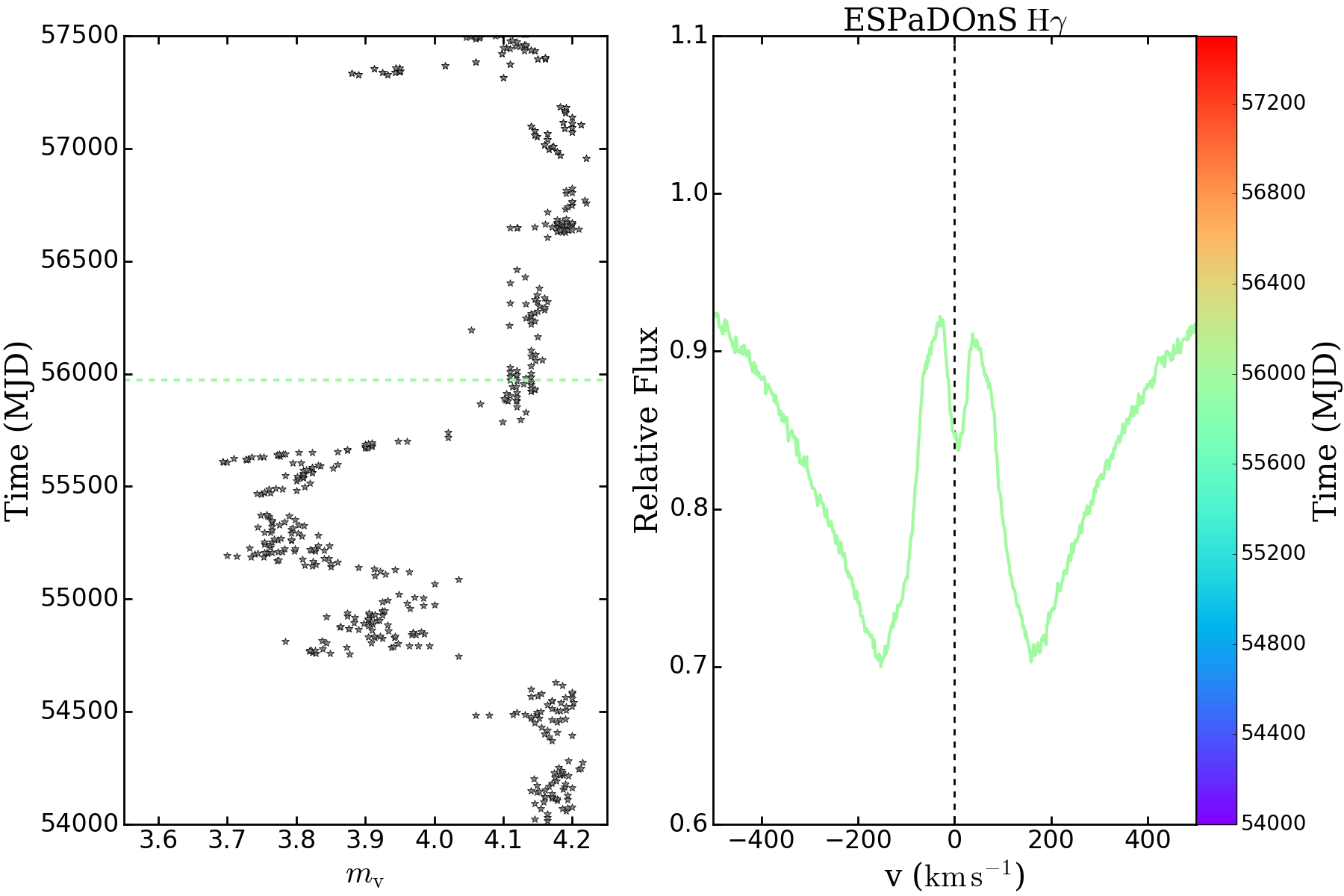}}
\end{minipage}%
\begin{minipage}{0.5\linewidth}
\centering
\subfloat[H$\delta$]{\includegraphics[width=1.0\linewidth]{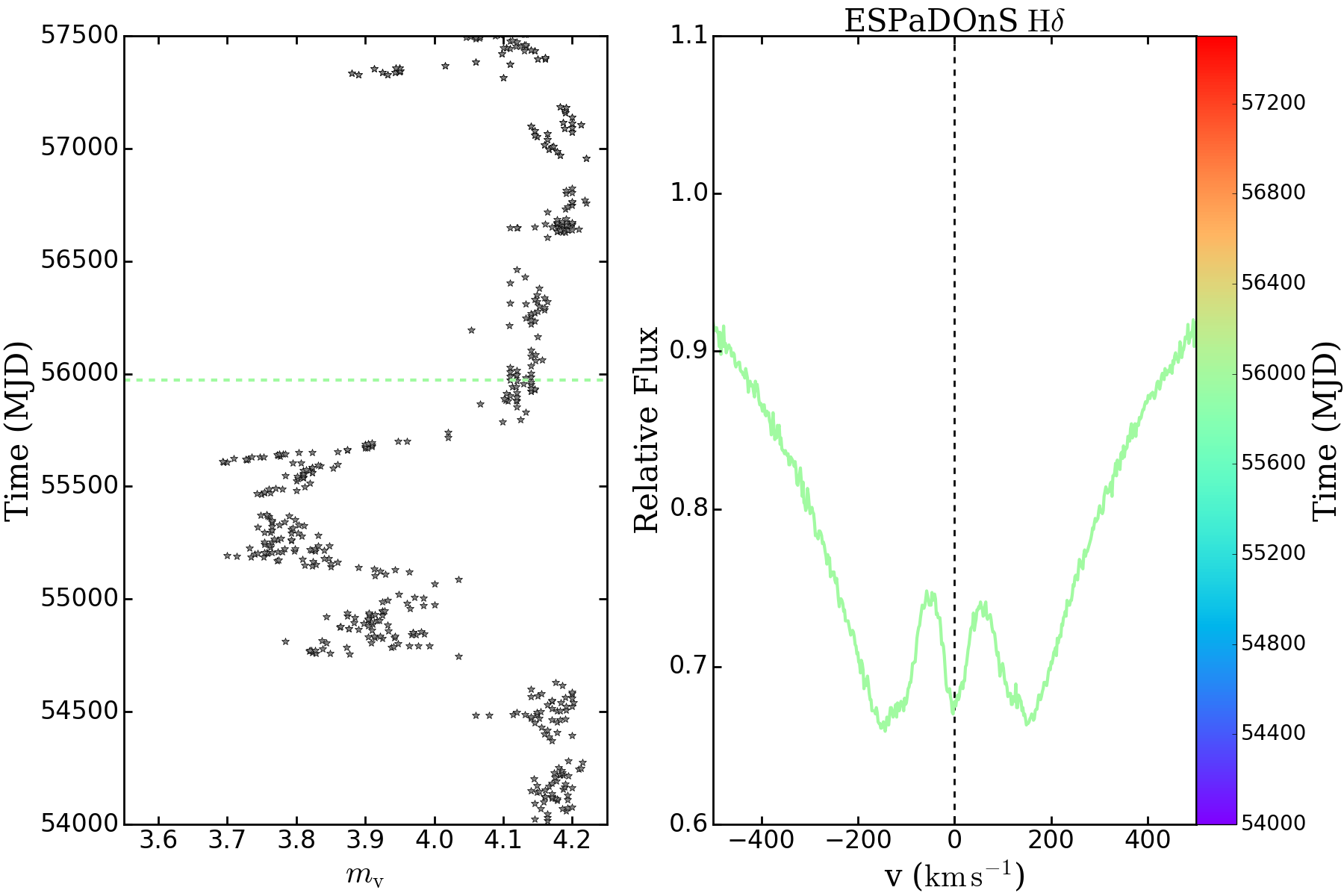}}
\end{minipage}%
\caption[Spectra of $\omega$ CMa observed by ESPaDOnS]{Hydrogen lines of $\omega$ CMa observed by ESPaDOnS.}
\label{fig:espadons}
\end{figure}


\begin{figure}
\begin{minipage}{0.5\linewidth}
\centering
\subfloat[H$\alpha$]{\includegraphics[width=1.0\linewidth]{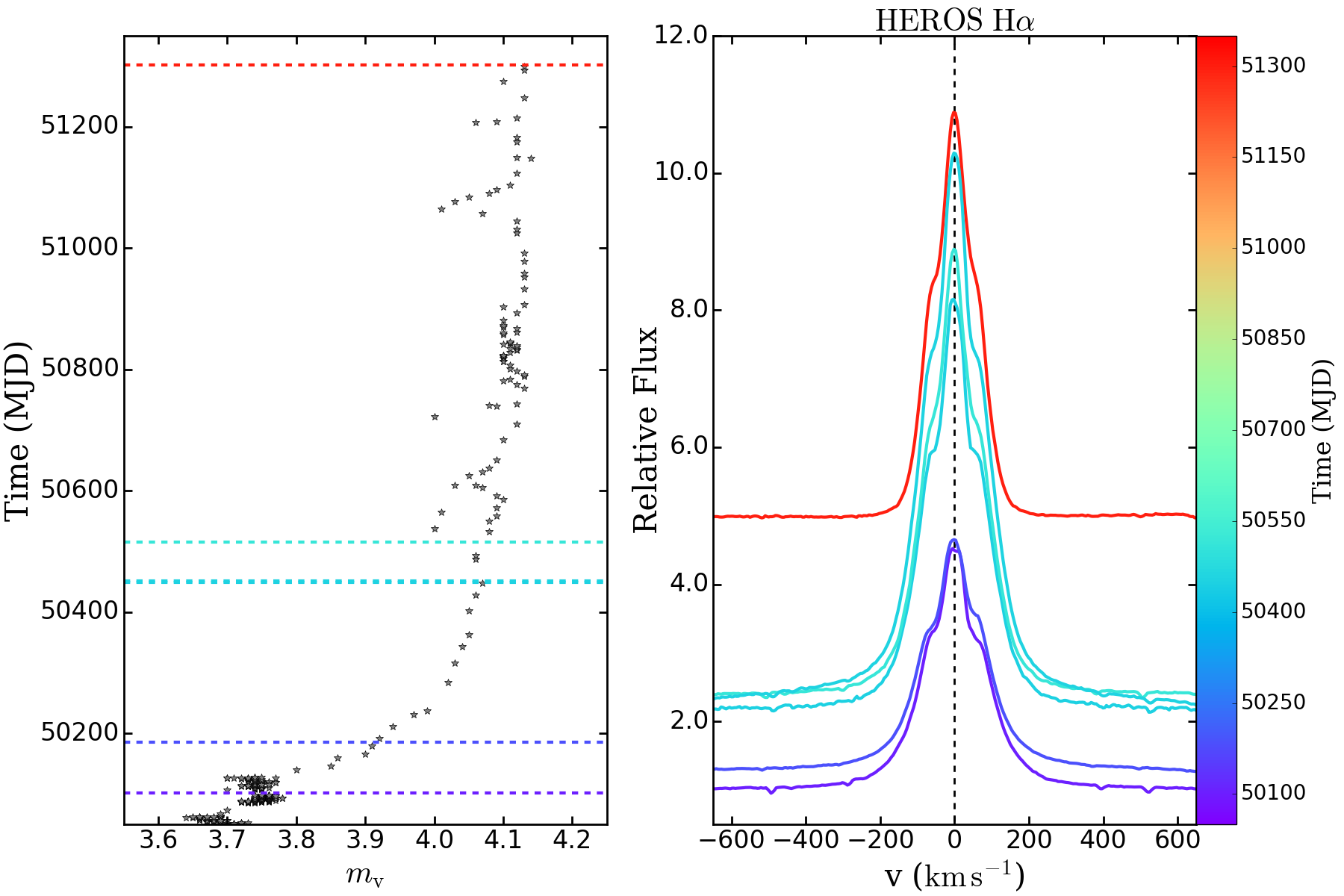}}
\end{minipage}%
\begin{minipage}{0.5\linewidth}
\centering
\subfloat[H$\beta$]{\includegraphics[width=1.0\linewidth]{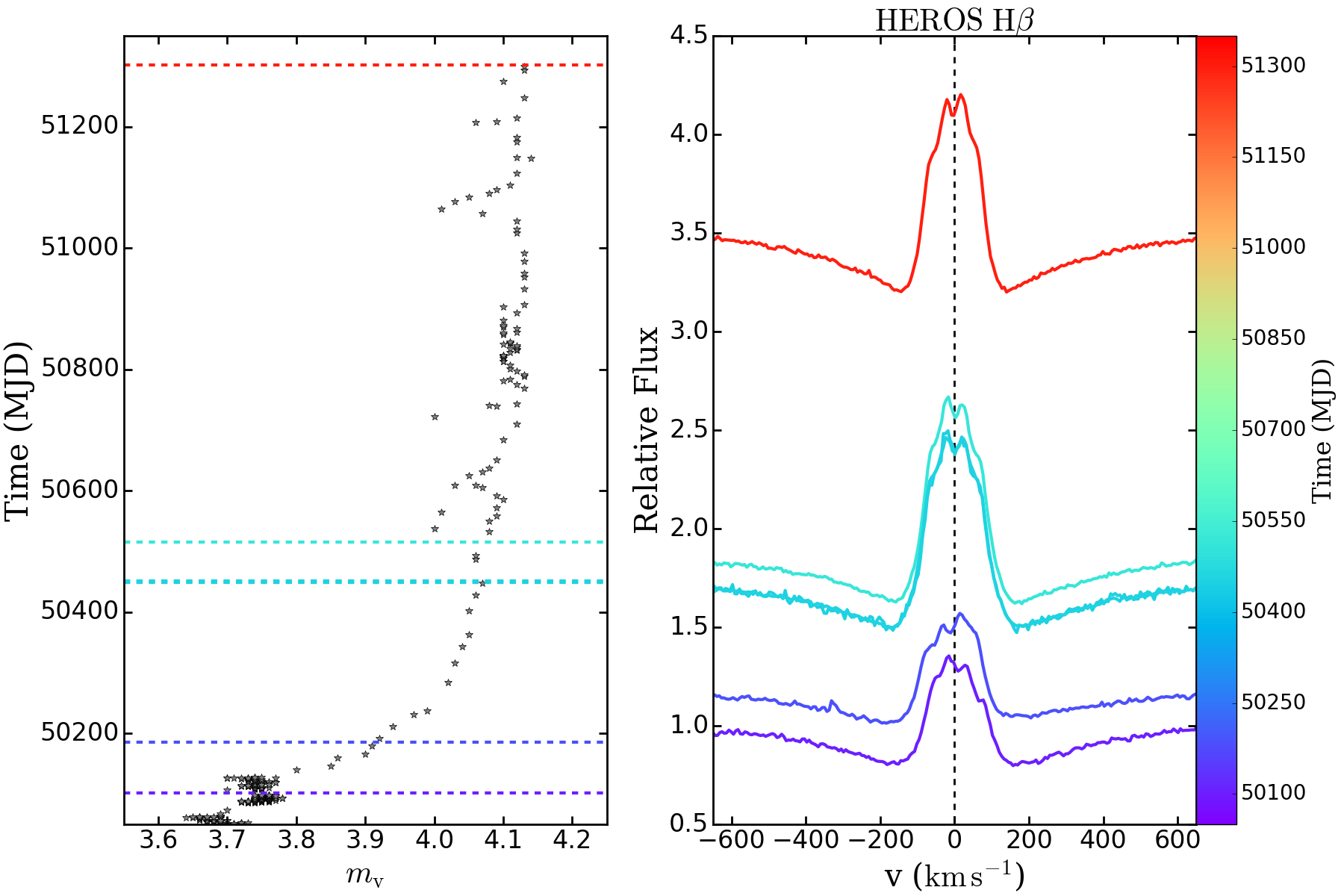}}
\end{minipage}\par\medskip
\begin{minipage}{0.5\linewidth}
\centering
\subfloat[H$\gamma$]{\includegraphics[width=1.0\linewidth]{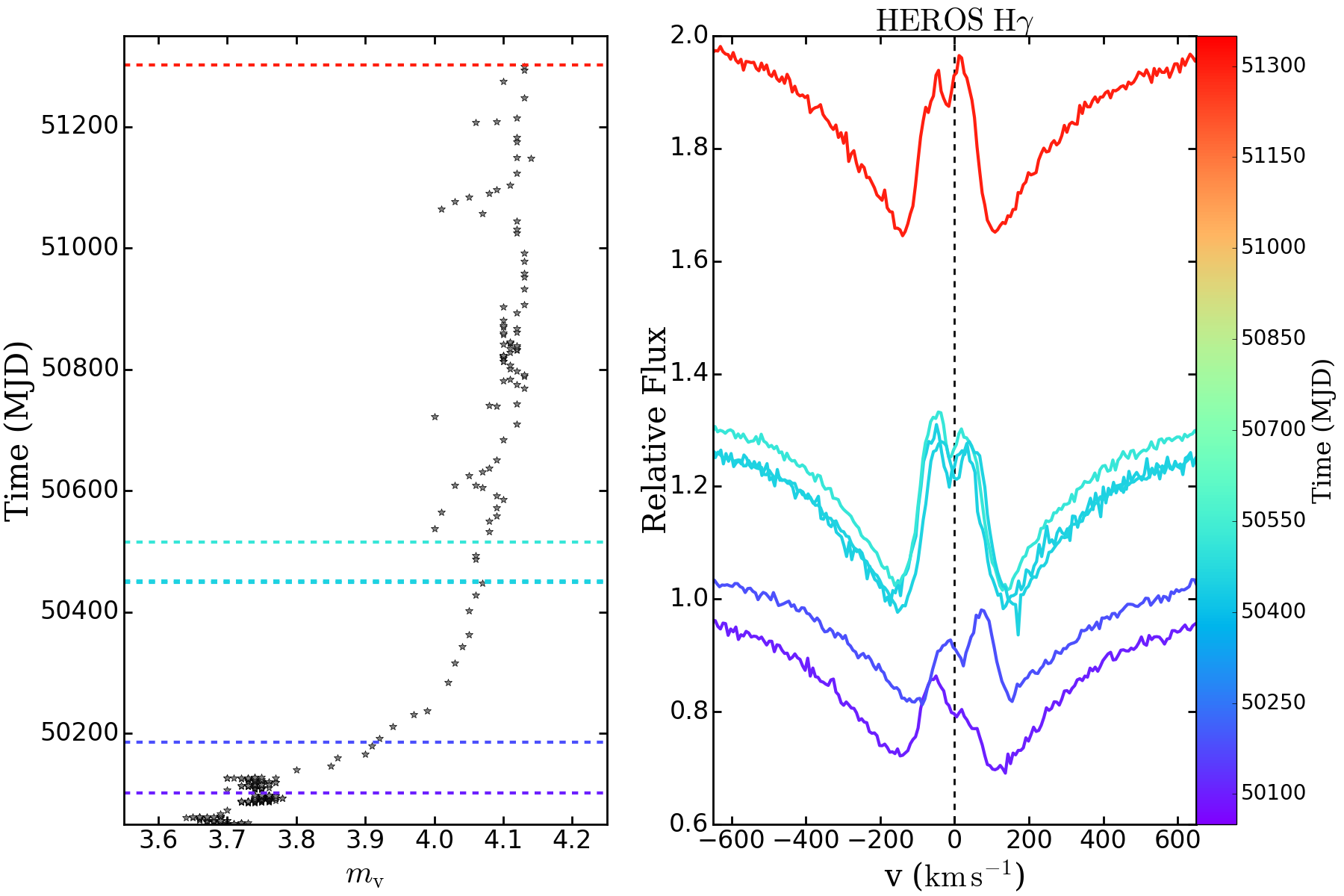}}
\end{minipage}%
\begin{minipage}{0.5\linewidth}
\centering
\subfloat[H$\delta$]{\includegraphics[width=1.0\linewidth]{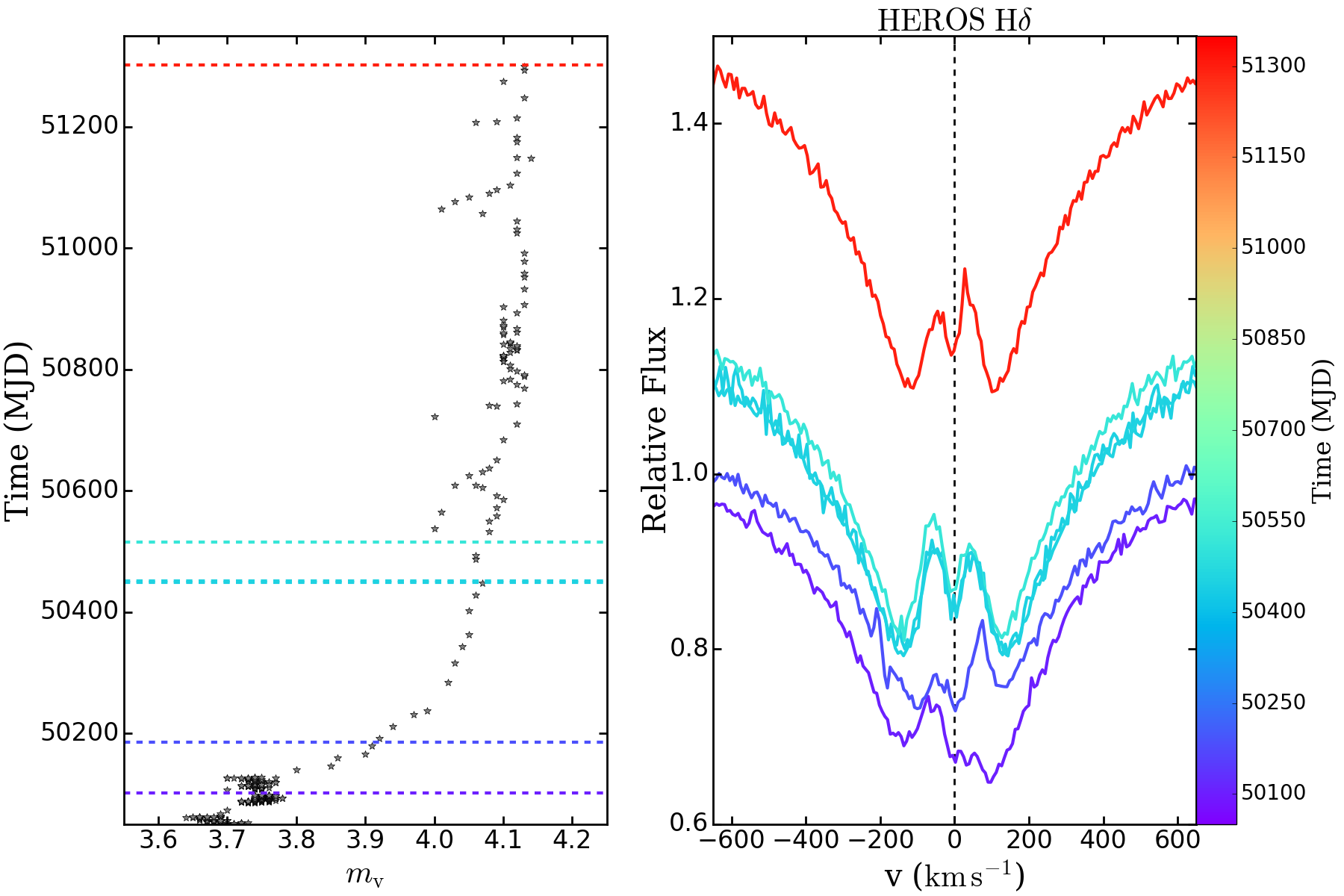}}
\end{minipage}%
\caption[Spectra of $\omega$ CMa observed by HEROS]{Hydrogen lines of $\omega$ CMa observed by HEROS.}
\label{fig:heros}
\end{figure}


\begin{figure}
\begin{center}
\setcaptionmargin{1cm}
\includegraphics[width=1.0 \columnwidth,angle=0]{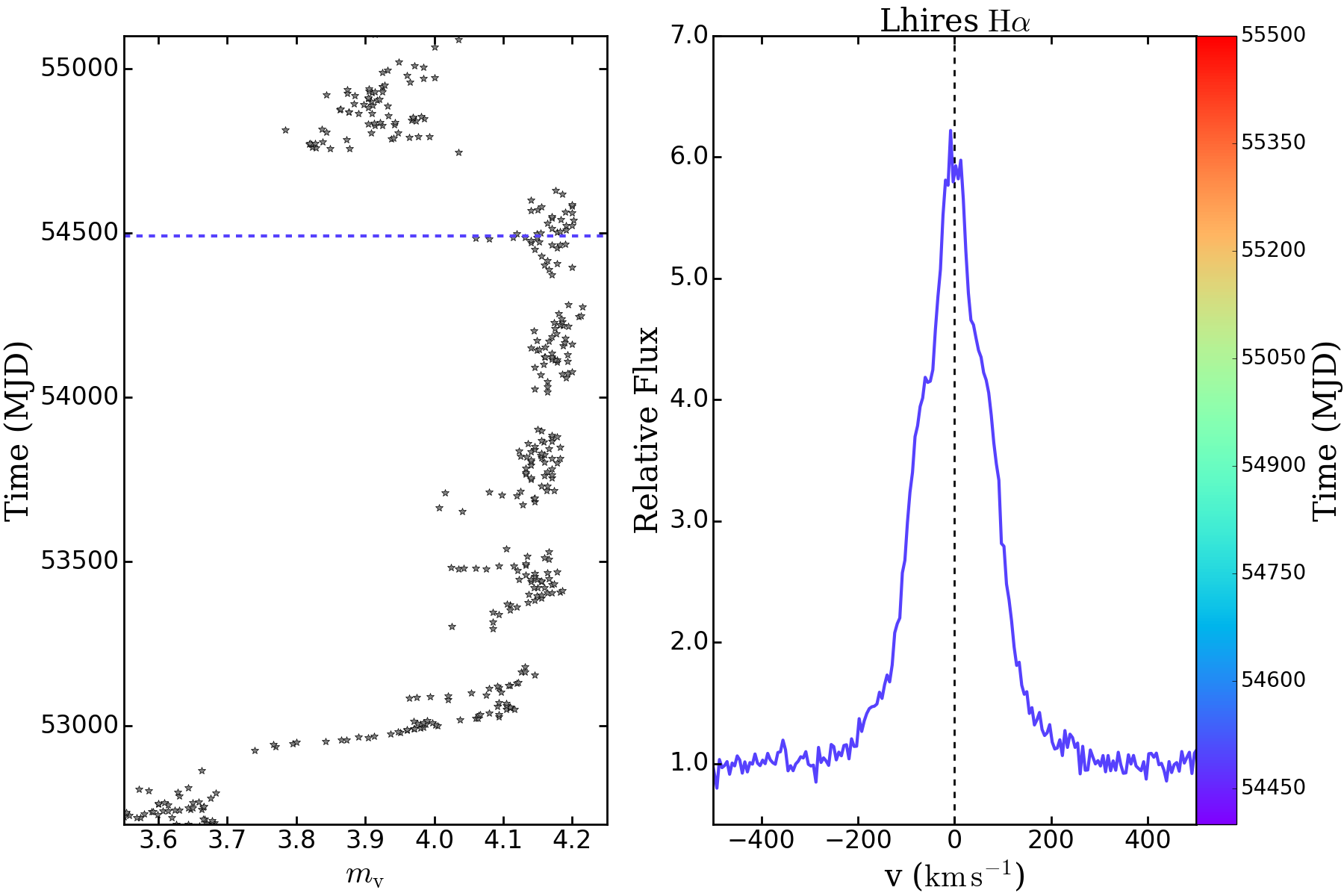}
\caption[Spectra of $\omega$ CMa observed by Lhires]{Hydrogen lines of $\omega$ CMa observed by Lhires.} 
\label{fig:lhires}
\end{center}
\end{figure}


\begin{figure}
\begin{minipage}{1.0\linewidth}
\centering
\subfloat[H$\alpha$]{\includegraphics[width=1.0\linewidth]{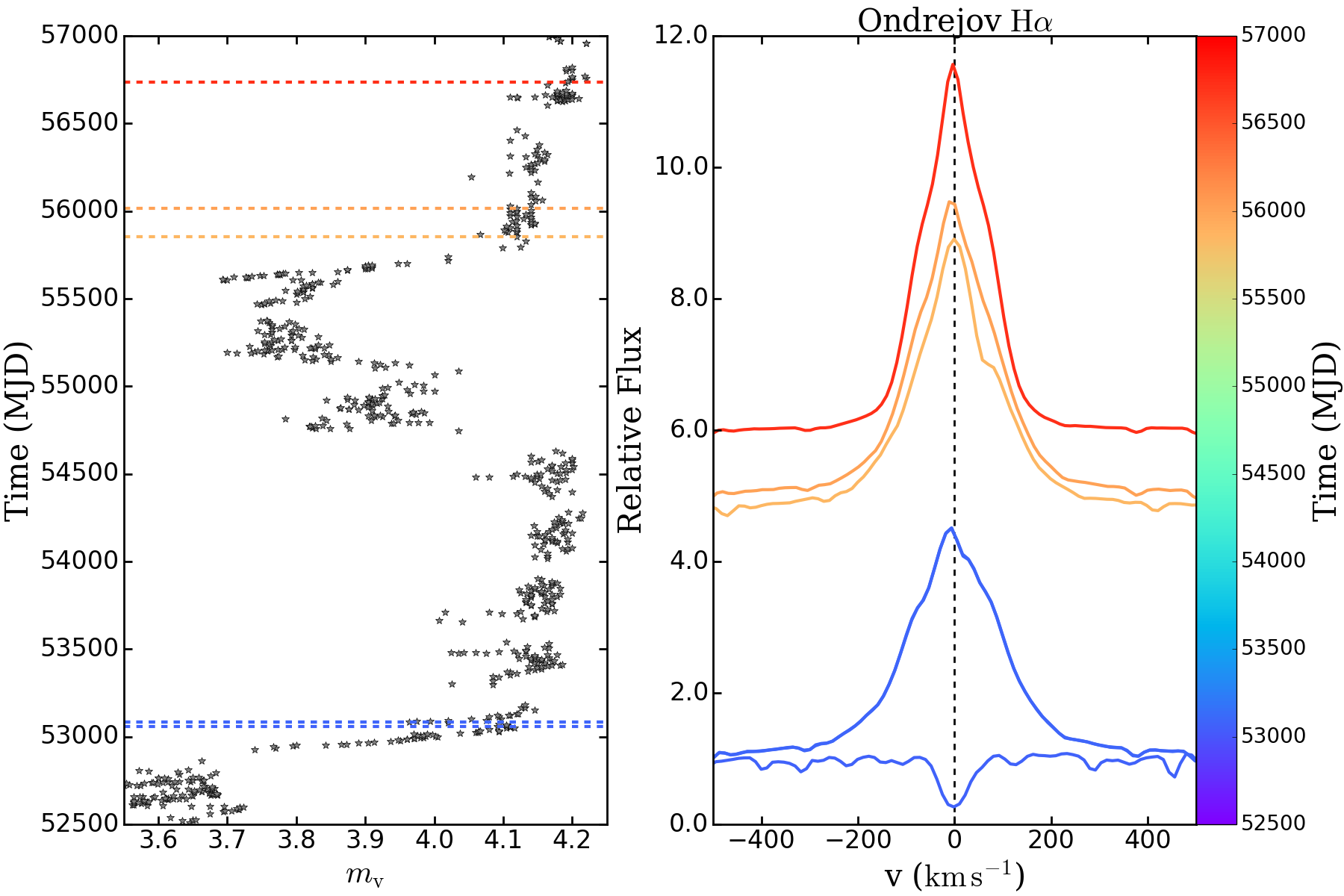}}
\end{minipage}\par\medskip
\begin{minipage}{1.0\linewidth}
\centering
\subfloat[H$\beta$]{\includegraphics[width=1.0\linewidth]{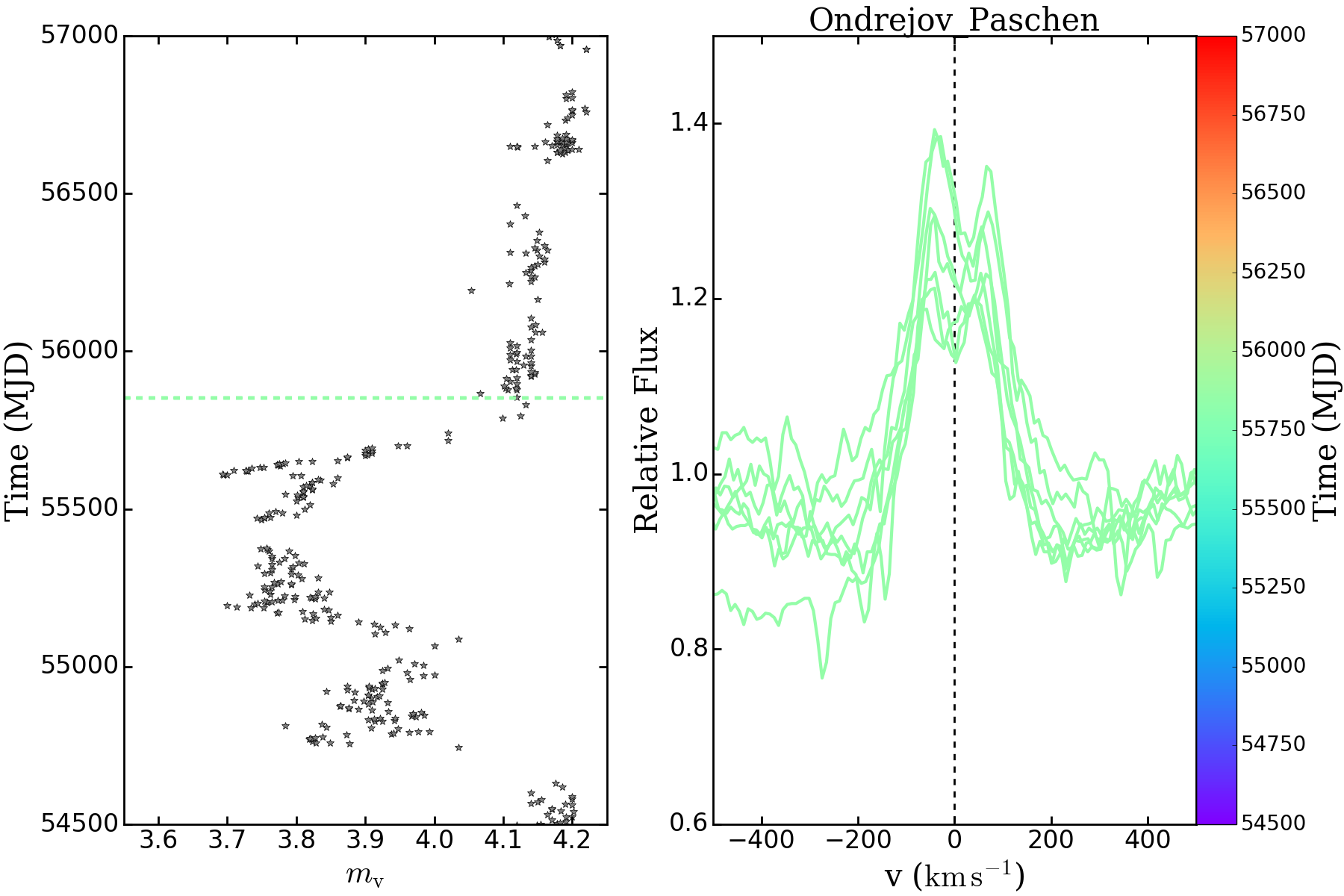}}
\end{minipage}%
\caption[Spectra of $\omega$ CMa observed by Ondrejov]{H$\alpha$ and Paschen lines of $\omega$ CMa observed by Ondrejov.}
\label{fig:ondrejov}
\end{figure}


\begin{figure}
\begin{minipage}{1.0\linewidth}
\centering
\subfloat[H$\alpha$]{\includegraphics[width=1.0\linewidth]{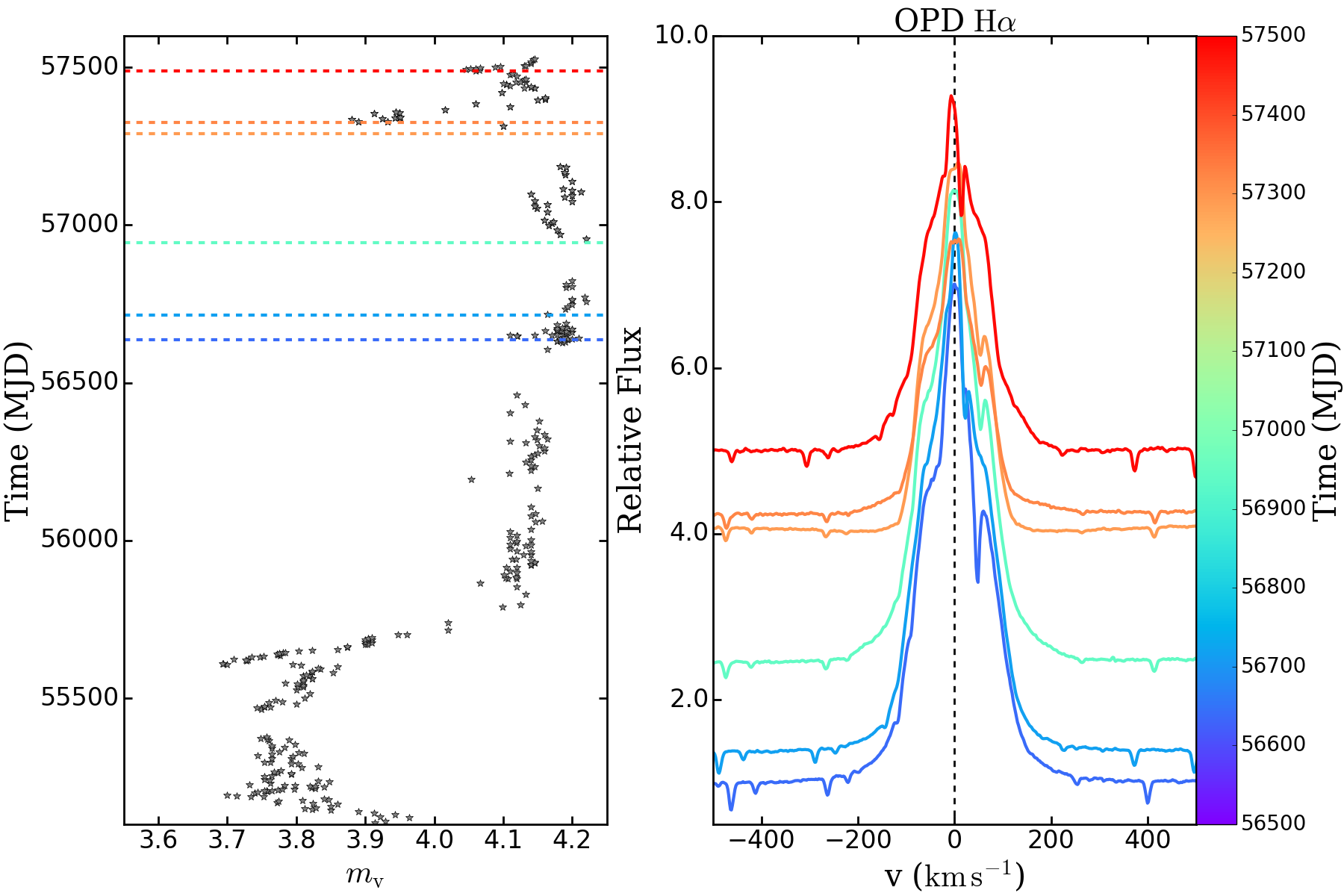}}
\end{minipage}\par\medskip
\begin{minipage}{0.5\linewidth}
\centering
\subfloat[H$\beta$]{\includegraphics[width=1.0\linewidth]{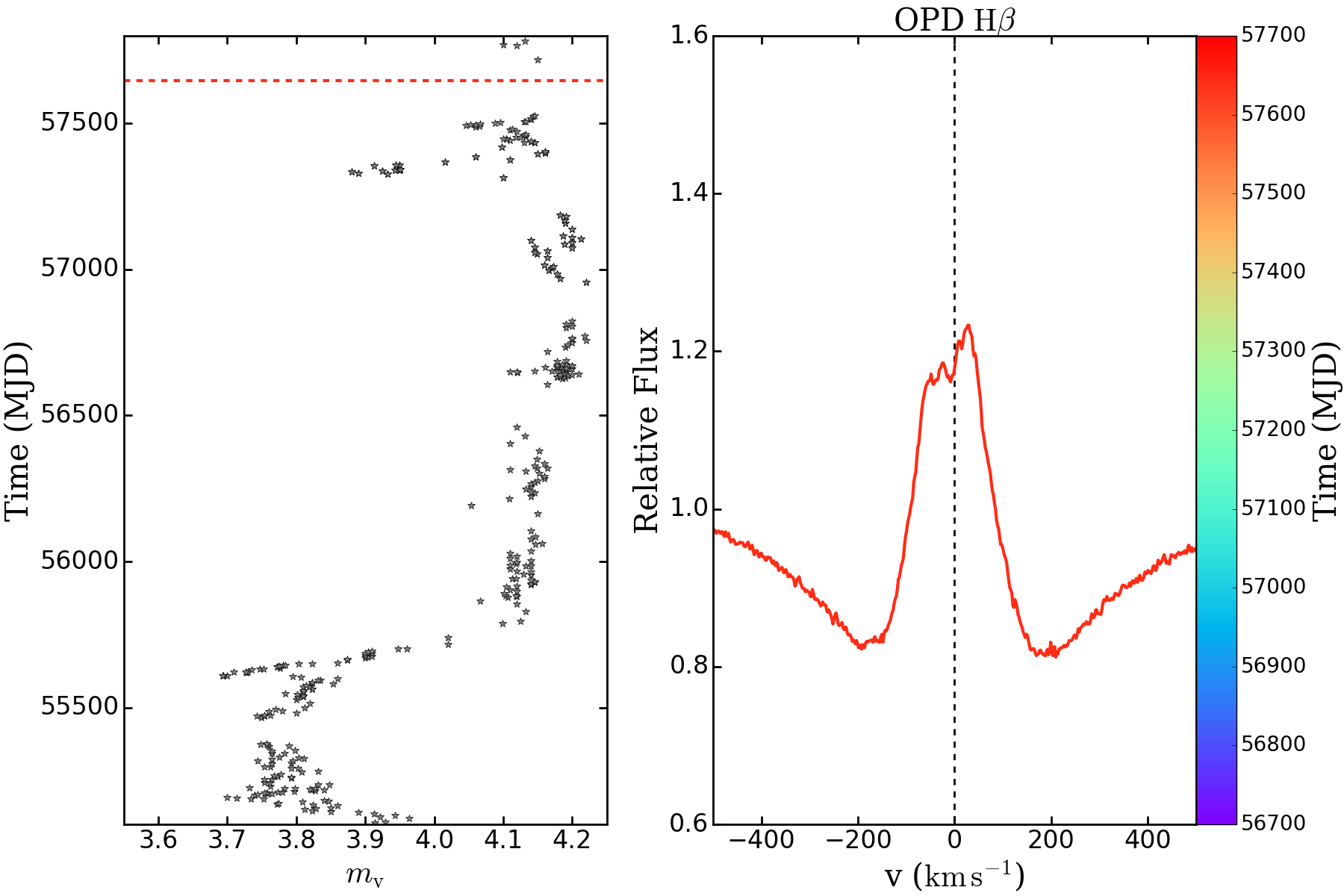}}
\end{minipage}%
\begin{minipage}{0.5\linewidth}
\centering
\subfloat[H$\gamma$]{\includegraphics[width=1.0\linewidth]{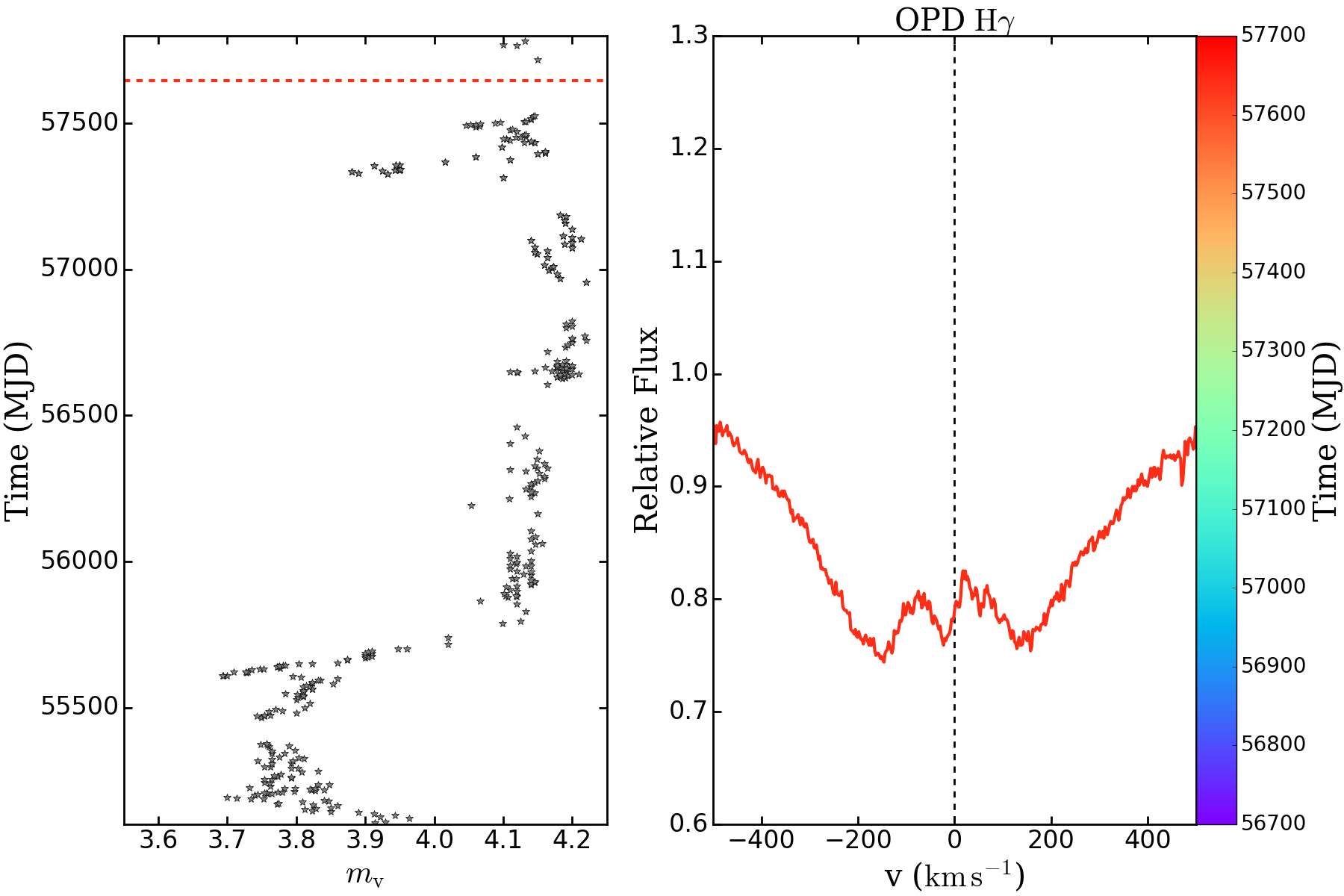}}
\end{minipage}%
\caption[Spectra of $\omega$ CMa observed by OPD]{Hydrogen lines of $\omega$ CMa observed by OPD.}
\label{fig:opd}
\end{figure}


\begin{figure}
\begin{center}
\setcaptionmargin{1cm}
\includegraphics[width=1.0 \columnwidth,angle=0]{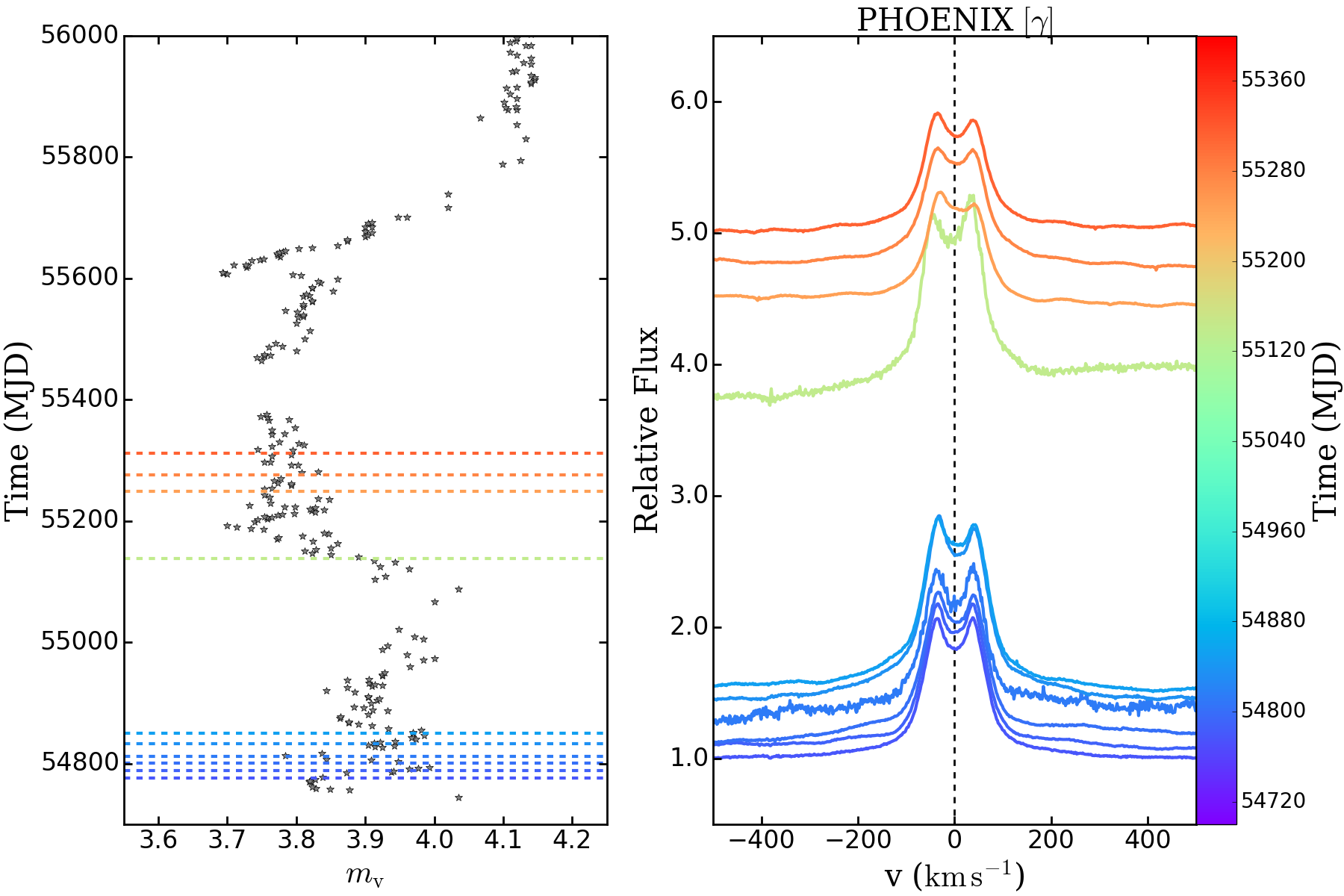}
\caption[Spectra of $\omega$ CMa observed by PHOENIX]{[$\gamma$] lines of $\omega$ CMa observed by PHOENIX.} 
\label{fig:phoenix}
\end{center}
\end{figure}


\begin{figure}
\begin{center}
\setcaptionmargin{1cm}
\includegraphics[width=1.0 \columnwidth,angle=0]{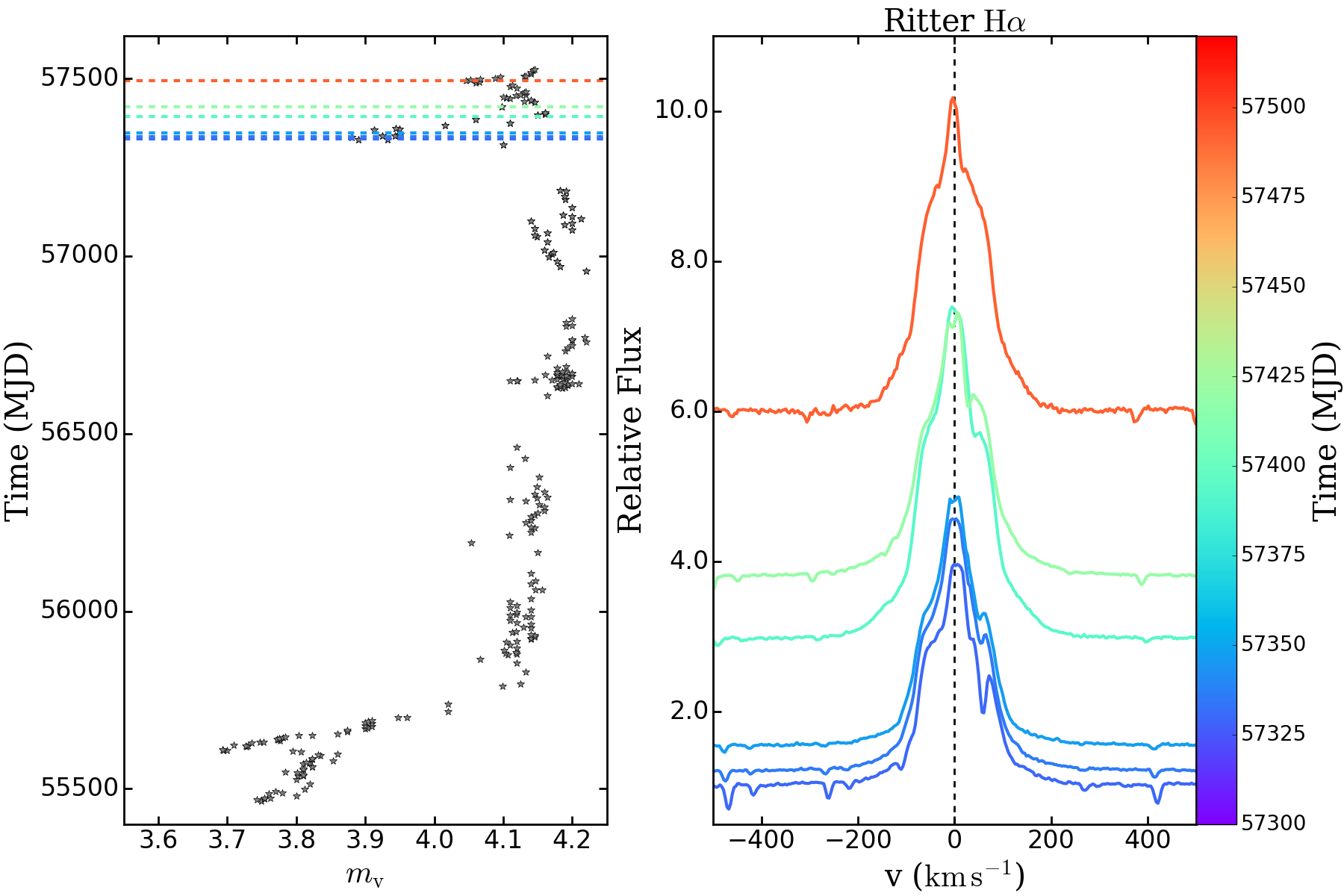}
\caption[Spectra of $\omega$ CMa observed by Ritter]{Hydrogen lines of $\omega$ CMa observed by Ritter.} 
\label{fig:ritter}
\end{center}
\end{figure}


\begin{figure}
\begin{minipage}{0.5\linewidth}
\centering
\subfloat[H$\alpha$]{\includegraphics[width=1.0\linewidth]{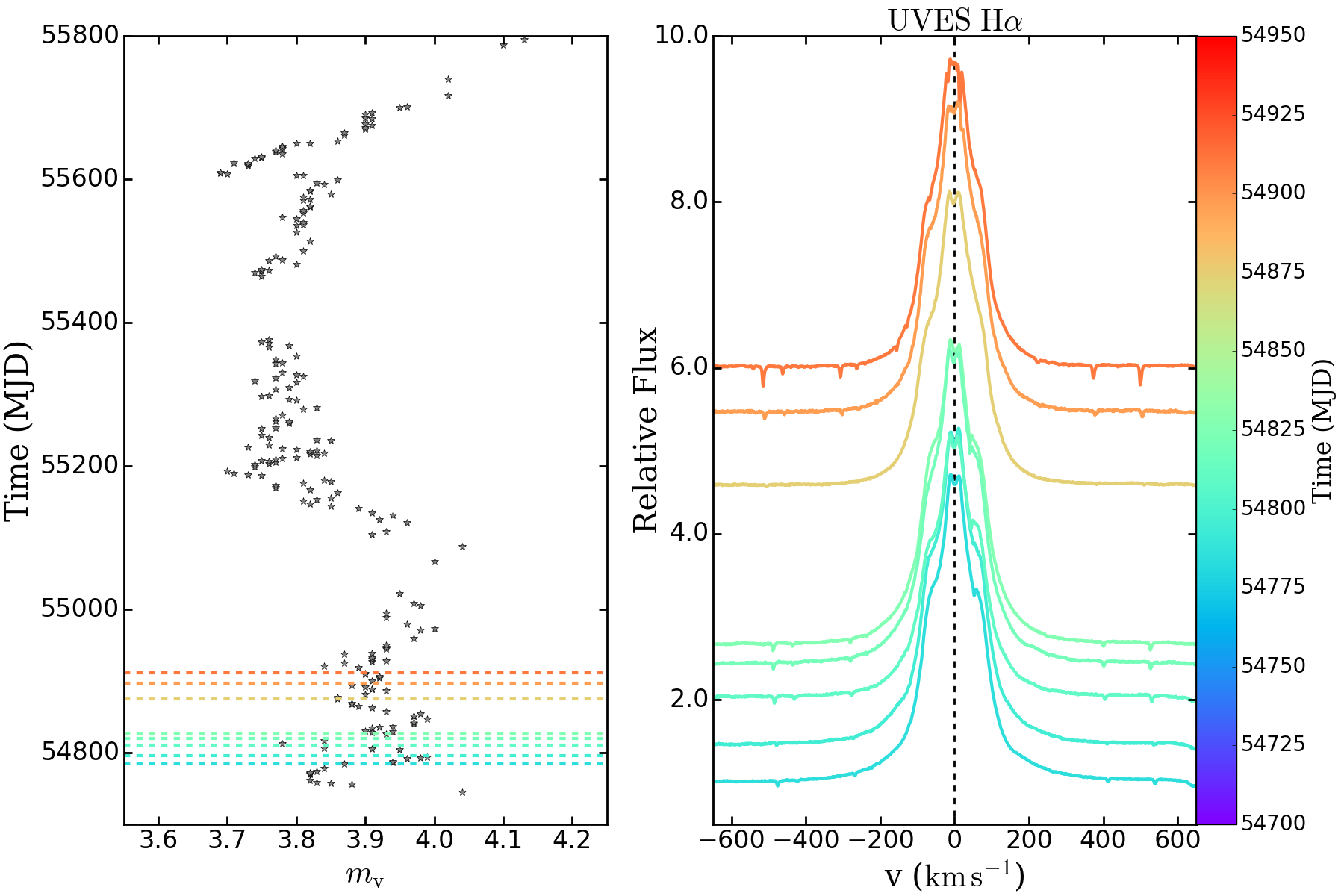}}
\end{minipage}%
\begin{minipage}{0.5\linewidth}
\centering
\subfloat[H$\beta$]{\includegraphics[width=1.0\linewidth]{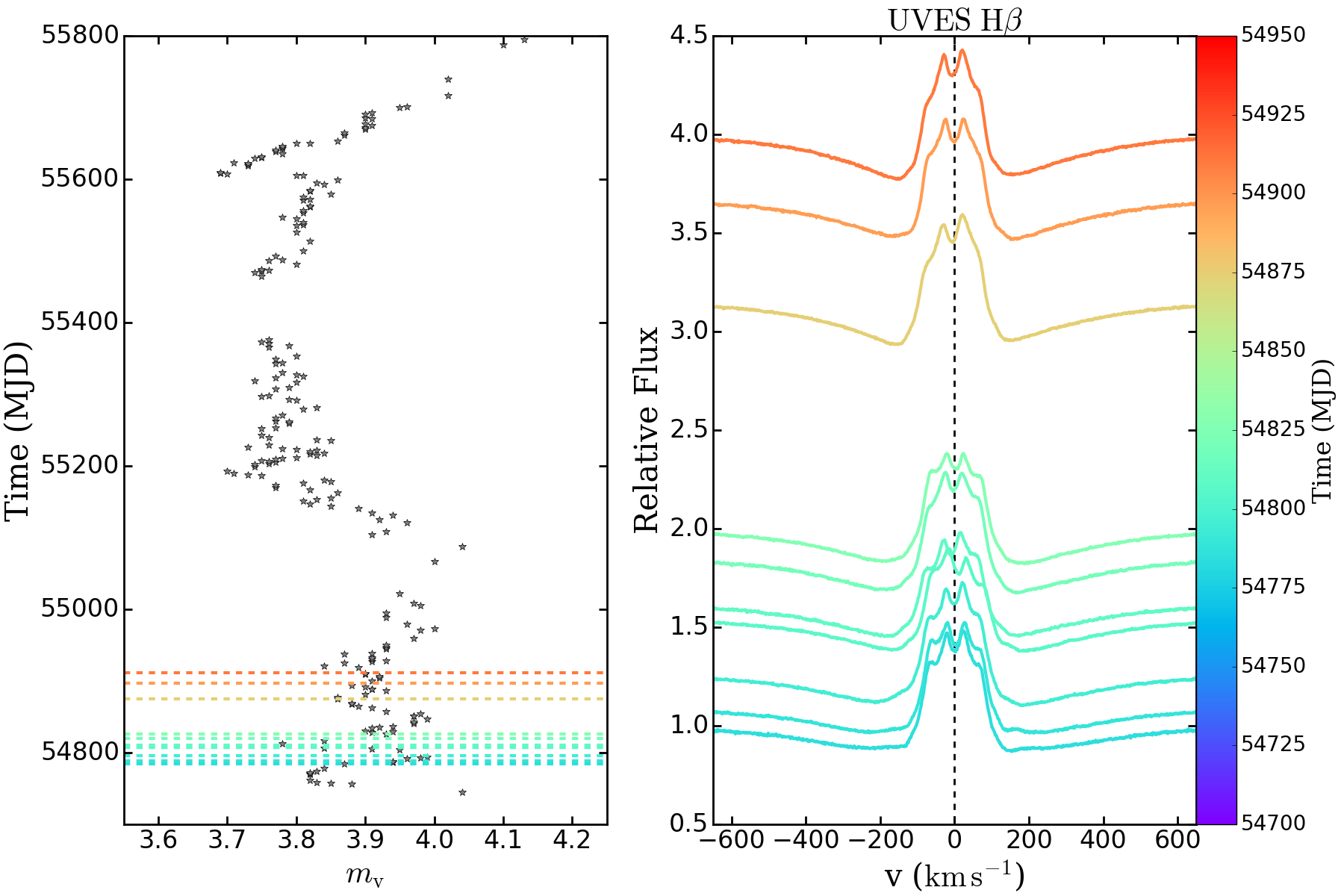}}
\end{minipage}\par\medskip
\begin{minipage}{0.5\linewidth}
\centering
\subfloat[H$\gamma$]{\includegraphics[width=1.0\linewidth]{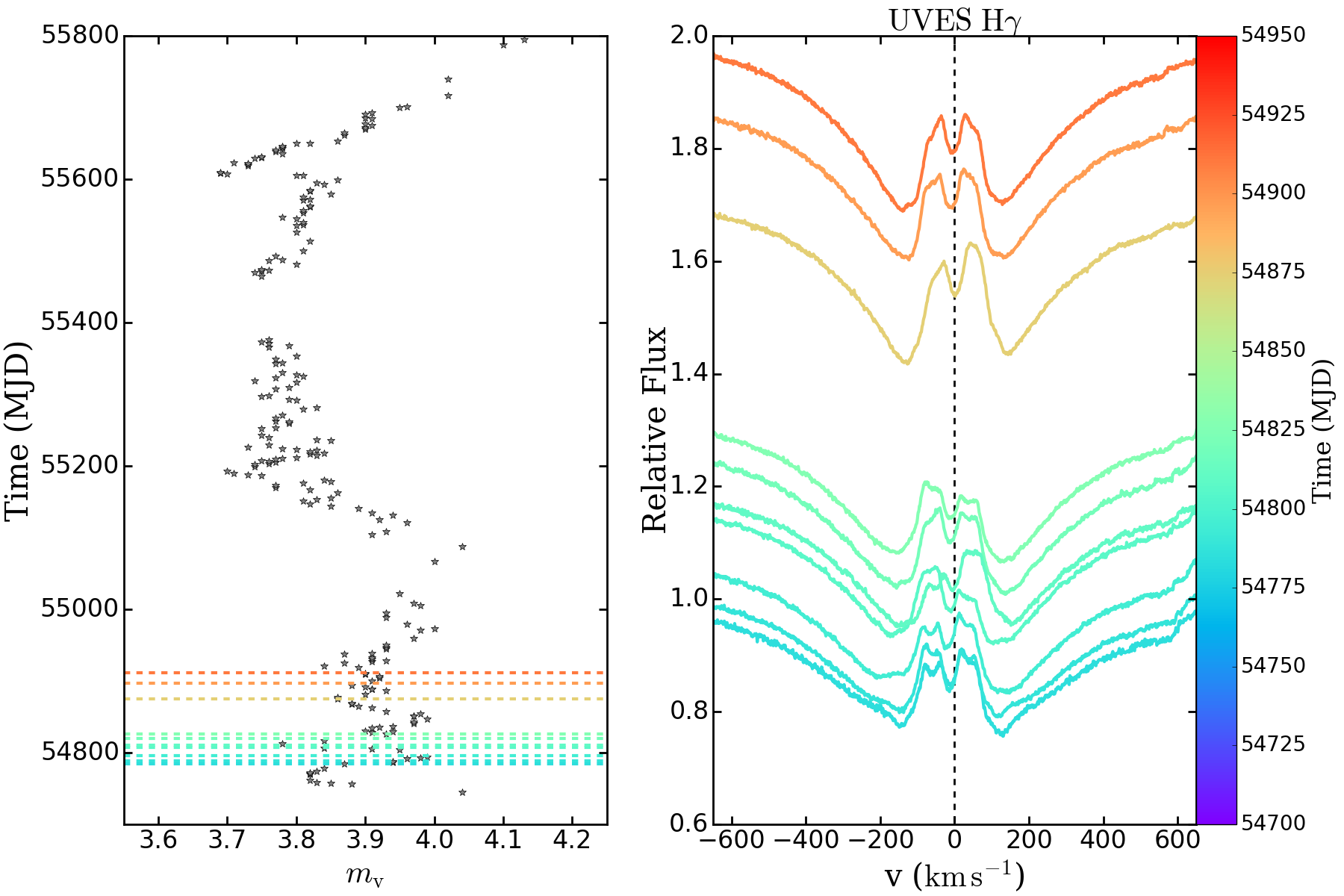}}
\end{minipage}%
\begin{minipage}{0.5\linewidth}
\centering
\subfloat[H$\delta$]{\includegraphics[width=1.0\linewidth]{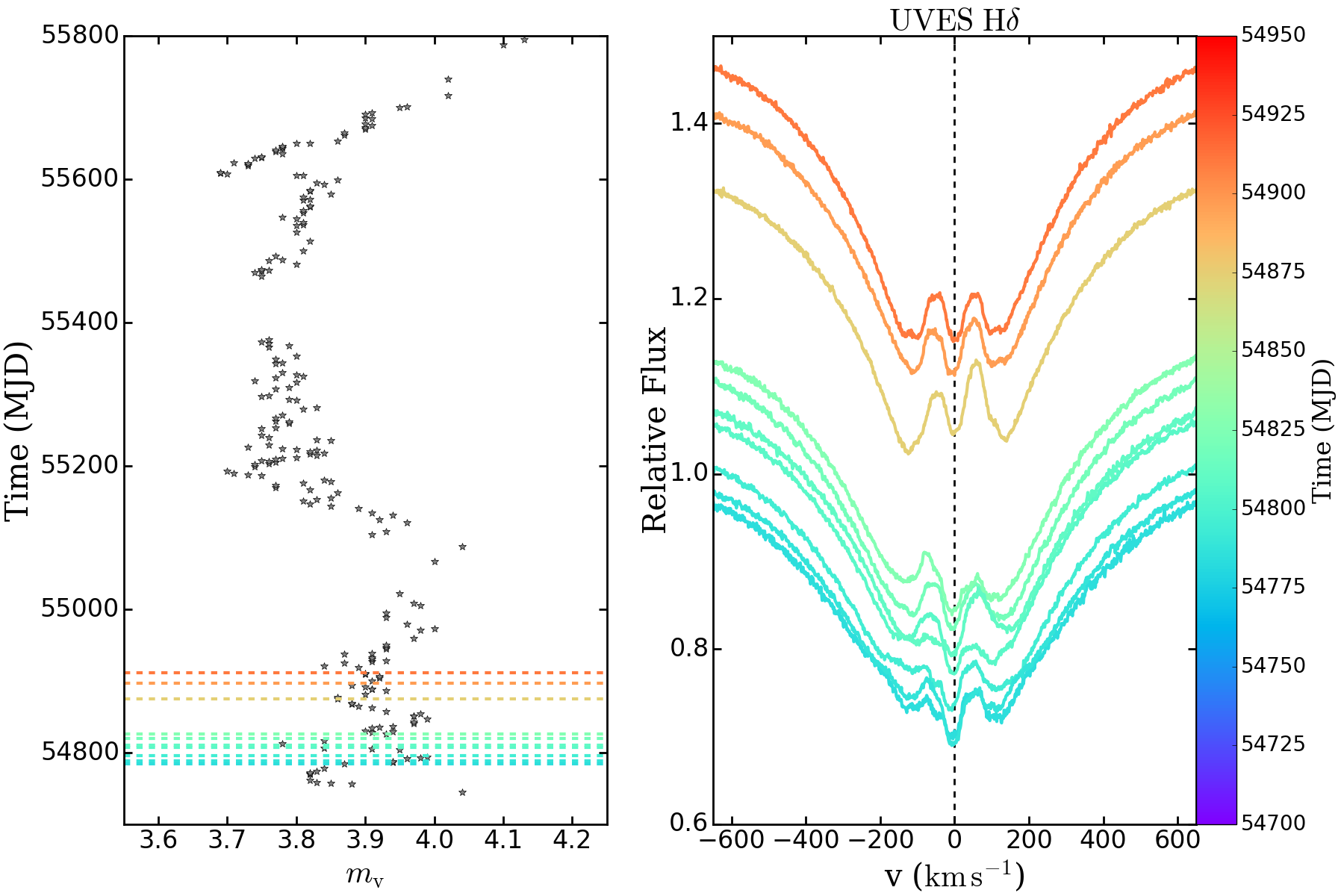}}
\end{minipage}%
\caption[Spectra of $\omega$ CMa observed by UVES]{Hydrogen lines of $\omega$ CMa observed by UVES.}
\label{fig:uves}
\end{figure}

\clearpage

\section{Data logs}\label{ap_a_2}

\begin{table}[!h]
\begin{center}
\caption[Photometric Data Logs]{Photometric Data Logs}
\begin{tabular}{@{}ccc}
\hline
\hline
Reference & Number of Points & Time Coverage (MJD) \\
\hline
& $V$-band & \\
\hline
Feinstein, 1975 & 11 & 38101 -- 41742 \\
Johnson, 1975 & 8 & 38383 -- 38722 \\
Van Hoof, 1975 & 16 & 40591 -- 41772 \\
Baade, 1976 & 16 & 43097 -- 43116 \\
Baade, 1982 & 2 & 44960 -- 45362 \\
Balona, 1987 & 657 & 46374 -- 46787 \\
Stagg, 1987 & 39 & 45419 -- 45756 \\
Dachs, 1988 & 2 & 43805 -- 45014 \\
Mennickent et al. 1994 & 6 & 46478 -- 46498 \\
{\v S}tefl et al. 2000 & 4 & 51579 -- 51585 \\
Edalati et al. 1989 & 398 & 45086 -- 51536 \\
Hipparcos & 185 & 47901 -- 49051 \\
LTPV & 104 & 45312 -- 48292 \\
{\v S}tefl, Geneva & 154 & 50044 -- 50127 \\
Biruni Observatory & 11 & 57452 -- 57526 \\
Mohammad R. Ghoreyshi & 13 & 57434 -- 57516 \\
Sebastian Otero & 963 & 50537 -- 58226 \\
\hline
& $JHKL$-bands & \\
\hline
Juan Fabregat (SAAO)\footnotemark & 8 & 54822 -- 55291 \\
\hline
& $uvby$-bands & \\
\hline
LTPV & 104 & 45312 -- 48292 \\
\hline
\end{tabular}
\label{table:phot_data_log}
\end{center}
\end{table}
\footnotetext{South African Astronomical Observatory}


\begin{table}
\begin{center}
\caption[Spectroscopic Data Logs]{Spectroscopic Data Logs}
\begin{tabular}{@{}ccc}
\hline
\hline
Reference & Number of Points & Time Coverage (MJD) \\
\hline
BeSS & 24 & 54083 -- 57465 \\
CES & 2 & 52659 -- 52660 \\
ESPaDOnS & 16 & 55971 -- 55971 \\
FEROS & 444 & 52277 -- 54822 \\
HEROS & 435 & 50102 -- 51301 \\
IUE & 12 & 43833 -- 44975 \\
Lhires & 1 & 54491 -- 54491 \\
Ondrejov & 7 & 53060 -- 56737 \\
OPD & 8 & 56636 -- 57645 \\
PHOENIX & 10 & 54776 -- 55311 \\
Ritter & 20 & 57329 -- 57496 \\
UVES & 141 & 54784 -- 54913 \\  
\hline
\end{tabular}
\label{table:spec_data_log}
\end{center}
\end{table}


\begin{table}
\begin{center}
\caption[Polarimetric Data Logs]{Polarimetric Data Logs}
\begin{tabular}{@{}ccc}
\hline
\hline
Reference & Number of Points & Time Coverage (MJD) \\
\hline
OPD & 57 & 54505 -- 57626\\ 
\hline
\end{tabular}
\label{table:pol_data_log}
\end{center}
\end{table}		
\end{apendice}				
						%
\end{document}